\documentclass[]{aastex63}
\usepackage{multirow}

\usepackage[utf8x]{inputenc}
\usepackage{graphics,times,subfigure}
\usepackage{booktabs}
\usepackage{graphicx}
\usepackage{longtable}
\usepackage{threeparttable}
\usepackage{bm}

\received{April 26, 2021}
\revised{August 26, 2021}
\accepted{August 30, 2021}
\submitjournal{ApJ}

\shorttitle{In search of infall motion in molecular clumps III}
\shortauthors{Yang et al.}

\begin{document}

\title{In search of infall motion in molecular clumps III: \\
HCO$^+$ (1-0) and H$^{13}$CO$^+$ (1-0) mapping observations toward the confirmed infall sources}

\correspondingauthor{Zhibo Jiang}
\email{zbjiang@pmo.ac.cn}


\author{Yang Yang}
\affiliation{Purple Mountain Observatory, Chinese Academy of Sciences, Nanjing 210023, China}
\affiliation{Key Laboratory of Radio Astronomy, Purple Mountain Observatory, Chinese Academy of Sciences, Nanjing 210023, China}
\affiliation{University of Science and Technology of China, Chinese Academy of Sciences, Hefei 230026, China}

\author{Zhibo Jiang}
\affiliation{Purple Mountain Observatory, Chinese Academy of Sciences, Nanjing 210023, China}
\affiliation{Key Laboratory of Radio Astronomy, Purple Mountain Observatory, Chinese Academy of Sciences, Nanjing 210023, China}
\affiliation{Center astronomy and Space Sciences, China Three Gorges University, Yichang 443002, China}

\author{Zhiwei Chen}
\affiliation{Purple Mountain Observatory, Chinese Academy of Sciences, Nanjing 210023, China}
\affiliation{Key Laboratory of Radio Astronomy, Purple Mountain Observatory, Chinese Academy of Sciences, Nanjing 210023, China}

\author{Yiping Ao}
\affiliation{Purple Mountain Observatory, Chinese Academy of Sciences, Nanjing 210023, China}
\affiliation{Key Laboratory of Radio Astronomy, Purple Mountain Observatory, Chinese Academy of Sciences, Nanjing 210023, China}
\affiliation{University of Science and Technology of China, Chinese Academy of Sciences, Hefei 230026, China}

\author{Shuling Yu}
\affiliation{Purple Mountain Observatory, Chinese Academy of Sciences, Nanjing 210023, China}
\affiliation{Key Laboratory of Radio Astronomy, Purple Mountain Observatory, Chinese Academy of Sciences, Nanjing 210023, China}
\affiliation{University of Science and Technology of China, Chinese Academy of Sciences, Hefei 230026, China}



\begin{abstract}

The study of infall motion helps us to understand the initial stages of star formation. In this paper, we use the IRAM 30-m telescope to make mapping observations of 24 infall sources confirmed in previous work. The lines we use to track gas infall motions are HCO$^+$ (1-0) and H$^{13}$CO$^+$ (1-0). All 24 sources show HCO$^+$ emissions, while 18 sources show H$^{13}$CO$^+$ emissions. The HCO$^+$ integrated intensity maps of 17 sources show clear clumpy structures; for the H$^{13}$CO$^+$ line, 15 sources show clumpy structures. We estimated the column density of HCO$^+$ and H$^{13}$CO$^+$ using the RADEX radiation transfer code, and the obtained [HCO$^+$]/[H$_2$] and [H$^{13}$CO$^+$]/[HCO$^+$] of these sources are about $10^{-11}$--$10^{-7}$ and $10^{-3}$--1, respectively. Based on the asymmetry of the line profile of the HCO$^+$, we distinguish these sources: 19 sources show blue asymmetric profiles, and the other sources show red profiles or symmetric peak profiles. For eight sources that have double-peaked blue line profiles and signal-to-noise ratios greater than 10, the RATRAN model is used to fit their HCO$^+$ (1-0) lines, and to estimate their infall parameters. The mean $V_{in}$ of these sources are 0.3 -- 1.3 km s$^{-1}$, and the $\dot M_{in}$ are about $10^{-3}$ -- $10^{-4}$ M$_{\odot}$ yr$^{-1}$, which are consistent with the results of intermediate or massive star formation in previous studies. The $V_{in}$ estimated from the Myers model are 0.1 -- 1.6 km s$^{-1}$, and the $\dot M_{in}$ are within $10^{-3}$ -- $10^{-5}$ M$_{\odot}$ yr$^{-1}$. In addition, some identified infall sources show other star-forming activities, such as outflows and maser emissions. Especially for those sources with a double-peaked blue asymmetric profile, most of them have both infall and outflow evidence.

\end{abstract}

\keywords{Star formation --- Interstellar medium --- Molecular clouds --- Collapsing clouds}


\section{Introduction} \label{sec:intro}

The gravitational infall in molecular clouds mainly occurs in the beginning stage of star formation \citep[e.g.][]{Bachiller+1996}, and it is an important part of the star-forming process. Stars are believed to form through the inside-out gravitational collapse of dense molecular cloud cores \citep[e.g.][]{Shu+etal+1987}. In the subsequent evolution, the accretion flow has been maintained to feed the forming stars \citep[e.g.][]{Allen+etal+2004}. The study of this process helps us to understand the initial stage of the star-formation process. There exist several studies focusing on the infall motions: \citet{Mardones+etal+1997} used the IRAM 30-m, SEST 15-m, and Haystack 37-m radio telescopes to observe some candidate protostars, and reported 15 infall candidates. Most of them are associated with clumps and have been found to have turbulent motions. \citet{Klaassen+etal+2008} observed seven massive star-forming regions (SFRs), and found that four of them have infall signatures based on their HCO$^+$ emission profiles. \citet{He+etal+2015} selected more than 400 compact sources from the MALT90 survey, and used the HCO$^+$ (1-0), HNC (1-0), and N$_2$H$^+$ (1-0) data to identify 131 infall candidates. Spitzer's mid-infrared data were also used to limit the evolutionary stages of the candidates as a supplement, and to analyze the relationship between the infall motion and evolution stage of these candidates. They found that the infall detection rate in the cores with protostars is relatively high, followed by the detection rate in prestellar cores. In addition, the infall detection rate in ultracompact HII (UC HII) clumps is higher than expected, which may indicate that many UC HII regions are still accreting matter. \citet{Calahan+etal+2018} identified six infall candidates from massive starless sources in the Bolocam Galactic Plane Survey, and calculated that their mass infall rates were about 500 to 2000 M$_{\odot}$ Myr$^{−1}$. \citet{Li+etal+2019} compared the infall detection rate of massive SFRs with and without outflow, and found that the former have a lower infall detection rate. Most of these studies have focused on the SFRs and compact sources. Nevertheless, the small number of infall samples prevents a comprehensive understanding of the collapse stage. Therefore, the large-scale search and identification of molecular clumps with infall motion signatures in the Milky Way will help us to reveal the initial stages of star formation.

In \citet{Yang+etal+2020}, we used the Milky Way Imaging Scroll Painting (MWISP)\footnote{This project is an on-going, unbiased CO survey along the Galactic plane, observing $^{12}$CO (1-0) and its isotopes $^{13}$CO (1-0) and C$^{18}$O (1-0) lines simultaneously \citep[e.g.][]{Su+etal+2019}.} CO data to search the molecular clouds with infall motion signatures. According to the infalling clumps model \citep[e.g.][]{Leung+Brown+1977}, the envelope material falls toward the center, resulting in a redshifted self-absorption profile for the optically thick line. This causes the optically thick line to produce a double-peaked profile, with the blue peak stronger than the red one. This kind of line profile is commonly referred to as a blue profile \citep[e.g.][]{Mardones+etal+1997}. However, multiple velocity components may also produce a similar line profile. An additional optically thin line is required to track the central radial velocity of the clump. The combination of optically thick and thin lines can distinguish whether the blue profile of the optically thick line is caused by infall motions or multiple velocity components. If the optically thin line shows a single-peaked profile with the peak located between the double peaks of the thick optical line or slightly skewed to the red peak, the double-peaked profile of the optically thick line is more likely caused by self-absorption than by multiple velocity components.

Based on the MWISP CO data, two pairs of lines, i.e. $^{12}$CO/$^{13}$CO and $^{13}$CO/C$^{18}$O, were used to search molecular clouds with infall motion signatures. We carried out a blind search of all available MWISP CO data with the two pairs of lines. About 3,200 molecular clouds with blue profiles were initially identified and checked up to the end of 2019 (Jiang et al. 2020, in prep). From this large sample, we selected 133 candidates with significant blue profiles in the $^{12}$CO line and C$^{18}$O line emission greater than 1 K for follow-up observations with the DLH 13.7-m telescope in the two optically thick lines of HCO$^+$ (1-0) and HCN (1-0). The HCO$^+$ and HCN lines trace denser regions than CO lines, thus, are often used as tracers for infall motions in molecular clouds \citep[e.g.][]{Vasyunina+etal+2011, Peretto+etal+2013, Yuan+etal+2018, Zhang+etal+2018}. According to the observation results, 56 of the 133 observed sources were confirmed as infall motion candidates with higher confidence \citep{Yang+etal+2020}. However, the single-point observations lack spatial information. It is necessary to map these confirmed sources with observations of higher spatial resolution, because some observations indicate that the blue profile exists at low resolution, but disappears at higher resolution \citep[e.g.][]{Wu+etal+2007}. On the other hand, not only infall but also outflow and rotation can cause blue profiles in some environments \citep[e.g.][]{Mardones+etal+1997, He+etal+2015}, which makes it difficult to distinguish infall, outflow, and rotation from single-point observations. For these 56 sources, we proposed mapping observations with the IRAM 30-m telescope to obtain the optically thick HCO$^+$ (1-0) line and optically thin H$^{13}$CO$^+$ (1-0) line emission maps. At present, we have completed the observations for 24 of them. 

In this paper, we present the results of mapping observations with the IRAM 30-m telescope and analyses of the physical properties of these 24 sources that are derived from the obtained data. In Section \ref{sec:obs}, we briefly introduce the mapping observations. The sources' morphology and clump properties based on these two lines are given in Section \ref{sec:results}. In Section \ref{sec:infall}, we identify the candidates with an infall profile out of 24 observed sources, and used two models to fit the HCO$^+$ lines of the sources with significant infall profiles, to estimate their infall physical parameters. In Section \ref{sec:sfa}, we confirmed that some infall candidates not only have infall characteristics but also outflow, and some other star-forming activities. A simple classification of the evolution stages of these sources is also presented in this section. Finally, Section \ref{sec:summary} summarizes the results and analyses of this work.

\section{Observations} \label{sec:obs}


\begin{table}
\begin{center}
  \caption{Source list.}\label{Tab:src-catalog}
 \setlength{\tabcolsep}{1mm}{
\begin{tabular}{cccccc}
  \hline\noalign{\smallskip}
Source  &  RA  &  Dec  & Vlsr &  Distance  &  Extent of map  \\ 
Name &	(J2000)	&	(J2000)	&	(km s$^{-1}$)	&	(kpc)	&		\\
  \hline\noalign{\smallskip}
G028.97+3.35     &     18:32:16.6     &     -01:59:27     & 7.31 & 0.42(0.02)\tablenotemark{$^1$} & $3\arcmin\times3\arcmin$ \\
G029.06+4.58     &     18:28:04.2     &     -01:20:47     & 7.48 & 0.49 & $2.5\arcmin\times2.5\arcmin$ \\
G029.60-0.63\tablenotemark{*}     &     18:47:36.1     &     -03:15:09     & 76.52 & 4.31 & $3\arcmin\times3\arcmin$ \\
G030.17+3.68     &     18:33:18.1     &     -00:46:31     & 9.02 & 0.61 & $2\arcmin\times2\arcmin$ \\
G031.41+5.25     &     18:29:58.6     &     +01:02:31     & 8.28 & 0.55 & $2\arcmin\times2\arcmin$ \\
G033.42+0.00     &     18:52:19.7     &     +00:26:04     & 10.6 & 0.69 & $3\arcmin\times3\arcmin$ \\
G039.33-1.03     &     19:06:49.4     &     +05:13:09     & 12.78 & 0.82 & $3\arcmin\times3\arcmin$ \\
G053.13+0.09\tablenotemark{*}    &     19:29:11.8     &     +17:56:19     & 21.71 & 1.67 & $4\arcmin\times6\arcmin$ \\
G077.91-1.16     &     20:34:12.6     &     +38:17:43     & -0.21 & 0.15 & $2.5\arcmin\times2.5\arcmin$ \\
G079.24+0.53\tablenotemark{*}     &     20:31:15.8     &     +40:22:09     & 0.79 & 0.39 & $2\arcmin\times2\arcmin$ \\
G079.71+0.14\tablenotemark{*}     &     20:34:22.8     &     +40:30:55     & 1.19 & 0.66 & $2.5\arcmin\times2.5\arcmin$ \\
G081.72+0.57\tablenotemark{*}     &     20:39:01.0     &     +42:22:41     & -3.22 & 1.50(0.08)\tablenotemark{$^2$} & $2.5\arcmin\times2.5\arcmin$ \\
G082.21-1.53\tablenotemark{*}     &     20:49:30.3     &     +41:27:35     & 3.07 & 1.13 & $3\arcmin\times3\arcmin$ \\
G107.50+4.47\tablenotemark{*}     &     22:28:32.1     &     +62:58:40     & -1.27 &     0.25\tablenotemark{$^3$}     & $3\arcmin\times3\arcmin$ \\
G109.00+2.73\tablenotemark{*}     &     22:47:17.6     &     +62:11:44     & -10.29 & 0.79 & $3.5\arcmin\times3.5\arcmin$ \\
G110.31+2.53     &     22:58:11.6     &     +62:35:36     & -11.5 & 0.9 & $3\arcmin\times3\arcmin$ \\
G110.40+1.67     &     23:01:57.9     &     +61:50:55     & -11.17 & 0.82 & $2.5\arcmin\times2.5\arcmin$ \\
G121.31+0.64\tablenotemark{*}     &     00:36:54.5     &     +63:27:57     & -17.54 & 0.89(0.05)\tablenotemark{$^4$} & $3\arcmin\times3\arcmin$ \\
G121.34+3.42     &     00:35:39.0     &     +66:14:34     & -5.18 & 0.19 & $3\arcmin\times3\arcmin$ \\
G126.53-1.17\tablenotemark{*}     &     01:21:39.1     &     61:29:25     & -12.26 & 0.72 & $2\arcmin\times2\arcmin$ \\
G143.04+1.74\tablenotemark{*}     &     03:33:49.5     &     +58:07:29     & -8.8 & 0.45\tablenotemark{$^5$} & $3\arcmin\times3\arcmin$ \\
G154.05+5.07     &     04:47:09.0     &     +53:03:26     & 4.58 &     0.17\tablenotemark{$^6$}     & $3\arcmin\times3\arcmin$ \\
G193.01+0.14\tablenotemark{*}     &     06:14:25.1     &     +17:43:12     & 7.59 & 1.91 & $3\arcmin\times3\arcmin$ \\
G217.30-0.05\tablenotemark{*}     &     06:59:14.3     &     -03:54:35     & 26.83 & 2.34 & $2\arcmin\times2\arcmin$ \\

  \hline\noalign{\smallskip}
\end{tabular}}
\end{center}
\tablecomments{$^*$ Source is associated with Class 0/I YSOs \citep{Yang+etal+2020}.\\
The distance of the sources are obtained from: $^1$ \citet{Konyves+etal+2015}; $^2$ \citet{Rygl+etal+2012}; $^3$ \citet{Bailer-Jones+etal+2018}; $^4$ \citet{Rygl+etal+2008}; $^5$ \citet{Stassun+etal+2018}; $^6$ \citet{Montillaud+etal+2015}, where $^1$, $^2$, and $^4$ are the parallax distances.
}
\end{table}


\begin{table}
\begin{center}
  \caption{Observation parameters}\label{Tab:obs}  
 \setlength{\tabcolsep}{1mm}{
\begin{tabular}{cccccccc}
  \hline\noalign{\smallskip}
Line & Frequency & Receiver & Backend & Bandwidth & Frequency resolution & Velocity resolution & RMS  \\
  & (GHz) &  &  & (GHz) & (kHz) & (km s$^{-1}$) & (K) \\
  \hline\noalign{\smallskip}
HCO$^+$(1-0) & 89.19 & EMIR & VESPA & 0.2 & 40 & 0.13 & $\textless$ 0.3 \\
H$^{13}$CO$^+$(1-0) & 86.75 & EMIR & FTS & 8$\times$4 & 195 & 0.67 & $\textless$ 0.2 \\ 
  \hline\noalign{\smallskip}
\end{tabular}}
\end{center}
\tablecomments{The values in columns 8 are the rms range corresponding to the rms values. }
\end{table}

We used the IRAM 30-m telescope to observe the HCO$^+$ (1-0) and H$^{13}$CO$^+$ (1-0) lines for 24 sources that are classified as infall candidates by \citet{Yang+etal+2020}. Table \ref{Tab:src-catalog} lists the observation sources, where the distance values are obtained from the literature. For sources whose distance value cannot be found in literature, we adopt the kinematic distance obtained by the \citet{Reid+etal+2014} model. If these sources have kinematic distance ambiguity, then we choose the nearer one as the source distance.

The observations were conducted with the Eight Mixer Receivers (EMIR) receiver of the IRAM 30-m telescope from February 20, 2008 to March 27, 2018 (project code: 134-18) and from June 11 to 19, 2019 (project code: 032-19). The back-end of the EMIR is the VESPA and FTS, with frequency resolutions of 40 kHz and 195 kHz, respectively (as shown in Table \ref{Tab:obs}). We used the OTF PSW observing mode to cover each observation area. The size of these areas depends on the MWISP $^{13}$CO (1-0) and C$^{18}$O (1-0) mapping data. The angular resolution of IRAM 30-m can achieve $28\arcsec$ in the 3-mm band, corresponding to a range of $0.02-0.60$ pc for our sources at the adopted distances. The main beam efficiency is approximately 0.81 at these frequencies, and the typical system temperature ($T_{sys}$) is between 120 and 130 K. For observation areas of different sizes, the on-source integration time is also different target by target. Including the time on pointing, focusing, calibrating and instrumental deadtimes, it takes about 0.7 to 3 h to observe one source. The total observing time is about 33 h. We used the CLASS of the GILDAS package\footnote{\url{http://www.iram.fr/IRAMFR/GILDAS/}}to reduce the raw data. The resultant RMS values of H$^{13}$CO$^+$ and HCO$^+$ lines are less than 0.2 K and 0.3 K, at their velocity resolutions. The ranges for rms corresponding to the rms values is 0.14 K and 0.23 K, respectively. We analyzed the morphology and physical properties of the sources based on the observed mapping data and chose some spectral line data with significant infall profiles from them for further analysis.

\begin{figure}
\centering
\renewcommand{\thesubfigure} \makeatletter
\subfigure[G053.13+0.09]{
\begin{minipage}[b]{0.45\textwidth}
\includegraphics[width=1.1\textwidth]{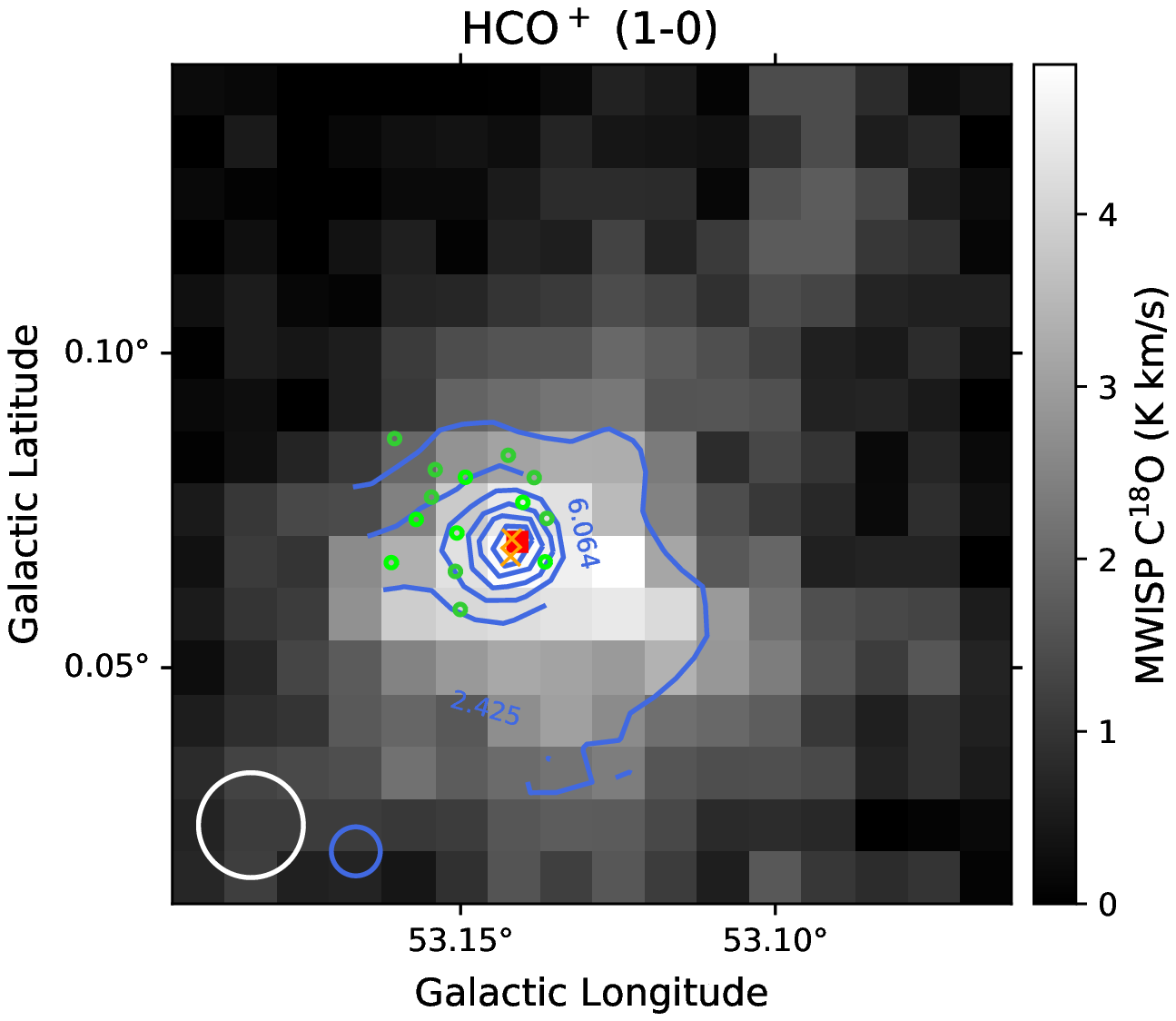} 
\end{minipage}
\begin{minipage}[b]{0.45\textwidth}
\includegraphics[width=1.1\textwidth]{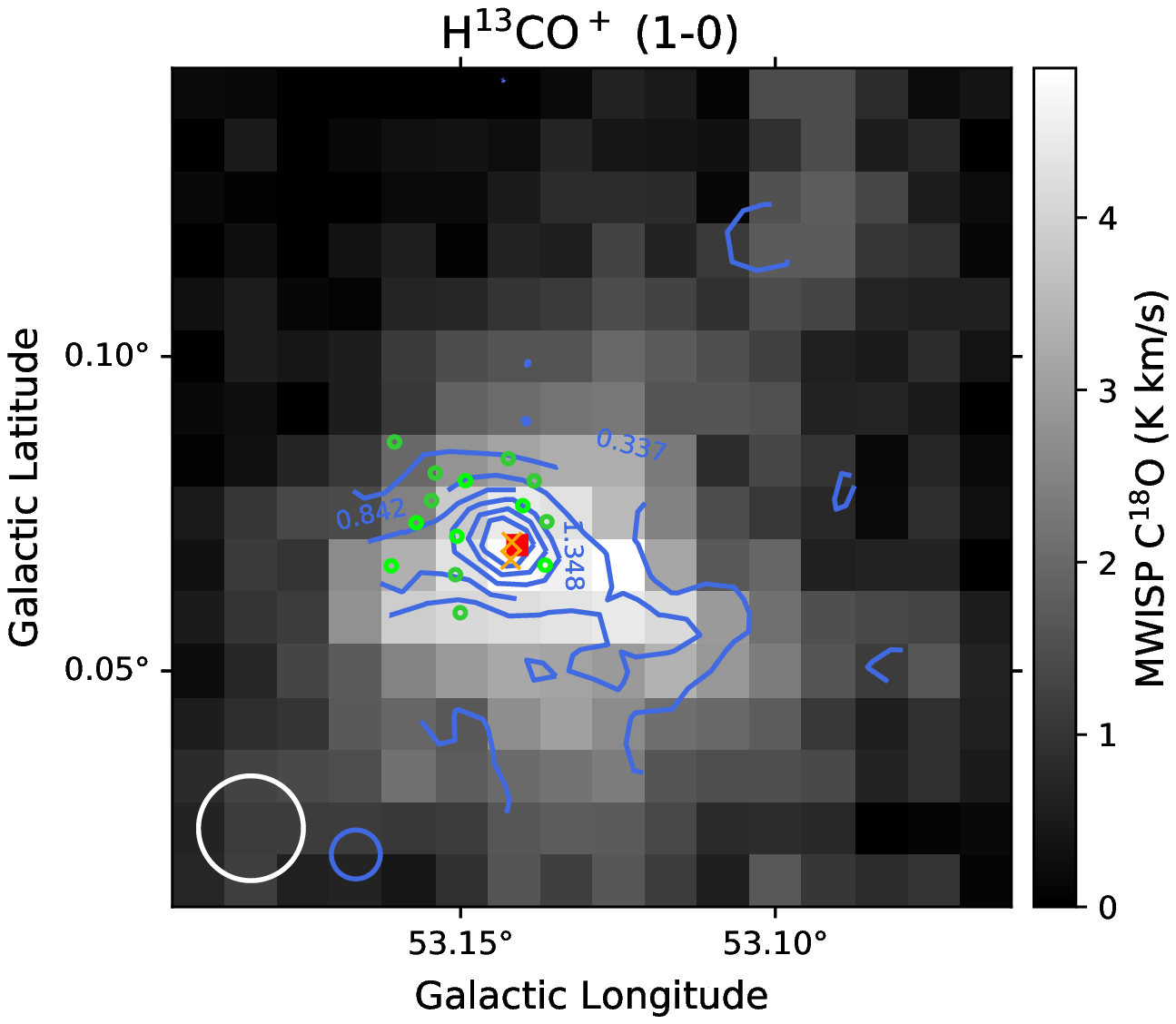}
\end{minipage}
}
\caption{Example of source G053.12+0.08: HCO$^+$ (1-0) and H$^{13}$CO$^+$ (1-0) integrated intensity contour (blue) maps superposed on the C$^{18}$O (1-0) integrated intensity image (C$^{18}$O data from the MWISP project). The white and blue circles on the lower left indicate the beam sizes of the DLH 13.7-m telescope and the IRAM 30-m telescope, respectively. The green circles denote YSOs from AllWISE data. The magenta circles denote IRAS sources. The red squares denote the star-forming regions. The orange crosses denote the maser sources.}
\label{fig:map_eg}
\end{figure}

\begin{figure*}[h]  
  \begin{minipage}[t]{0.5\linewidth}
  \centering
   \includegraphics[width=100mm]{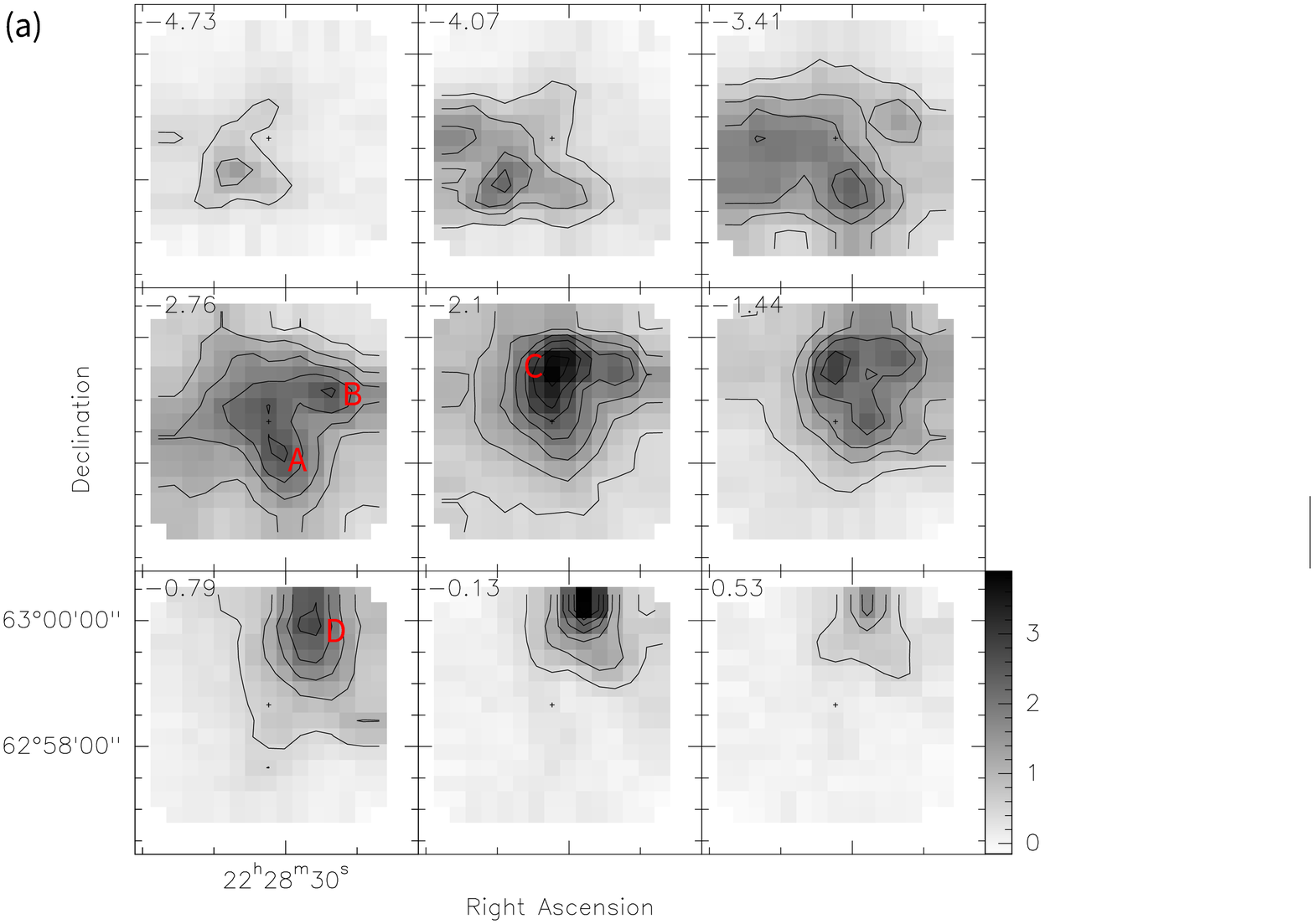}
  \end{minipage}%
  \begin{minipage}[t]{0.5\linewidth}
  \centering
   \includegraphics[width=100mm]{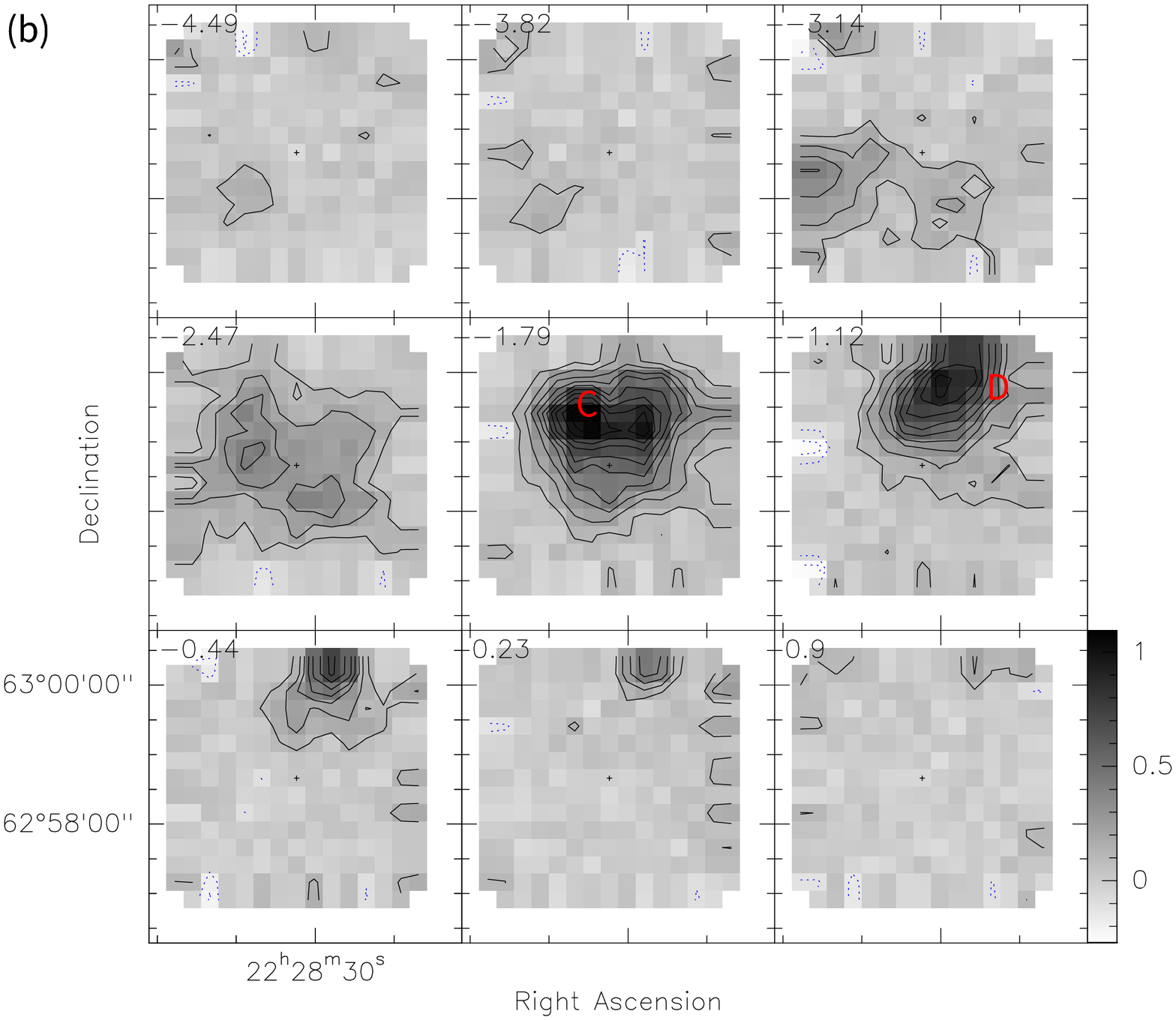}
  \end{minipage}%
  \gridline{\fig{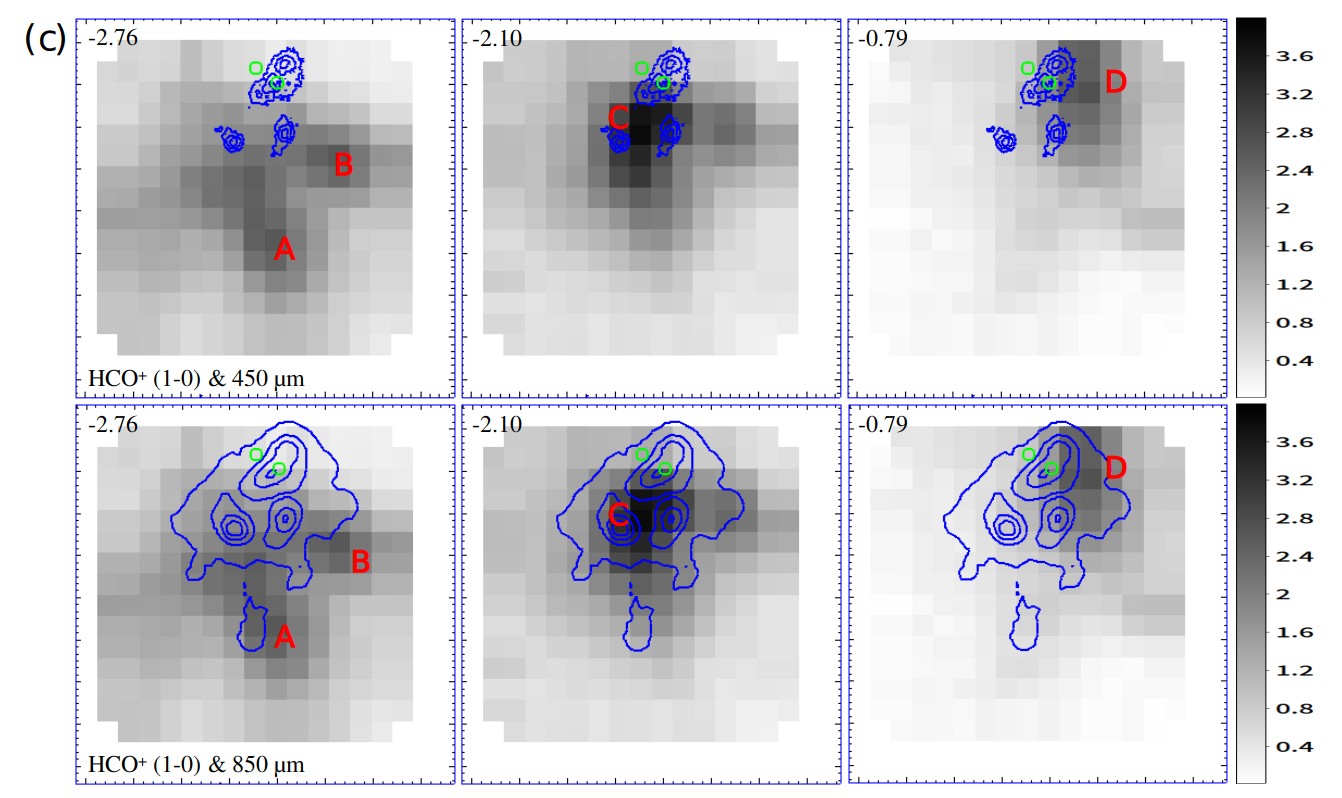}{0.55\textwidth}{}} 
\caption{(a) panel: Channel maps of G107.50+4.47 HCO$^+$ (1-0) lines from -4.73 km s$^{-1}$ to 0.53 km s$^{-1}$. (b) panel: Channel maps of G107.50+4.47 H$^{13}$CO$^+$ (1-0) lines from -4.49 km s$^{-1}$ to 0.90 km s$^{-1}$. (c) panel: The 450 $\mu$m and 850 $\mu$m continuum data from SCUBA-2, superposed on the HCO$^+$ channel maps (velocity are -2.76, -2.10, and -0.79 km s$^{-1}$, respectively. The green circles denote Class 0/I YSOs from AllWISE data.).
\label{fig:G107_1}}
\end{figure*}

\section{Mapping Results} \label{sec:results}

HCO$^+$ (1-0) emissions are detected from all 24 sources, while H$^{13}$CO$^+$ (1-0) emissions are detected from 18 sources (listed in Table \ref{Tab:result}). Although most sources have detected the emissions of these two spectral lines, due to some reasons such as insufficient signal-to-noise ratios and resolution of the lines, not all sources show clear clumpy structures in the mapping image. For the sources with compact structures, we analyzed their morphology and clump properties in this section, while for other sources, we only analyzed their spectral line profiles (the analysis of line profiles will be given in Section \ref{sec:infall}).

\subsection{Morphology} \label{subsec:mappting}

The HCO$^+$ integrated intensity maps of 17 sources show clumpy structures, and the H$^{13}$CO$^+$ integrated intensity maps of 15 sources show clumpy structures (which are listed in Table \ref{Tab:result}). The HCO$^+$ and H$^{13}$CO$^+$ integrated intensity maps of sources with clumpy structures superposed on the C$^{18}$O (1-0) data are presented in Section \ref{sec:Appendix1} (Figure \ref{fig:map_eg} shows an example of source G053.12+0.08). The C$^{18}$O mapping data come from the MWISP project, with spatial resolutions of about 52$^{\prime\prime}$. In addition, the maps also show the catalog positions of SFR \citep{Avedisova+etal+2002}, maser sources, as well as the young stellar objects (YSOs), which are  associated with the clumps.

Among them, 14 sources show clear clumpy structures in both HCO$^+$ (1-0) and H$^{13}$CO$^+$ (1-0) lines. For G029.06+4.58, G039.33-1.03, and G110.31+2.53, only HCO$^+$ maps show clear clumps. HCO$^+$ emissions have also been found in other sources, but due to insufficient angular resolution and/or sensitivity of the telescope, the integrated intensity maps do not show obvious clumps. G028.97+3.35 is different, i.e., its H$^{13}$CO$^+$ integrated intensity map shows a compact clump. However the HCO$^+$ map does not have sufficient signal-to-noise ratio, and no clear clump is found in this area.

Although the observation targets are selected based on the MWISP CO mapping data, the HCO$^+$ spatial distributions of some sources seem to be slightly different from the CO distributions, and deviate from the center position of the observation area. This may be caused by HCO$^+$ and CO tracing different density regions. Similarly, although HCO$^+$ and H$^{13}$CO$^+$ lines are isotopic molecular lines, the areas they traced may also be different. In most cases, the maps of HCO$^+$ and H$^{13}$CO$^+$ display similar compact regions. However, there are some exceptions: the map of HCO$^+$ emissions in G029.60-0.63 shows two clumps located in the north and south positions, while the H$^{13}$CO$^+$ emission map shows only one clump at the same velocity region, and the clump located slightly east of the two clumps from the HCO$^+$ data. The HCO$^+$ and H$^{13}$CO$^+$ maps of G109.00+2.73 also display some differences. The compact clump shown in the H$^{13}$CO$^+$ map is located northeast of the clumps in the HCO$^+$ emission map. This may be related to the different distribution of HCO$^+$ and H$^{13}$CO$^+$ in the clumps or the different optical depths of these two lines.

Another source with a relatively complicated structure is G107.50+4.47. Our mapping results show that both HCO$^+$ and H$^{13}$CO$^+$ integrated intensity maps of this source have complex geometric shapes. We made channel maps to analyze further the structure of this source. Figure \ref{fig:G107_1} (a) and (b) panels show the channel maps of HCO$^+$ and H$^{13}$CO$^+$, respectively. We found that there may be four clumps in this area, which named clump A, B, C, and D (as shown in the Figure \ref{fig:G107_1}). Their velocity ranges are approximately [-4.7, -1.4], [-3.4, -1.4], [-2.8, -1.4], and [-1.0, 0.4] km s$^{-1}$, respectively. Among them, clump D is associated with two YSOs. Figure \ref{fig:G107_1} (c) presents the HCO$^+$ (1-0) channel maps superposed on the SCUBA-2 450 $\mu$m and 850 $\mu$m continuum contour maps\footnote{\url{http://www.cadc-ccda.hia-iha.nrc-cnrc.gc.ca/en/search/}} \citep{Geach+etal+2017}. The continuum data also show that there is a complex clump in this area. Some structures in the north and south of the continuum maps seem to correspond to the molecular lines' emissions. However, due to the limitation of the angular resolution of the telescope, the more detailed structure and properties of the clumps in this area are still unknown. We hope to use a telescope with higher resolution and sensitivity to map this target in follow-up study.


\begin{table}
\begin{center}
  \caption{Spectral line parameters}\label{Tab:result}  
 \setlength{\tabcolsep}{1mm}{
\begin{tabular}{cccccccccc}
  \hline\noalign{\smallskip}
Source&  \multicolumn{4}{c}{H$^{13}$CO$^+$ (1-0)}  &  \multicolumn{4}{c}{HCO$^+$ (1-0)} \\
\cmidrule(lr){2-6}  \cmidrule(lr){7-10}
Name & V$_{peak}$ & $\sigma$ & T$_{peak}$ & RMS & Clump & V$_{peak}$ & T$_{peak}$ & RMS & Clump \\
 & (km s$^{-1}$) & (km s$^{-1}$) & (K) & (K) & & (km s$^{-1}$) & (K) & (K) & \\
  \hline\noalign{\smallskip}
G028.97+3.35 & 7.31(0.01) & 0.34(0.02) & 0.8(0.02)  & 0.13  & yes & 5.72 & 0.5  & 0.21  & no\\ 
G029.06+4.58 & 7.48(0.02) & 0.35(0.03) & 0.4(0.01)  & 0.16  & no & 7.14 & 1.1  & 0.26  & yes\\ 
G029.60-0.63 & 76.52(0.03) & 1.28(0.07) & 0.3(0.01)  & 0.19  & yes & 75.01  & 3.0  & 0.32  & yes\\ 
G030.17+3.68 & 9.02(0.01) & 0.32(0.03) & 0.3(0.03)  & 0.13  & no & 8.57 & 0.8  & 0.21  & no\\ 
G031.41+5.25 & Nd & Nd & Nd & 0.12  & no & 7.01 & 0.6  & 0.22  & no\\ 
G033.42+0.00 & Nd & Nd & Nd & 0.18  & no & 10.58 & 0.9  & 0.32  & no\\ 
G039.33-1.03 & Nd & Nd & Nd & 0.14  & no & 13.19 & 1.3  & 0.23  & yes\\ 
G053.13+0.09 & 21.71(0.01) & 1.06(0.02) & 1.6(0.01)  & 0.15  & yes & 21.02 & 10.1  & 0.25  & yes\\ 
G077.91-1.16 & -0.21(0.01) & 0.57(0.03) & 0.6(0.01)  & 0.13  & yes & 0.57 & 0.6  & 0.22  & yes\\ 
G079.24+0.53 & 0.79(0.01) & 0.44(0.02) & 0.6(0.01)  & 0.12  & yes & -0.62 & 0.6  & 0.20  & yes\\ 
G079.71+0.14 & 1.19(0.01) & 0.36(0.09) & 0.5(0.04)  & 0.15  & yes & 0.61 & 1.2  & 0.25  & yes\\ 
G081.72+0.57 & -3.22(0.01) & 1.80(0.01) & 2.5(0.01)  & 0.13  & yes & -5.20 & 13.9  & 0.22  & yes\\ 
G082.21-1.53 & 3.07(0.01) & 0.41(0.01) & 1.1(0.01)  & 0.14  & yes & 2.31 & 2.1  & 0.23  & yes\\ 
G107.50+4.47 & -1.27(0.01) & 0.53(0.01) & 1.0(0.01)  & 0.14  & yes & -1.84 & 4.6  & 0.23  & yes\\ 
G109.00+2.73 & -10.29(0.02) & 0.57(0.04) & 0.4(0.01)  & 0.14  & yes & -11.01 & 1.2  & 0.24  & yes\\ 
G110.31+2.53 & Nd & Nd & Nd & 0.12  & no & -11.43 & 1.5  & 0.19  & yes\\ 
G110.40+1.67 & Nd & Nd & Nd & 0.12  & no & -12.78 & 0.4  & 0.20  & no\\ 
G121.31+0.64 & -17.54(0.01) & 1.02(0.01) & 2.3(0.01)  & 0.18  & yes & -18.35 & 13.7  & 0.30  & yes\\ 
G121.34+3.42 & -5.18(0.01) & 0.34(0.02) & 0.5(0.01)  & 0.12  & no & -6.04 & 0.7  & 0.21  & no\\ 
G126.53-1.17 & -12.26(0.04) & 0.30(0.09) & 0.6(0.2)  & 0.10  & yes & -13.16 & 1.5  & 0.14  & yes\\ 
G143.04+1.74 & -8.8(0.01) & 0.55(0.03) & 0.5(0.01)  & 0.14  & yes & -9.05 & 1.7  & 0.23  & yes\\ 
G154.05+5.07 & Nd & Nd & Nd & 0.15 & no & 4.04 & 1.1  & 0.25  & no\\ 
G193.01+0.14 & 7.59(0.01) & 1.10(0.05) & 0.7(0.01)  & 0.12  & yes & 7.11 & 7.5  & 0.20  & yes\\ 
G217.30-0.05 & 26.83(0.02) & 0.60(0.07) & 0.3(0.01)  & 0.13 & yes & 26.21 & 2.3  & 0.21  & yes\\ 
  \hline\noalign{\smallskip}
\end{tabular}}
\end{center}
\tablecomments{
Nd denotes non-detection with H$^{13}$CO$^+$ (1-0) emission. H$^{13}$CO$^+$ line parameters are derived from Gaussian fitting. The values in parentheses are the uncertainty of the Gaussian fitting.}
\end{table}

\subsection{Physical Parameters of the Clumps} \label{subsec:clumps}


\begin{table}
\begin{center}
\caption{Physical parameters of the clumps.}\label{Tab:clump}
\setlength{\tabcolsep}{0.3mm}{
\begin{tabular}{cccccccccccc}   
  \hline\noalign{\smallskip}
Source & Radius & $T_{kin}$ & log($\frac{N({\rm H}_2)}{{\rm cm}^{-2}}$) & log($\frac{\rho}{{\rm cm}^{-3}}$) & log($\frac{M}{M_{\odot}}$) & $T_{ex}$(HCO$^+$) & $T_{ex}$(H$^{13}$CO$^+$) & log($\frac{N({\rm HCO}^+)}{{\rm cm}^{-2}}$)  & log($\frac{N({\rm H^{13}CO}^+)}{{\rm cm}^{-2}}$)  & [HCO$^+$]/ & [H$^{13}$CO$^+$]/ \\
Name  & (pc) & (K) &  &  &  & (K) & (K) &  &   & [H$_2$] & [HCO$^+$] \\ 
  \hline\noalign{\smallskip} 
G028.97+3.35 &0.03(0.01) & 12.4(0.4) &22.4(0.3) & 4.8(0.7) & 1 &$9.1_{-1.0}^{+2.1}$  &$9.8_{-1.1}^{+2.3}$  &$11.6_{-0.1}^{+0.2}$ &$11.7_{-0.1}^{+0.2}$ & $1.5\times10^{-11}$ & $1.3\times10^{0}$ \\
G029.06+4.58 &0.05 & 8.7(0.3) & 21.6(0.1) & 4.3 & 1 &$4.0_{-0.01}^{+0.01}$ &$3.6_{-0.01}^{+0.03}$ &$12.6_{-0.1}^{+0.1}$&$12.1_{-0.1}^{+0.1}$& $9.7\times10^{-10}$ & $5.6\times10^{-1}$ \\
G029.60-0.63 &0.56 & 13.9(0.4) & 22.6(0.5) & 4.2 & 3 &$6.1_{-0.02}^{+0.6}$ &$3.6_{-0.5}^{+1.7}$ &$13.8_{-0.5}^{+0.5}$&$12.5_{-0.3}^{+0.5}$& $1.9\times10^{-9}$ & $3.9\times10^{-2}$ \\
G030.17+3.68 & - & 10.4(0.4) &21.8(0.2) & - & - & - & - & - & - & - & - \\
G031.41+5.25 & - & 13.8(0.6) &22.0(0.3) & - & - & - & Nd &  - & Nd & - & Nd \\
G033.42+0.00 & - & 8.8(0.4) &21.6(0.1) & - & - & - & Nd &  - & Nd & - & Nd \\
G039.33-1.03 &0.06 & 8.8(0.3) &21.8(0.1) & 4.4 & 1 &$4.1_{-0.01}^{+0.01}$ & Nd &$12.4_{-0.1}^{+0.1}$& Nd & $4.0\times10^{-10}$ & Nd \\
G053.13+0.09 &0.20 & 16.4(0.4) &22.0(0.2)& 4.1 & 2 &$16.3_{-0.3}^{+0.1}$ &$4.4_{-0.01}^{+0.1}$ &$15.2_{-0.9}^{+1.1}$&$13.3_{-0.2}^{+0.2}$& $1.6\times10^{-7}$ & $1.3\times10^{-2}$ \\
G077.91-1.16 &0.01 & 14.3(0.3) &22.0(0.2) & 5.6 & 1 &$15.8_{-3.8}^{+6.0}$ &$17.1_{-4.3}^{+6.5}$ &$12.2_{-0.01}^{+0.1}$&$11.8_{-0.01}^{+0.1}$& $1.0\times10^{-10}$ & $4.2\times10^{-1}$ \\
G079.24+0.53 &0.03 & 10.4(0.3) &22.4(0.4) & 5.3 & 1 &$7.6_{-2.5}^{+3.3}$ &$8.0_{-2.7}^{+3.3}$ &$12.0_{-0.01}^{+0.1}$&$11.8_{-0.01}^{+0.1}$& $3.6\times10^{-11}$ & $6.6\times10^{-1}$ \\
G079.71+0.14 &0.07 & 9.1(0.3) &21.7(0.1) & 4.2 & 1 &$4.0_{-0.1}^{+0.1}$ &$3.5_{-0.1}^{+0.1}$ &$12.8_{-0.1}^{+0.1}$&$12.2_{-0.1}^{+0.1}$& $1.4\times10^{-9}$ & $2.3\times10^{-1}$ \\
G081.72+0.57 &0.22(0.02) & 38.0(0.4) &23.2(1.3) & 5.0(0.2) & 3 &$29.6_{-6.1}^{+5.1}$ &$23.8_{-1.9}^{+3.9}$ &$13.9_{-0.1}^{+1.4}$&$13.0_{-0.01}^{+0.1}$& $5.2\times10^{-10}$ & $1.1\times10^{-1}$ \\
G082.21-1.53 &0.23 & 12.7(0.3) &22.0(0.2) & 4.0 & 2 &$6.8_{-0.02}^{+0.01}$ &$3.9_{-0.1}^{+0.03}$ &$13.4_{-0.2}^{+0.2}$&$12.8_{-0.2}^{+0.2}$& $2.4\times10^{-9}$ & $2.6\times10^{-1}$ \\
G107.50+4.47 &0.03 & 18.8(0.4) &22.4(0.4) & 5.3 & 1 &$16.3_{-5.7}^{+9.4}$ &$18.5_{-0.1}^{+0.1}$ &$13.0_{-0.01}^{+0.2}$&$12.0_{-0.01}^{+0.1}$& $4.3\times10^{-10}$ & $1.0\times10^{-1}$ \\
G109.00+2.73 &0.06 & 17.4(0.3) &22.4(0.3) & 5.0 & 1 &$8.3_{-2.7}^{+6.2}$ &$8.7_{-3.1}^{+7.7}$ &$12.1_{-0.1}^{+0.1}$&$11.7_{-0.03}^{+0.1}$& $5.2\times10^{-11}$ & $4.0\times10^{-1}$ \\
G110.31+2.53 & - & 22.8(0.4) &22.2(0.3) & - & - & - & Nd & - & Nd & - & Nd  \\
G110.40+1.67 & - & 12.6(0.6) &21.8(0.1) & - & - & - & Nd & - & Nd & - & Nd  \\
G121.31+0.64 &0.12(0.01) & 16.6(0.4) &22.2(0.2)& 4.4(0.1) & 2 &$15.9_{-0.4}^{+0.4}$ &$5.4_{-0.4}^{+0.9}$ &$15.4_{-1.4}^{+1.4}$&$13.1_{-0.2}^{+0.2}$& $1.5\times10^{-7}$ & $7.9\times10^{-3}$ \\
G121.34+3.42 & - & 7.2(0.3) &22.0(0.2) & - & - & - & - & - & - & $5.5\times10^{-10}$ & $5.9\times10^{-1}$  \\
G126.53-1.17 &0.05 & 11.9(0.4) &22.0(0.3)& 4.7 & 1 &$5.1_{-0.6}^{+1.5}$ &$4.8_{-0.9}^{+1.8}$ &$12.3_{-0.2}^{+0.3}$&$11.8_{-0.1}^{+0.2}$& $2.2\times10^{-10}$ & $2.8\times10^{-1}$ \\
G143.04+1.74 &0.05 & 14.6(0.4) &22.0(0.2)& 4.7 & 1 &$5.5_{-0.7}^{+1.2}$ &$5.1_{-0.8}^{+1.4}$ &$12.6_{-0.1}^{+0.2}$&$11.9_{-0.1}^{+0.1}$& $3.9\times10^{-10}$ & $2.2\times10^{-1}$ \\
G154.05+5.07 & - & 10.3(0.4) &21.5(0.1)& - & - & - & Nd &-& Nd & - & Nd \\
G193.01+0.14 &0.31 & 20.4(0.4) &22.2(0.4)& 4.1 & 2&$12.4_{-0.01}^{+0.01}$ &$3.8_{-0.3}^{+1.1}$ &$14.4_{-0.4}^{+0.4}$&$12.7_{-0.3}^{+0.4}$& $1.5\times10^{-8}$ & $2.2\times10^{-2}$ \\
G217.30-0.05 &0.33 & 22.1(0.4) &22.4(0.5)& 4.3 & 2 &$6.6_{-0.03}^{+1.6}$ &$4.0_{-0.8}^{+2.9}$ &$13.4_{-0.5}^{+0.5}$&$11.8_{-0.3}^{+0.4}$& $9.7\times10^{-10}$ & $2.9\times10^{-2}$ \\
  \hline\noalign{\smallskip}

\end{tabular}}
\end{center}
\tablecomments{
Nd denotes non-detection with H$^{13}$CO$^+$ (1-0) emissions. The values in parentheses are the uncertainty of parameters.
}
\end{table}

Using HCO$^+$ and H$^{13}$CO$^+$ data, we can estimate the physical parameters of these clumps. Table \ref{Tab:clump} lists the parameters, including the radius, density and mass of these clumps, the excitation temperatures, and the column densities of HCO$^+$ and H$^{13}$CO$^+$ lines. Among them, the radius is obtained by the 2-dimensional Gaussian fitting of the of HCO$^+$ integrated intensity maps (for G028.97+3.35, the H$^{13}$CO$^+$ map was used). If the source is non-spherical, we adopt the average radius $r = \sqrt{a \times b}$, where a and b indicate the long and short axes of the clump. The angular sizes have been converted to linear sizes. The H$_2$ column densities are obtained from the previous study \citep{Yang+etal+2020}. Meanwhile, we roughly estimate the mass of these clumps by $M=\pi R^2 \mu m_H N(H_2)$, where R is the radius, $\mu$ is the mean molecular weight of the interstellar medium (i.e., 2.8), $m_H$ is the mass of a hydrogen atom, and $N(H_2)$ is the H$_2$ column density. However, because the beam size of the DLH telescope is about twice that of these observations, the H$_2$ density and clump mass may be underestimated due to beam dilution. In addition, due to the influence of distance and radius uncertainty, the clump mass error may also be large. What is given here is an estimate on the order of magnitude.

The excitation temperatures and column densities of HCO$^+$ and H$^{13}$CO$^+$ are estimated from the 1D non-LTE RADEX radiative transfer code \citep{vanderTak+etal+2007}. The input parameters of this program are the spectral range for the output, the cosmic microwave background temperature (i.e., 2.73 K), the kinetic temperature (estimated from the $T_{ex}$ of $^{12}$CO, while the $T_{ex}$ is calculated from $^{12}CO$ data given by the MWISP project \citep{Yang+etal+2020}) and the H$_2$ volume density (assuming that the source is a uniform sphere, it is estimated from the H$_2$ column density) of the cloud, and the molecular line width.

We iterate through the molecular column density in a loop, to match the model line intensities with the observed line intensities' peak values. The obtained results show that the excitation temperatures of HCO$^+$ and H$^{13}$CO$^+$ are about a few to twenty Kelvin. The HCO$^+$ column densities are about $10^{11}$ to $10^{15}$ cm$^{-2}$, while the H$^{13}$CO$^+$ densities are about $10^{11}$ to $10^{13}$ cm$^{-2}$. Combining the H$_2$ column density, we can estimate the abundance ratios of H$_2$, HCO$^+$, and H$^{13}$CO$^+$ in these clumps: [HCO$^+$]/[H$_2$] is about $10^{-11}$ to $10^{-7}$, and [H$^{13}$CO$^+$]/[HCO$^+$] is about $10^{-3}$ to 1.3. Except for the H$^{13}$CO$^+$ density of G028.97+3.35, which is slightly higher than for HCO$^+$, the HCO$^+$ abundances of the remaining clumps are all greater than H$^{13}$CO$^+$. In most clumps, the abundances of HCO$^+$ are 10 to 100 times those of H$^{13}$CO$^+$. We changed the input parameters and performed many calculations to estimate the uncertainty values of the results caused by the input parameter errors. However, due to the radius uncertainty caused by the kinematic distance error, and the local thermal equilibrium assumption and the uncertainty of H$_2$ abundance ratio we used when estimated the H$_2$ density, the uncertainty values of the results may be greater than the values we give.

\begin{figure*}[h]
  \begin{minipage}[t]{0.24\linewidth}
  \centering
   \includegraphics[width=50mm]{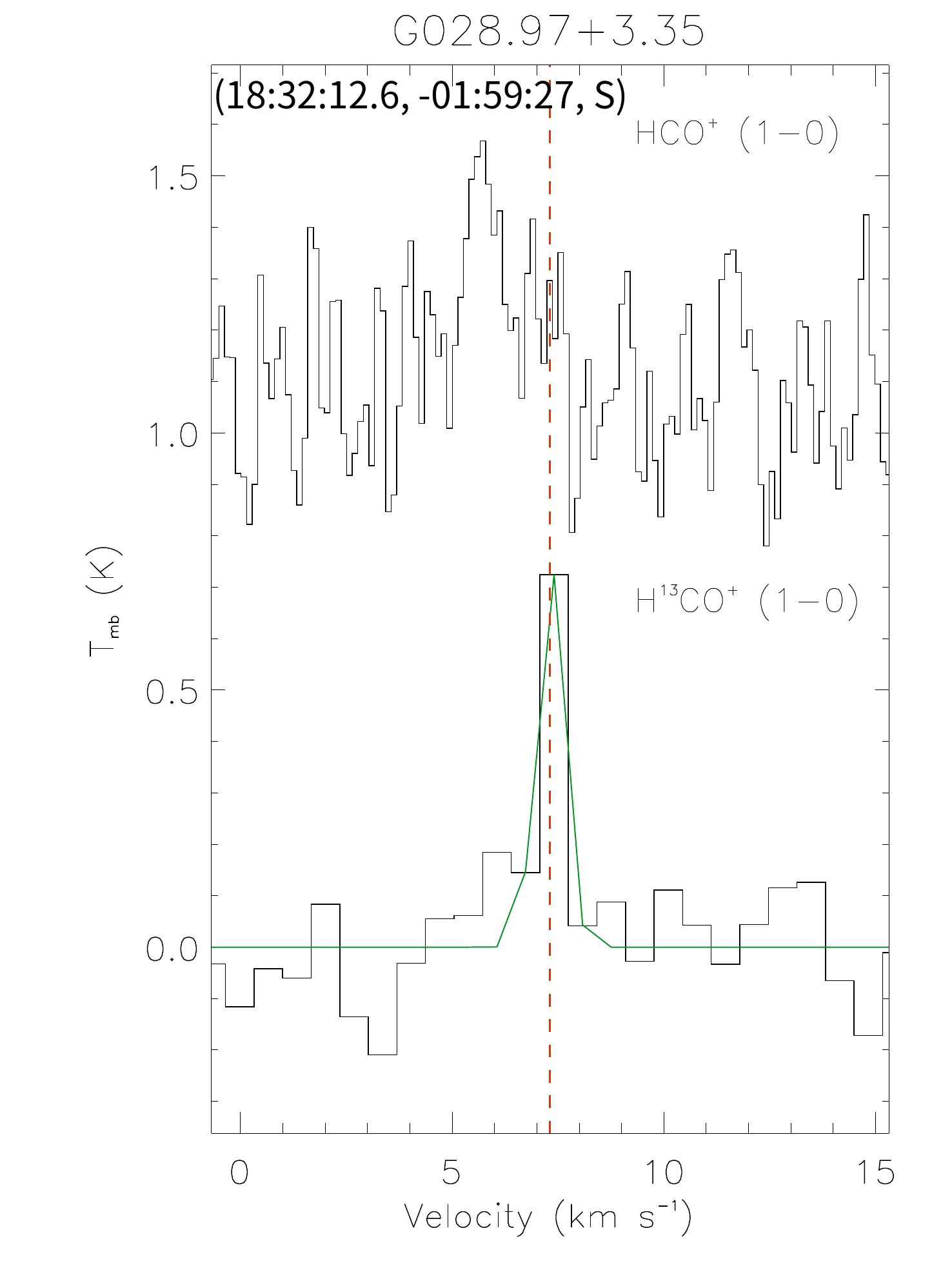}
  \end{minipage}%
  \begin{minipage}[t]{0.24\linewidth}
  \centering
   \includegraphics[width=50mm]{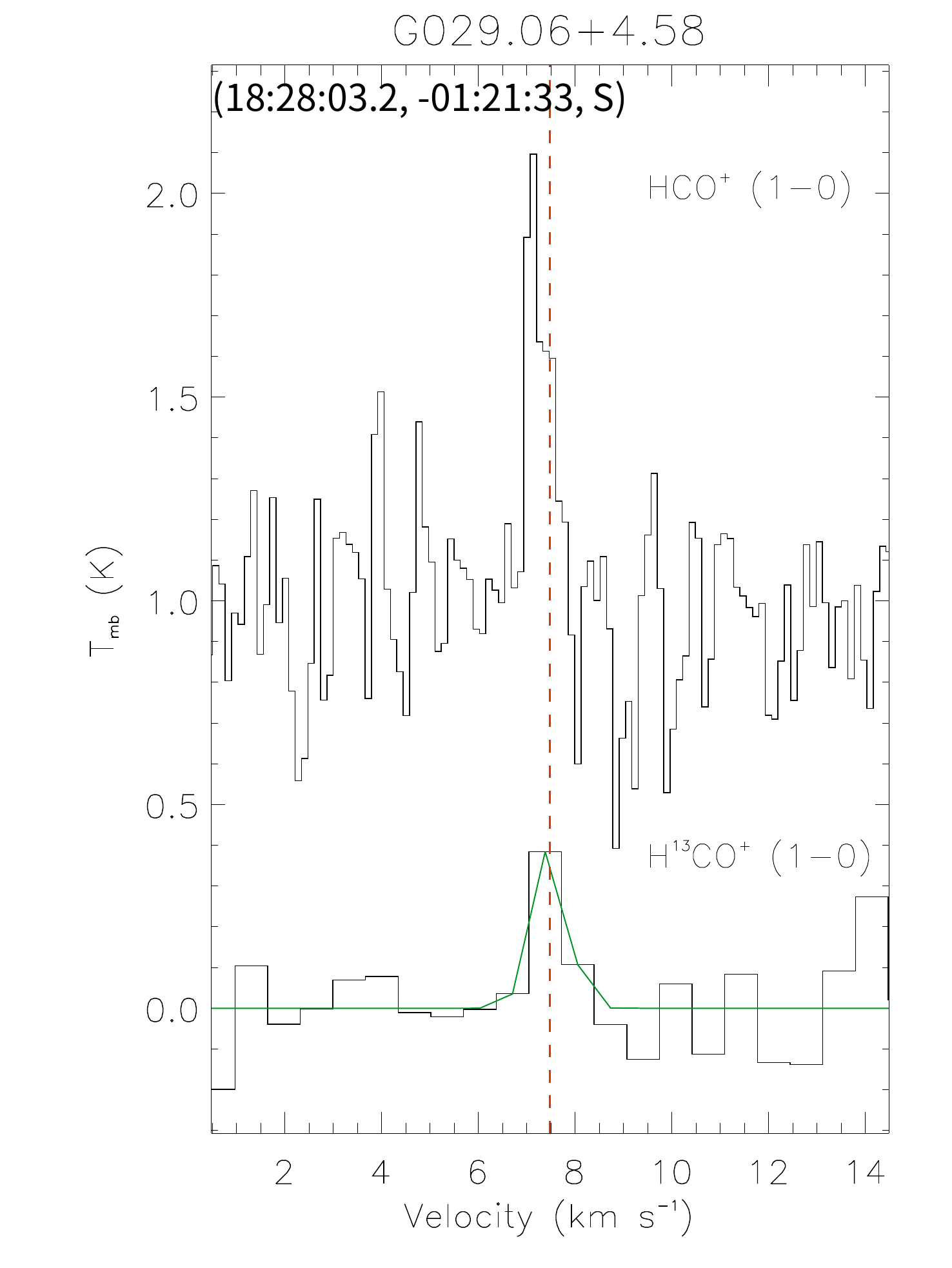}
  \end{minipage}%
  \begin{minipage}[t]{0.24\linewidth}
  \centering
   \includegraphics[width=50mm]{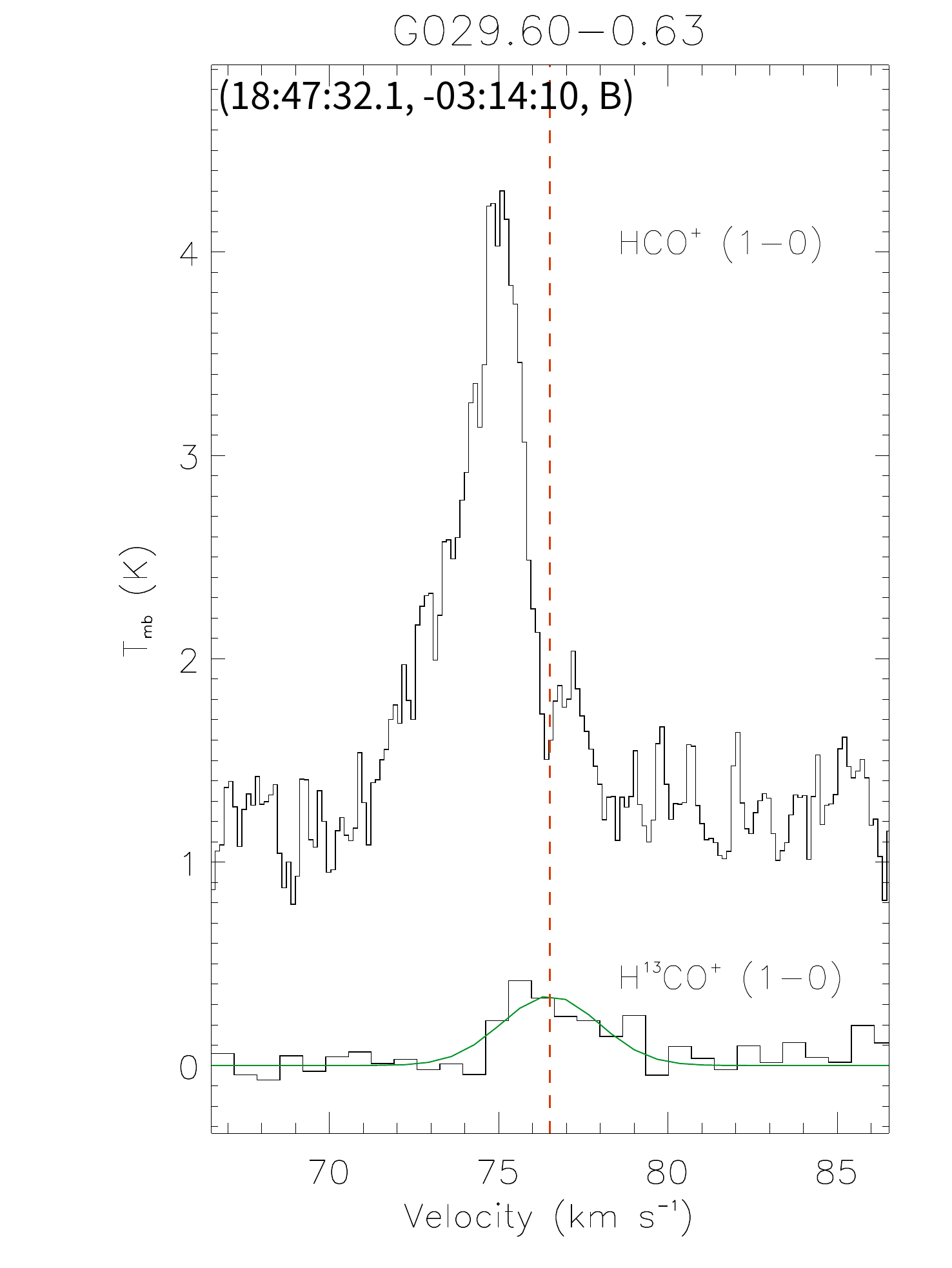}
  \end{minipage}%
  \begin{minipage}[t]{0.24\linewidth}
  \centering
   \includegraphics[width=50mm]{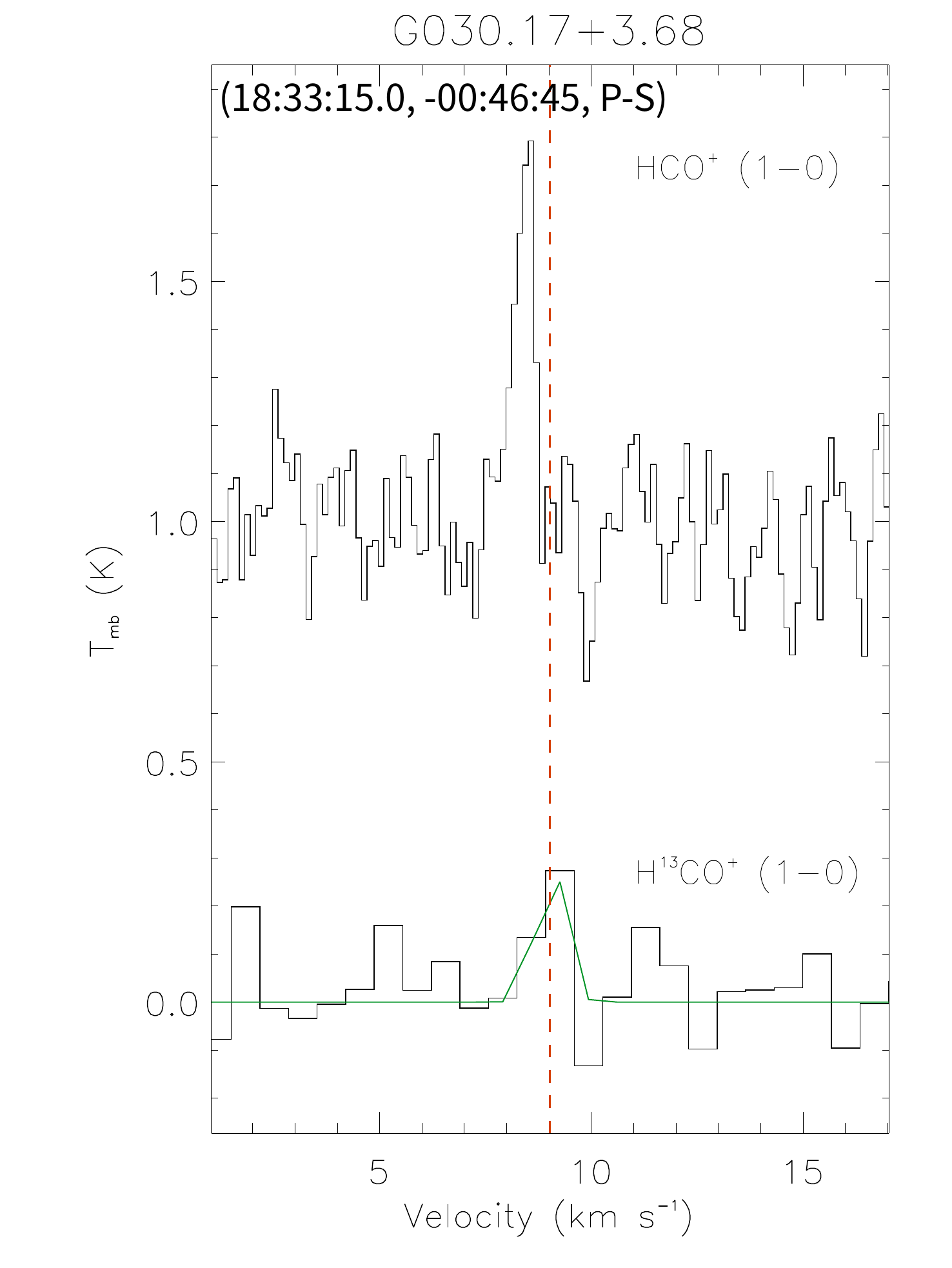}
  \end{minipage}%
  
  \begin{minipage}[t]{0.24\linewidth}
  \centering
   \includegraphics[width=50mm]{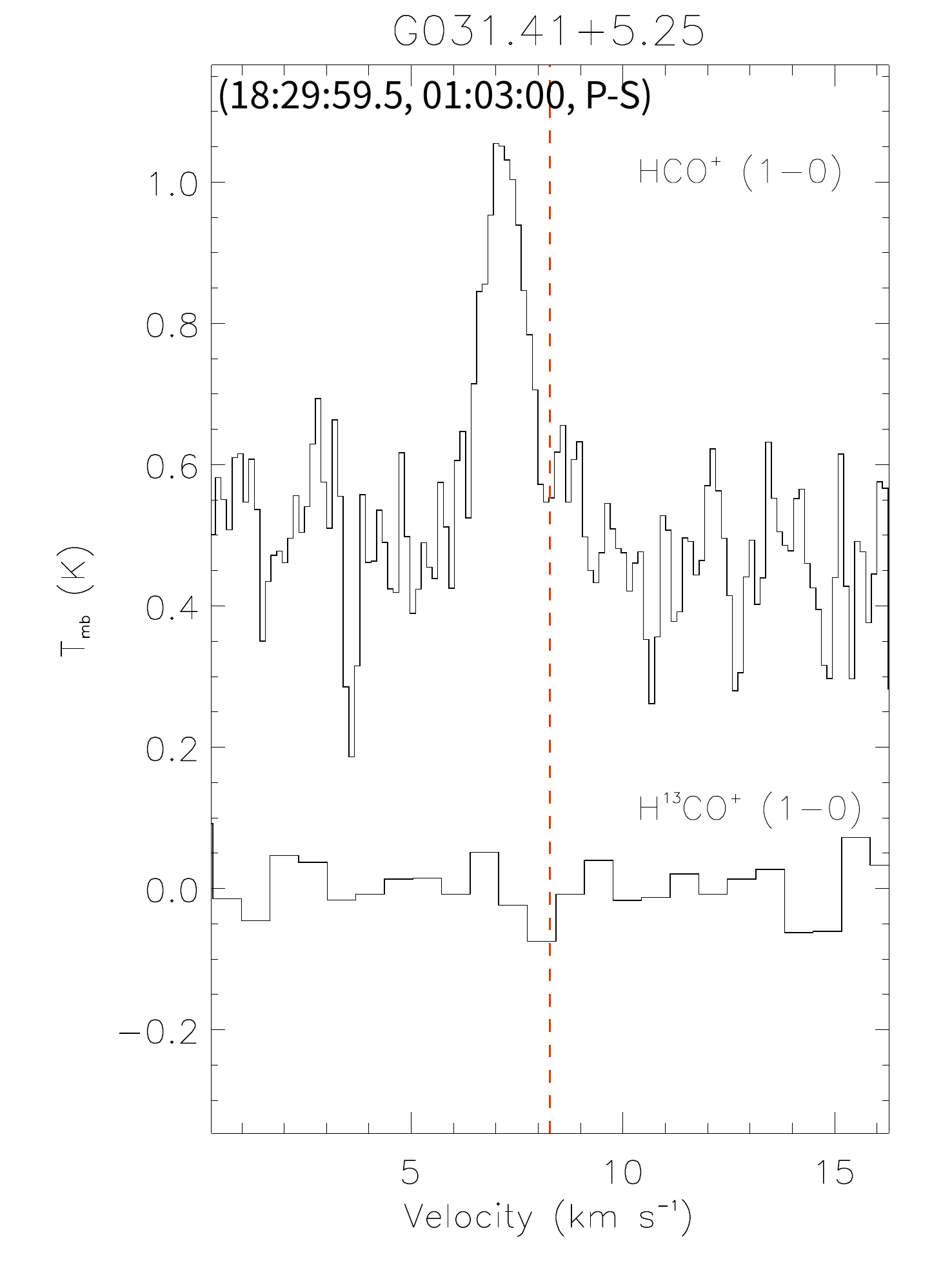}
  \end{minipage}%
  \begin{minipage}[t]{0.24\linewidth}
  \centering
   \includegraphics[width=50mm]{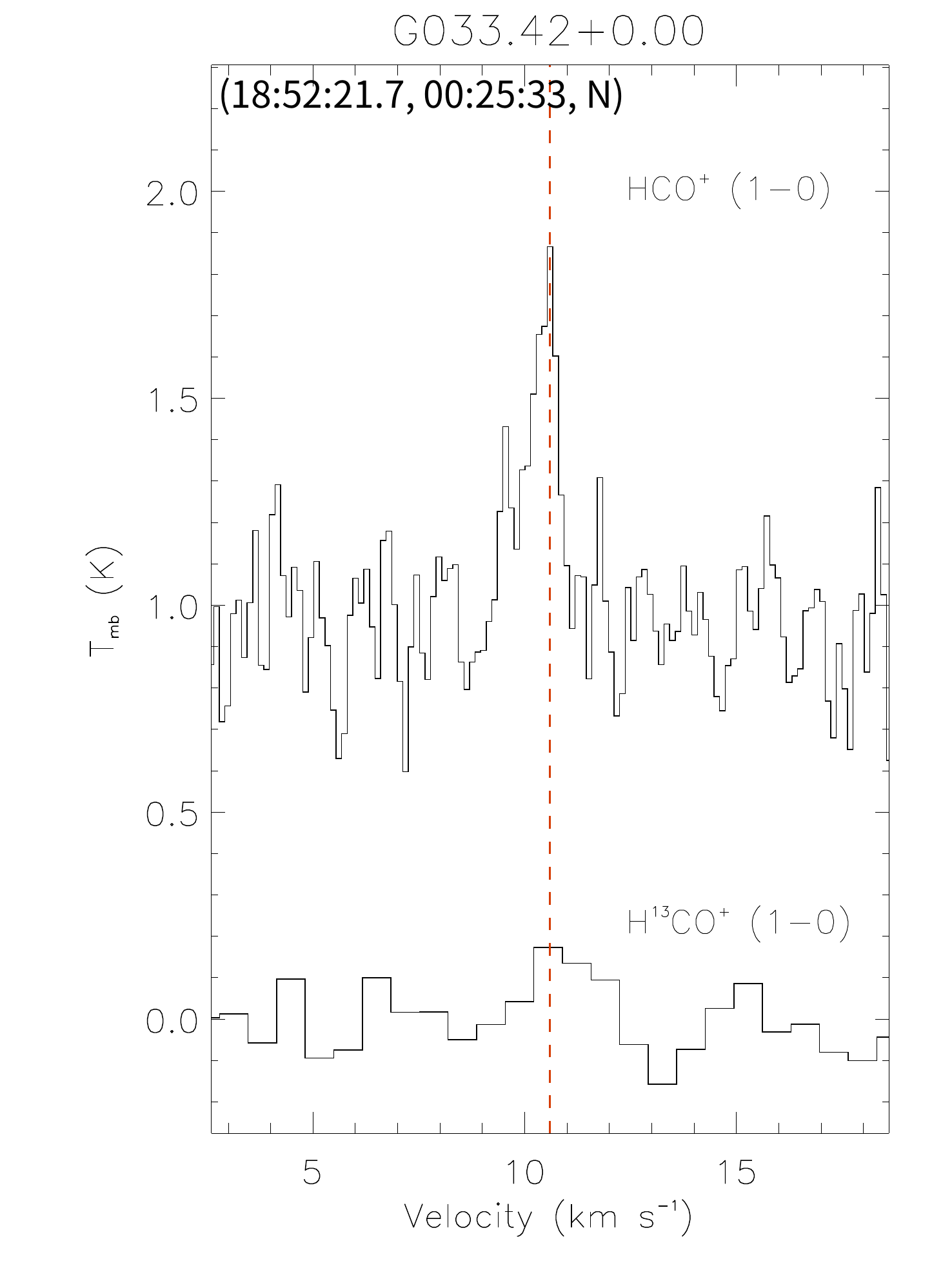}
  \end{minipage}%
  \begin{minipage}[t]{0.24\linewidth}
  \centering
   \includegraphics[width=50mm]{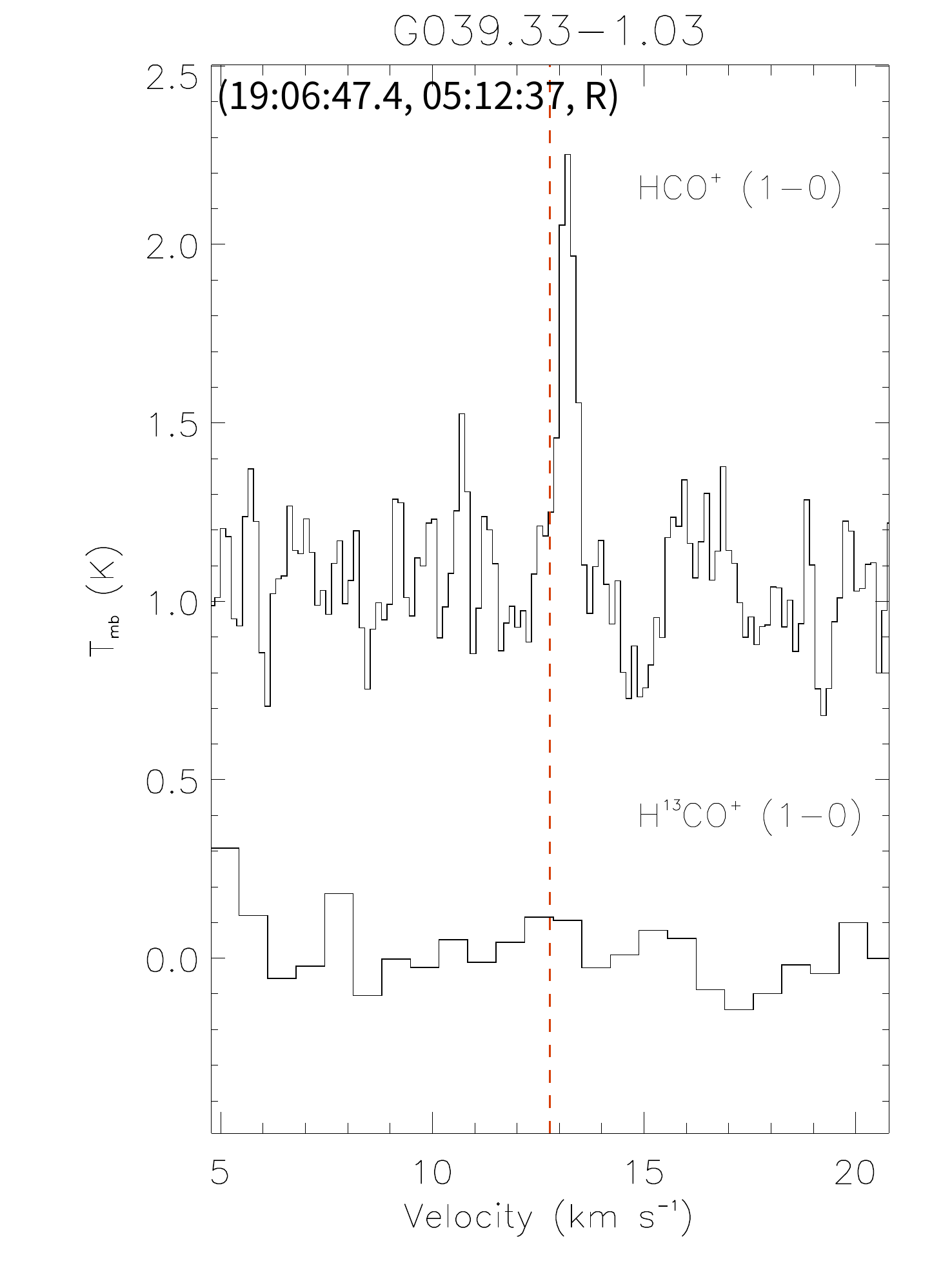}
  \end{minipage}%
  \begin{minipage}[t]{0.24\linewidth}
  \centering
   \includegraphics[width=50mm]{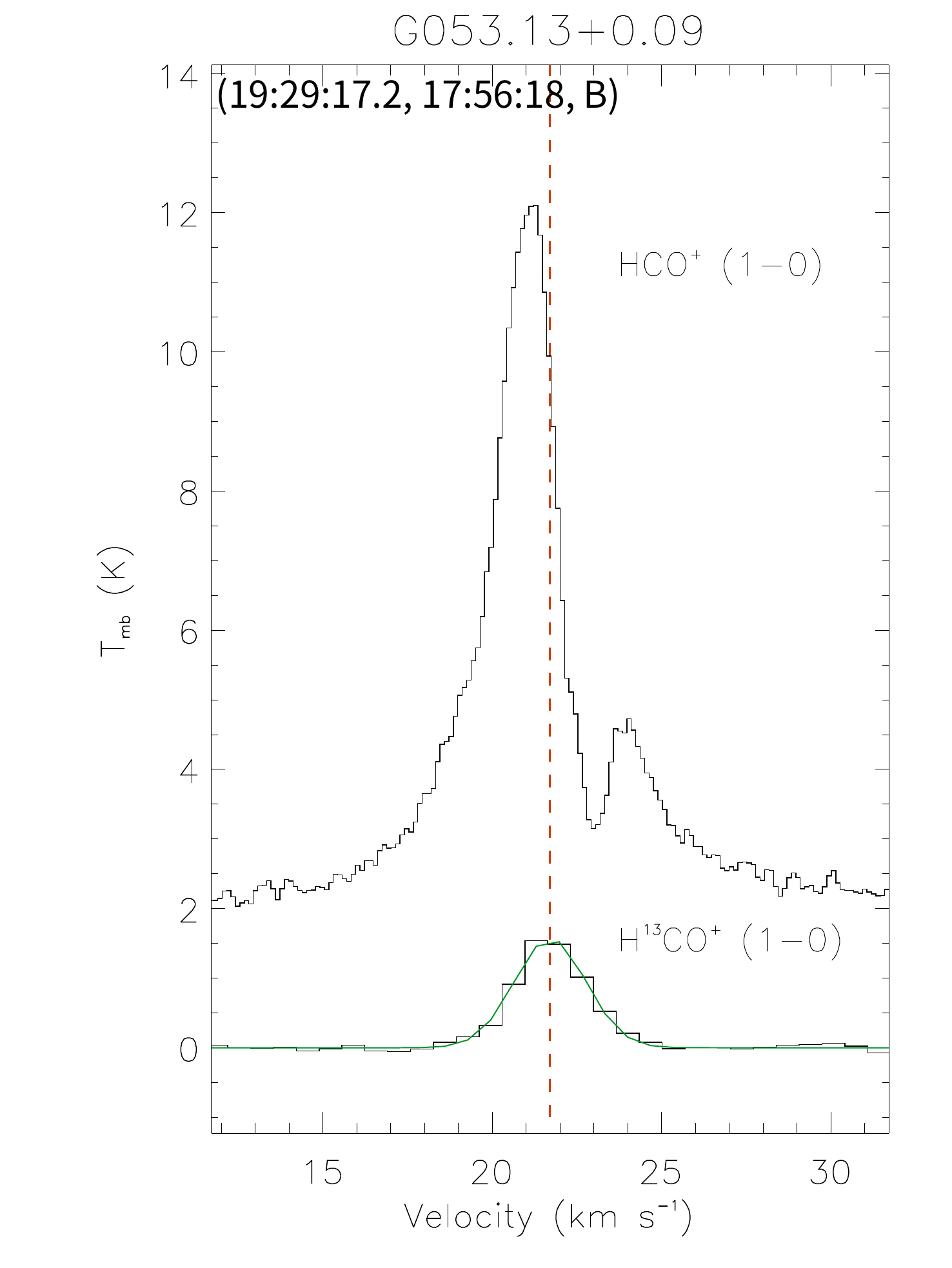}
  \end{minipage}%
  
  \begin{minipage}[t]{0.24\linewidth}
  \centering
   \includegraphics[width=50mm]{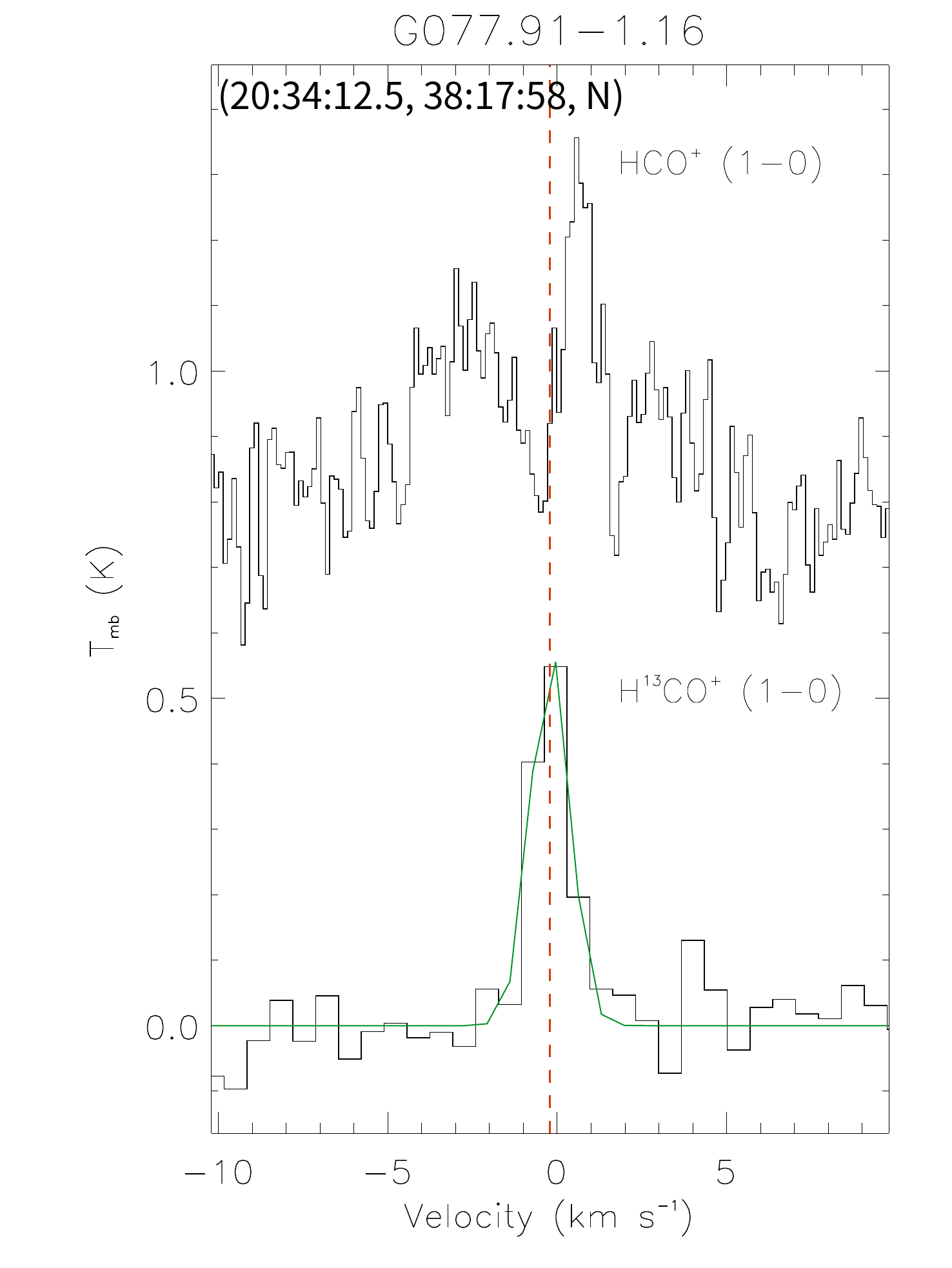}
  \end{minipage}%
  \begin{minipage}[t]{0.24\linewidth}
  \centering
   \includegraphics[width=50mm]{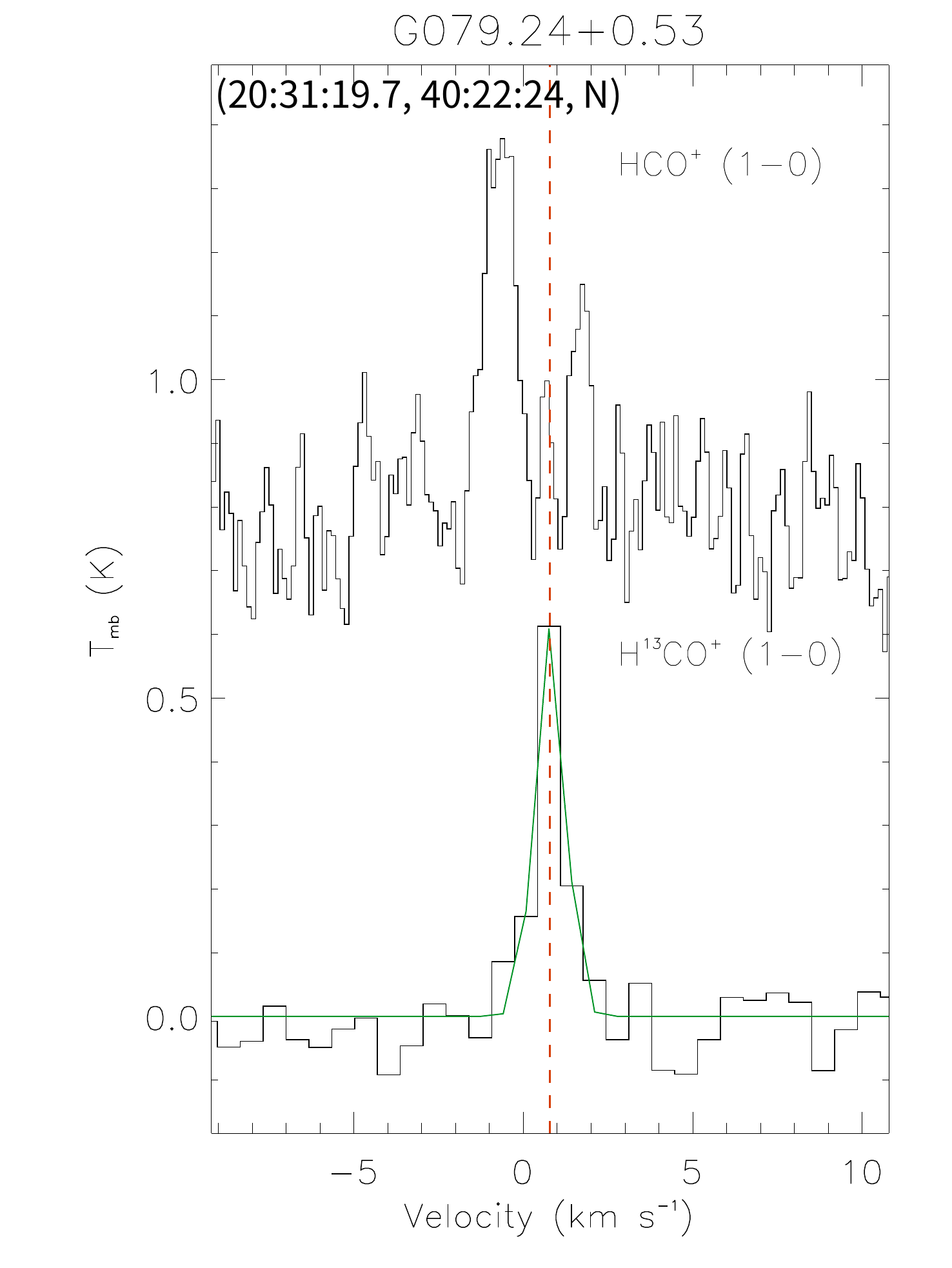}
  \end{minipage}%
  \begin{minipage}[t]{0.24\linewidth}
  \centering
   \includegraphics[width=50mm]{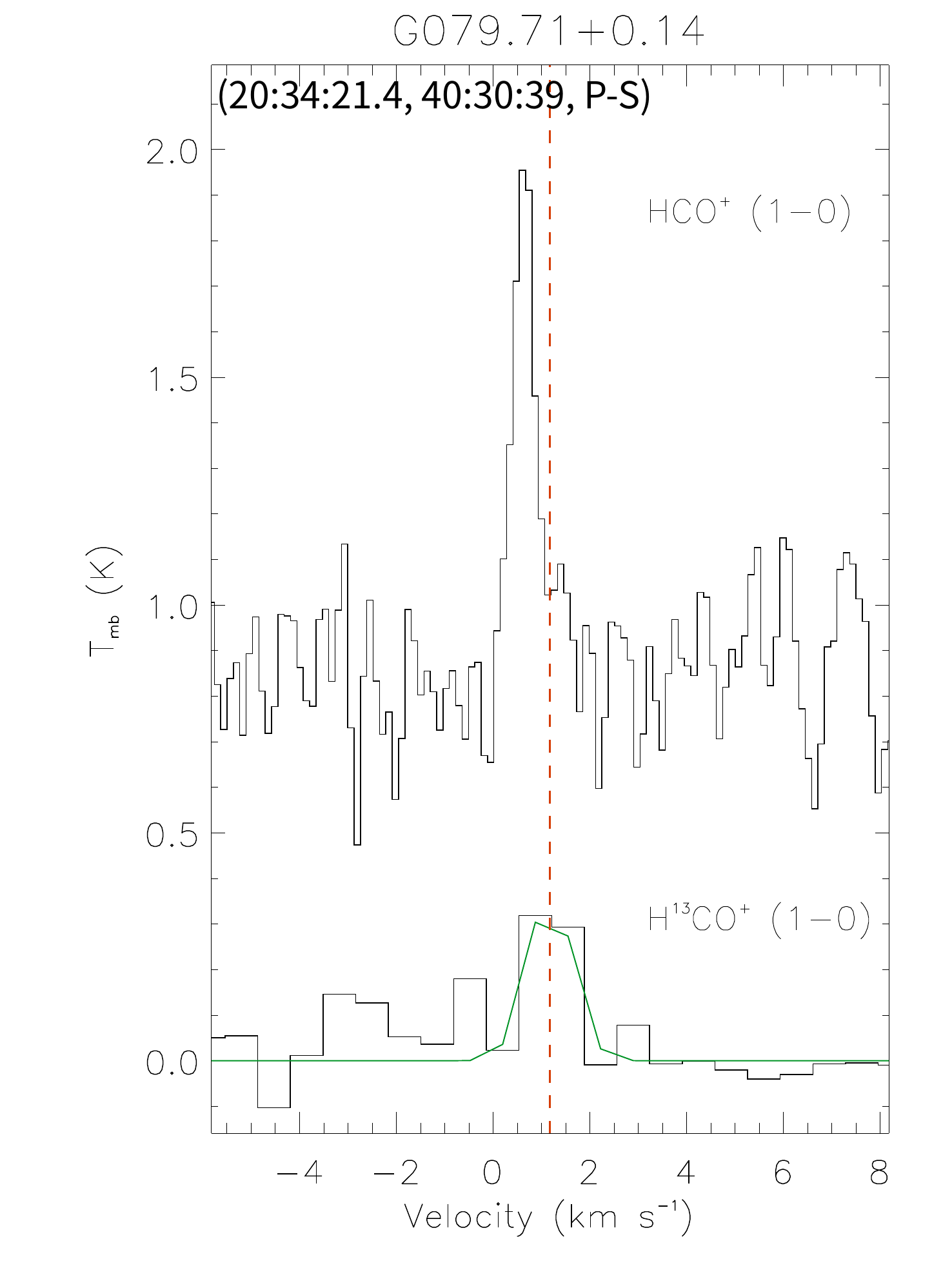}
  \end{minipage}%
  \begin{minipage}[t]{0.24\linewidth}
  \centering
   \includegraphics[width=50mm]{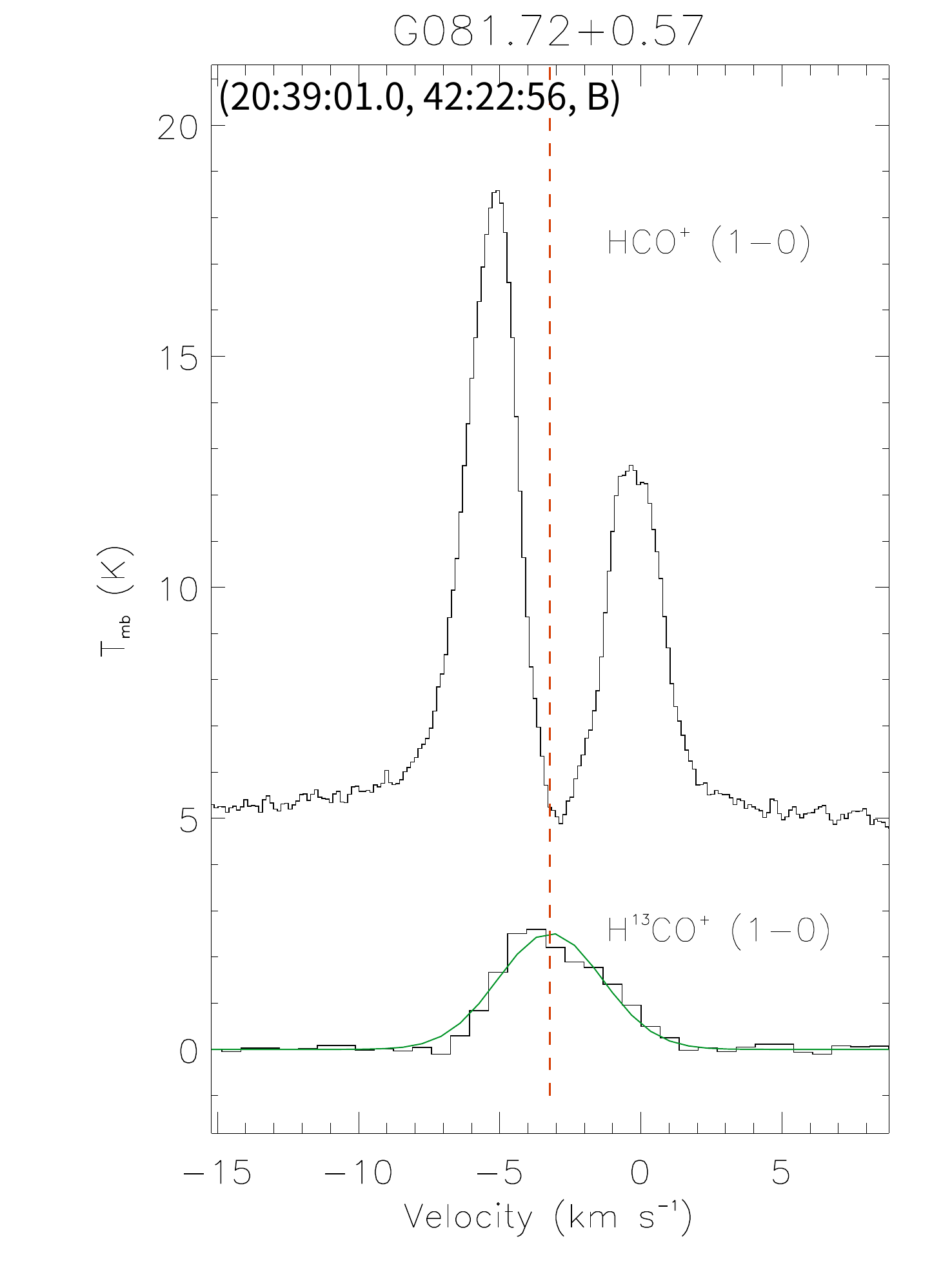}
  \end{minipage}%
\caption{The extracted spectra of HCO$^+$ (1-0) and H$^{13}$CO$^+$ (1-0) from the mapping images. The selected positions are the positions where the asymmetry of the line profiles are obvious (the extraction point coordinates and line profile types are marked on the upper left corner of each figures). The green line is the result of Gaussian fitting of H$^{13}$CO$^+$ (1-0), and the dashed red line indicates the central radial velocity of H$^{13}$CO$^+$ (1-0) (If H$^{13}$CO$^+$ emissions are not detected, we use C$^{18}$O data to track the central radial velocity). The HCO$^+$ (1-0) profiles are evaluated: B denotes blue asymmetric double-peaked profile; P-S denotes peak-shoulder profile; S denotes single-peaked profile with the peak skewed to the blue; R denotes red profile; N denotes symmetric line profile or multi-peaked profile.
\label{fig:lines}}
\end{figure*}

\begin{figure*}[h]
\addtocounter{figure}{-1} 
  \begin{minipage}[t]{0.24\linewidth}
  \centering
   \includegraphics[width=50mm]{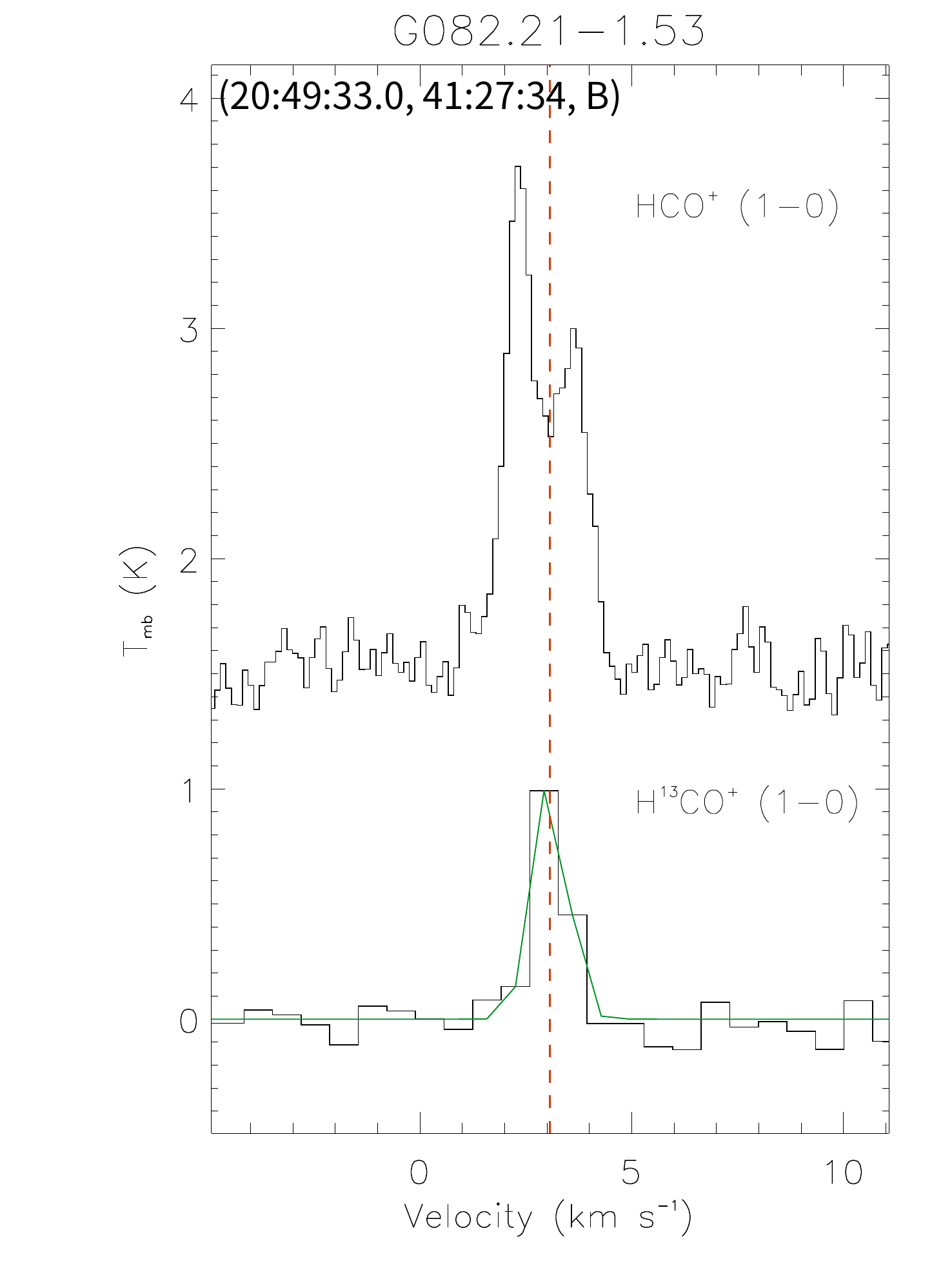}
  \end{minipage}%
  \begin{minipage}[t]{0.24\linewidth}
  \centering
   \includegraphics[width=50mm]{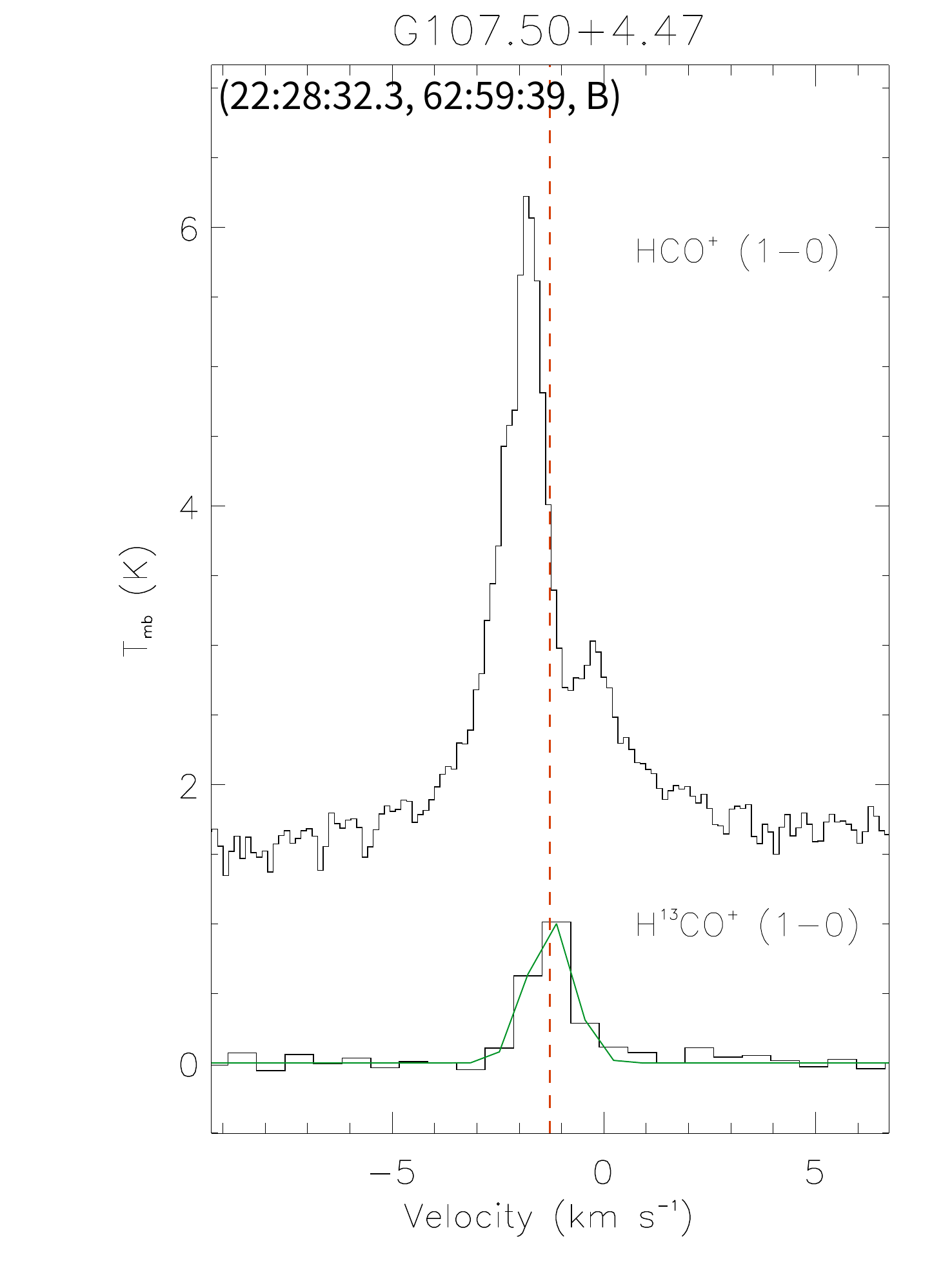}
  \end{minipage}%
  \begin{minipage}[t]{0.24\linewidth}
  \centering
   \includegraphics[width=50mm]{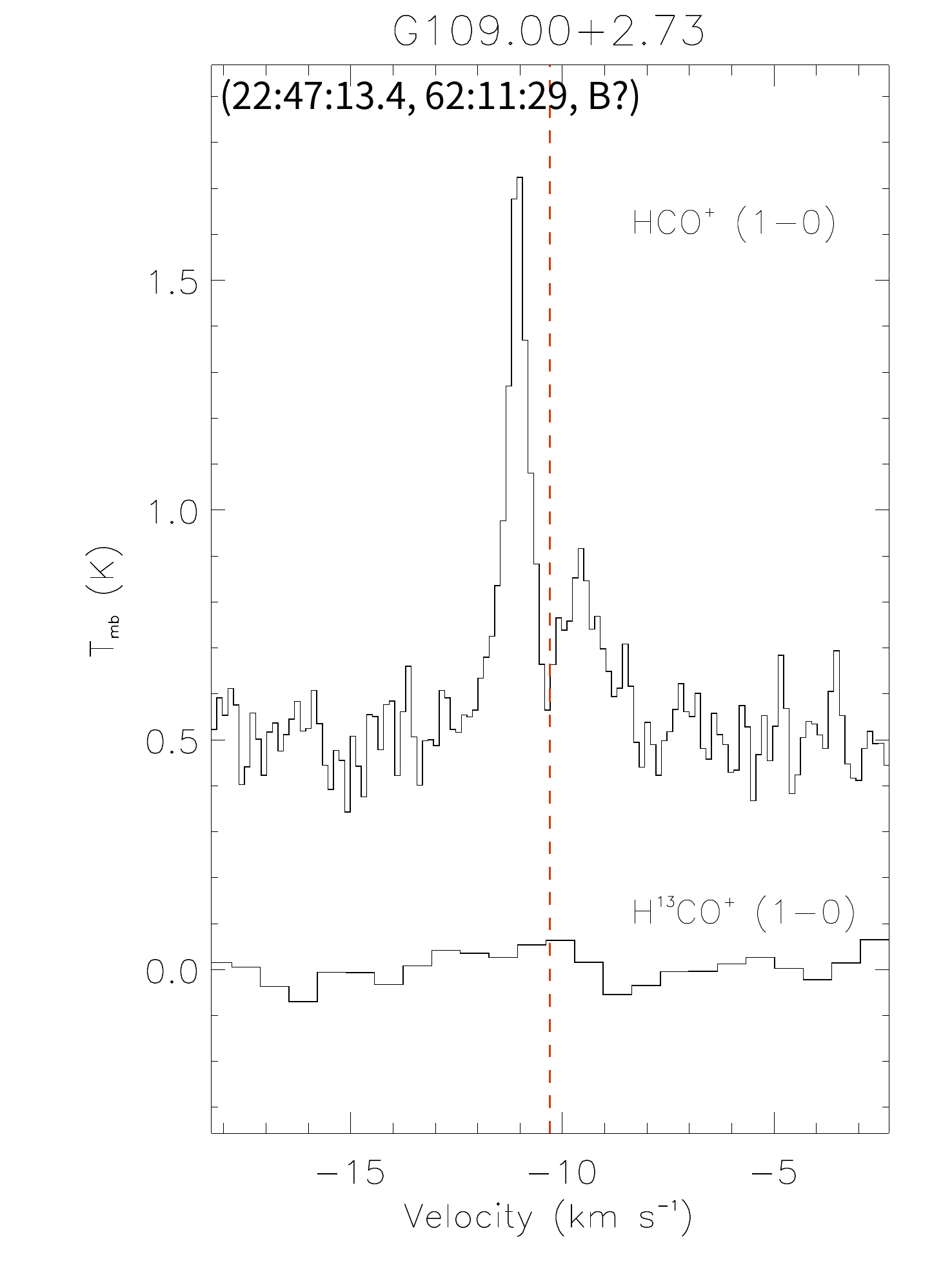}
  \end{minipage}%
  \begin{minipage}[t]{0.24\linewidth}
  \centering
   \includegraphics[width=50mm]{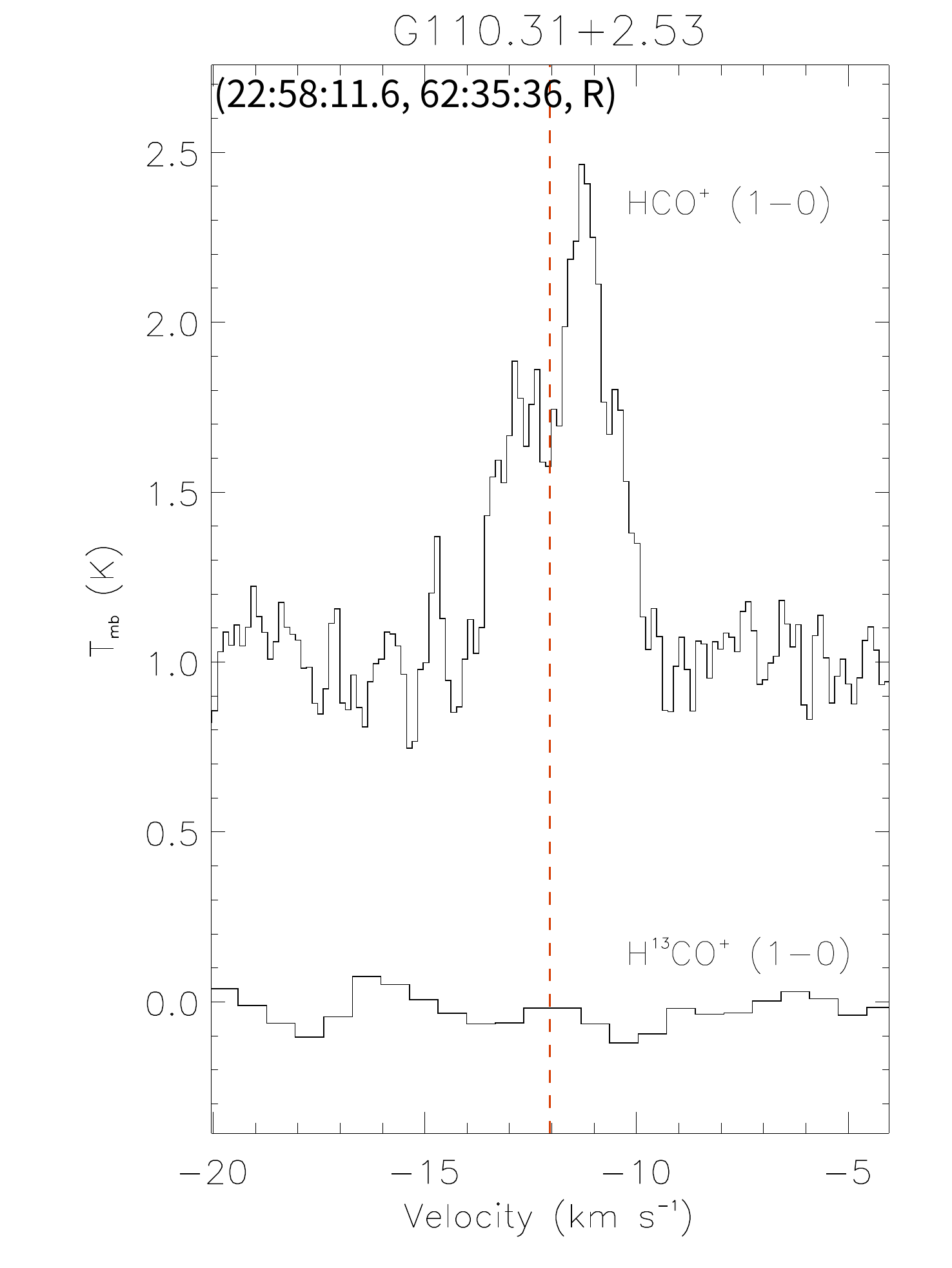}
  \end{minipage}%
  
  \begin{minipage}[t]{0.24\linewidth}
  \centering
   \includegraphics[width=50mm]{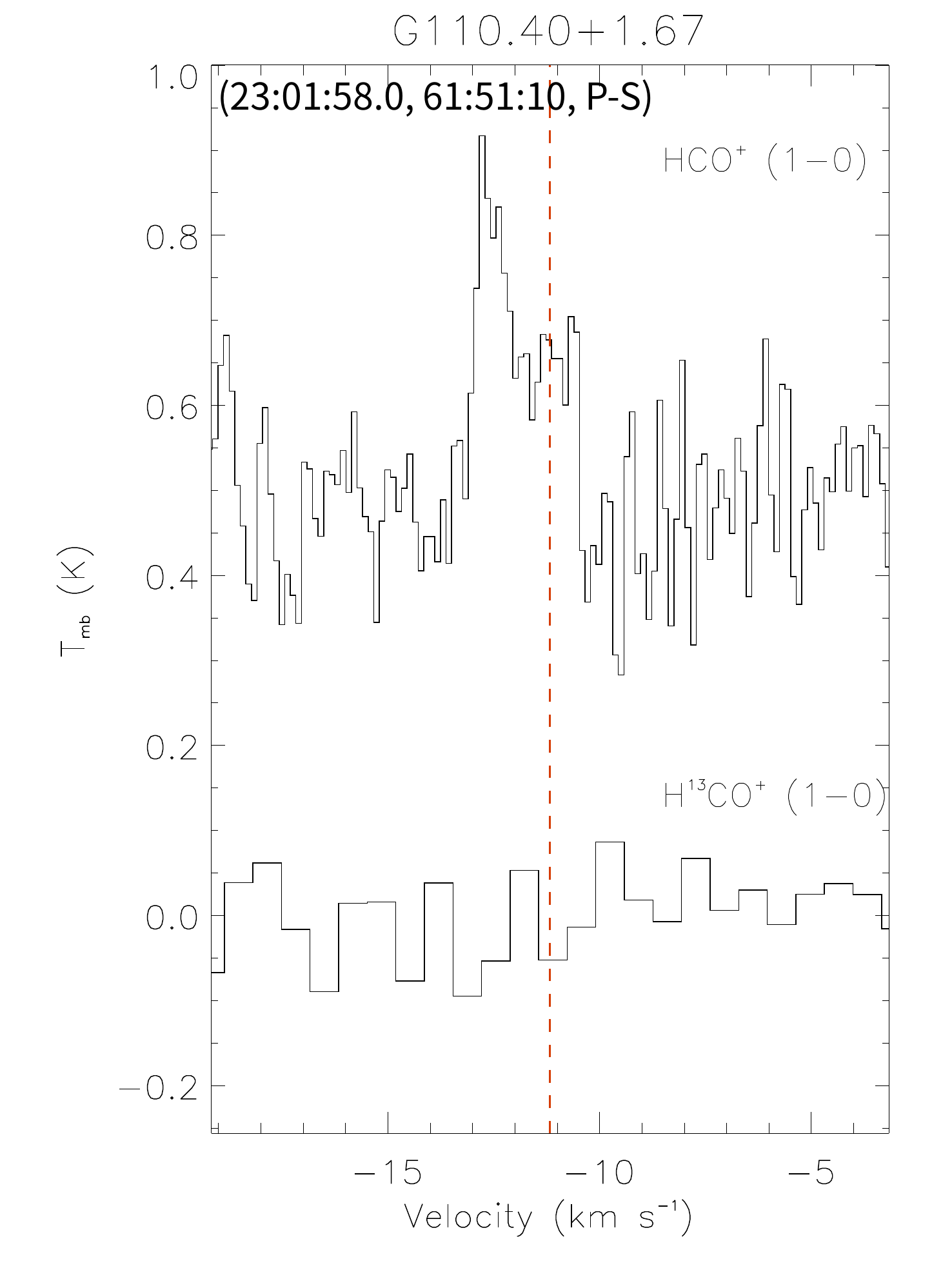}
  \end{minipage}%
  \begin{minipage}[t]{0.24\linewidth}
  \centering
   \includegraphics[width=50mm]{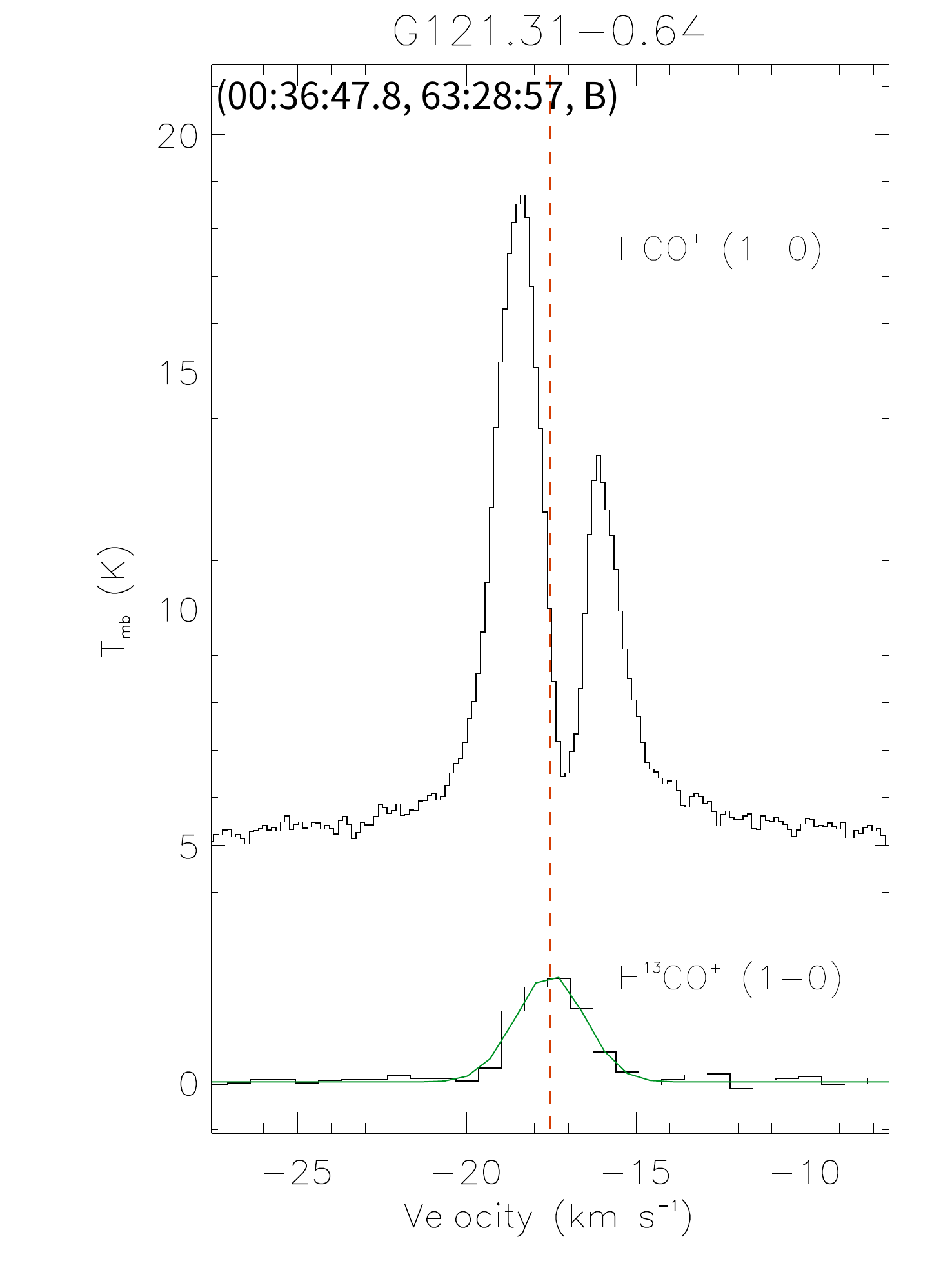}
  \end{minipage}%
  \begin{minipage}[t]{0.24\linewidth}
  \centering
   \includegraphics[width=50mm]{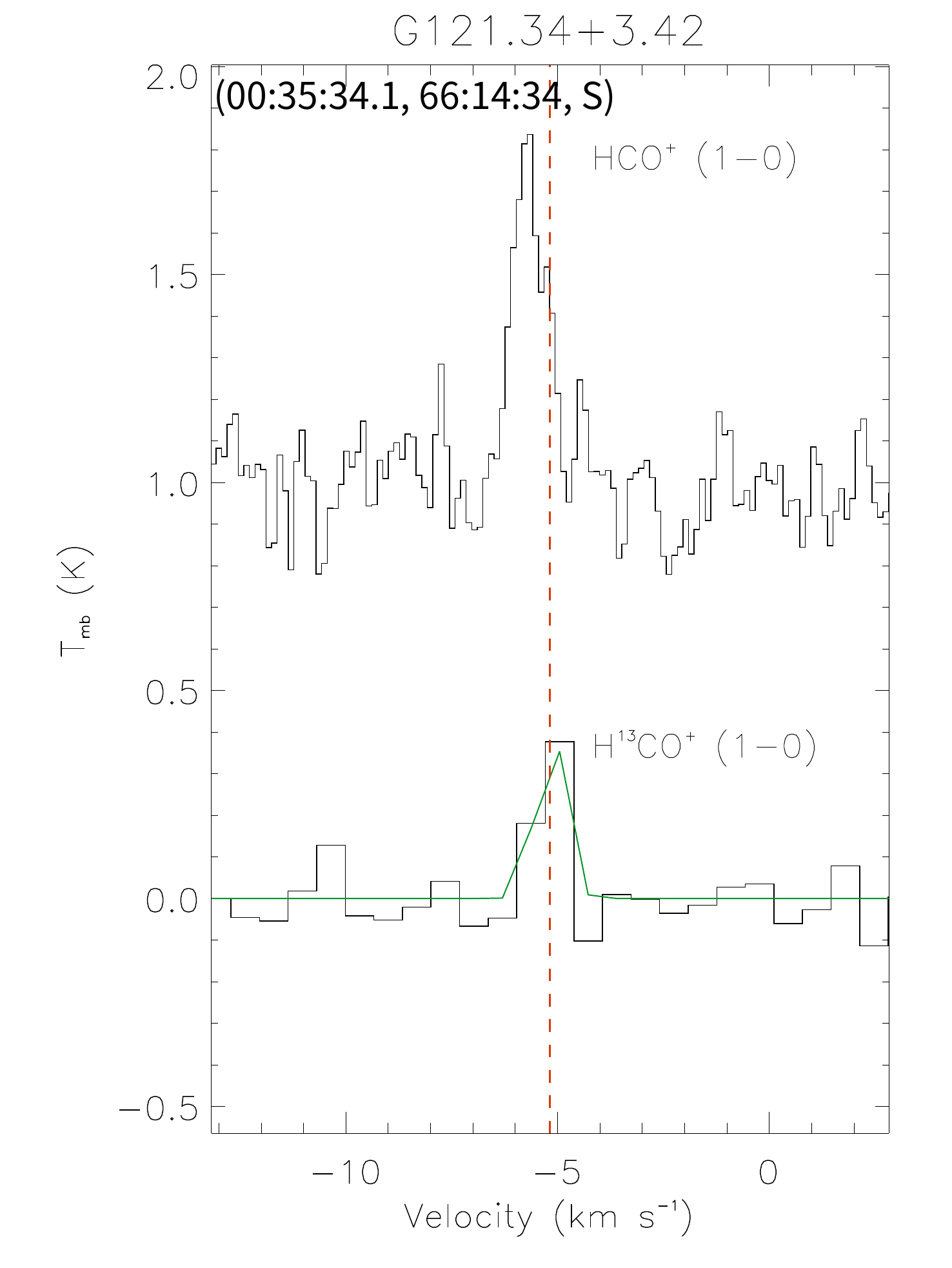}
  \end{minipage}%
  \begin{minipage}[t]{0.24\linewidth}
  \centering
   \includegraphics[width=50mm]{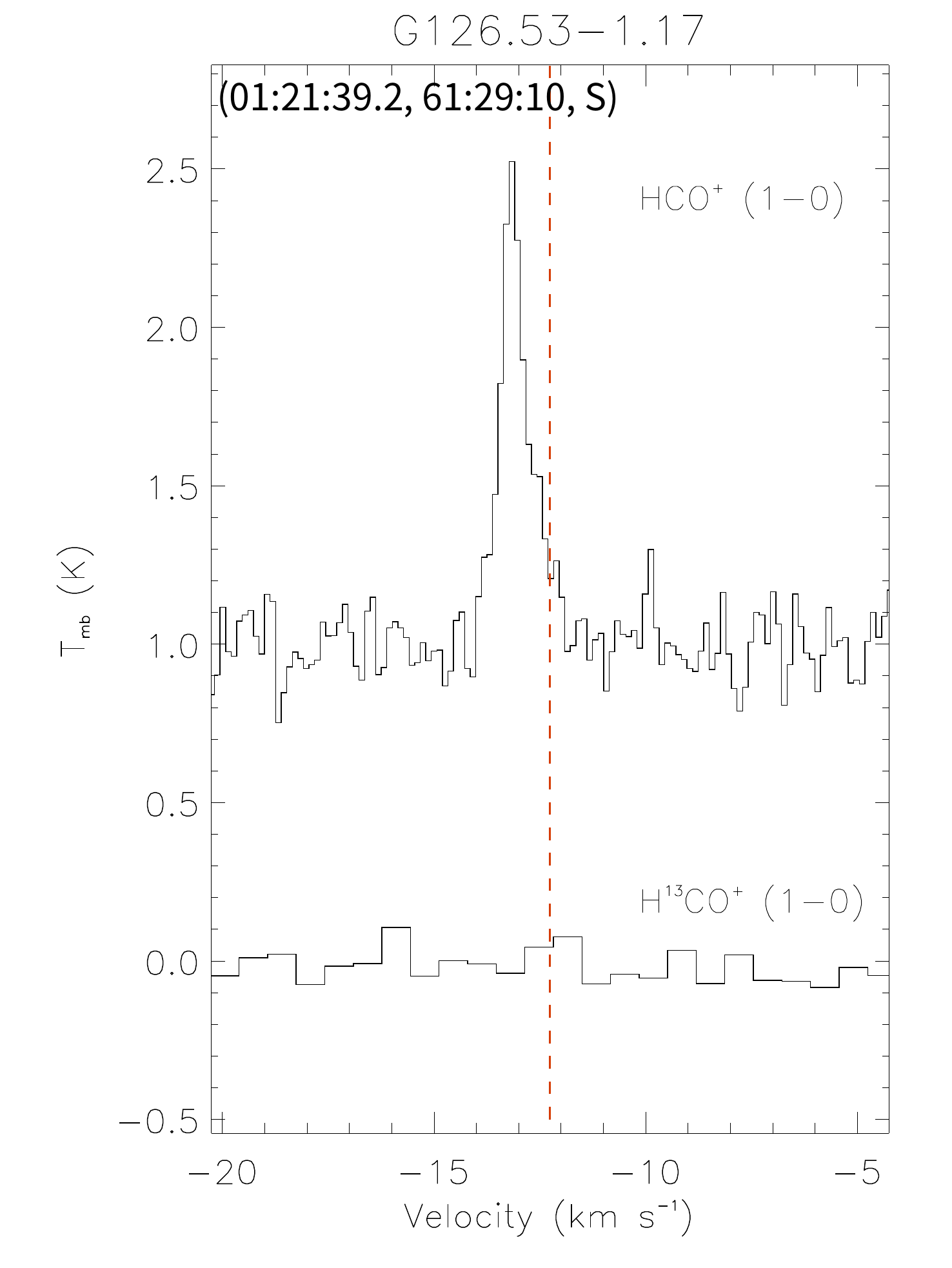}
  \end{minipage}%

  \begin{minipage}[t]{0.24\linewidth}
  \centering
   \includegraphics[width=50mm]{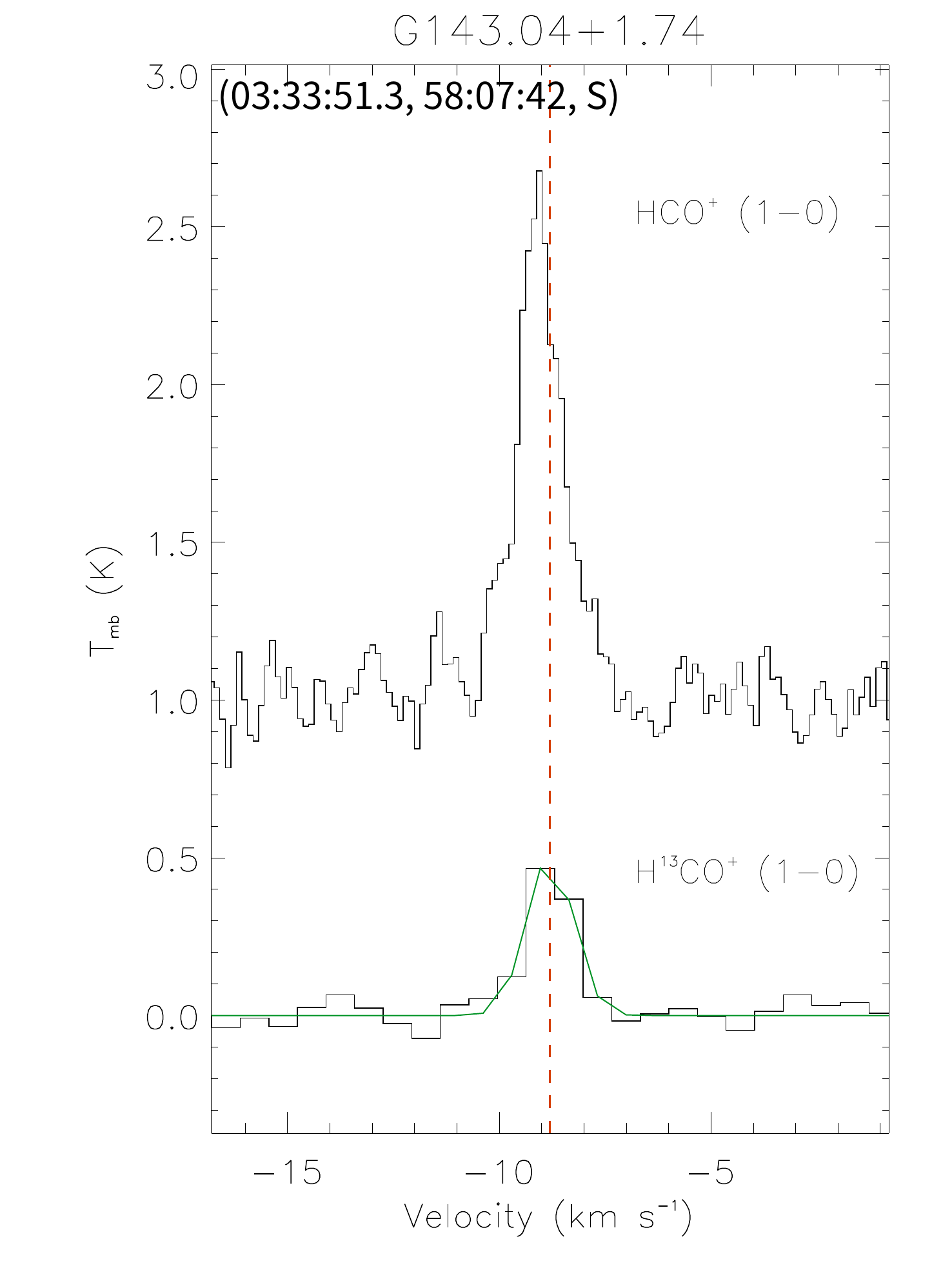}
  \end{minipage}%
  \begin{minipage}[t]{0.24\linewidth}
  \centering
   \includegraphics[width=50mm]{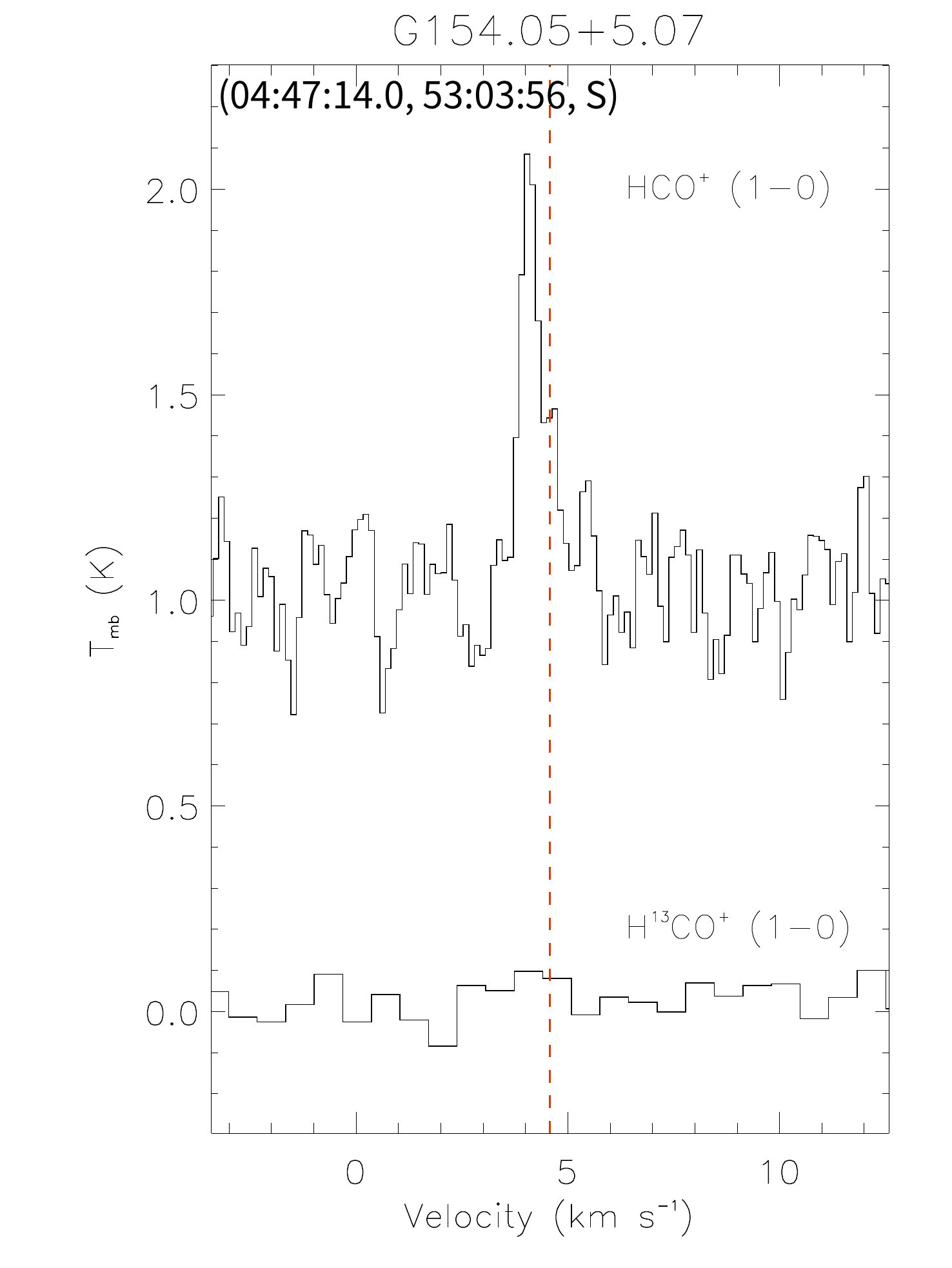}
  \end{minipage}%
  \begin{minipage}[t]{0.24\linewidth}
  \centering
   \includegraphics[width=50mm]{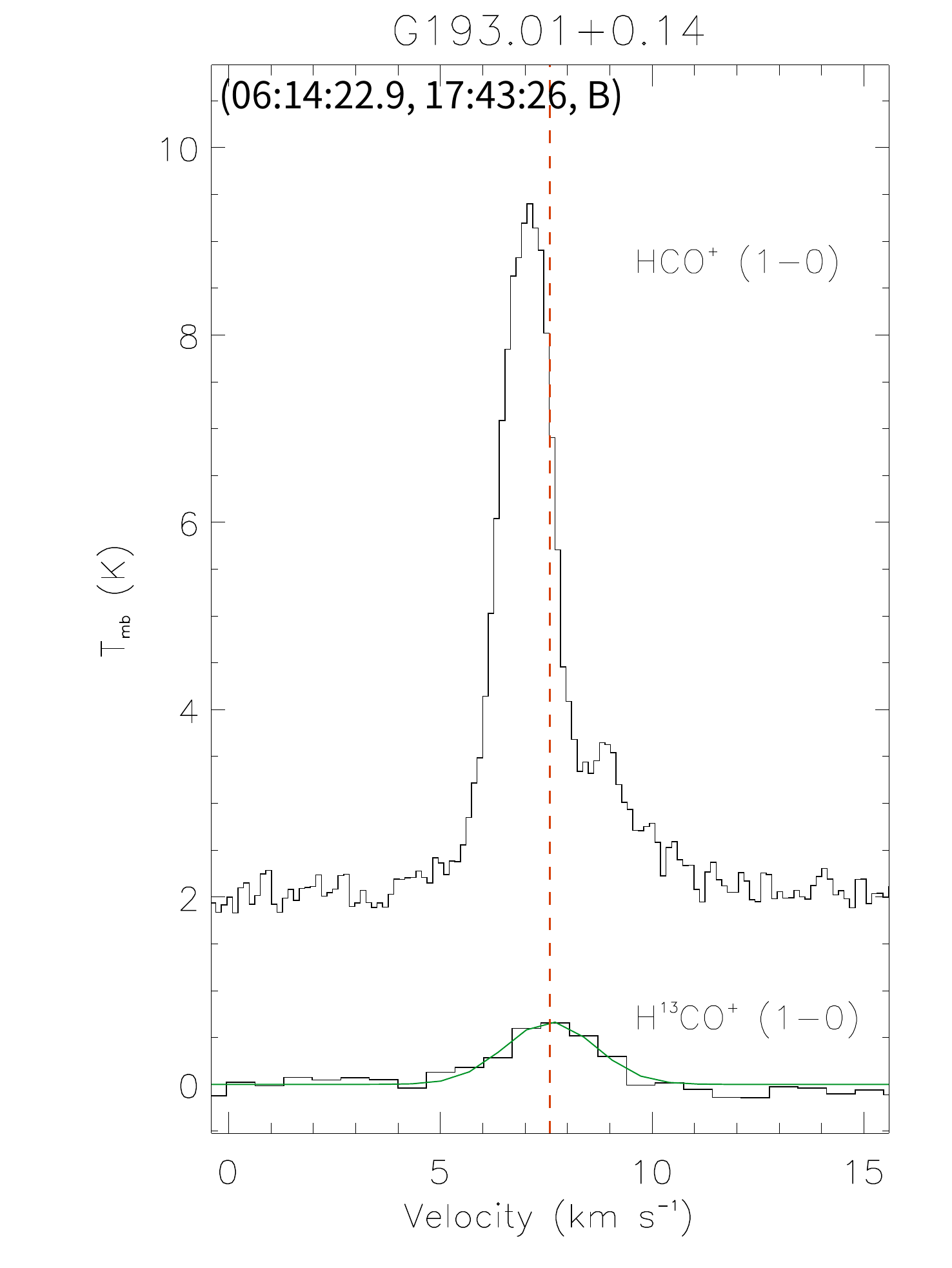}
  \end{minipage}%
  \begin{minipage}[t]{0.24\linewidth}
  \centering
   \includegraphics[width=50mm]{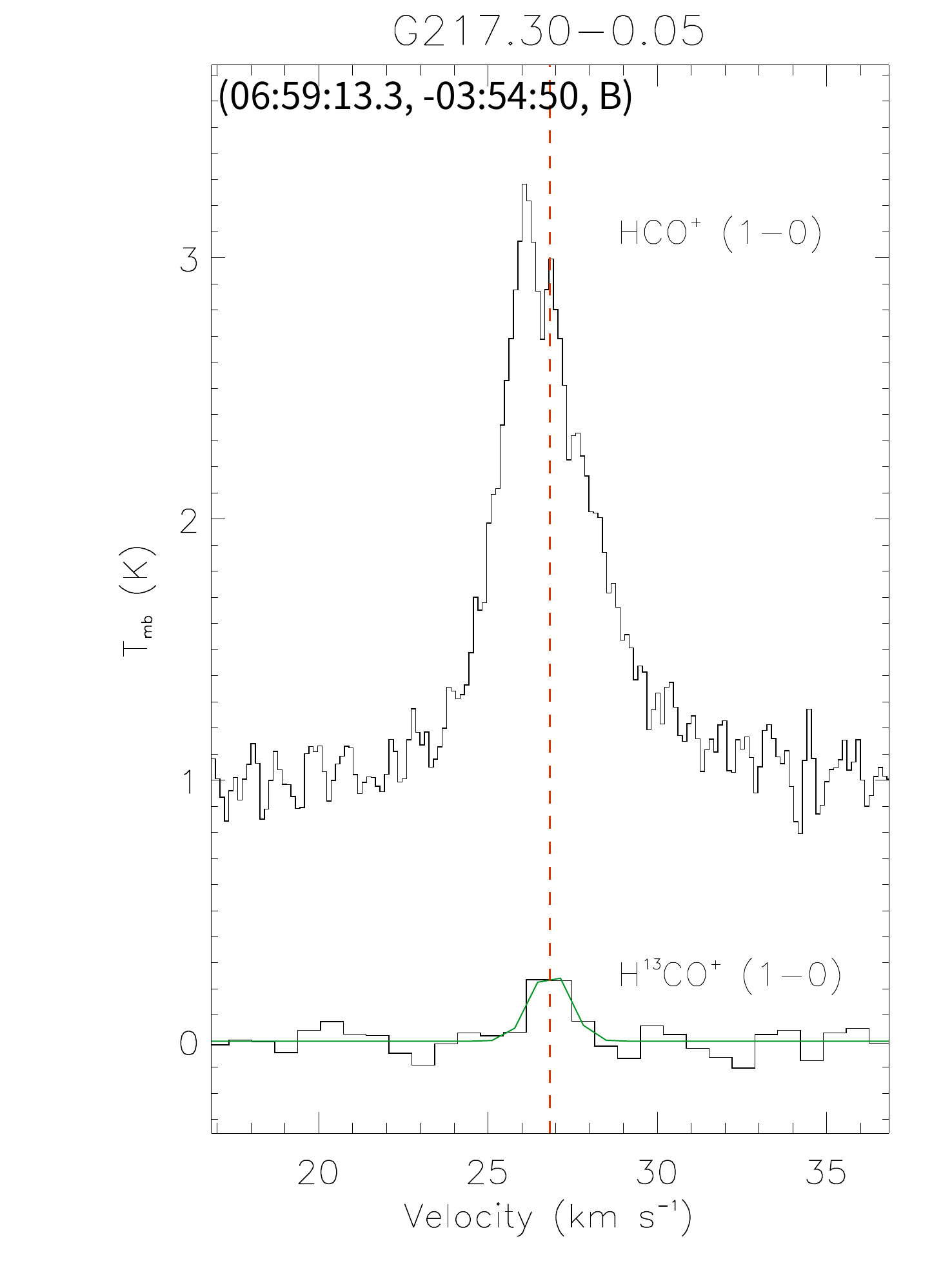}
  \end{minipage}%
\caption{Continued.
\label{fig:lines}}
\end{figure*}

\section{Infall Candidates} \label{sec:infall}

The combination of HCO$^+$ and H$^{13}$CO$^+$ lines is generally used to detect the infall motion in molecular clouds. In this section, we will use these two lines to identify the gas infall motion. For the confirmed infall sources, through the asymmetric line profile of the HCO$^+$ line relative to the optically thin line H$^{13}$CO$^+$, we can further estimate the infall velocities and the mass infall rates.

\subsection{Line Profiles and Asymmetries} \label{subsec:profiles}

To obtain the spatial details of the line profiles of these sources, we use the mapping data to form the HCO$^+$ (1-0) map grids of these sources, which are gridded to one beam size (28$^{\prime\prime}$). The map grids of all 24 sources are shown in Appendix \ref{sec:Appendix2}. Five sources, G029.60-0.63, G053.13+0.09, G081.72+0.57, G121.31+0.64, and G193.01+0.14, show significant blue asymmetric double-peaked profiles over the entire clump areas, which may indicate that these sources are global infall sources. Among them, G081.72+0.57 is located in the famous SFR DR21, which has been detected to have gas infall motion in previous studies \citep[e.g.][]{Schneider+etal+2010}. G053.13+0.09 and G121.31+0.64 also have been confirmed to have a trend of global gas collapse \citep[e.g.][]{Zhang+etal+2017, Pirogov+etal+2016}. Other sources, i.e., G082.21-1.53, G107.50+4.47, G109.00+2.73, and G217.30-0.05, show both blue and red line profiles in the observed areas, showing that these sources have more complex local motions. For the remaining sources, the optically thick line does not show clear double-peaked profiles. Therefore, we need to quantify the asymmetry of HCO$^+$ lines through a dimensionless parameter, to confirm further the line profiles of other sources.

We selected some single-point spectra from the mapping data, and then gave their line parameters. In most cases, the positions of the single-point spectra we selected are the positions where the HCO$^+$ lines have the strongest emissions and the most significant asymmetric profiles (excluding strong emissions from other sources at the edge of the mapping area). For G028.97+3.35, its HCO$^+$ emission of this source is weak, the single-point spectrum we selected is located at the position where the emission of the H$^{13}$CO$^+$ line is strongest. For the sources with complex line components in optically thick lines, we also select the lines at the strongest emission positions of optically thin lines. The position coordinates of these single-point spectra are listed in Table \ref{Tab:line} columns 2 and 3, and the line profiles of all 24 sources are presented in Figure \ref{fig:lines}.

According to the asymmetric profiles of the optically thick lines, we can divide these sources into three kinds: sources with blue profiles, red profiles and neither (including multipeaked profiles). \citet{Mardones+etal+1997} proposed that a dimensionless parameter, $\delta V$ can be used to quantify the asymmetry of the line profile:

\begin{equation}
  \delta V = \frac{V_{thick} - V_{thin}}{\Delta V_{thin}}
\label{eq:equ1}
\end{equation}

where $V_{thick}$ and $V_{thin}$ are the peak velocities of optically thick and thin lines, respectively, and $\Delta V_{thin}$ is the FWHM of the optically thin line. The $V_{thick}$ values are taken from the brightest emission peak positions of HCO$^+$ (1-0) lines, and the $V_{thin}$ and $\Delta V_{thin}$ are taken from the Gaussian fitting of H$^{13}$CO$^+$ (1-0). If no H$^{13}$CO$^+$ emissions were detected in the sources, the C$^{18}$O (1-0) lines from \cite{Yang+etal+2020} were used as optically thin lines (these sources are marked in Table \ref{Tab:line}). For G109.00+2.73 and G126.53-1.17, because no H$^{13}$CO$^+$ emissions are detected at their positions, we used H$^{13}$CO$^+$ line data at approximately 32 arcsec and 54 arcsec offsets to determine the central radial velocities. The dimensionless parameter $\delta V$ less than 0 indicates that the HCO$^+$ line shows a blue asymmetric profile, $\delta V$ greater than 0 indicates that the HCO$^+$ line shows a red profile. If $\delta V$ is approximately equal to 0, it indicates that the HCO$^+$ line does not show obvious asymmetry profiles.

In addition to the nine sources mentioned above that show significant blue profiles, there are sources without typical infall line characteristics, but may still be infall sources. Because in some cases, the self-absorption of the optically thick lines is weak, this makes the double-peaked profile not so obvious. In such a situation, the optically thick line may show a peak-shoulder profile or a single-peaked one with the peak skewed to the blue. In generally, we also categorize this profile as a blue profile. Four sources show a peak-shoulder profile, while another six show a single-peaked profile with the peak positions skewed to the blue. All these sources are marked as P-S and S in Table \ref{Tab:line}.

Among the remaining sources, the HCO$^+$ lines of G039.33-1.03 and G110.31+2.53 show a red profile. No H$^{13}$CO$^+$ emission is detected in G033.42+0.00, so its central radial velocity is based on C$^{18}$O data. Compared with the center radial velocity, we found that the HCO$^+$ line of this source did not show an asymmetric profile. The HCO$^+$ lines of G077.91-1.16 and G079.24+0.53 show multiple peaks, and optically thin lines have only one velocity component. However, because of the low velocity resolution of H$^{13}$CO$^+$ and the insufficient signal-to-noise ratios at the dip of the HCO$^+$ line, we cannot determine whether the multiple peaks are caused by multiple velocity components or systematic motions.


\begin{table}
\begin{center}
\caption{The derived line parameters and profiles of sources.}\label{Tab:line}
 \setlength{\tabcolsep}{1mm}{
\begin{tabular}{cccccccc}   
  \hline\noalign{\smallskip}
Source & \multicolumn{2}{c}{Extraction point} & $V_{thick}$ & $V_{thin}$ & $\Delta V_{thin}$ & $\delta V$ & Line \\
Name & Ra & Dec & & & & & Profile \\
  & (J2000) & (J2000) & (km s$^{-1}$) & (km s$^{-1}$) & (km s$^{-1}$) & &  \\
  \hline\noalign{\smallskip} 
G028.97+3.35  &     18:32:12.6    &     -01:59:27 \tablenotemark{$\dagger$}& 5.72 & 7.31(0.01) & 0.76(0.02) & -2.09&  S \\
G029.06+4.58  &     18:28:03.2    &     -01:21:33  & 7.14 & 7.48(0.02) & 0.83(0.03)& -0.41&  S \\
G029.60-0.63  &     18:47:32.1    &     -03:14:10  & 75.01& 76.52(0.03)& 3.00(0.07)& -0.50&  B \\
G030.17+3.68  &     18:33:15.0    &     -00:46:45  & 8.57 & 9.02(0.01) & 0.74(0.03)& -0.61&  P-S \\
G031.41+5.25  &     18:29:59.5    &      01:03:00  & 7.01 &  8.28\tablenotemark{*} &  0.82\tablenotemark{*} & -1.55&  P-S \\
G033.42+0.00  &     18:52:21.7    &      00:25:33  & 10.58&  10.60\tablenotemark{*} &  0.55\tablenotemark{*} & -0.04&  N \\
G039.33-1.03  &     19:06:47.4    &      05:12:37  & 13.19&  12.90\tablenotemark{*} &  0.72\tablenotemark{*} & 0.40&  R \\
G053.13+0.09  &     19:29:17.2    &      17:56:18  & 21.02& 21.71(0.01)& 2.49(0.01)& -0.28&  B \\
G077.91-1.16  &     20:34:12.5    &      38:17:58 \tablenotemark{$\dagger$}& Multiple peaks & -0.21(0.01) & 1.35(0.03)& - &  N \\
G079.24+0.53  &     20:31:19.7    &      40:22:24 \tablenotemark{$\dagger$}& Multiple peaks & 0.79(0.01) & 1.03(0.02)&  - &  N \\
G079.71+0.14  &     20:34:21.4    &      40:30:39  & 0.61 & 1.19(0.01) & 0.85(0.09)& -0.68&  P-S \\
G081.72+0.57  &     20:39:01.0    &      42:22:56  & -5.2 & -3.22(0.01) & 4.24(0.01)& -0.47&  B \\
G082.21-1.53  &     20:49:33.0    &      41:27:34 \tablenotemark{$\dagger$}& 2.31 & 3.07(0.01) & 0.96(0.01)& -0.79&  B \\
G107.50+4.47  &     22:28:32.3    &      62:59:39  & -1.84& -1.27(0.01) & 1.25(0.01)& -0.46&  B \\
G109.00+2.73&     22:47:13.4    &      62:11:29  &-11.01&  -10.29(0.02)\tablenotemark{**} &  1.34(0.04)\tablenotemark{**} & -0.54&  B? \\
G110.31+2.53  &     22:58:11.6    &      62:35:36  &-11.43&  -12.04\tablenotemark{*} &  1.18\tablenotemark{*} & 0.52&  R \\
G110.40+1.67  &     23:01:58.0    &      61:51:10  &-12.78&  -11.17\tablenotemark{*} &  0.80\tablenotemark{*} & -2.01&  P-S \\
G121.31+0.64  &     00:36:47.8    &      63:28:57  &-18.35& -17.54(0.01)& 2.38(0.01)& -0.34&  B \\
G121.34+3.42  &     00:35:34.1    &      66:14:34 \tablenotemark{$\dagger$}& -6.04& -5.18(0.01) & 0.75(0.02)& -1.15&  S \\
G126.53-1.17&       01:21:39.2    &      61:29:10  &-13.16& -12.26(0.04)\tablenotemark{**}& 0.69(0.18)\tablenotemark{**}& -1.30&  S \\
G143.04+1.74  &     03:33:51.3    &      58:07:42  & -9.05& -8.80(0.01)& 1.28(0.03)& -0.2&  S \\
G154.05+5.07  &     04:47:14.0    &      53:03:56  & 4.04 &  4.58\tablenotemark{*} &  0.44\tablenotemark{*} & -1.23&  S \\
G193.01+0.14  &     06:14:22.9    &      17:43:26  & 7.11 & 7.59(0.01)& 2.51(0.03)& -0.19&  B \\
G217.30-0.05  &     06:59:13.3    &     -03:54:50  & 26.21& 26.83(0.02)& 1.32(0.06)& -0.47&  B \\
  \hline\noalign{\smallskip}

\end{tabular}}
\end{center}
\tablecomments{ 
The HCO$^+$ (1-0) profiles are evaluated: B denotes blue asymmetric double-peaked profile; P-S denotes peak-shoulder profile; S denotes single-peaked profile with the peak skewed to the blue; R denotes red profile; N denotes symmetric line profile or multi-peaked profile. \tablenotetext{\dagger}{ The position where the spectrum is extracted is the position of the strongest emission of the H$^{13}$CO$^+$ line.} \tablenotetext{*}{ The optically thin line data of the source comes from the C$^{18}$O (1-0) data of \citet{Yang+etal+2020}.} \tablenotetext{**}{ The optically thin line data of the source comes from the H$^{13}$CO$^+$ (1-0) data from other position of the same source.} }
\end{table}

\subsection{Infall Velocities} \label{subsec:Vin}


\begin{table}
  \begin{center}
    \small
\caption{Infall velocities and mass infall rates of nine infall candidates with double-peaked profile.}\label{Tab:infall}
 \setlength{\tabcolsep}{2mm}{
\begin{tabular}{cccccc}   
  \hline\noalign{\smallskip}
Source & $\sigma$ & $V_{in}$ & $V_{in}$ & $\dot M_{in}$ & $\dot M_{in}$ \\
   &  RAT  & RAT$^*$ & Myers & RAT & Myers \\
  Name & ($\mathrm{km\,s}^{-1}$) & ($\mathrm{km\,s}^{-1}$) & ($\mathrm{km\,s}^{-1}$) & ($\times10^{-4} M_{\odot}\,\mathrm{yr}^{-1}$) & ($\times10^{-4} M_{\odot}\,\mathrm{yr}^{-1}$) \\
  \hline\noalign{\smallskip} 
G029.60-0.63 & 0.5(0.2) & 1.3(0.2) &  1.6(0.9) & 45 & 55 \\
G053.13+0.09 & 1.8(0.1) & 0.9(0.1) &  0.7(0.5) & 2.8 & 2.2 \\
G081.72+0.57 & 1.0(0.2) & 0.5(0.2) &  0.4(0.6) & 27 &   22\\
G082.21-1.53 & 0.7(0.3) & 0.5(0.1) &  0.1(0.3) & 2.9 & 0.57 \\
G107.50+4.47 & 0.7(0.2) & 0.6(0.1) &  0.3(0.2) & 0.97 & 0.49 \\
G109.00+2.73 &  -  &  -  &  0.3(0.8) & - & 1.0  \\   
G121.31+0.64 & 1.0(0.1) & 0.6(0.2) &  0.3(0.9) & 2.4 & 1.2 \\
G193.01+0.14 & 0.5(0.2) &  0.7(0.2) &  1.5(0.8) & 5.3 & 11 \\
G217.30-0.05 & 0.4(0.2) & 0.3(0.2) &  0.2(0.1) & 3.8 & 2.6 \\
  \hline\noalign{\smallskip}

\end{tabular}}
\end{center}
\tablecomments{$^*$ $V_{in}(RAT)$ is the mean velocity value (weighted by the shell's width) of the clump obtained by the RATRAN model.}
\end{table}

\begin{figure*}[h]
\gridline{\fig{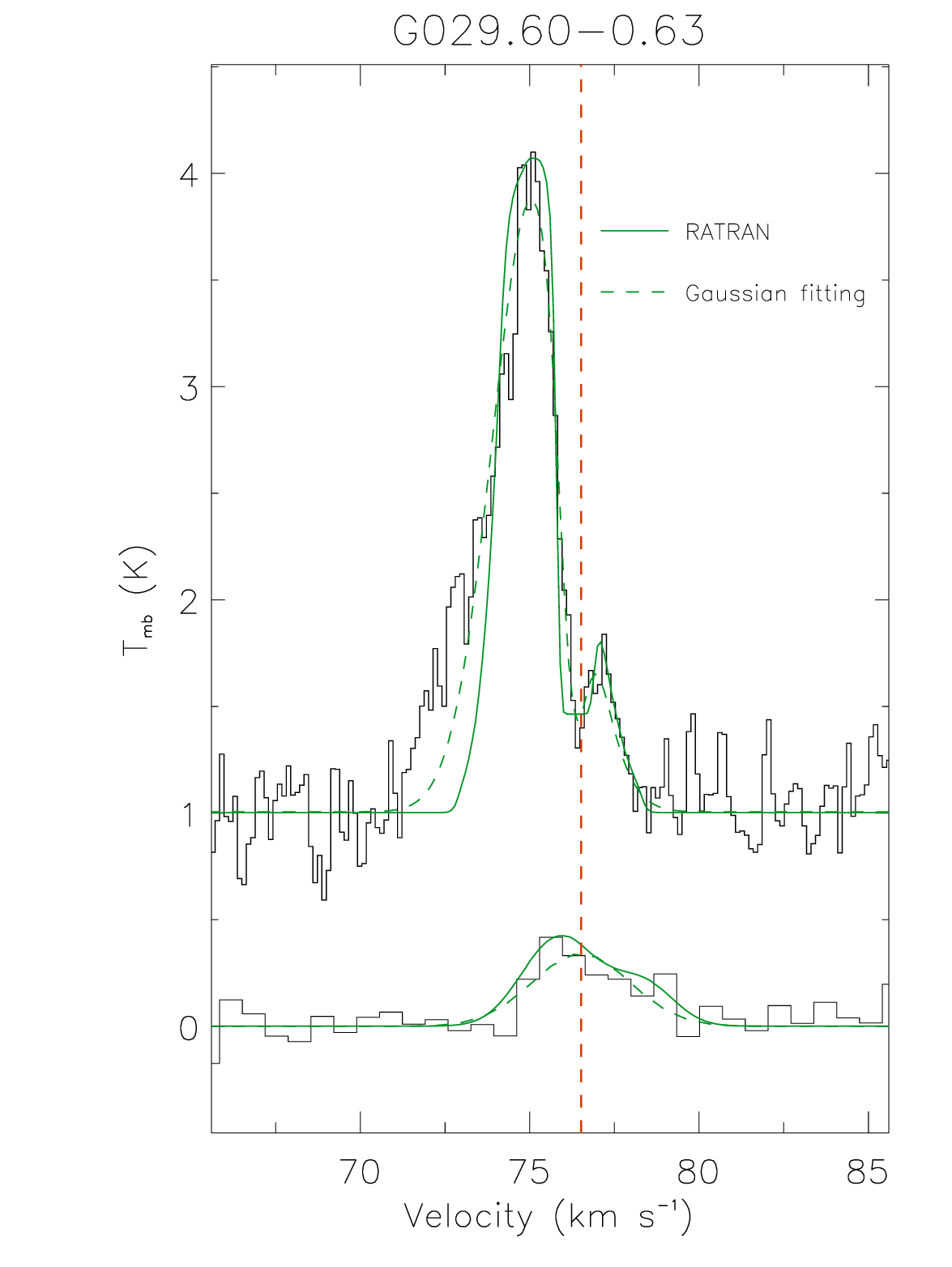}{0.28\textwidth}{}
          \fig{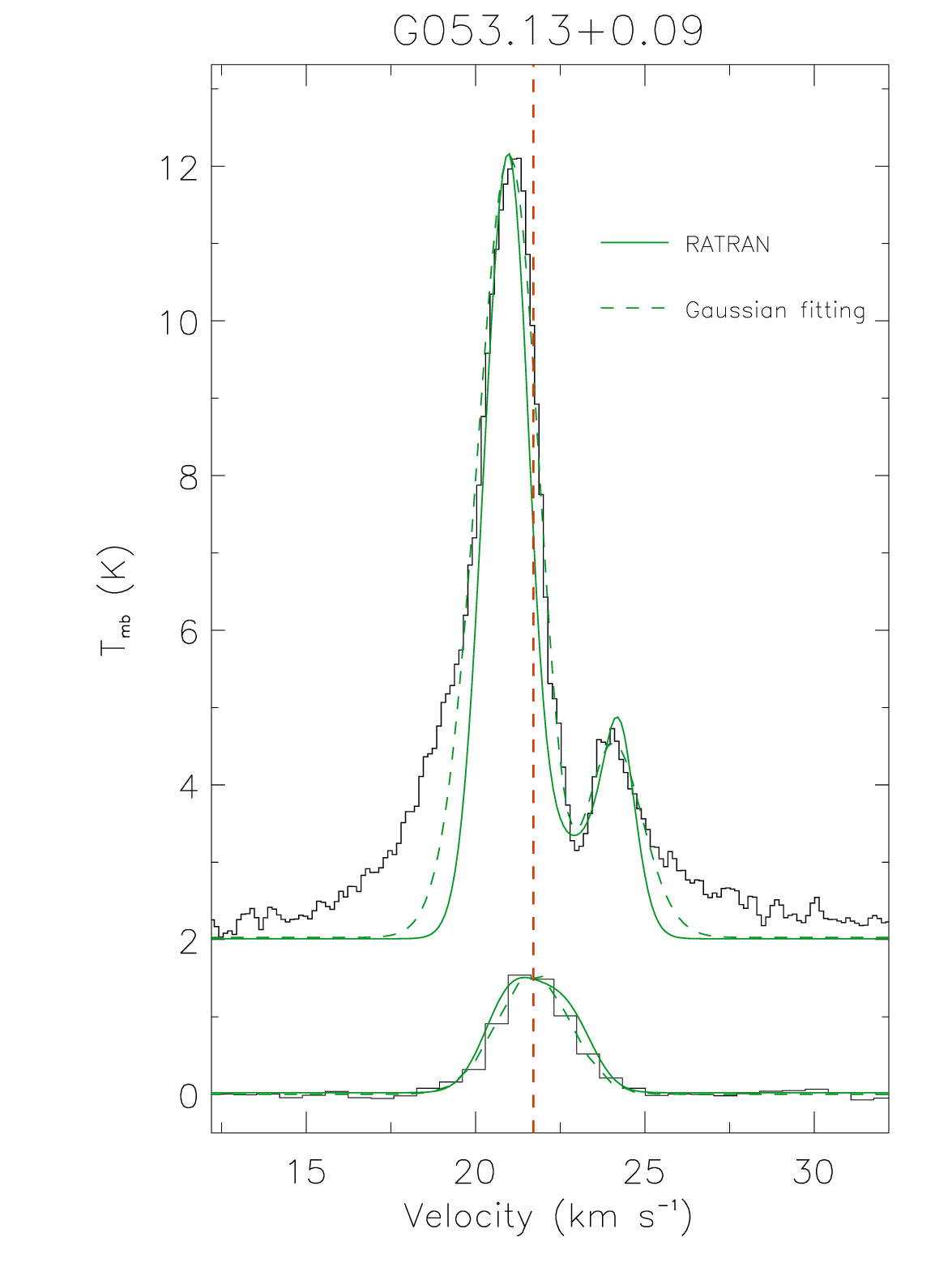}{0.28\textwidth}{}
          \fig{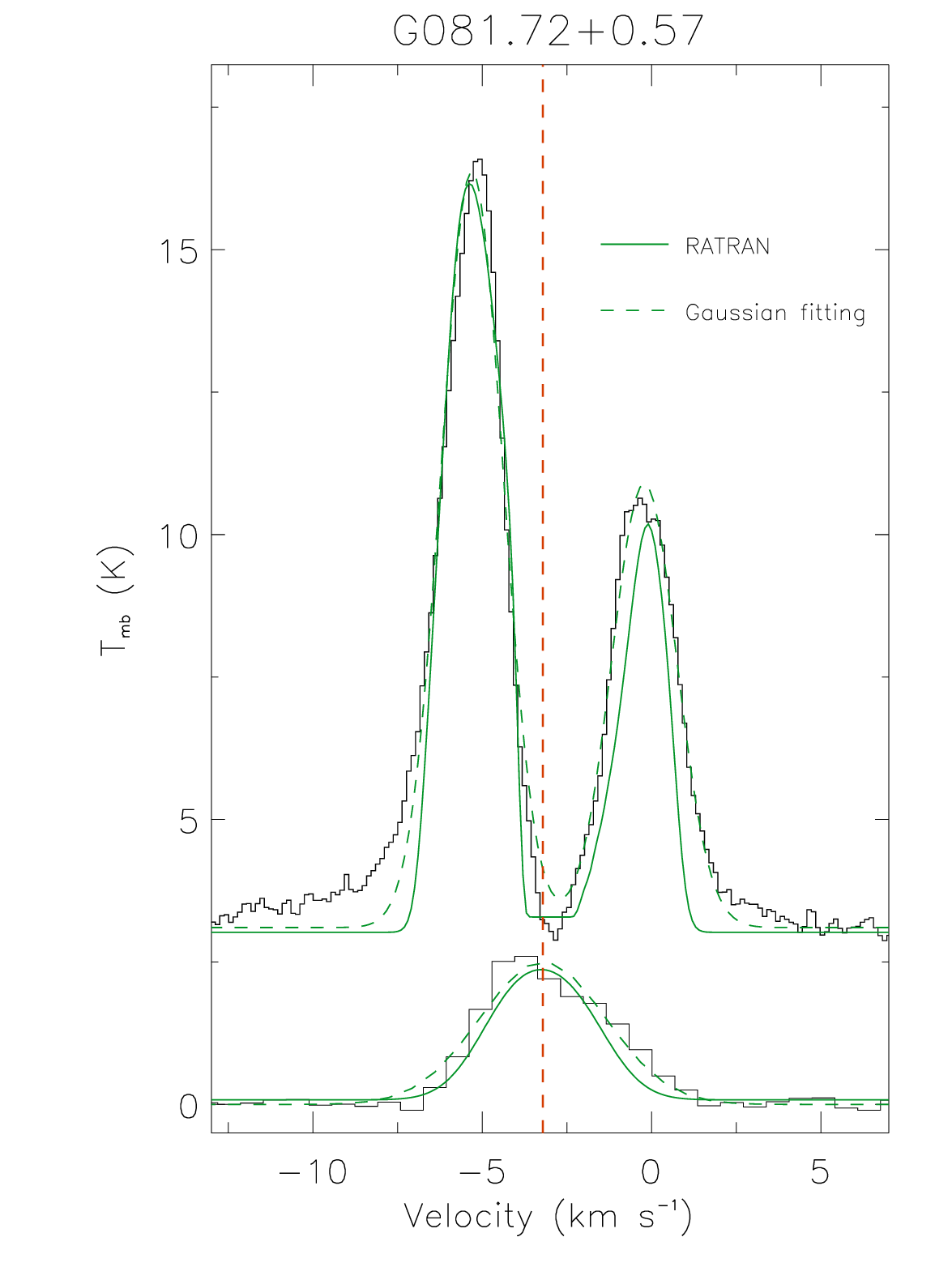}{0.28\textwidth}{}
          }
\gridline{\fig{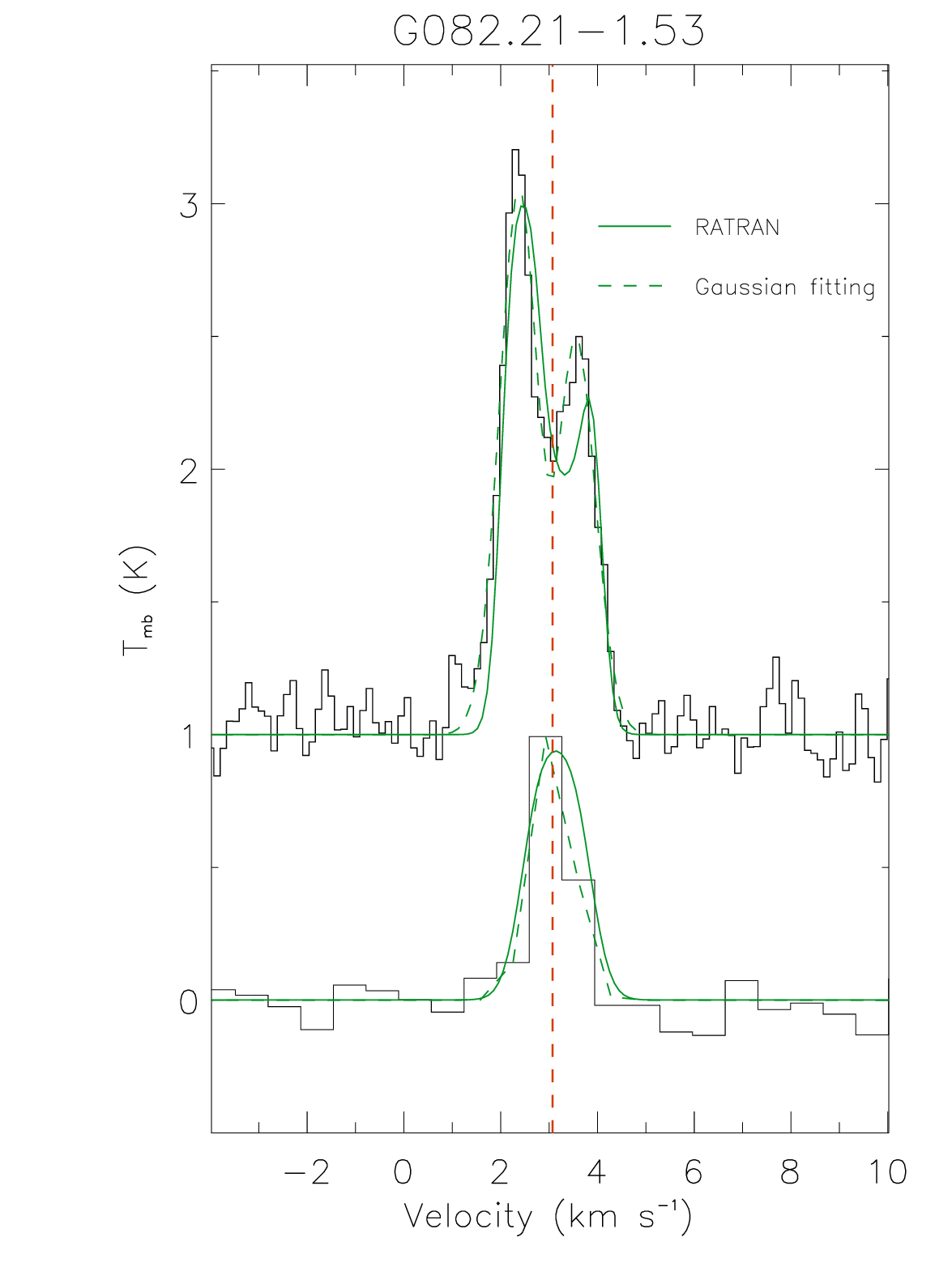}{0.28\textwidth}{}
          \fig{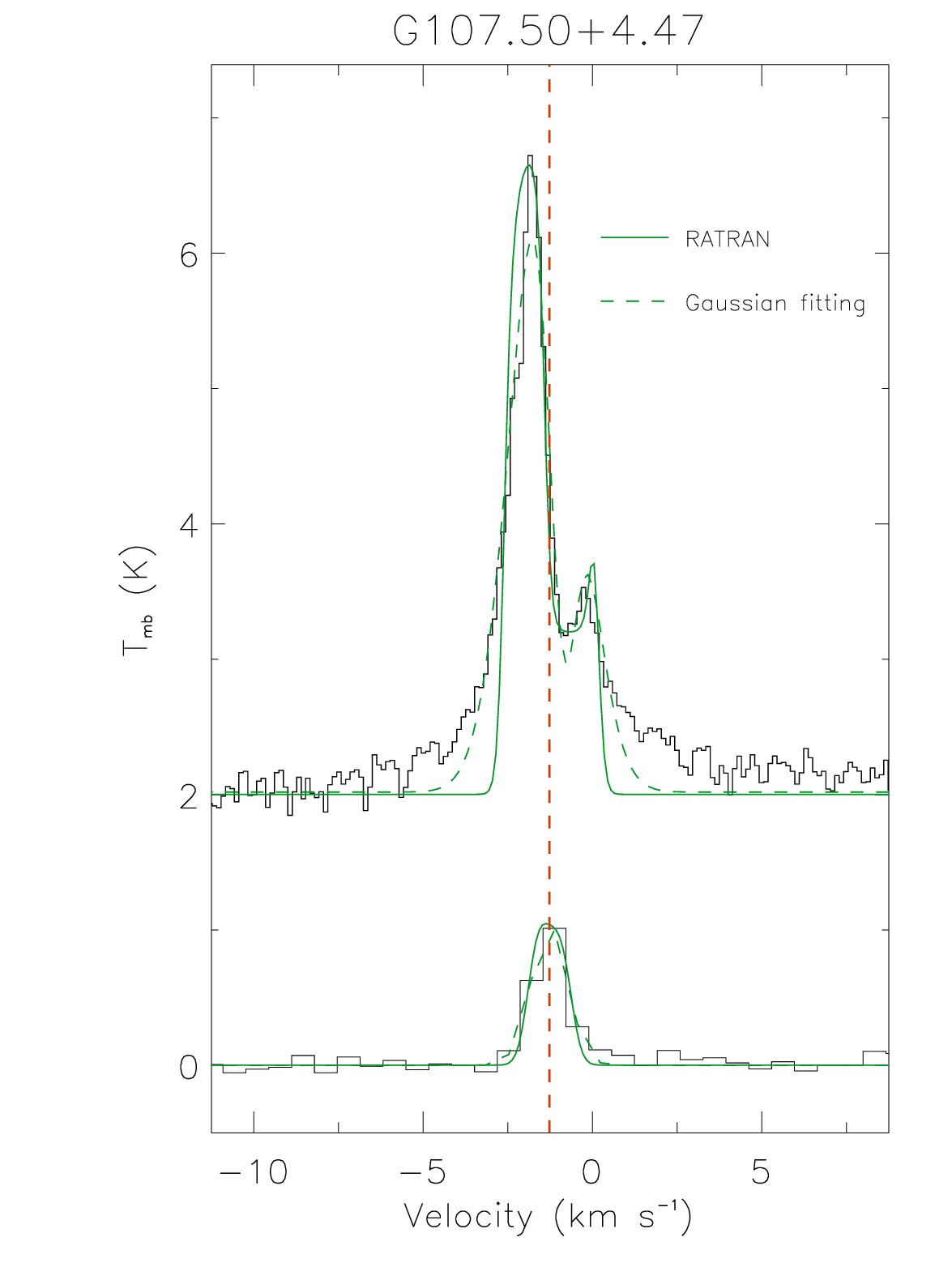}{0.28\textwidth}{}
          \fig{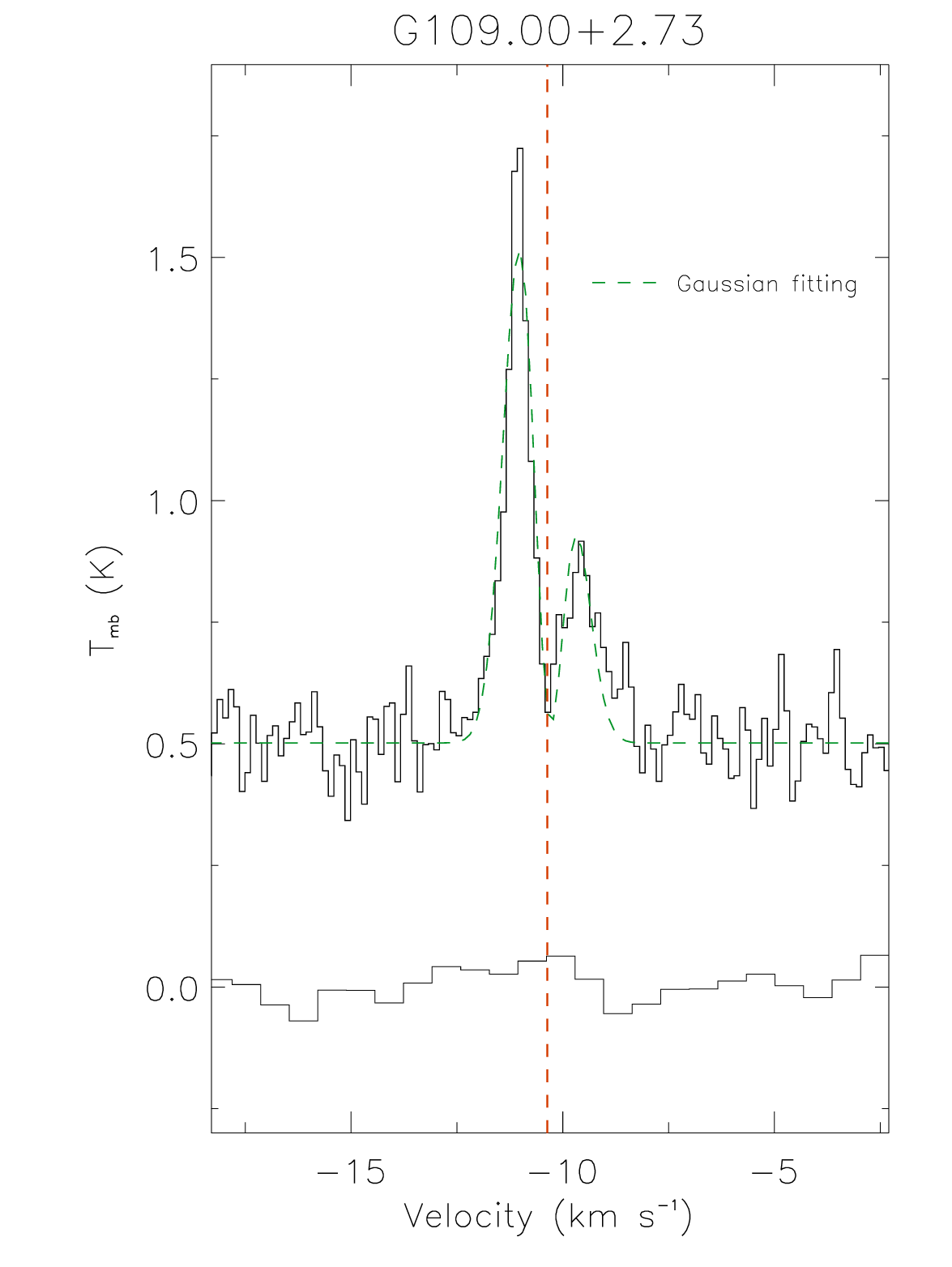}{0.28\textwidth}{}
          }
\gridline{\fig{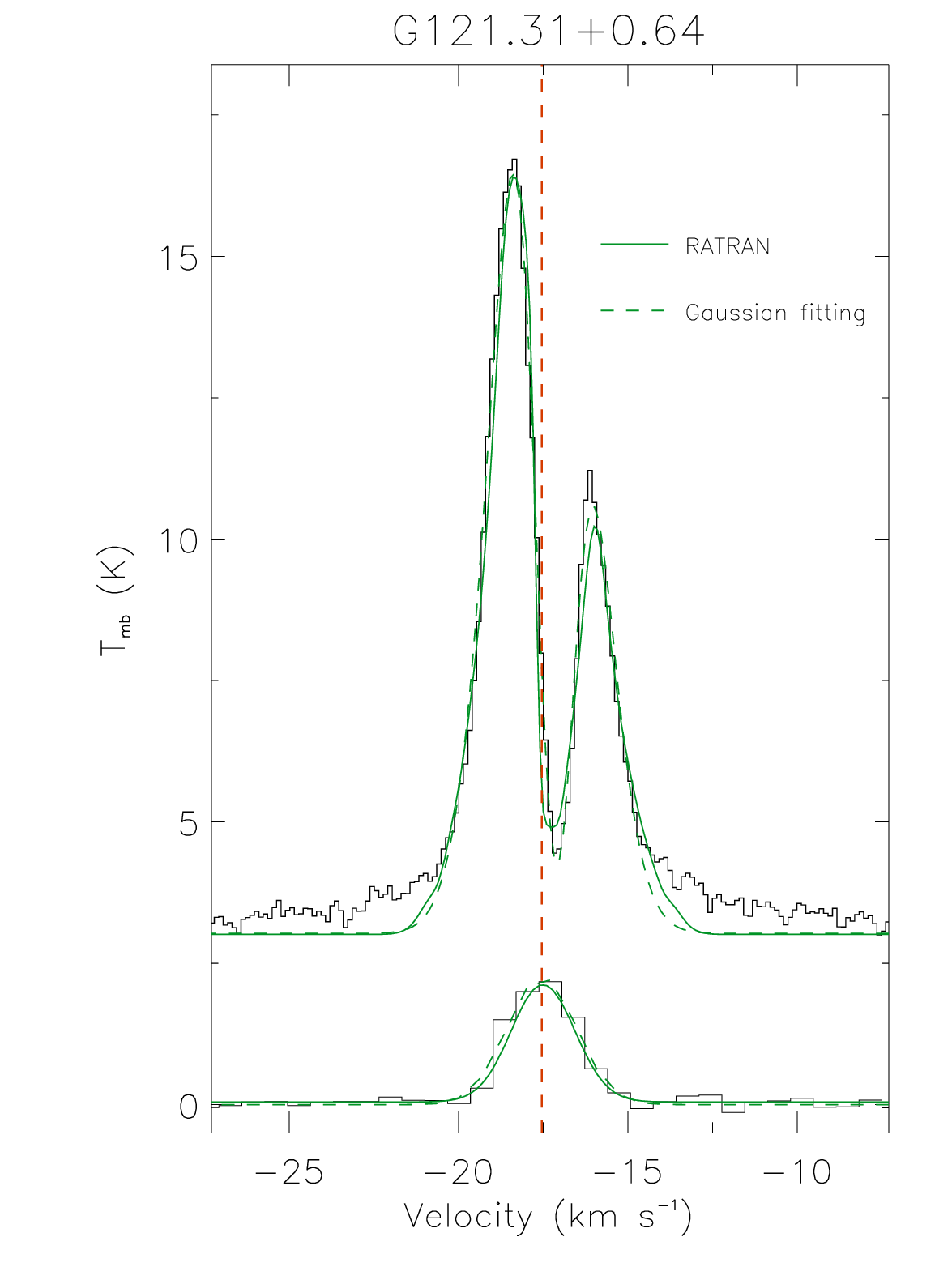}{0.28\textwidth}{}
          \fig{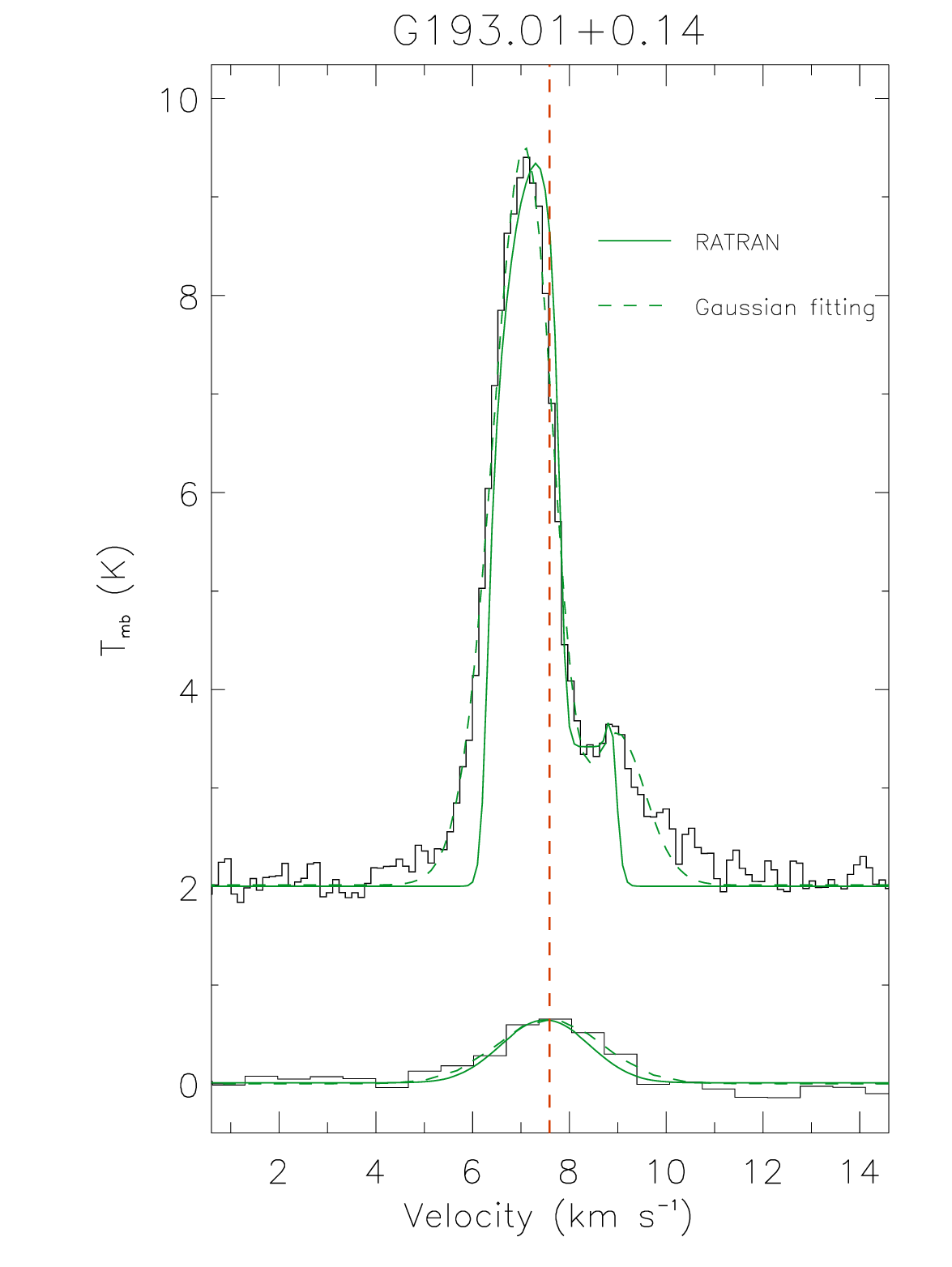}{0.28\textwidth}{}
          \fig{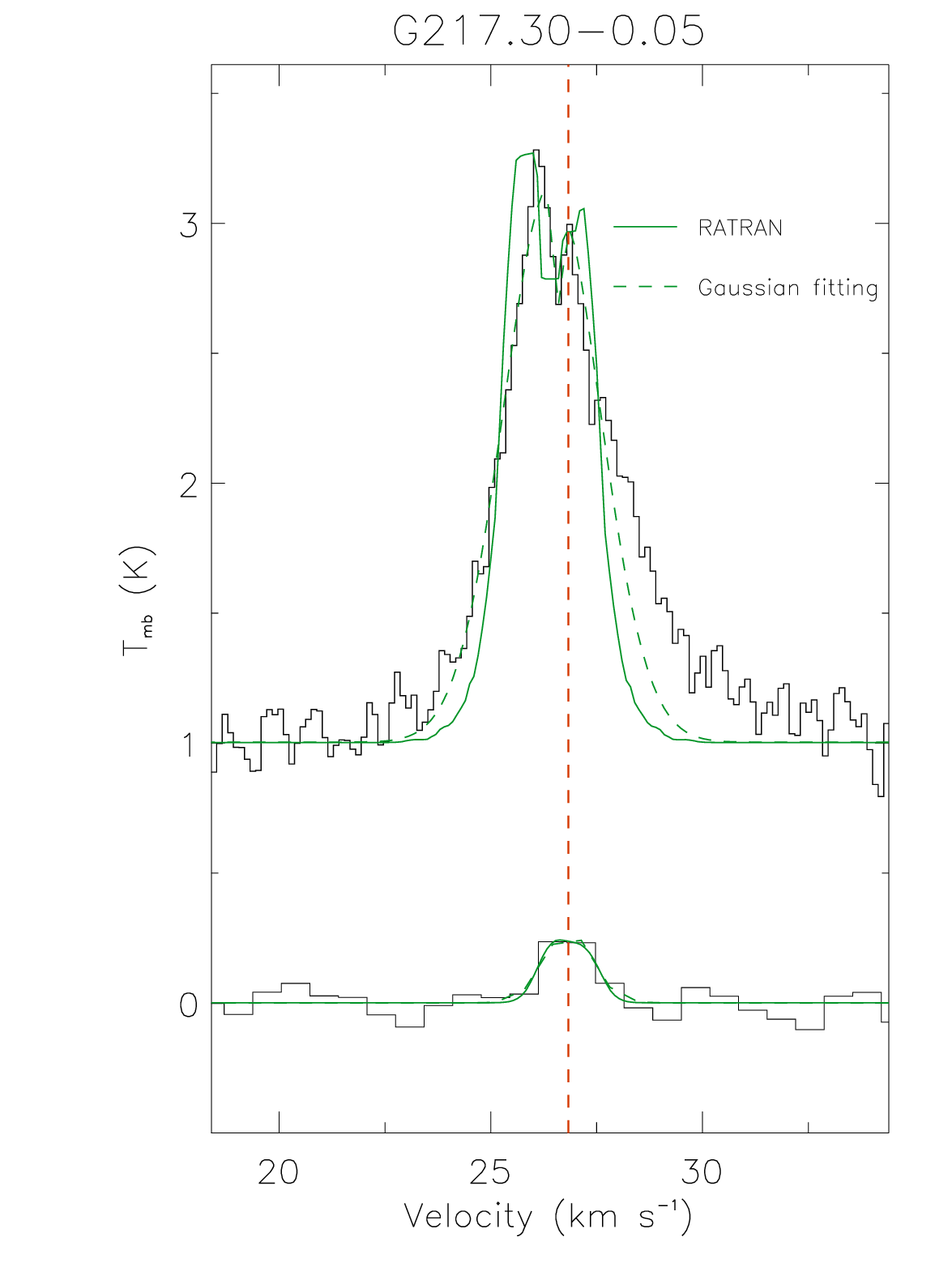}{0.28\textwidth}{}
          }
\caption{HCO$^+$ (1−0) spectra obtained from RATRAN modelling of a collapsing cloud (see text). The observed HCO$^+$ (1−0) spectrum is plotted in black, and the position of the spectrum is shown in column 2 and 3 of Table \ref{Tab:line}. The dashed green line is the spectrum obtained by RATRAN model, and the green line is the result of Gaussian fitting of HCO$^+$ (1−0). The dashed red line indicates the central radial velocity of H$^{13}$CO$^+$. \label{fig:infall}}
\end{figure*}

In the previous subsection, we identified 19 sources that have gas infall motions. Among them, the HCO$^+$ lines of eight sources show a significant blue profile, and the signal-to-noise ratios are greater than 10. We use some models to analyze further the physical properties of these sources. The RATRAN (radiative transfer and molecular excitation) 1D Monte Carlo radiative transfer code model of a collapsing cloud \citep{Hogerheijde+etal+2000} is a model usually used to fit the optically thick lines to estimate the infall velocity of sources. Using this model, we fit the HCO$^+$ lines of their single-point spectra, and then estimated the infall velocities and the mass infall rates of these candidates. The input parameters of RATRAN are the shell number, the radius of the clump, the density distribution (cm$^{-3}$) and the abundance distribution of the line, the kinetic temperature distribution, turbulent velocity dispersion, and infall velocity distribution. Among them, the size and other input parameters of these sources we used are quoted in Table \ref{Tab:clump}. We use 10 concentric spherical shells with a power-law density and velocity distribution to model the envelope, with power index of -1.5 and -0.5, respectively. Based on Section \ref{subsec:clumps}, we fixed the HCO$^+$ abundance relative to H$_2$. Then, we ran the code varying the infall velocity and the velocity dispersion to get a series of models to fit the HCO$^+$ lines. The Kolmogorov-Smirnov (K-S) test is used to verify the similarity between the HCO$^+$ fitting results obtained by the model and the observation. We selected the one with the highest probability in the K-S test in a series of models as a reasonable fit of the HCO$^+$ line. For H$^{13}$CO$^+$ line, we also use the RATRAN model for fitting. However, compared with HCO$^+$, the variation of H$^{13}$CO$^+$ infall velocity and velocity dispersion has no significant effect on the fitting results. The model results of HCO$^+$ and H$^{13}$CO$^+$ are presented in Figure \ref{fig:infall}, and the parameters are listed in Table \ref{Tab:infall}.

An alternative way to estimate the infall velocity is \citet{Myers+etal+1996} model. This analytical model allows one to infer an infall velocity from the data of the optically thick and thin lines:
\begin{equation}
  V_{in}=\frac{\sigma^{2}}{V_{red}-V_{blue}}\ln\left(\frac{1+e(T_{bd}/T_{dip})}{1+e(T_{rd}/T_{dip})}\right)
\label{eq:equ2}
\end{equation}
where $\sigma$ is the velocity dispersion of the H$^{13}$CO$^+$ line. $V_{red}$ and $V_{blue}$ are the velocities of the HCO$^+$ line. $T_{dip}$ is the intensity of the self-absorption dip. $T_{bd}$ and $T_{rd}$ are the temperatures of the blue and red peaks above the dip, respectively. All these parameters come from a combination of an emission Gaussian peak and an absorption Gaussian peak fitting. As a comparison, we also used this approach to estimate the infall velocity of the sources with a blue profile, and the results are also listed in Table \ref{Tab:infall}. Compared with the infall velocities calculated by the same method in \citet{Yang+etal+2020}, the result for G121.31+0.64, G029.60-0.63 and G081.72+0.57 infall velocities reduced. This may be due to the infall line profile is quite sensitive to changes in spatial position. Therefore, the difference of $V_{in}$ may be related to factors such as the coordinate position of the extracted single-point line, the pointing accuracy and the spatial resolution of the telescope. Compared with the infall velocities calculated by RATRAN, the velocities of G082.21-1.53, G107.50+4.47, and G121.31+0.64 are lower, and the infall velocity of G193.01+0.14 is larger than the result of the RATRAN model. This may be caused by the difference between the two models. For other sources, the results are relatively close.

\subsection{Mass Infall Rate} \label{subsec:Min}

Using \citet{Lopez-Sepulcre+etal+2010} method, a rough estimate of the mass infall rate, $\dot M_{in}$, may be determined from: 
\begin{equation}
  \dot M_{in}=4 \pi R^2 V_{in} \rho
\label{eq:equ3}
\end{equation}
where R is the clump radius estimated from the integrated intensity map, $V_{in}$ is an estimate of infall velocity from the RATRAN and/or Myers model (see Section \ref{subsec:Vin}), and $\rho$ is the average density of the clump. Table \ref{Tab:infall} also lists the obtained mass infall rates. The results show that the $\dot M_{in}$ of these sources is between $10^{-3}$ and $10^{-5}$ M$_{\odot}$ yr$^{-1}$. For most sources, there is roughly  no difference in the magnitude of the mass infall rate values calculated by the two models. According to some previous studies, for massive star formation, the mass infall rates onto the host clumps are about $10^{-2}$ to $10^{-4}$ M$_{\odot}$ yr$^{-1}$ \citep[e.g.][]{Palla+Stahler+1993, Whitney+etal+1997, Kirk+etal+2005}. For low-mass star formation, the mass infall rates are about $10^{-5}$ to $10^{-6}$ M$_{\odot}$ yr$^{-1}$ \citep[e.g.][]{Rygl+etal+2013, He+etal+2015}. This suggests that intermediate or massive stars may be forming in some of the clumps. However, this estimation approach assumes that the clump is isotropic, which may cause the local mass infall rate to be overestimated or underestimated. To study the local mass infall rate of clumps further, higher-resolution observations are necessary.

\section{Other Star-forming Activities of the Infall Candidates} \label{sec:sfa}

\subsection{Association with Infrared Point Sources} \label{subsec:IR}

We also checked the correlation between infrared point sources and the identified infall sources to distinguish the evolutionary stages of these sources. If the infrared point sources are located near the clump centers, or near the spectral lines' emission positions, we consider that this source is associated with the infrared point sources. The infrared source catalogs we used are the AllWISE Sources Catalog \citep[][]{vizier:II/328} and the Spitzer catalog, GLIMPSE \citep{Benjamin+etal+2003, Churchwell+etal+2009}. After excluding the foreground sources, shock emission blobs, and resolved structure, all sources with infall signatures were associated with AllWISE and/or GLIMPSE sources, with the exception of G031.41+5.25 and G110.40+1.67.

Using the sources' classification criteria suggested by \citet{Koenig+etal+2012}, we have identified that 12 of them are associated with Class 0/I YSOs, which are marked in Table \ref{Tab:src-catalog}. Most of these sources are close to the radio sources in the SFR, as well as a variety of masers, including Class I and Class II methanol, OH, and/or H$_2$O masers, indicating that these sources have more active star-forming activities. We will analyze and discuss each source with other star-forming activities in the next subsection.  

The remaining sources, G028.97+3.35, G029.06+4.58, G030.17+3.68, G121.34+3.42, and G154.05+5.07 are not associated with Class 0/I YSOs, while G031.41+5.25 and G110.40+1.67 lack infrared sources associated with them. Thus, those data suggests that these sources may be at a very early stage of evolution. Unfortunately, the signal-to-noise ratios of these sources are relatively low, and the line profiles are mostly peak-shoulder profiles or single-peaked profiles with the peaks skewed to the blue. The model we currently use cannot fit these line profiles well. If there are better optically thick line data in the future, further research on these sources will help us better understand the physical process of the initial stage of star formation.

\subsection{Outflow and Other Star-forming Activities} \label{subsec:sfa} 

\begin{figure*}[h]
  \begin{minipage}[t]{0.5\linewidth}
  \centering
   \includegraphics[width=100mm]{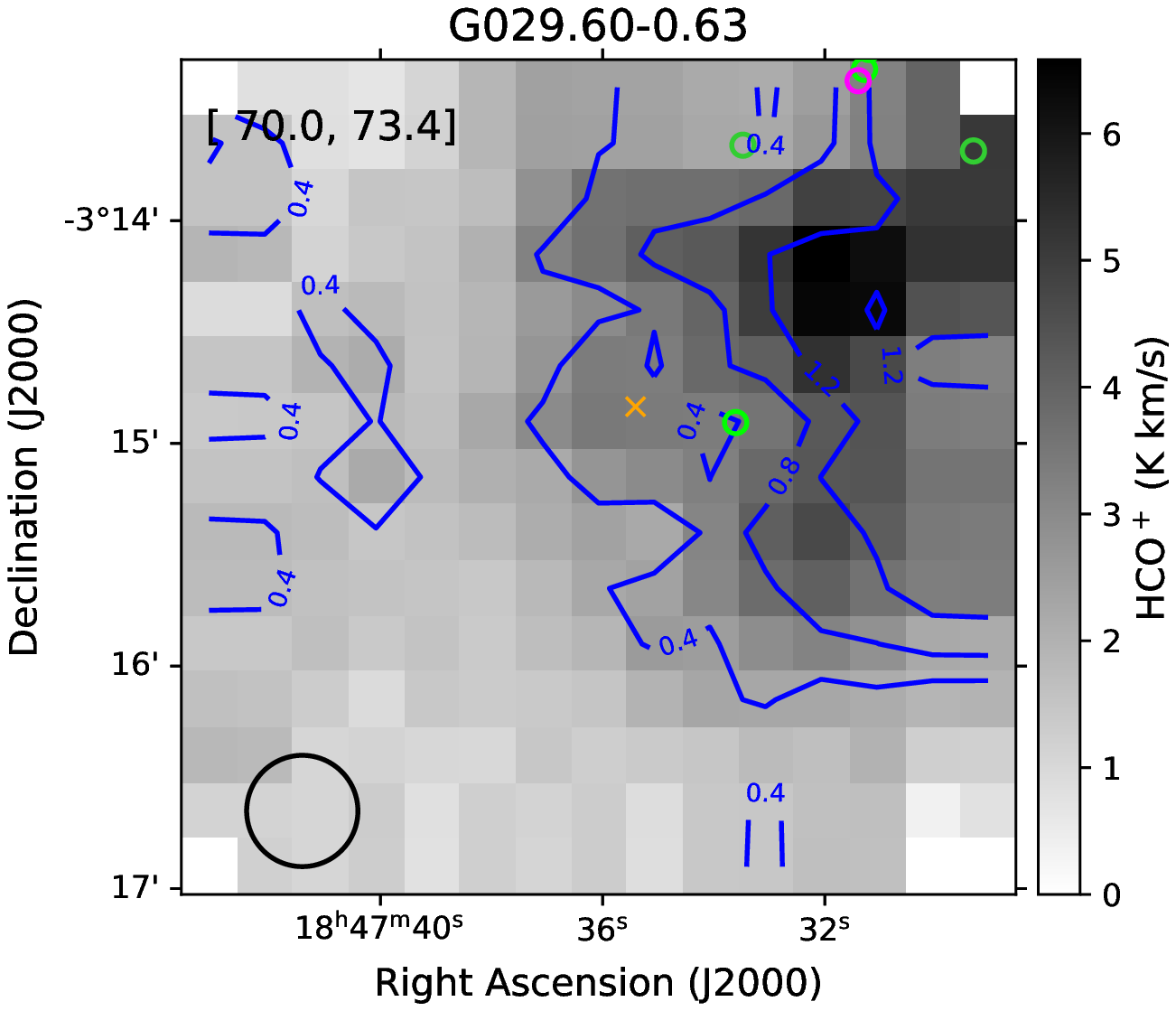}
  \end{minipage}%
  \begin{minipage}[t]{0.63\linewidth}
  \centering
   \includegraphics[width=100mm]{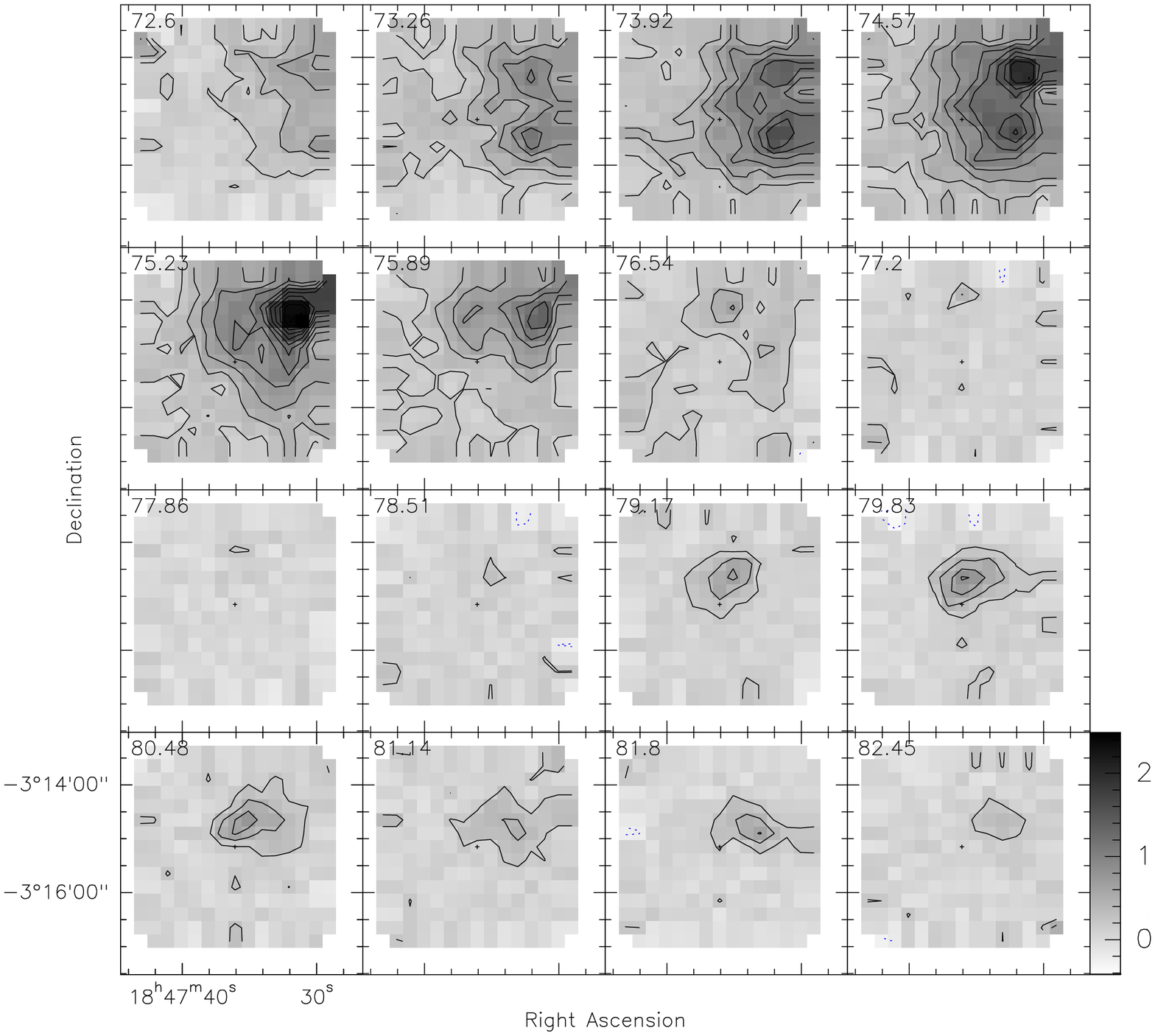}
  \end{minipage}%
    \begin{minipage}[t]{0.5\linewidth}
  \centering
   \includegraphics[width=100mm]{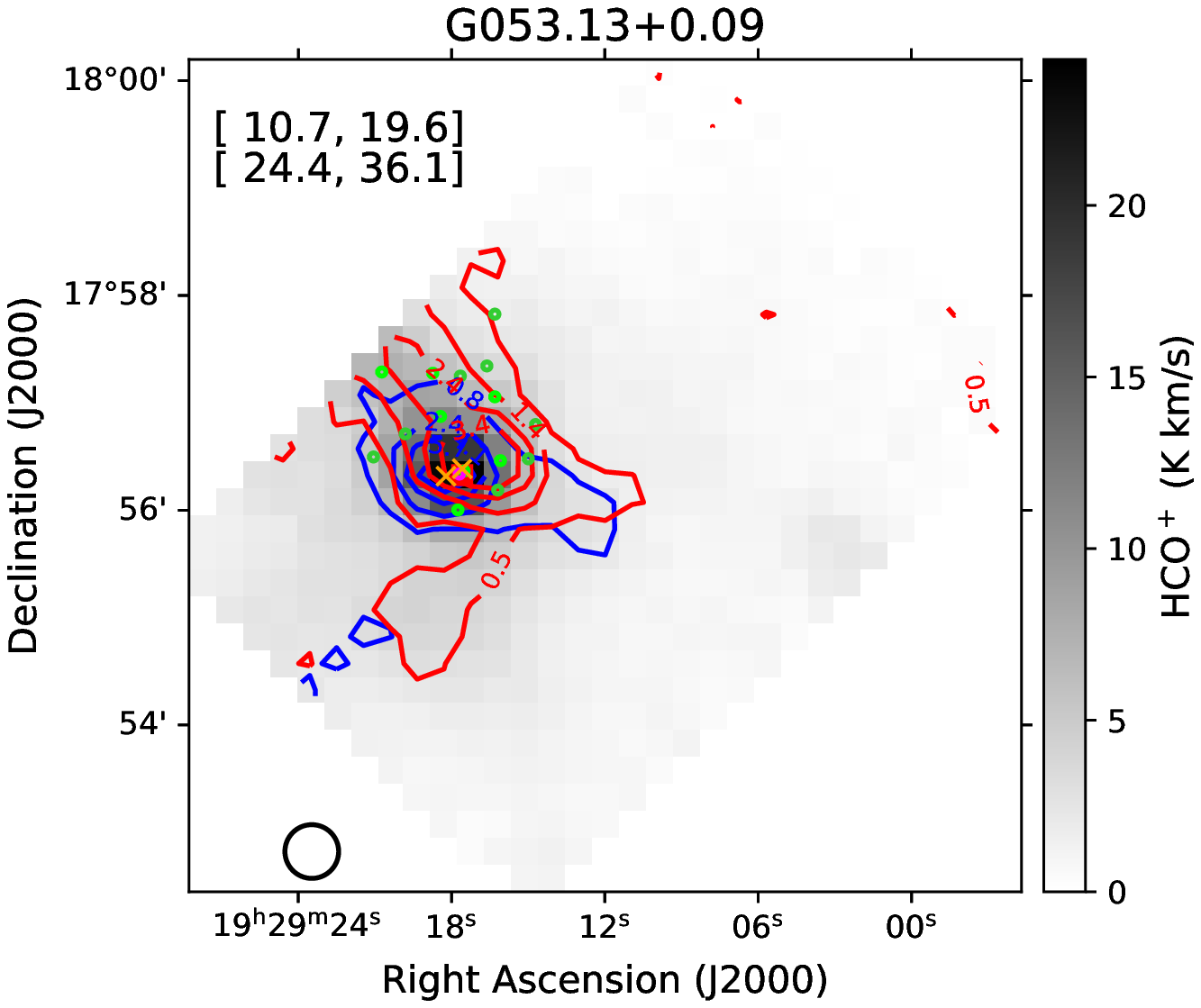}
  \end{minipage}%
  \begin{minipage}[t]{0.5\linewidth}
  \centering
   \includegraphics[width=100mm]{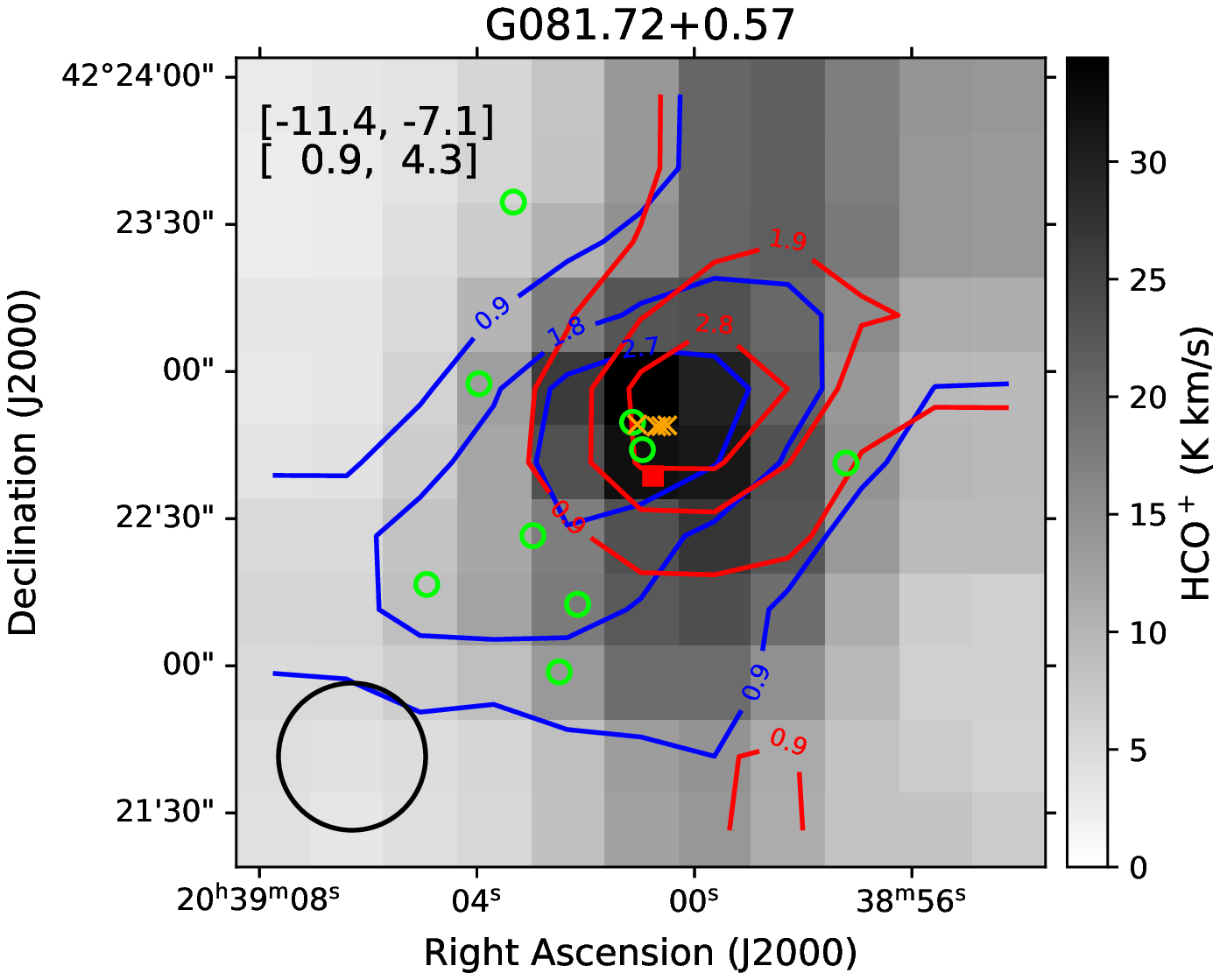}
  \end{minipage}%
  \begin{minipage}[t]{0.5\linewidth}
  \centering
   \includegraphics[width=100mm]{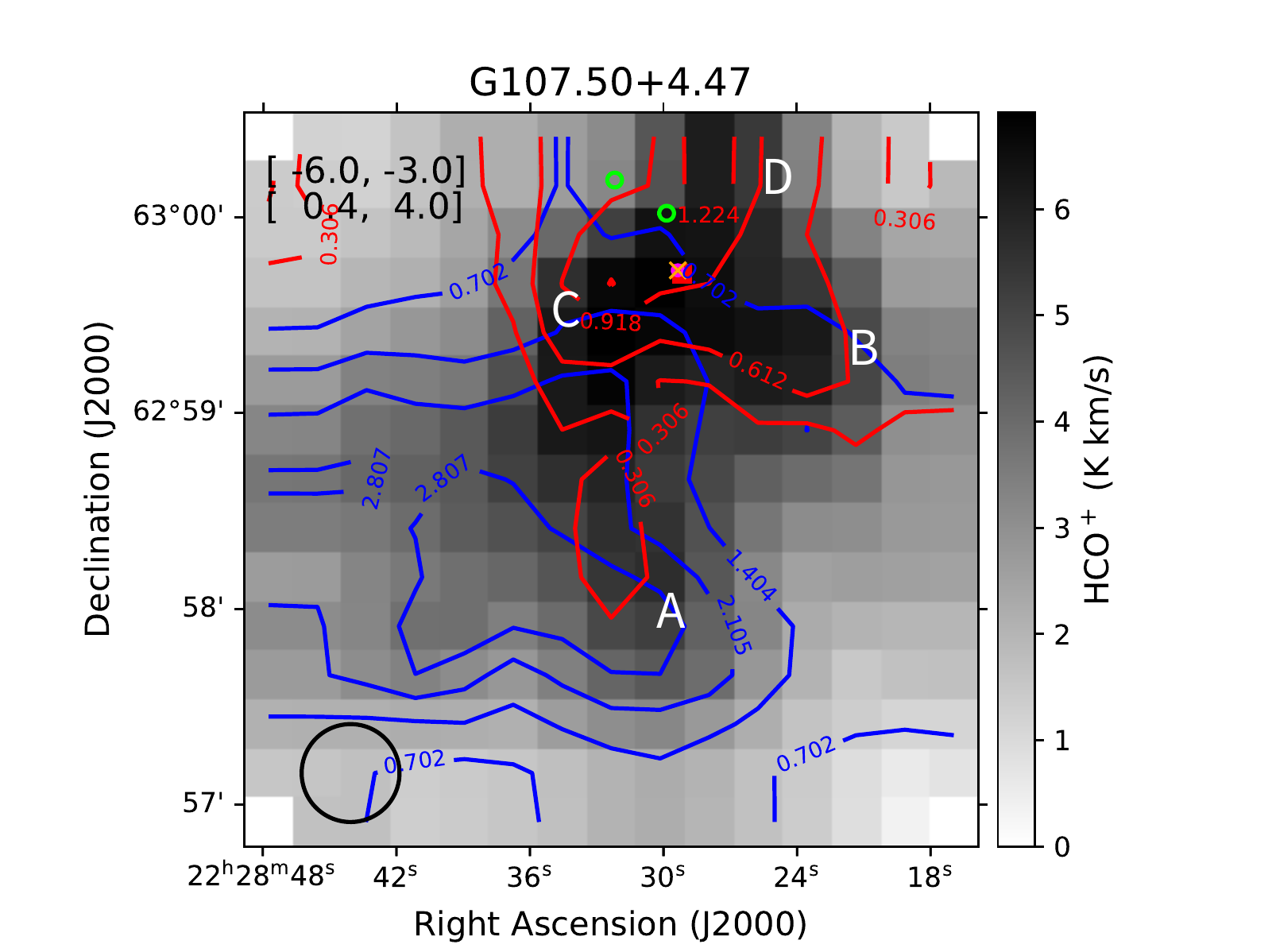}
  \end{minipage}%
  \begin{minipage}[t]{0.5\linewidth}
  \centering
   \includegraphics[width=100mm]{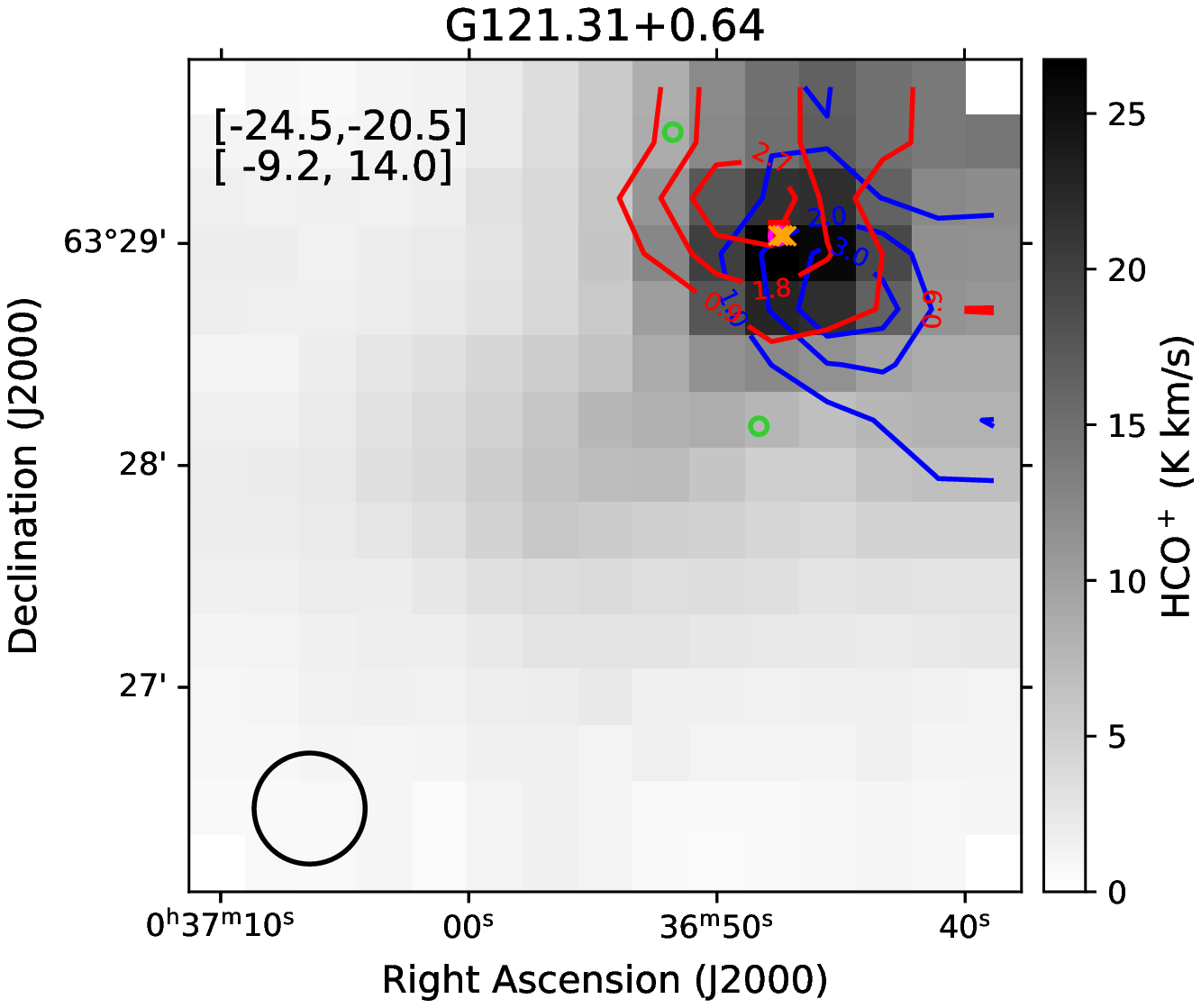}
  \end{minipage}%
\caption{The integrated intensity contour of the line wing of the sources with outflow characteristics, superposed on the HCO$^+$ integrated intensity image (The velocity range is marked on the images). The black circles indicate the beam sizes of the IRAM 30-m telescope. The green circles denote YSOs from AllWISE data. The magenta circles denote IRAS sources. The red squares denote the star-forming regions. The orange crosses denote the maser sources. Top right panel: Channel map of G029.60-0.63 HCO$^+$ (1-0) lines from 72.6 km s$^{-1}$ to 82.5 km s$^{-1}$.
\label{fig:outflow}}
\end{figure*}

\begin{figure*}[h]
\addtocounter{figure}{-1} 
  \begin{minipage}[t]{0.5\linewidth}
  \centering
   \includegraphics[width=100mm]{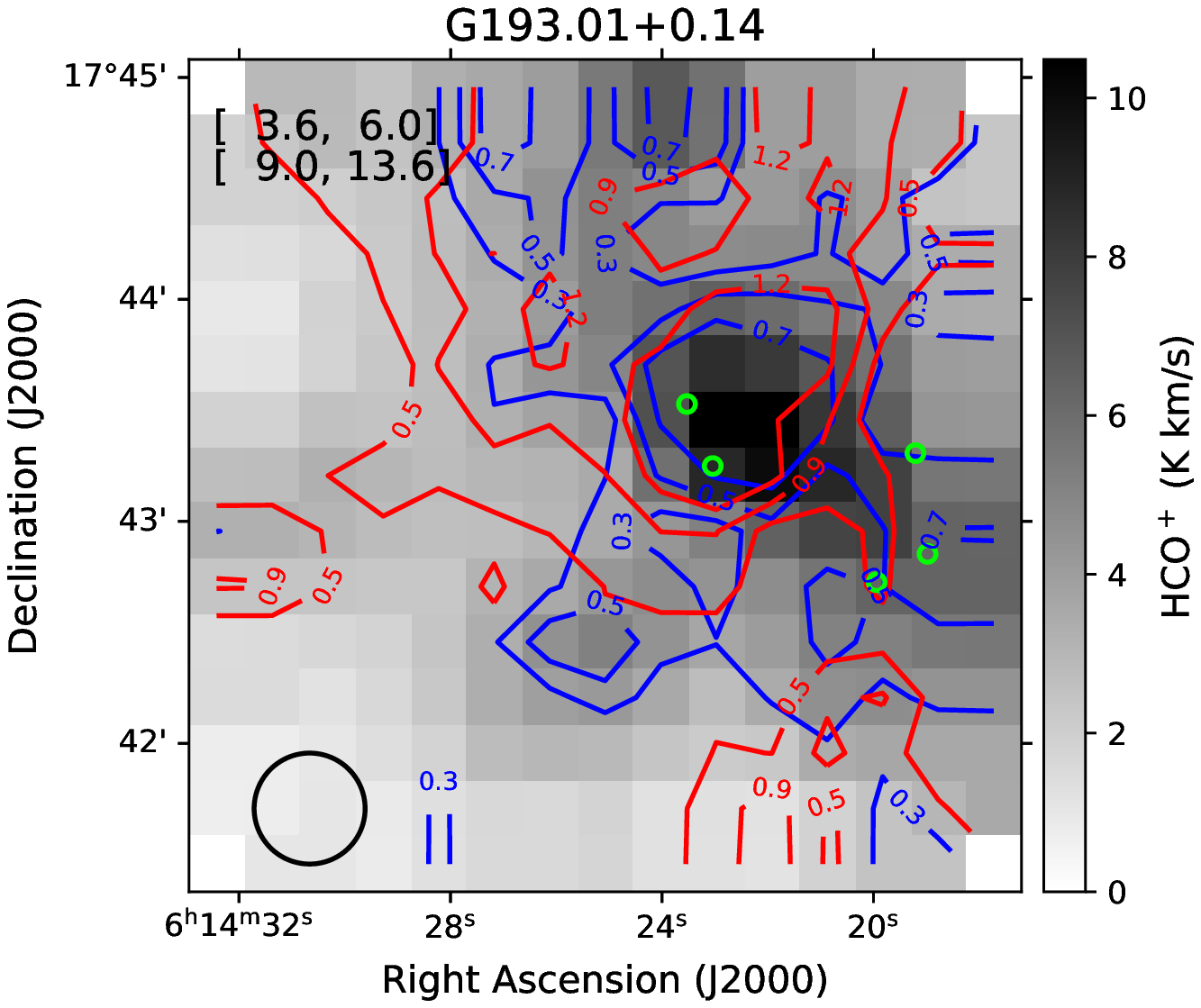}
  \end{minipage}%
  \begin{minipage}[t]{0.5\linewidth}
  \centering
   \includegraphics[width=100mm]{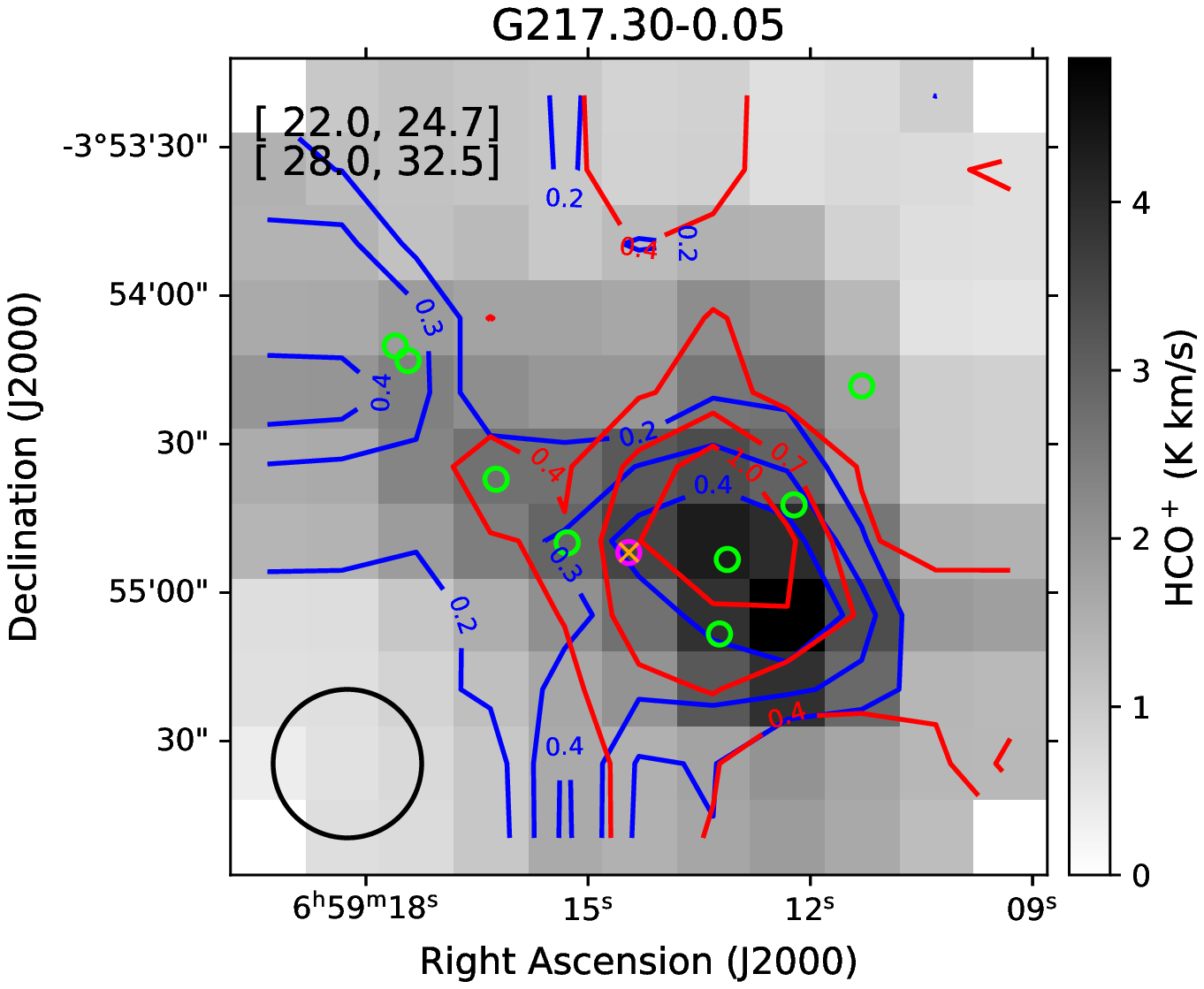}
  \end{minipage}%
\caption{Continued.
\label{fig:outflow}}
\end{figure*}

\begin{table}
\begin{center}
\caption{Outflow and other star-forming activities of infall candidates.}\label{Tab:sfa}
 \setlength{\tabcolsep}{1mm}{
\begin{tabular}{ccccccccccccc}   
  \hline\noalign{\smallskip}
Source & Association with & Outflow & \multicolumn{7}{c}{Methanol Maser} & H$_2$O Maser & OH Maser \\
Name & Class 0/I YSOs & Lobe & 6.7GHz & 36GHz & 44GHz & 84.5GHz & 95GHz & 107GHz & 133GHz & & \\
  \hline\noalign{\smallskip} 
G028.97+3.35 &   &     &   &  &  &  &  &  &  &  &  \\
G029.06+4.58 &   &     &   &  &  &  &  &  &  &  &  \\
G029.60-0.63 & + & b   & + &  &  &  & + &  &  & + &  \\
G030.17+3.68 &   &     &   &  &  &  &  &  &  &  &  \\
G031.41+5.25 &   &     &   &  &  &  &  &  &  &  &  \\
G053.13+0.09 & + & r,b & +  &  &  &  & + &  &  & + & + \\
G079.71+0.14 & + &     &   &  &  &  &  &  &  &  &  \\
G081.72+0.57 & + & r,b & + & + & + & + & + & + & + & + & + \\
G082.21-1.53 & + &     &   &  &  &  &  &  &  &  &  \\
G107.50+4.47 & + & r,b &   &  &  &  &  &  &  & + &  \\  
G109.00+2.73 & + &     &   &  &  &  &  &  &  &  &  \\
G110.40+1.67 &   &     &   &  &  &  &  &  &  &  &  \\
G121.31+0.64 & + & r,b & + &  & + &  & + &  &  & + &  \\
G121.34+3.42 &   &     &   &  &  &  &  &  &  &  &  \\
G126.53-1.17 & + &     &   &  &  &  &  &  &  &  &  \\
G143.04+1.74 & + &     &   &  &  &  &  &  &  &  &  \\
G154.05+5.07 &   &     &   &  &  &  &  &  &  &  &  \\
G193.01+0.14 & + & r,b &   &  &  &  &  &  &  &  &  \\
G217.30-0.05 & + & r,b &   &  &  &  &  &  &  & + &  \\
  \hline\noalign{\smallskip}

\end{tabular}}
\end{center}
\tablecomments{For outflow, if any, we mark out the lobe of the outflow, r denotes red lobe, and b denotes blue lobe.}
\end{table}

In addition to showing infall characteristics, some infall candidates also have evidence for outflow and other star-forming activities. In this paper, we only make a rough diagnosis, and a more detailed analysis is beyond the scope of this work and will be given in a future paper.

G029.60-0.63 spectra map (presented in Figure \ref{fig:figA25}) shows that the HCO$^+$ lines of the north clump have infall signatures with a blue line wing. We made an integrated intensity map of the line wing that is presented in the Figure \ref{fig:outflow}, top left panel, with a velocity range of 70.0--73.5 km s$^{-1}$. This source may be an outflow candidate with a blue lobe. Using the $^{13}$CO (3-2) and C$^{18}$O (3-2) data, a previous study also found outflows in this area \citep[e.g.][]{deVilliers+etal+2014}. However, what they found is a red lobe outflow, which is different from our result. Therefore, we made a velocity channel map of the HCO$^+$ with a velocity range of 72.6--82.5 km s$^{-1}$ to analyze the structure of this source (as shown in Figure \ref{fig:outflow} top right panel). It can be seen that the velocity range of the two clumps in the upper right corner of the map is about 72.6 to 76.5 km s$^{-1}$. However, in the range of 79.2 to 82.5 km s$^{-1}$, another clump with a slightly different location seems to exist. The red line wing found in previous CO data is likely to correspond to this velocity component. In addition, G029.60-0.63 is associated with an SFR, while several Class 0/I YSOs have been found in this region \citep[e.g.][]{Camarata+etal+2015}. However, the clump position of HCO$^+$ with infall line profile was found to deviate slightly from the positions of these YSOs. Other studies have detected a 22 GHz H$_2$O maser \citep[e.g.][]{Svoboda+etal+2016} and a variety of kinds of methanol masers \citep[such as 6.7 GHz Class II and 95 GHz Class I methanol masers, e.g.][]{deVilliers+etal+2014, Yang+etal+2017} near this clump. Among them, Class II methanol masers usually trace the high-mass SFRs \citep[e.g.][]{Billington+etal+2019}, while a H$_2$O maser is generated in the shock wave produced by outflows and stellar winds \citep[e.g.][]{Torrelles+etal+2005}. The evidence shows that the star-forming activity of this source is very active, and massive stars may be forming in this area.

The HCO$^+$ lines of G053.13+0.09 have obvious blue and red wing components, indicating that this source may have a bipolar outflow. From the lobes' integral diagram shown in Figure \ref{fig:outflow}, it can be seen that the direction of outflow is roughly from southeast to northwest. The outflow center basically coincides with the radio source 53.14+0.07 in the massive SFR. And several YSOs at the Class I stage have been detected in this region \citep[e.g.][]{Avedisova+etal+2002}. This is consistent with the results obtained by \citet{Zhang+etal+2017} using SiO lines to trace the shock gas associated with molecular outflows. The Class I YSOs here are likely to be the host sources of the infall and outflow motion, but further analysis still requires higher-resolution observations to support this conclusion. On the other hand, this source is also maser-rich, with 6.7 GHz Class II CH$_3$OH, 95 GHz Class I CH$_3$OH, H$_2$O and OH masers reported \citep[e.g.][]{Szymczak+etal1+2000, Hu+etal+2016, Svoboda+etal+2016, Yang+etal+2017, Beuther+etal+2019}. This indicates that the star-formation activity of this source is very active, and intermediate- or high-mass stars may be forming in this area.

G081.72+0.57 is associated with the well-known massive SFR DR21 \citep{Downes+Rinehart+1966}, which consists of a group of several compact HII regions \citep[e.g.][]{Harris+etal+1973} with energetic outflow \citep[e.g.][]{Garden+etal+1991a, Russell+etal+1992, Davis+Smith+1996}. We found significant line wings at its HCO$^+$ line, which are likely to be caused by outflow gas. The integral diagram of the red and blue lobes is shown in Figure \ref{fig:outflow}, which seems to show that the outflow from this source has a southeast--northwest direction, while some YSOs are close to the outflow center. However, because there are several YSOs in this area, we are not sure which sources host the molecular outflow and/or infall. A large number of previous studies have reported that there are many kinds of masers here: \citet{Genzel+Downes+1977} found 22 GHz H$_2$O maser, and other researchers have detected OH masers here \citep[e.g.][]{Norris+etal+1982, Argon+etal+2000, Szymczak+etal2+2000}. A large amount of research has shown that the methanol maser here is also rich, with a variety of Class I and Class II methanol masers, including 6.7, 36, 44, 84.5, 95, 107, and 133 GHz methanol masers \citep[e.g.][]{Haschick+Baan+1989, Haschick+etal+1990, Menten+etal+1991, Val'tts+etal+1995, Slysh+etal+1997, Kalenskii+etal+2001}. In addition, many studies have been focused on gas collapse in this area. Some of them have indicated a global gas infall motion here and tried to estimate the infall velocity of the gas. \citet{Schneider+etal+2010} used the simple \citet{Myers+etal+1996} model to infer an infall velocity from the HCO$^+$ line profiles of the DR21 filament to be about 0.2--1 km s $^{-1}$. Using the RATRAN and Myers model, the infall velocities we estimated are about 0.5 and 0.4 km s$^{-1}$, respectively, which are consistent with previous result.

G107.50+4.47 is also an outflow candidate. Its HCO$^+$ lines have some components that exceed the Gaussian profiles, and the lobes integral diagram of this source is shown in Figure \ref{fig:outflow}. In this source, we detected some HCO$^+$ lines with blue profiles at the positions of clumps A, B, and C (marked with white letters in the figure, the velocity ranges are slightly different, see Section \ref{subsec:mappting}), while in the southwest of the mapping area, some red profile lines are detected. A part of the mapping area shows a blue line profile, and the other shows a red line profile. The distribution of the two kinds of line profiles is almost symmetric, which may also be evidence of gas outflow \citep{He+etal+2015}. Using the CO (2-1) data, \citet{Kim+etal+2006} previously confirmed that there is gas outflow in this area. In the ensuing research, they observed HCO$^+$ (1-0) and SiO (2-1) on 18 YSOs, including a source in this area, and confirmed that there may be intermediate- or high-mass counterparts of Class 0 objects, and it is experiencing active accretion \citep{Kim+etal+2015}. Moreover, there is an SFR 107.43+4.54 nearby, with a velocity range close to these clumps. Clumps A, B, and C all show some infall line profiles, but clump C is the most obvious one with global infall characteristics. \citet{Valdettaro+etal+2001} detected H$_2$O maser emissions near clump C, while the clumps associated with the H$_2$O maser usually have other indicators of star-formation activity. Using the RATRAN and Myers model, we estimated the infall velocities of clump C to be about 0.6 and 0.3 km s$^{-1}$, respectively. The mass infall rates are about 10$^{-4}$ -- 10$^{-5}$ M$_{\odot}$ yr$^{-1}$. This indicates that there may be intermediate-mass stars forming in this clump, which is similar to previous observation results.

G121.31+0.64 is associated with SFR 121.20+0.62 (IRAS 00338+6312). Some previous studies have already detected signs of outflow and infall here: CO observation results indicate a bipolar outflow of CO gas \citep[e.g.][]{Snell+etal+1990, Yang+etal+1991}, and the HCO$^+$ and H$^{13}$CO$^+$ data taken by the 20-m radio telescope of the Onsala Space Observatory (Sweden) indicates infall motion in this source \citep{Pirogov+etal+2016}. Our observations also confirmed these results. The spectra map grids of this source show that the HCO$^+$ lines have a blue profile in the entire area, which indicates that this source may be a global collapse source, while the mass infall rate we estimated is about 10$^{-4}$ M$_{\odot}$ yr$^{-1}$. This means that G121.31+0.64 is likely to be gestating intermediate-mass stars. We also found some outflow evidence here. HCO$^+$ lines show both blue and red wings, and the integrated intensity map of the line wings shows a northeast--southwest outflow direction (as shown in Figure \ref{fig:outflow}), which is consistent with the results of previous studies \citep[e.g.][]{Juarez+etal+2019}. In addition, rich maser sources have been found in this source, including OH, H$_2$O, and a variety of methanol masers \citep[6.7 GHz Class II, 44 GHz and 95 GHz Class I methanol masers; e.g.][]{Wouterloot+etal+1993, Valdettaro+etal+2001, MacLeod+etal+1998, Gan+etal+2013, Yang+etal+2017}. OH and H$_2$O masers can be used to trace SFRs, while Class II methanol masers are usually found close to strong radiation sources, and are associated with high-mass SFRs. This is also strong evidence that massive stars are forming in this source.

The HCO$^+$ line profiles of G193.01+0.14 also have line wings components. Figure \ref{fig:outflow} shows the integrated intensity map of the line wings of this source. From the figure, it can be seen that the outflow direction of G193.01+0.14 seems to be toward the line of sight. Some infrared point sources have been found in the center of the bipolar outflow. These infrared sources may be the host source of the gas outflow and infall.

G217.30-0.05 is associated with IRAS 06567-0350. We have also found some WISE sources in the early stages of evolution, which are located around this SFR. In addition, there is an HII region BFS56 (Gl = 217.31$^{\circ}$ \& Gb = -0.05$^{\circ}$) associated with the molecular cloud complex in this area \citep[e.g.][]{Blitz+Fich+1982}. Some previous studies have detected H$_2$O masers \citep[e.g.][]{Han+etal+1998, Valdettaro+etal+2001}, but not the 95 GHz Class I methanol maser here \citep[e.g.][]{Gan+etal+2013}. \citet{Volp+etal+2000} confirms that IRAS 06567-0350 has very similar characteristics to other transitional YSOs. From the spectra map grids of G217.30-0.05 (Figure \ref{fig:figA25}), we found that optically thick line HCO$^+$ shows blue asymmetrical profiles in the northeast of the clump, while it shows red profiles in the southwest. Its HCO$^+$ profile also shows blue and red wings at the HCO$^+$ lines, which may indicate that this source has outflow motion. The integrated intensity map of the line wings is shown in Figure \ref{fig:outflow}, with the integration range being $22.0-24.7$ km s$^{-1}$ and $28.0-30.0$ km s$^{-1}$. It can be seen that the direction of outflow in G217.30-0.05 is roughly northeast--southwest. There are some class 0/I YSOs in the outflow center, which may be related to outflow and infall activities in this source.

It can be seen that out of the nine confirmed infall sources with a blue asymmetric double-peaked profile, seven have outflow characteristics. The percentage of double-peaked infall sources with outflow motion to the total candidates is higher than expected. Moreover, six of them are associated with one or more kinds of masers. All six sources are associated with SFRs. On the other hand, no confirmed infall sources with a peak-shoulder profile or single-peaked profile have been found with outflow evidence. It cannot be ruled out whether it is caused by the weak HCO$^+$ signals of these sources. Even if the effects of insufficient signal-to-noise ratios are excluded, there are still some confirmed sources that do not have outflow characteristics. These sources should be in the early stages of evolution, probably earlier than the Class 0 stage. At this time, the central object has not yet experienced molecular outflow.


\section{Summary} \label{sec:summary}

Using the IRAM 30-m telescope, we have observed the HCO$^+$ (1-0) and H$^{13}$CO$^+$ (1-0) lines of 24 sources. All these sources are infall candidates with high confidence, which are selected from the single-point observations in \citet{Yang+etal+2020}. We carried out mapping observations of optically thick and thin lines at the same time, and further analyzed the gas infall motions in these sources. The results are summarized as follows:

(i) All 24 sources have observed HCO$^+$ emission (RMS<0.3K) and 18 sources show H$^{13}$CO$^+$ (RMS<0.2K), with detected rates of 100\% and 75\%, respectively. Among them, the HCO$^+$ mapping observations of 17 sources show clear clumpy structures, and the H$^{13}$CO$^+$ mapping of 15 sources shows clumpy structures. 

(ii) Using RADEX radiative transfer code, we roughly estimated the column densities of HCO$^+$ and H$^{13}$CO$^+$ lines, and further calculated the abundance ratio of [HCO$ ^+$]/[H$_2$] and [H$^{13}$CO$^+$]/[HCO$ ^+$] in these sources. The abundance ratios [HCO$^+$]/[H$_2$] are about $10^{-11}$ to $10^{-7}$, and [H$^{13}$CO$^+$]/[HCO$^+$] are about $10^{-3}$ to 1. 

(iii) According to the classification of optically thick line profiles of the observed sources, nine sources show double-peaked blue profiles in HCO$^+$ lines. Another four sources show peak-shoulder profiles, and six sources show single-peaked profiles with the peak skewed to the blue. All these sources may be the confirmed infall sources. In addition, other sources show red profiles, symmetrical profiles, or multiple peaks.

(iv) The RATRAN model was used to analyze eight sources with significant blue profiles and signal-to-noise ratios greater than 10. By comparing the results of the model and observations, we estimated that the mean infall velocities of these sources are between 0.3 and 1.3 km s$^{-1}$, and the mass infall rates are about $10^{-3}$--$10^{-4}$ M$_{\odot}$ yr$^{-1}$, which are consistent with the results of intermediate or massive star formation in previous studies. We also used Myers model to estimate the infall velocity. Except for four sources, the infall velocities of other sources estimated by these two models are consistent. For most sources, the magnitude of the mass infall rate value calculated by the two models is not much different.

(v) We checked the association between the infrared point sources and confirmed infall sources. According to the AllWISE sources classification criteria, there are 12 sources associated with Class 0/I YSOs. The remaining seven sources are not associated with Class 0/I YSOs, which may be at a very beginning stage of evolution. In addition, seven confirmed infall sources show molecular outflow evidence, while six are associated with well-known SFRs. In addition, one or more kinds of maser sources have been detected. All these sources show double-peaked blue asymmetric line profiles.

\acknowledgments

We are grateful to the staff of the Institut de Radioastronomie Millim{\'e}trique (IRAM) for their assistance and support during the observations. We also have made use of data products from the Wide-field Infrared Survey Explorer and the observations made with the Spitzer Space Telescope.  

This work is supported by the National Key R\&D Program of China (Grant No. 2017YFA0402702), and the National Natural Science Foundation of China (NSFC, Grant Nos. 11873093 and U2031202), Z. Chen acknowledges the support from the NSFC general grant 11903083, and Y. Ao acknowledges support by the NSFC grant 11933011.

\vspace{5mm}

\facility{IRAM:30m}

\software{GILDAS \citep{Pety+2005, Gildas+2013}, RADEX \citep{vanderTak+etal+2007}, RATRAN \citep{Hogerheijde+etal+2000}}

%






\clearpage

\restartappendixnumbering

\appendix

\section{HCO$^+$ (1-0) and H$^{13}$CO$^+$ (1-0) Integrated Intensity Maps} \label{sec:Appendix1}

\begin{figure}[h]
\centering
\renewcommand{\thesubfigure} \makeatletter
\subfigure[G028.97+3.35]{
\begin{minipage}[b]{0.45\textwidth}
\includegraphics[width=1.1\textwidth]{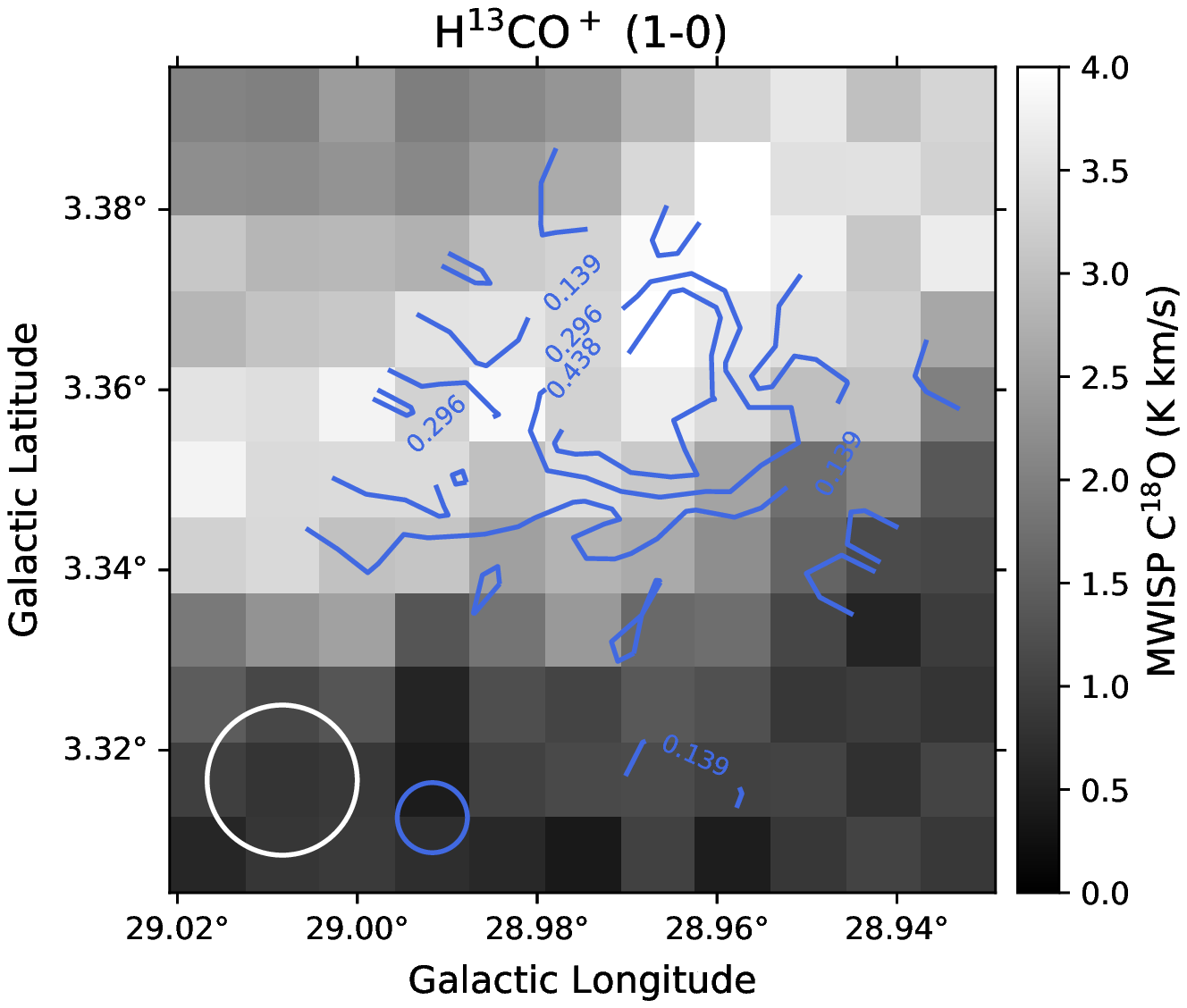}
\end{minipage}
}
\subfigure[G029.06+4.58]{
\begin{minipage}[b]{0.45\textwidth}
\includegraphics[width=1.1\textwidth]{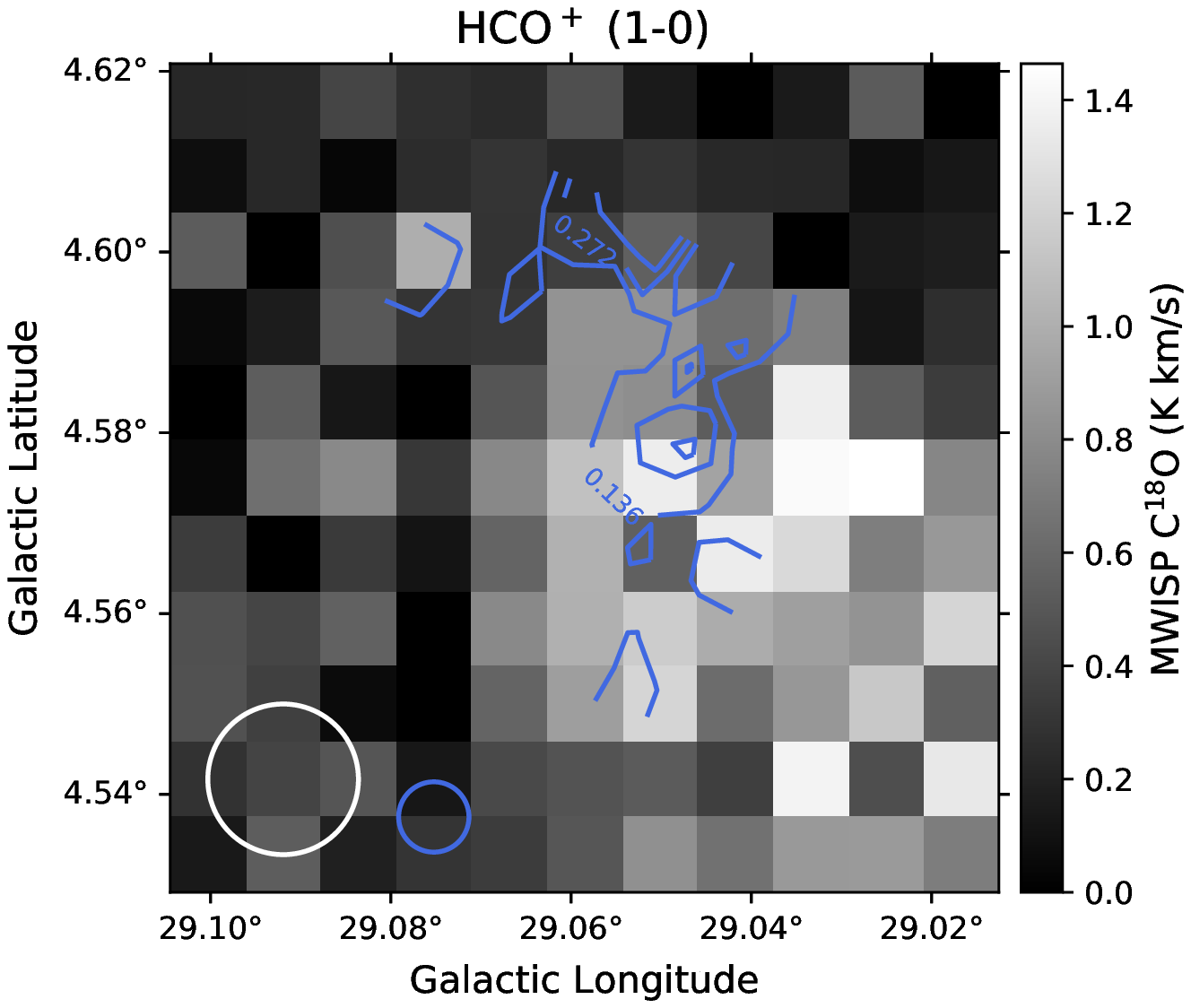}
\end{minipage}
}
\subfigure[G029.60-0.63]{
\begin{minipage}[b]{0.45\textwidth}
\includegraphics[width=1.1\textwidth]{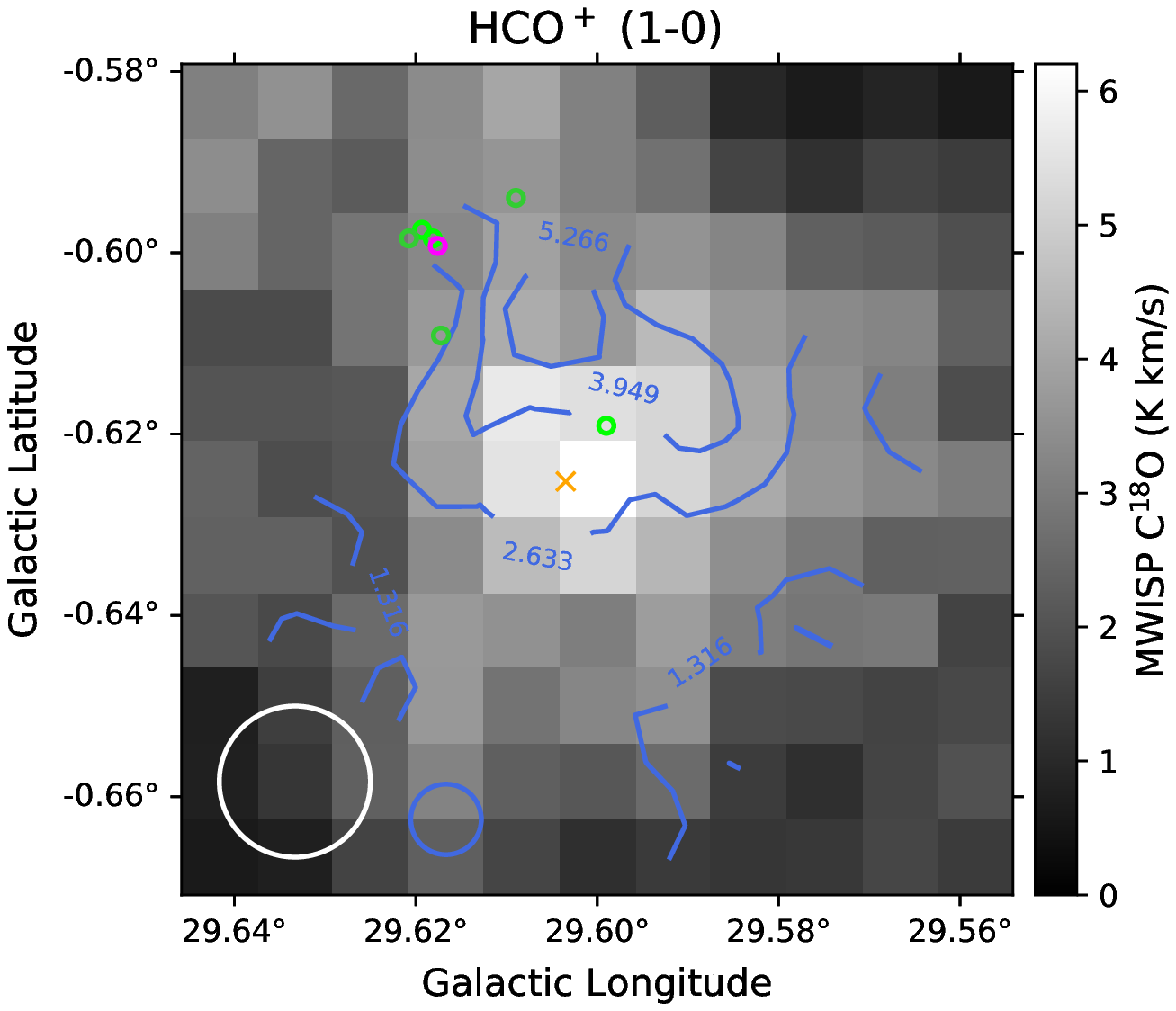}
\end{minipage}
\begin{minipage}[b]{0.45\textwidth}
\includegraphics[width=1.1\textwidth]{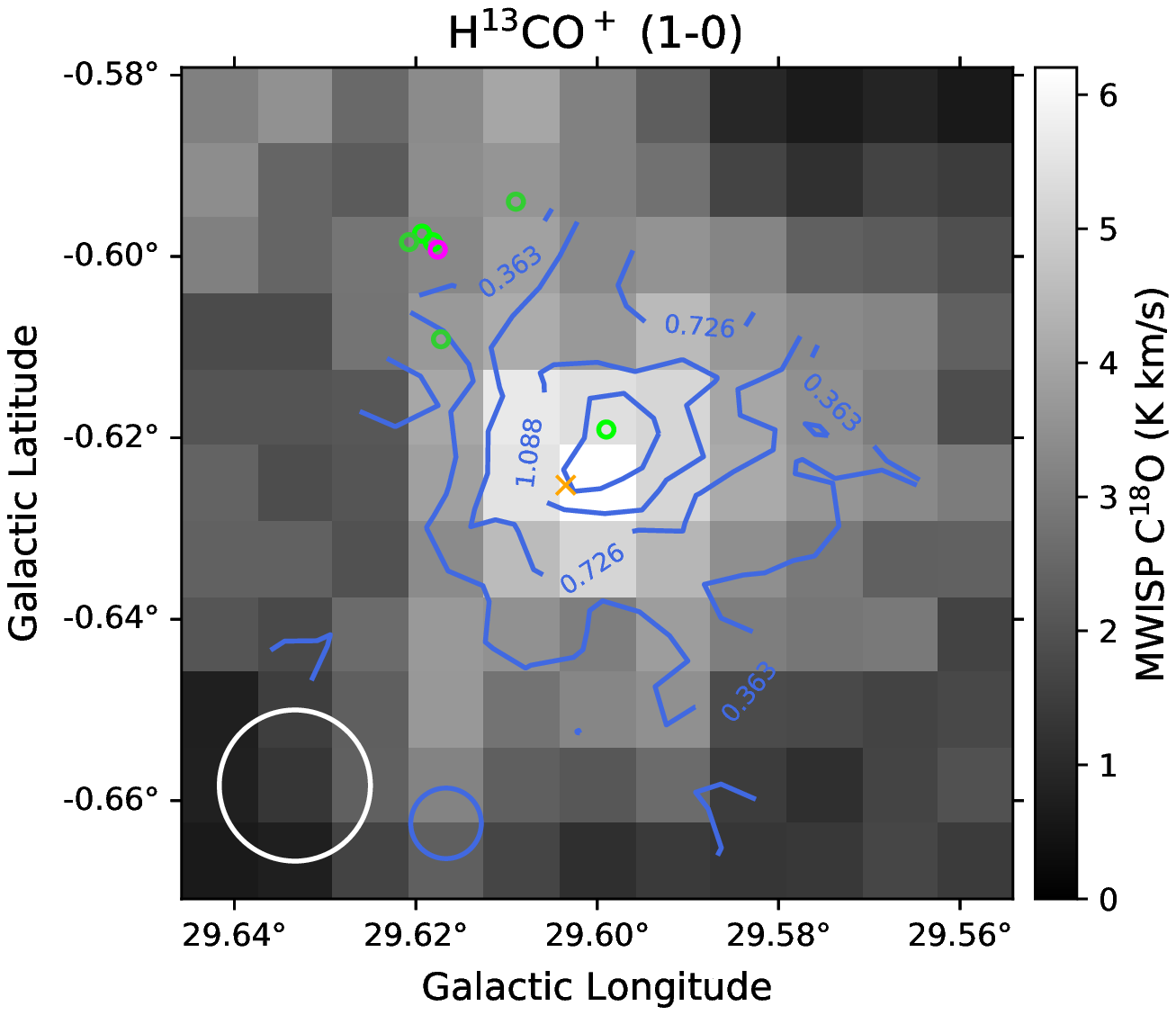}
\end{minipage}
}
\subfigure[G039.33-1.03]{
\begin{minipage}[b]{0.45\textwidth}
\includegraphics[width=1.1\textwidth]{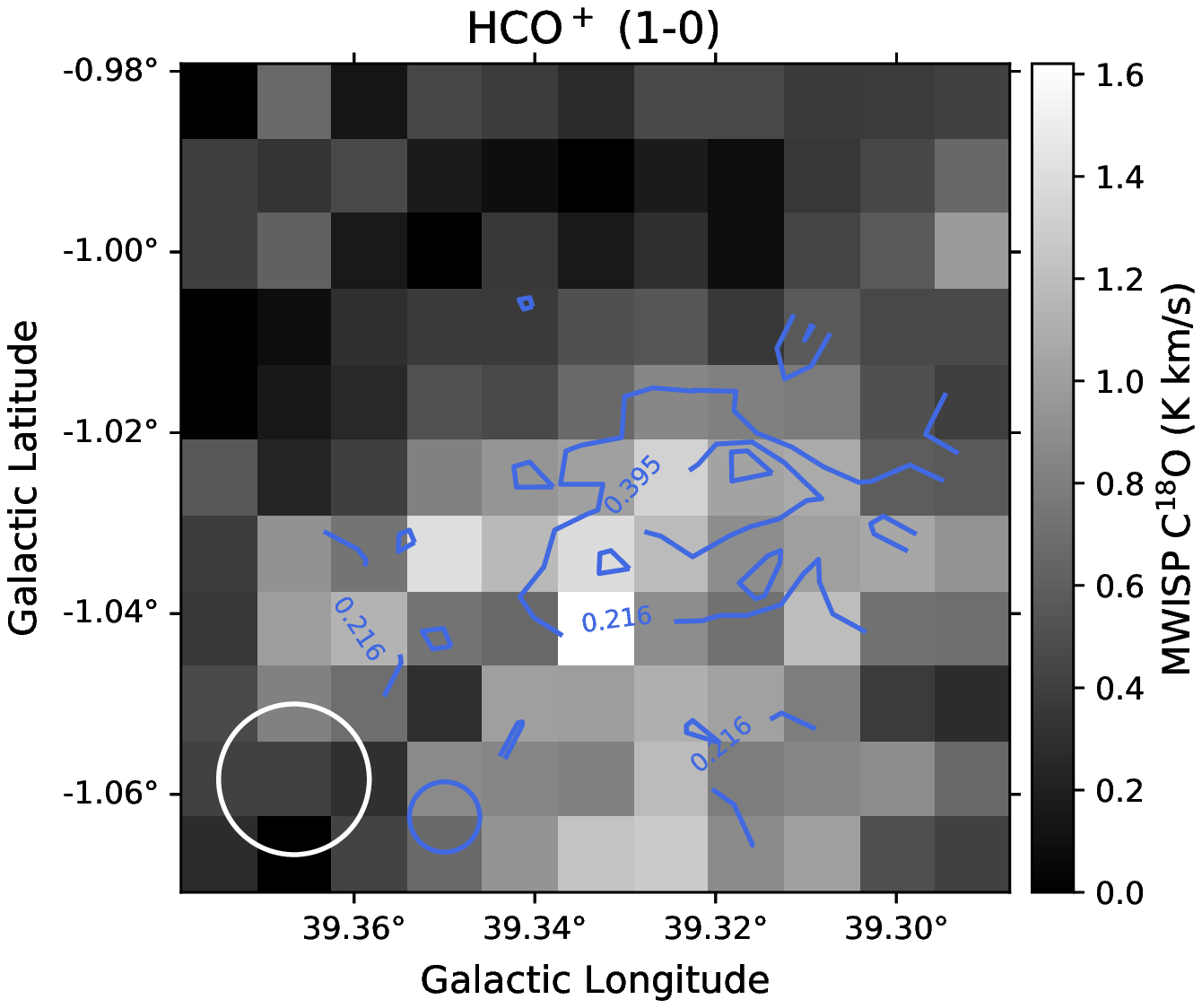}
\end{minipage}
}
\caption{HCO$^+$ (1-0) and H$^{13}$CO$^+$ (1-0) integrated intensity contour (blue) maps superposed on the C$^{18}$O (1-0) integrated intensity image( C$^{18}$O data from the MWISP project). The white and blue circles on the lower left indicate the beam sizes of the DLH 13.7-m telescope and the IRAM 30-m telescope, respectively. The green circles denote YSOs from AllWISE data. The magenta circles denote IRAS sources. The red squares denote the star-forming regions. The orange crosses denote the maser sources.}
\label{fig:map}
\end{figure}

\begin{figure}
\centering
\addtocounter{figure}{-1} 
\renewcommand{\thesubfigure} \makeatletter
\subfigure[G053.13+0.09]{
\begin{minipage}[b]{0.45\textwidth}
\includegraphics[width=1.1\textwidth]{./fig_map_eg_1.eps} 
\end{minipage}
\begin{minipage}[b]{0.45\textwidth}
\includegraphics[width=1.1\textwidth]{./fig_map_eg_2.eps}
\end{minipage}
}
\subfigure[G077.91-1.16]{
\begin{minipage}[b]{0.45\textwidth}
\includegraphics[width=1.1\textwidth]{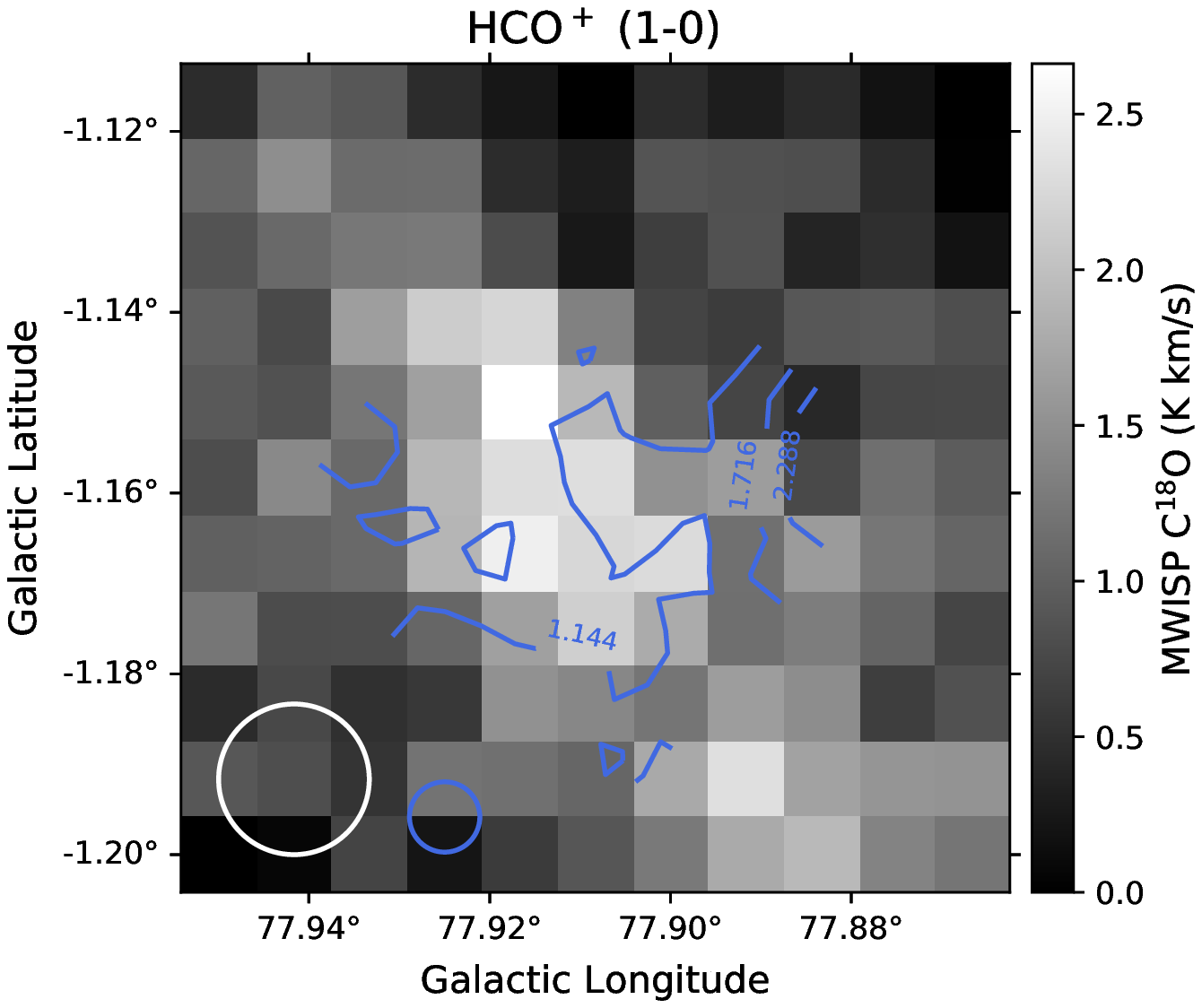} 
\end{minipage}
\begin{minipage}[b]{0.45\textwidth}
\includegraphics[width=1.1\textwidth]{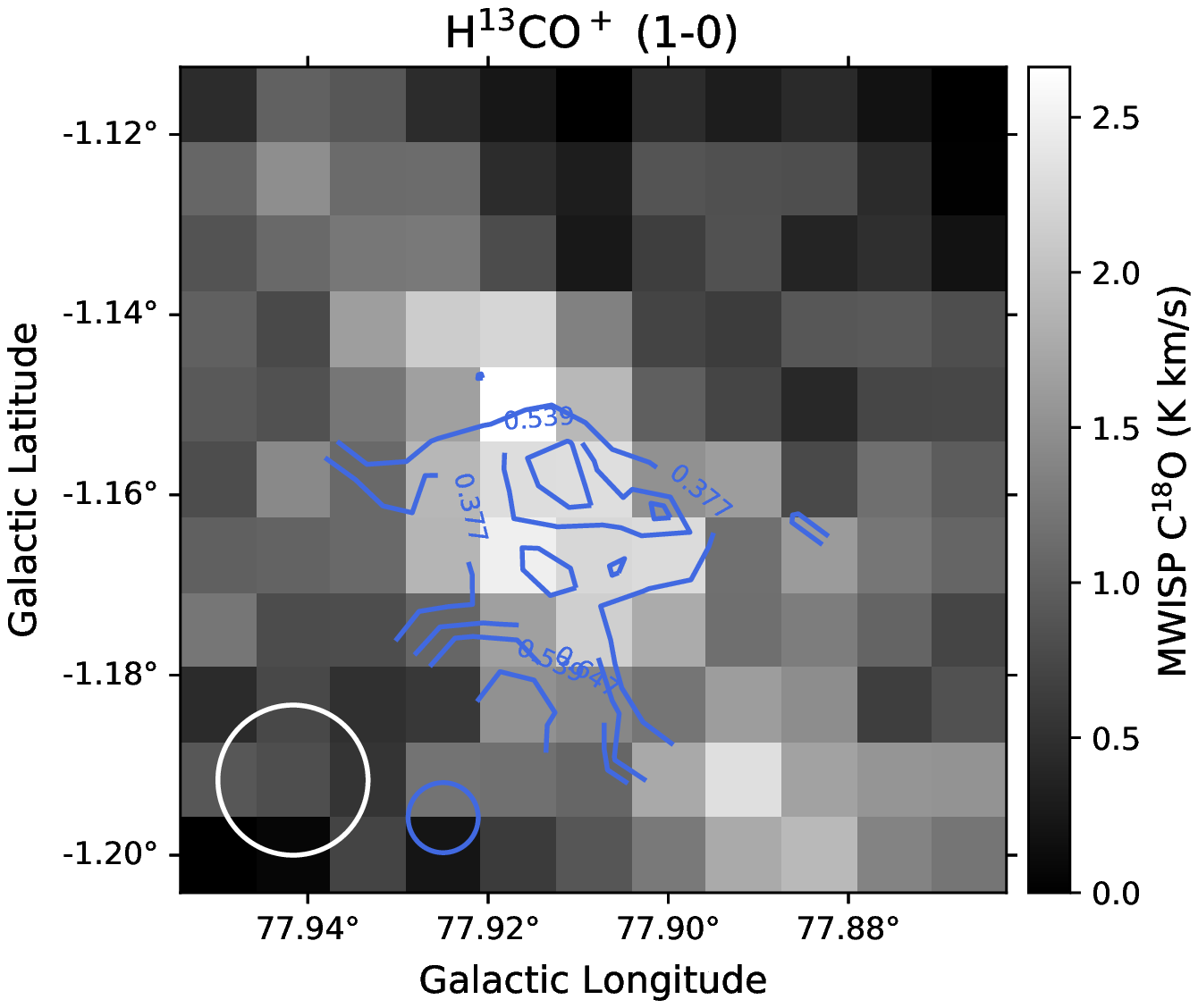}
\end{minipage}
}
\subfigure[G079.24+0.53]{
\begin{minipage}[b]{0.45\textwidth}
\includegraphics[width=1.1\textwidth]{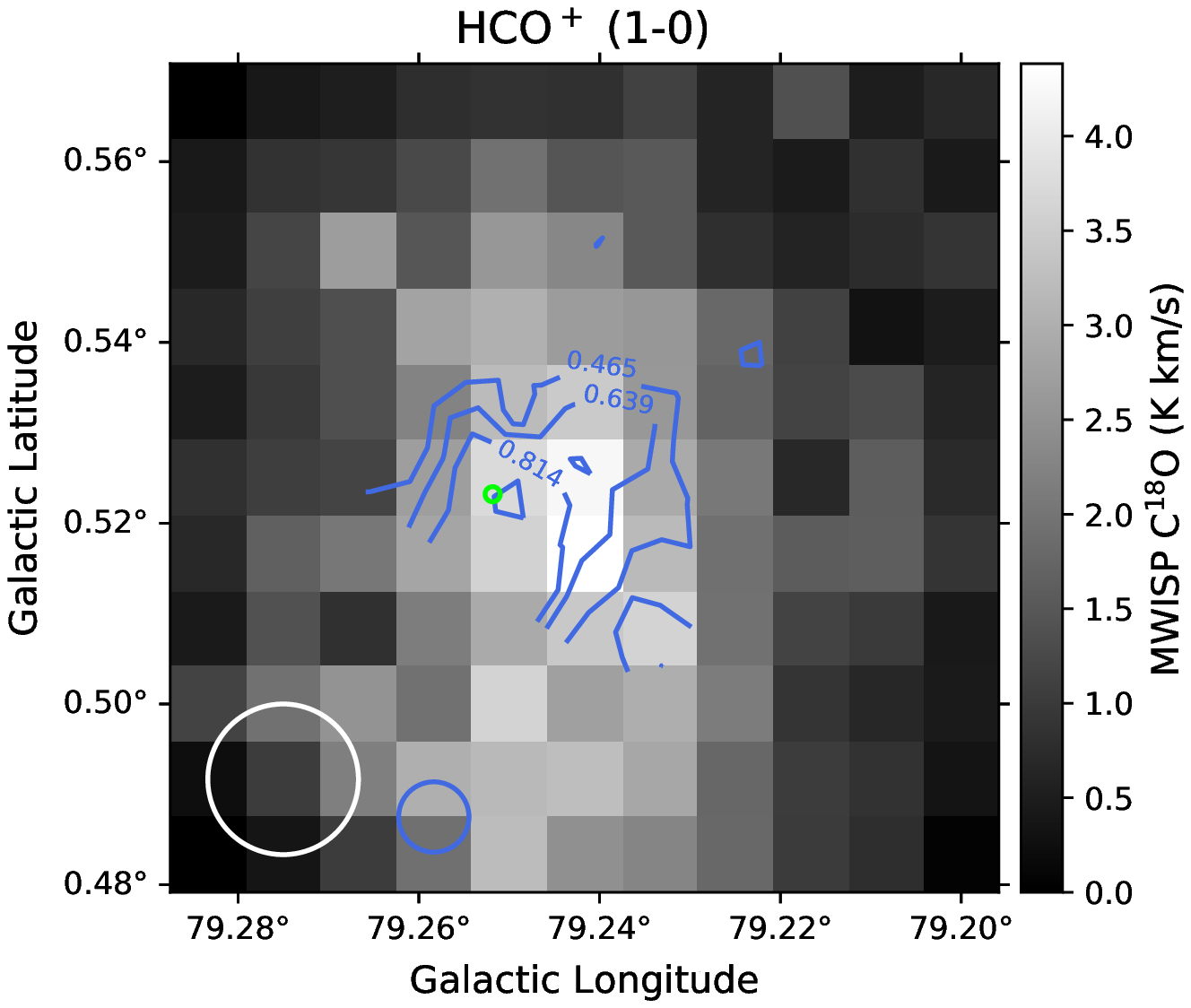} 
\end{minipage}
\begin{minipage}[b]{0.45\textwidth}
\includegraphics[width=1.1\textwidth]{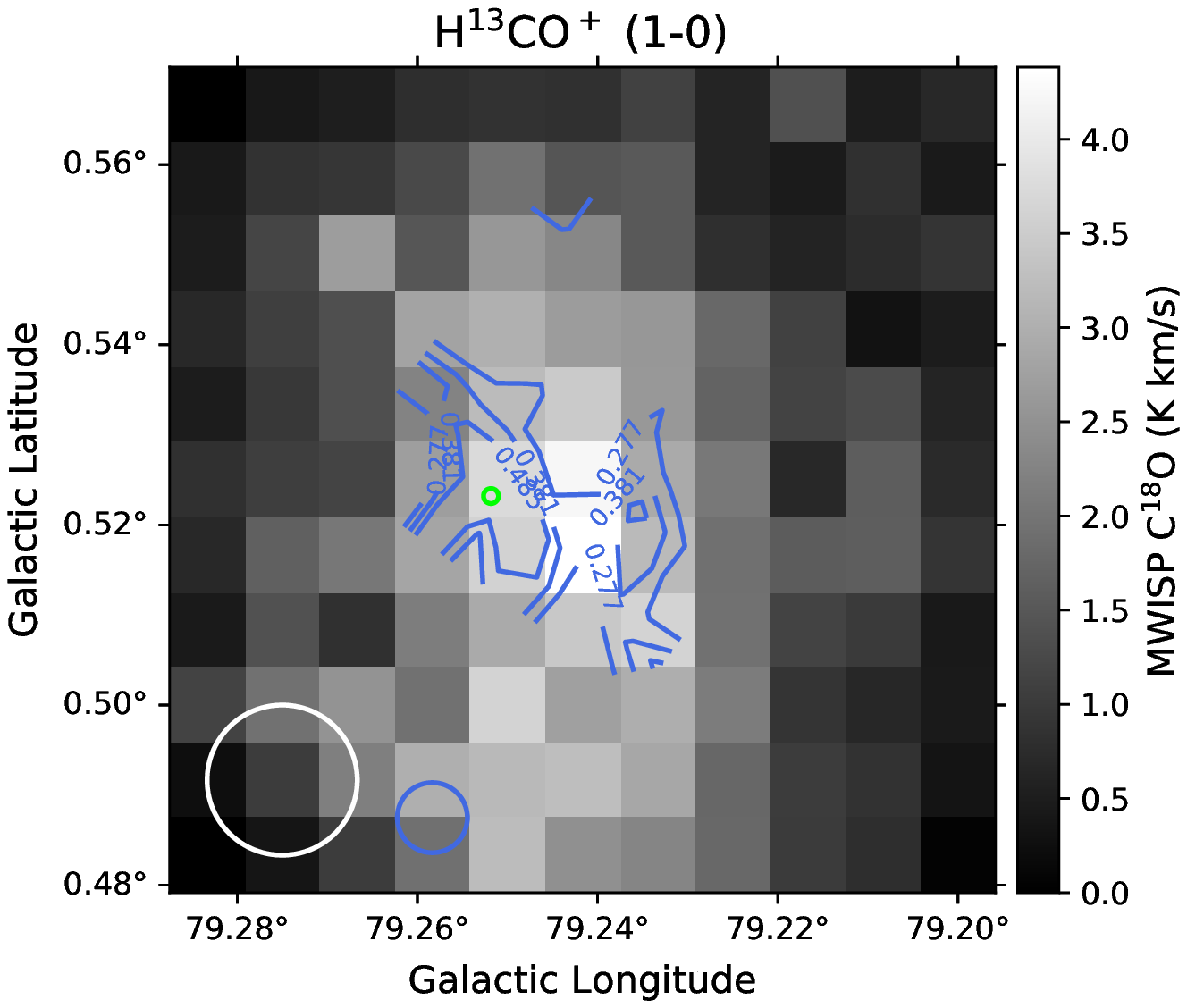}
\end{minipage}
}
\caption{Continued.}
\label{fig:map}
\end{figure}

\begin{figure}
\centering
\addtocounter{figure}{-1} 
\renewcommand{\thesubfigure} \makeatletter
\subfigure[G079.71+0.14]{
\begin{minipage}[b]{0.45\textwidth}
\includegraphics[width=1.1\textwidth]{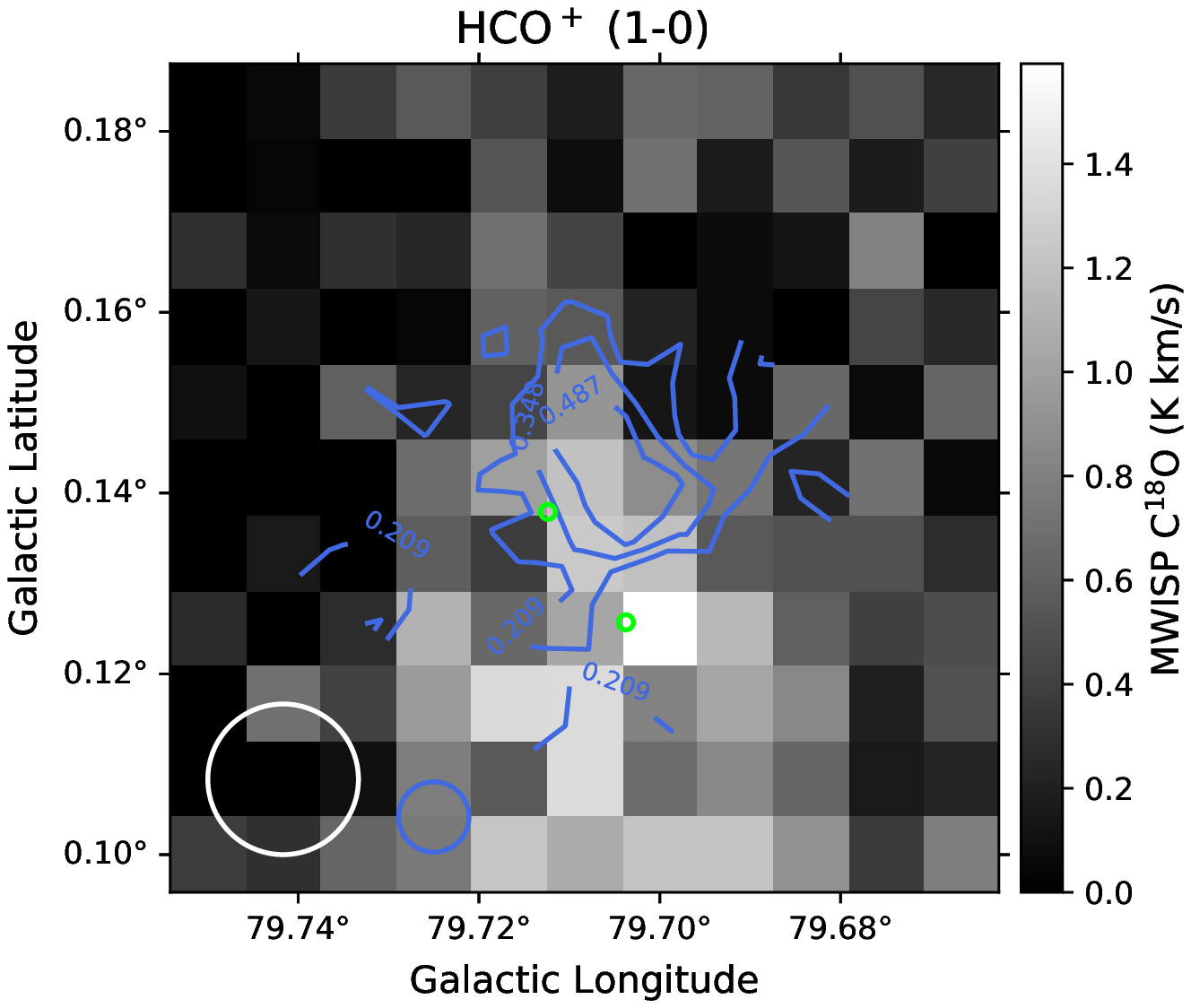} 
\end{minipage}
\begin{minipage}[b]{0.45\textwidth}
\includegraphics[width=1.1\textwidth]{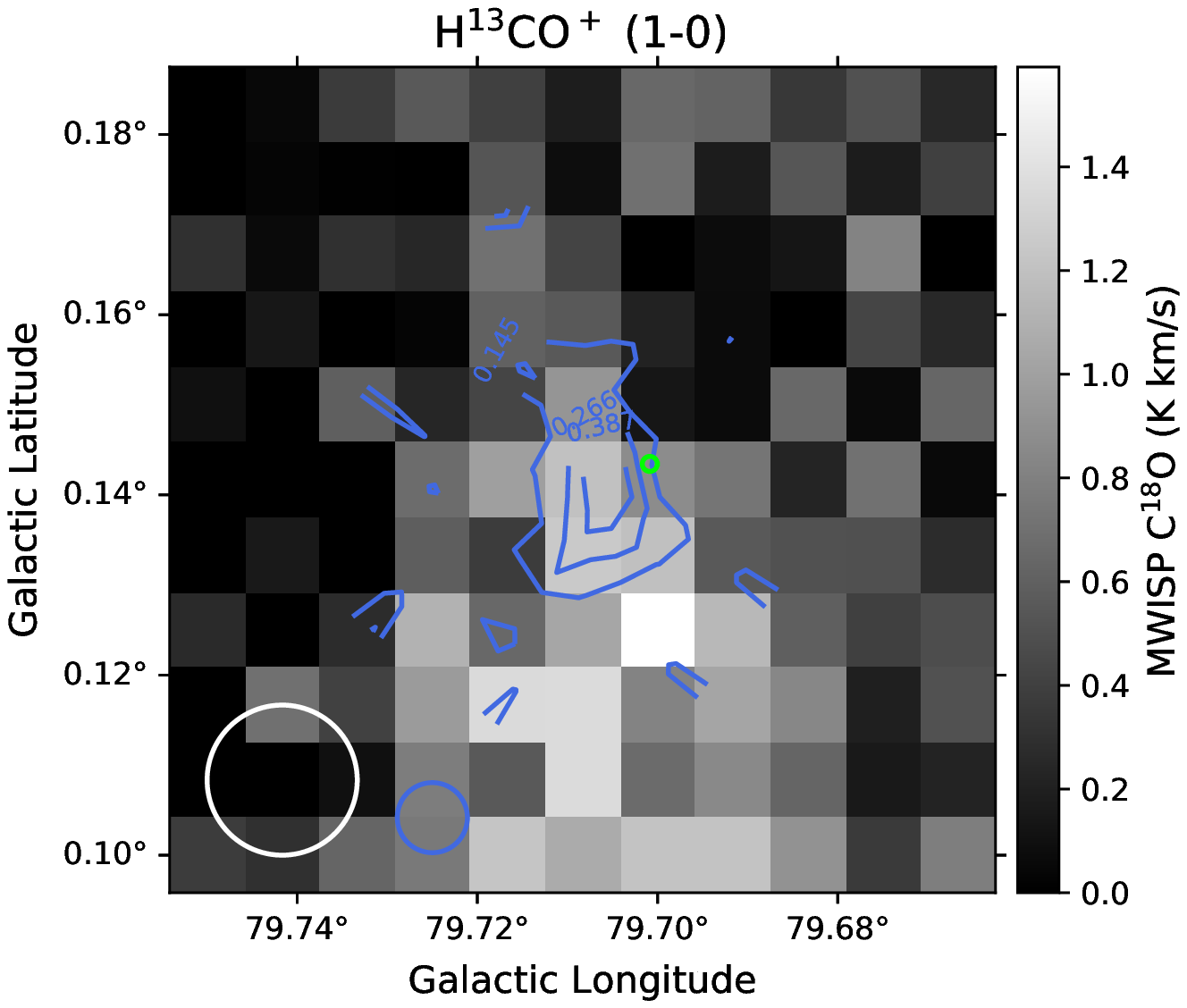}
\end{minipage}
}
\subfigure[G081.72+0.57]{
\begin{minipage}[b]{0.45\textwidth}
\includegraphics[width=1.1\textwidth]{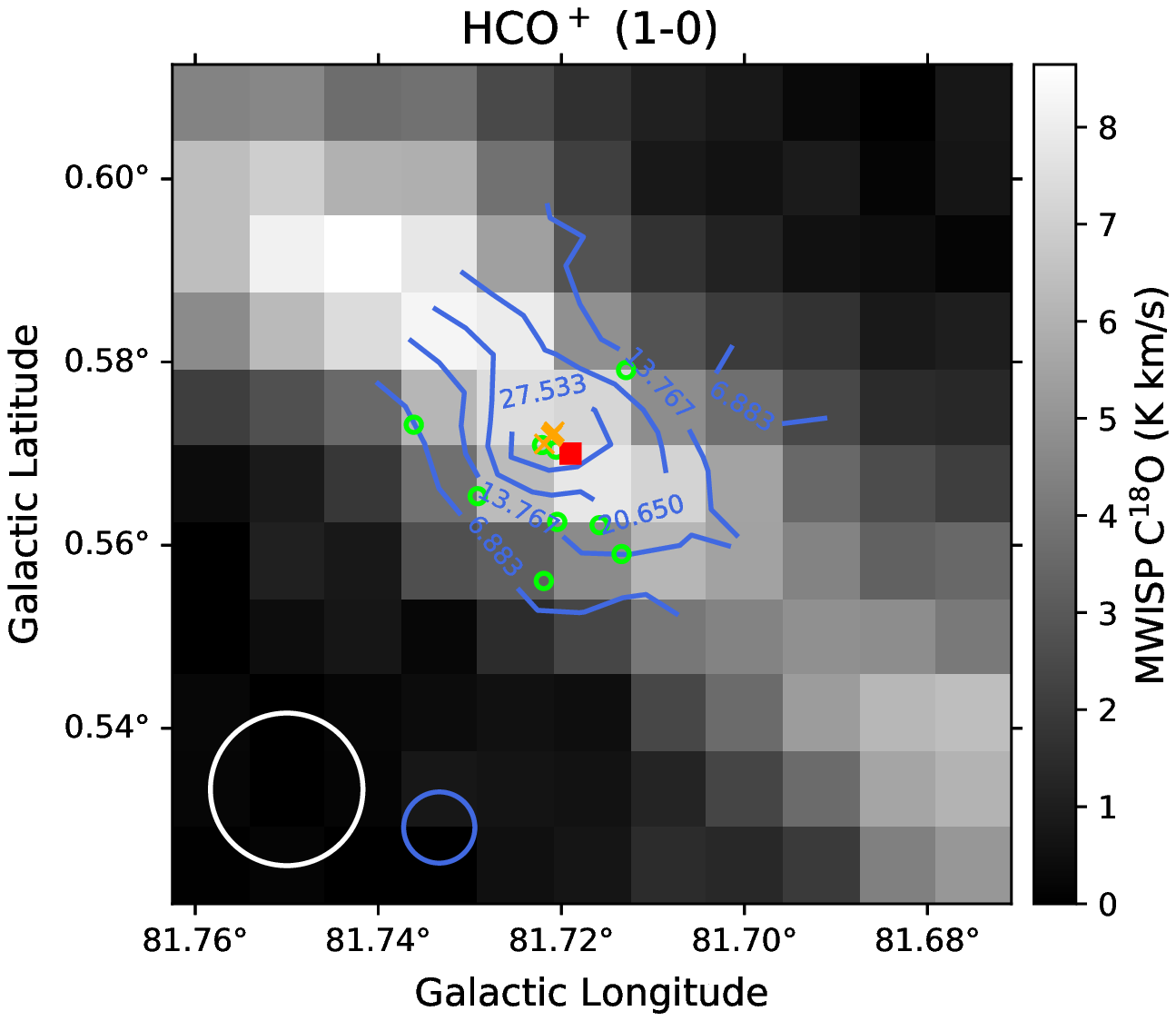} 
\end{minipage}
\begin{minipage}[b]{0.45\textwidth}
\includegraphics[width=1.1\textwidth]{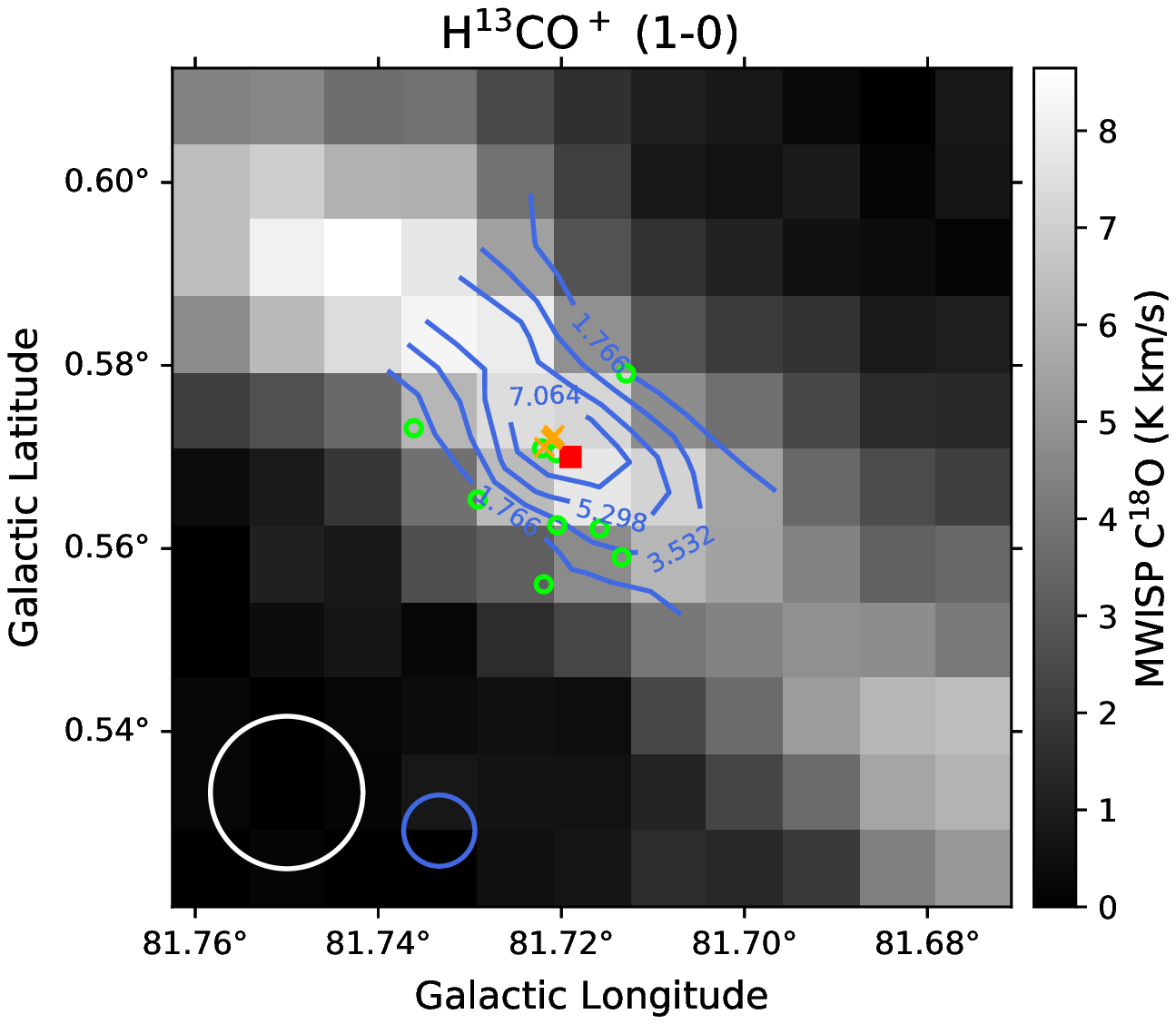}
\end{minipage}
}
\subfigure[G082.21-1.53]{
\begin{minipage}[b]{0.45\textwidth}
\includegraphics[width=1.1\textwidth]{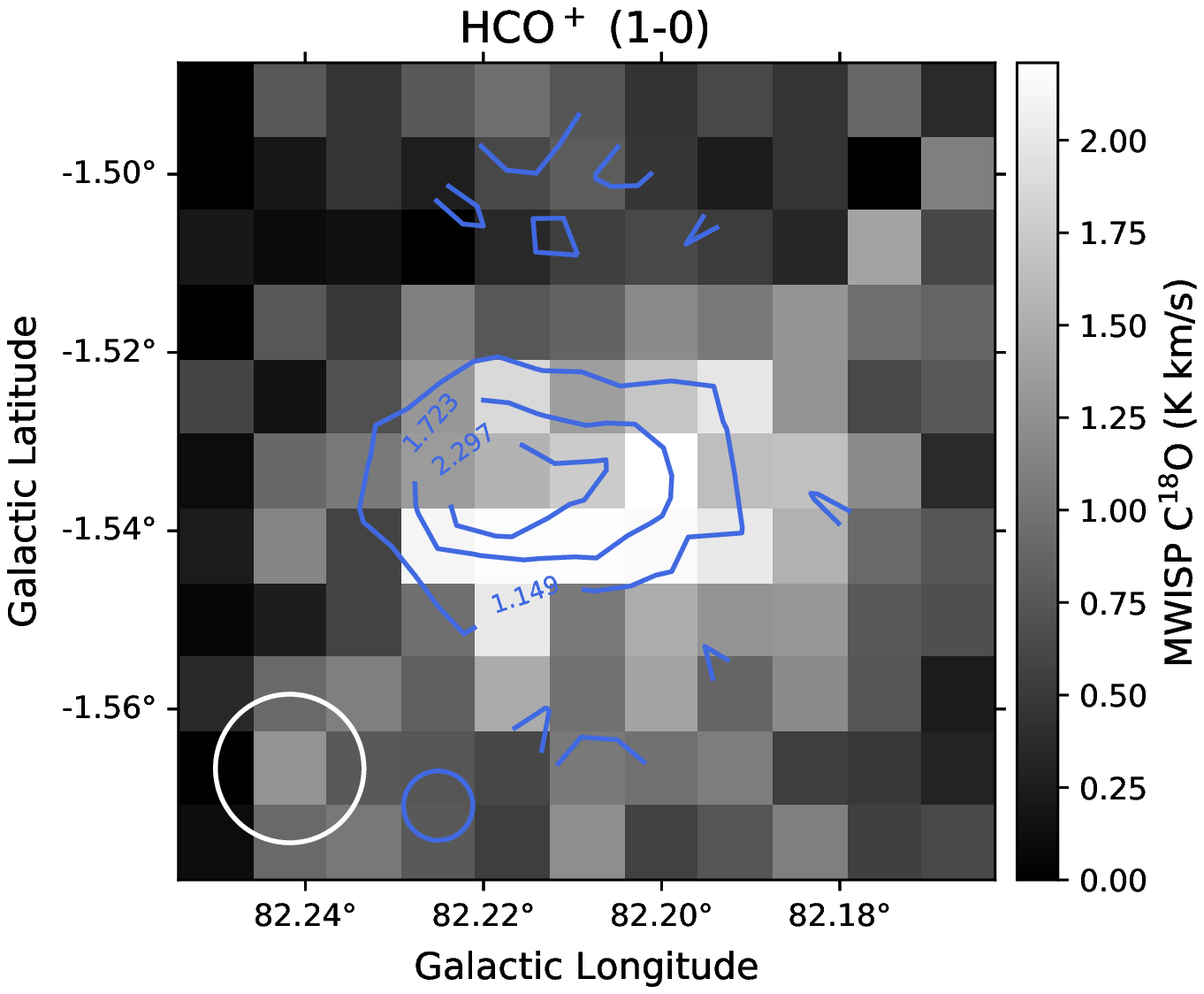} 
\end{minipage}
\begin{minipage}[b]{0.45\textwidth}
\includegraphics[width=1.1\textwidth]{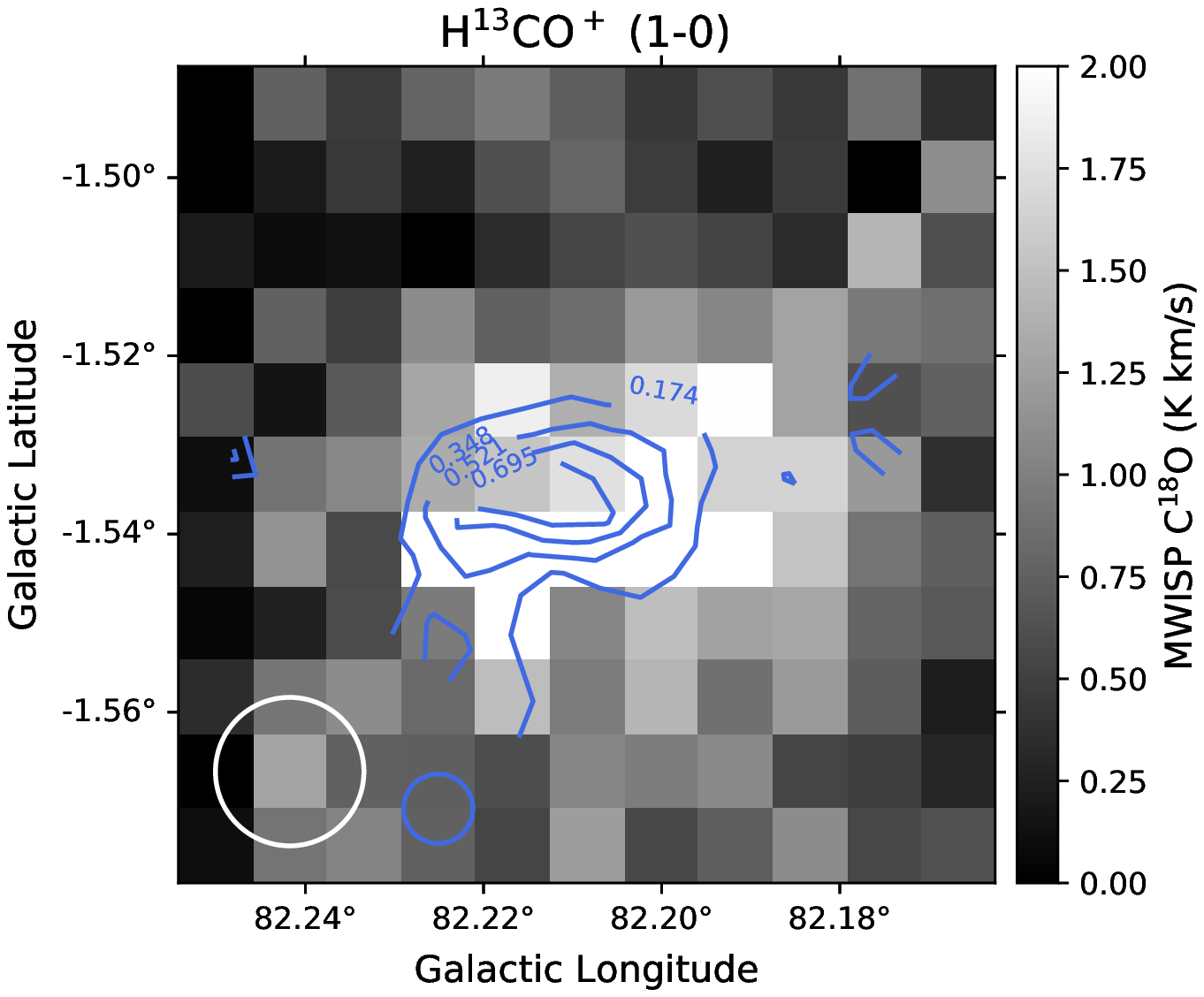}
\end{minipage}
}
\caption{Continued.}
\label{fig:map}
\end{figure}

\begin{figure}
\centering
\addtocounter{figure}{-1} 
\renewcommand{\thesubfigure} \makeatletter
\subfigure[G107.50+4.47]{
\begin{minipage}[b]{0.45\textwidth}
\includegraphics[width=1.1\textwidth]{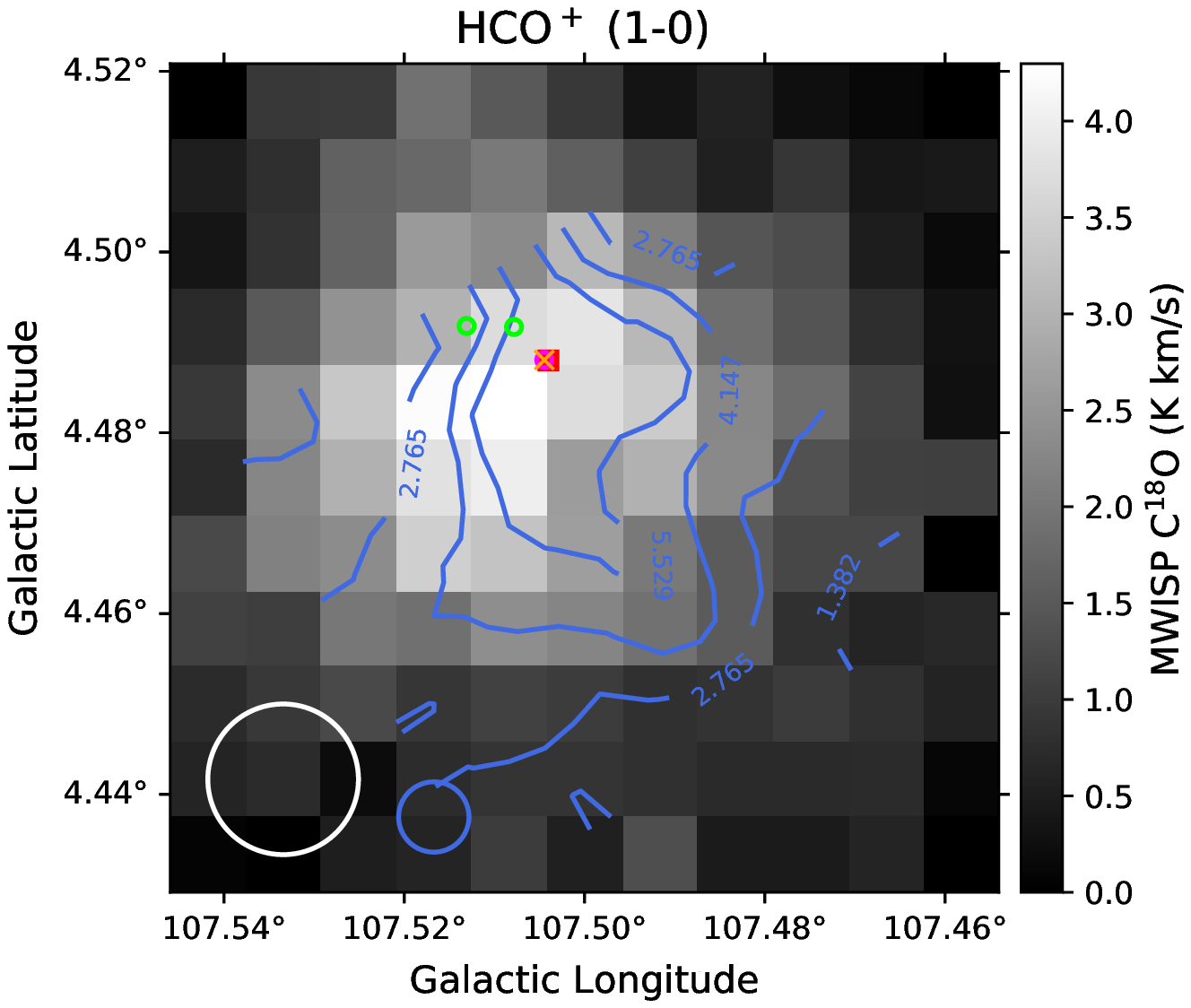} 
\end{minipage}
\begin{minipage}[b]{0.45\textwidth}
\includegraphics[width=1.1\textwidth]{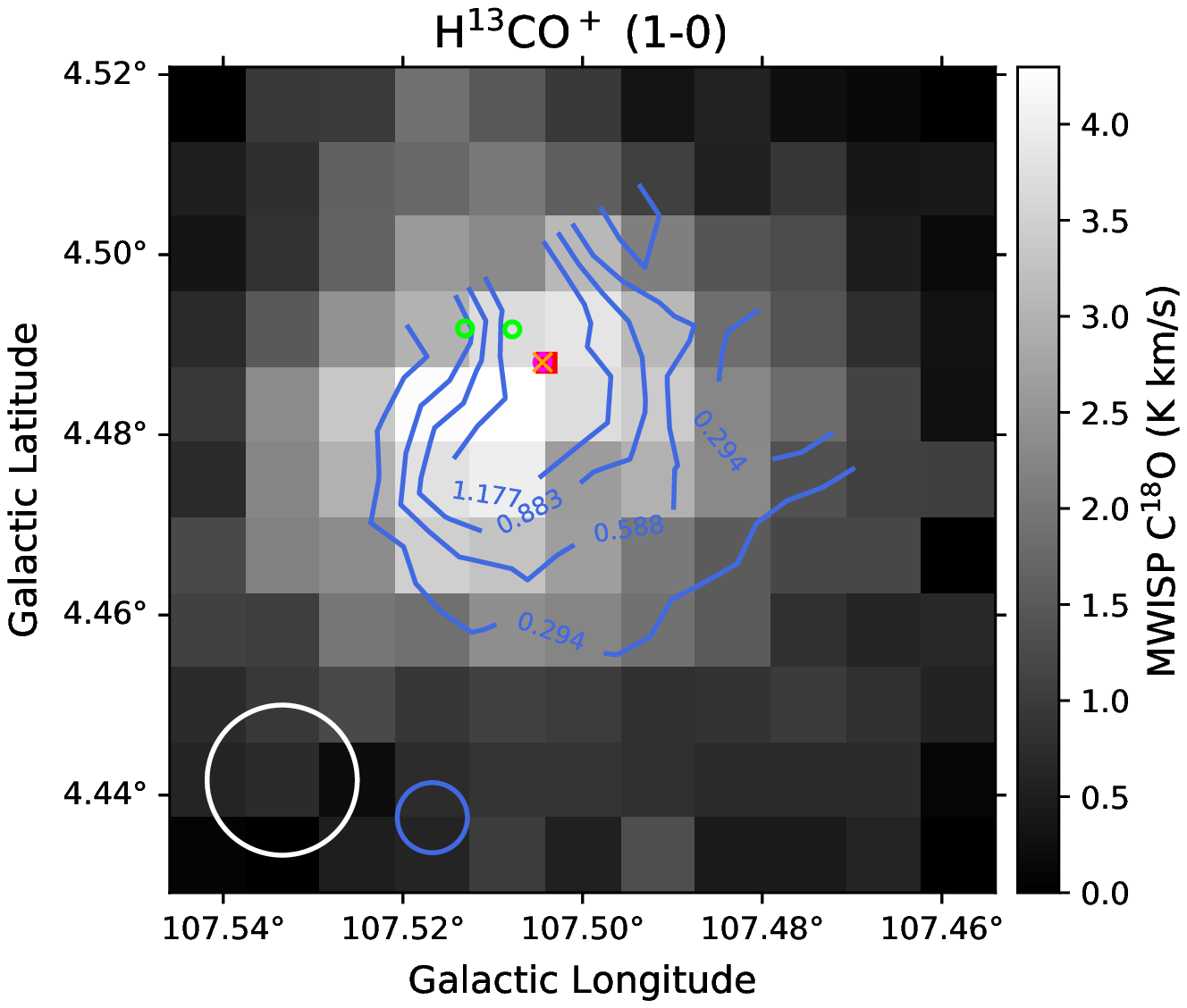}
\end{minipage}
}
\subfigure[G109.00+2.73]{
\begin{minipage}[b]{0.45\textwidth}
\includegraphics[width=1.1\textwidth]{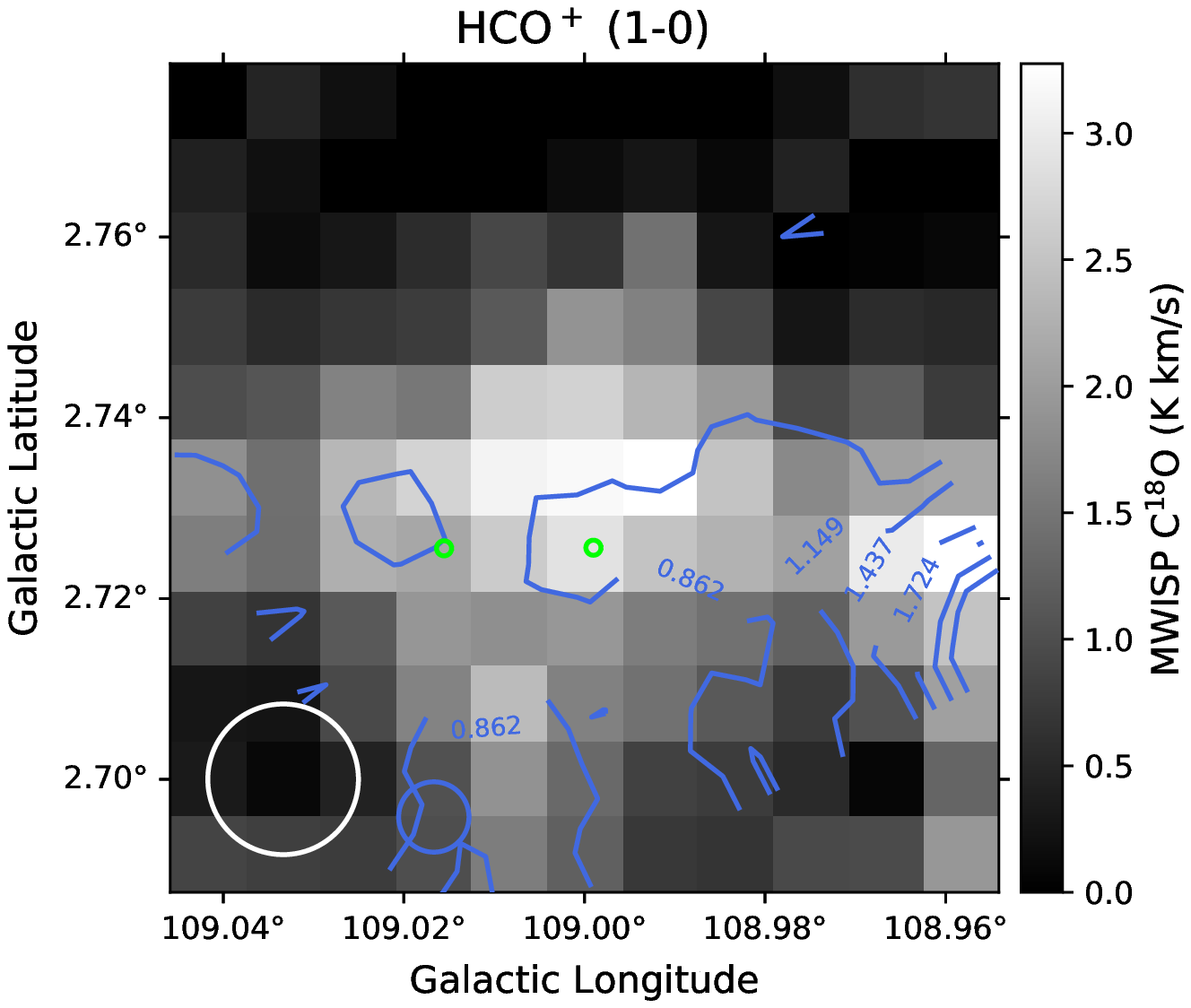}
\end{minipage}
\begin{minipage}[b]{0.45\textwidth}
\includegraphics[width=1.1\textwidth]{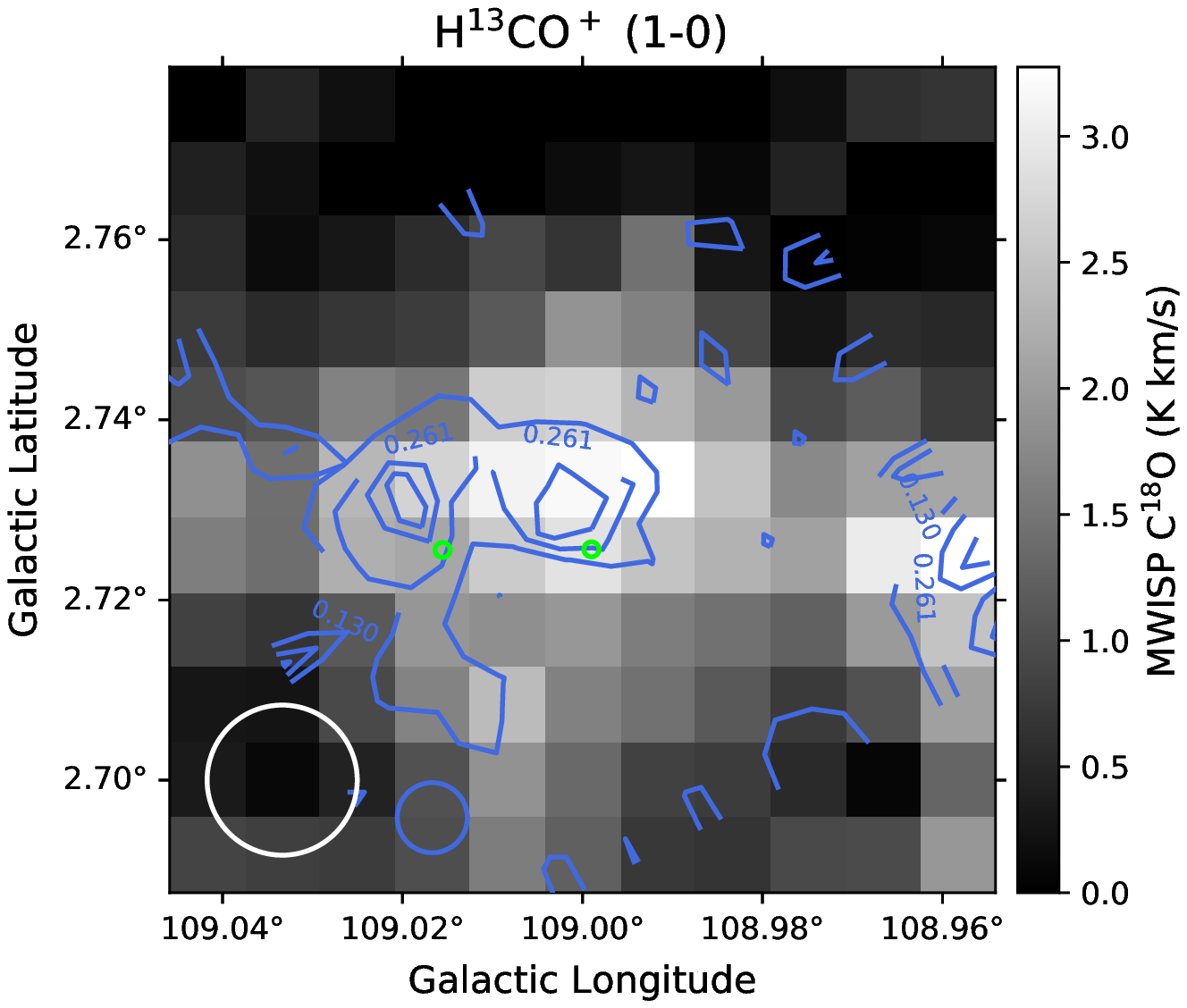}
\end{minipage}
}
\subfigure[G110.31+2.53]{
\begin{minipage}[b]{0.45\textwidth}
\includegraphics[width=1.1\textwidth]{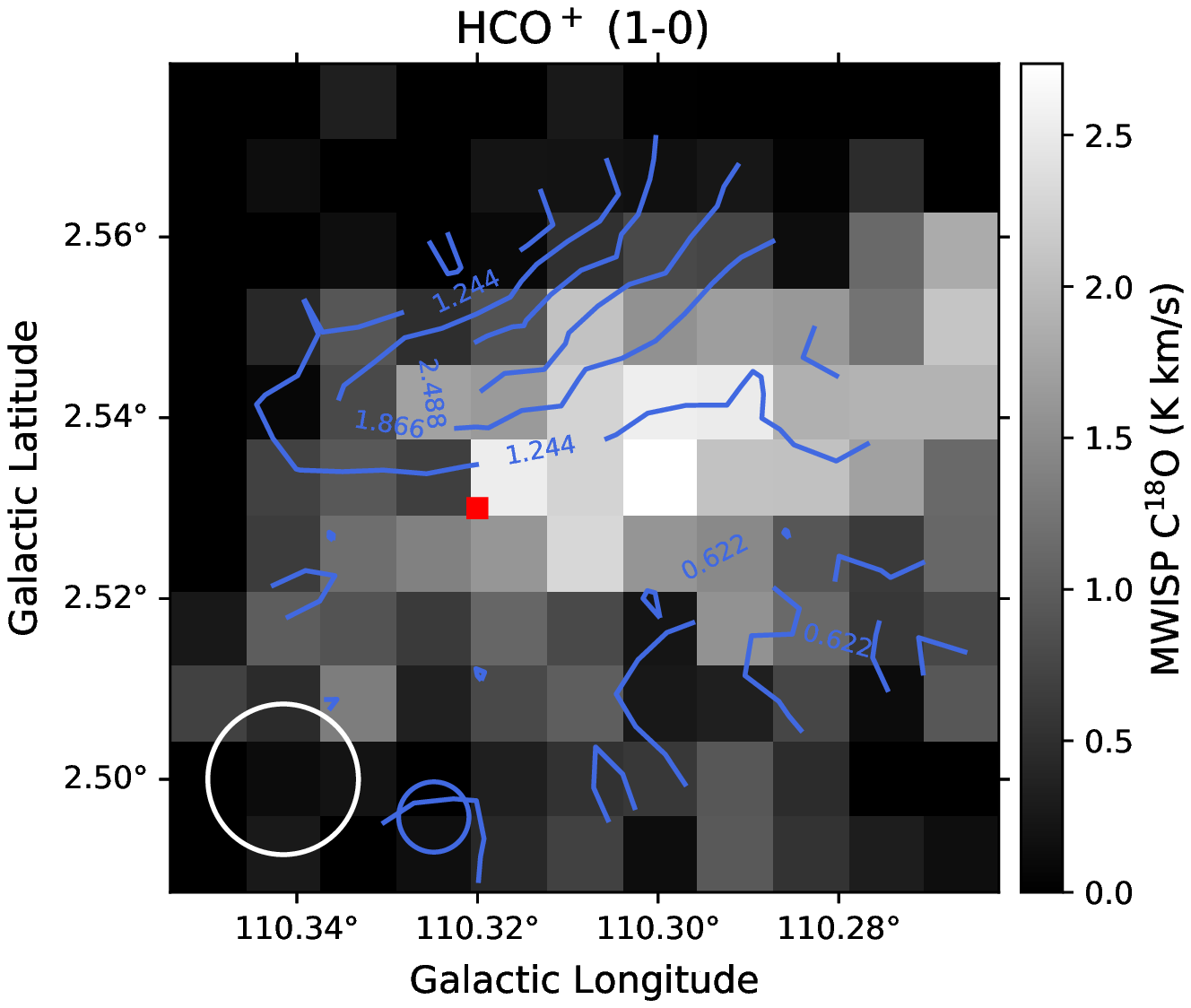} 
\end{minipage}
}
\caption{Continued.}
\label{fig:map}
\end{figure}

\begin{figure}
\centering
\addtocounter{figure}{-1} 
\renewcommand{\thesubfigure} \makeatletter
\subfigure[G121.31+0.64]{
\begin{minipage}[b]{0.45\textwidth}
\includegraphics[width=1.1\textwidth]{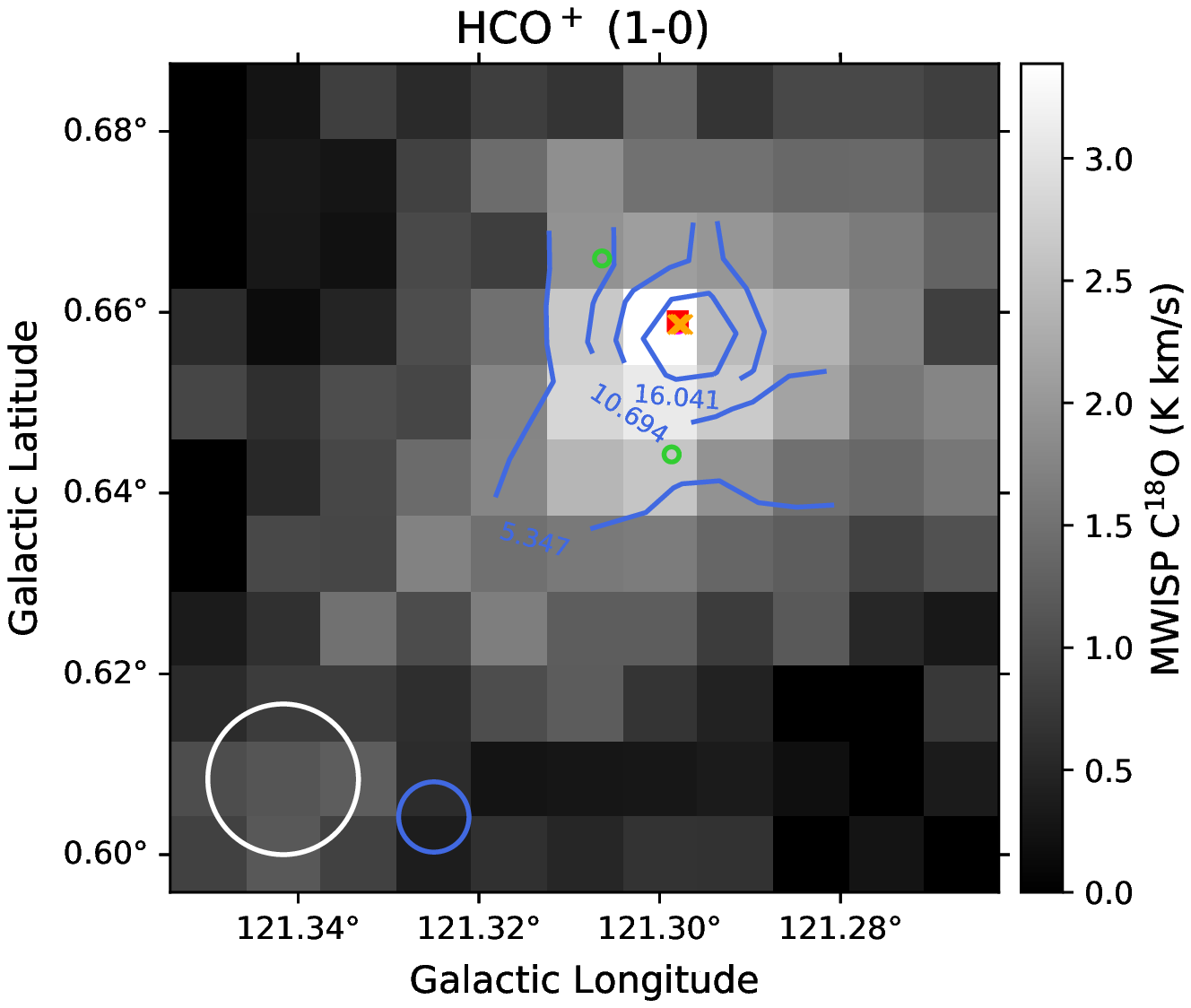} 
\end{minipage}
\begin{minipage}[b]{0.45\textwidth}
\includegraphics[width=1.1\textwidth]{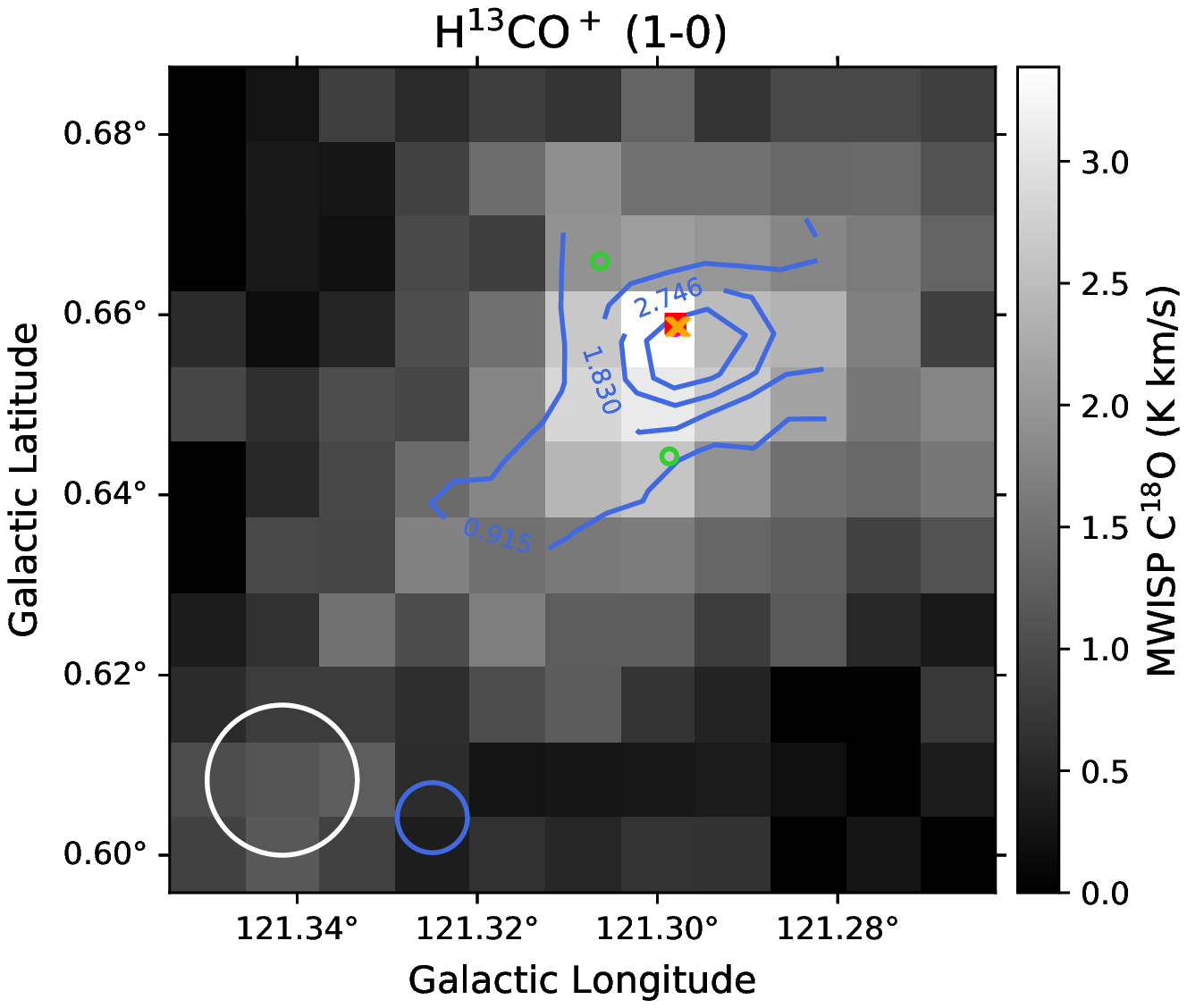}
\end{minipage}
}
\subfigure[G126.53-1.17]{
\begin{minipage}[b]{0.45\textwidth}
\includegraphics[width=1.1\textwidth]{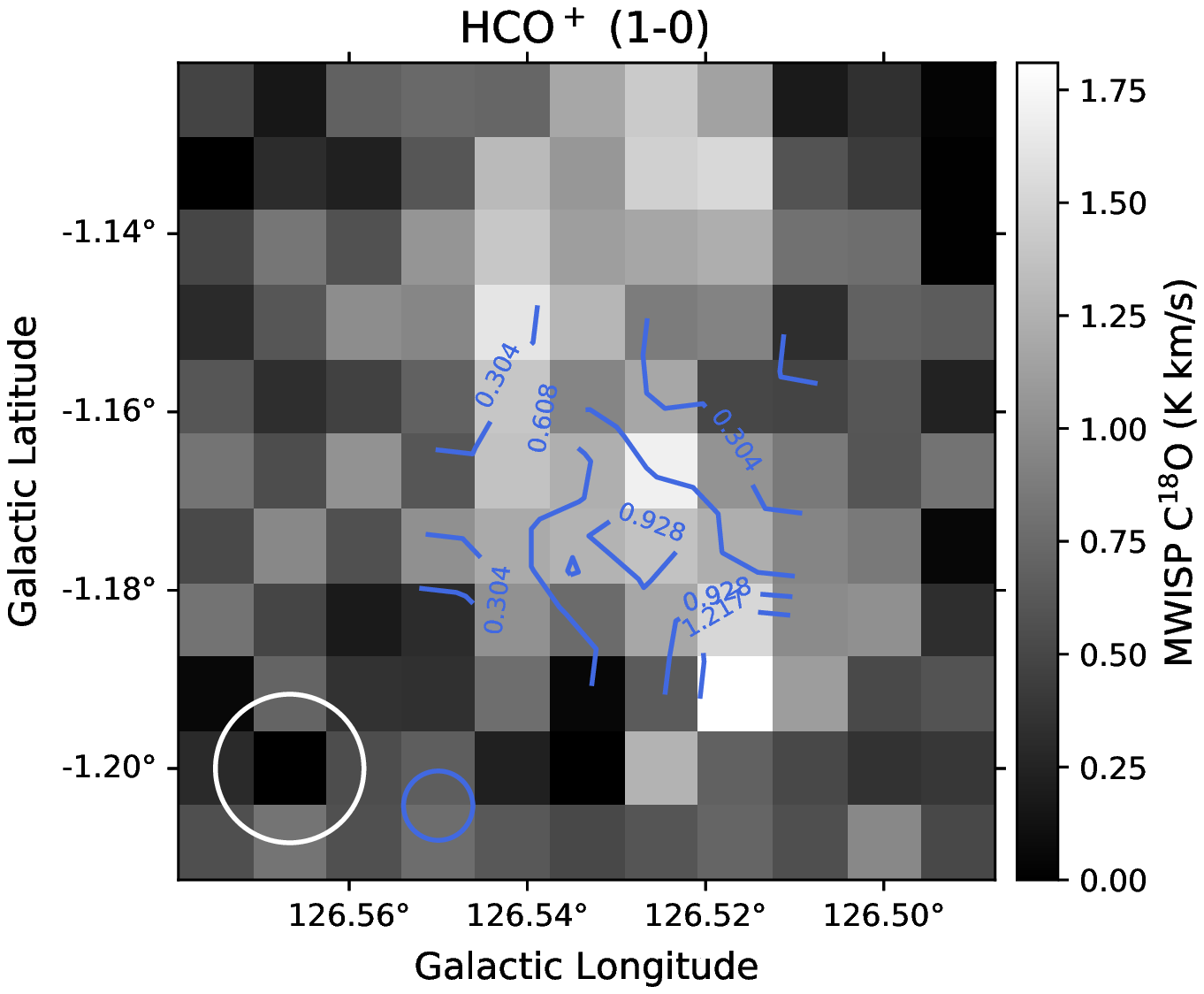} 
\end{minipage}
\begin{minipage}[b]{0.45\textwidth}
\includegraphics[width=1.1\textwidth]{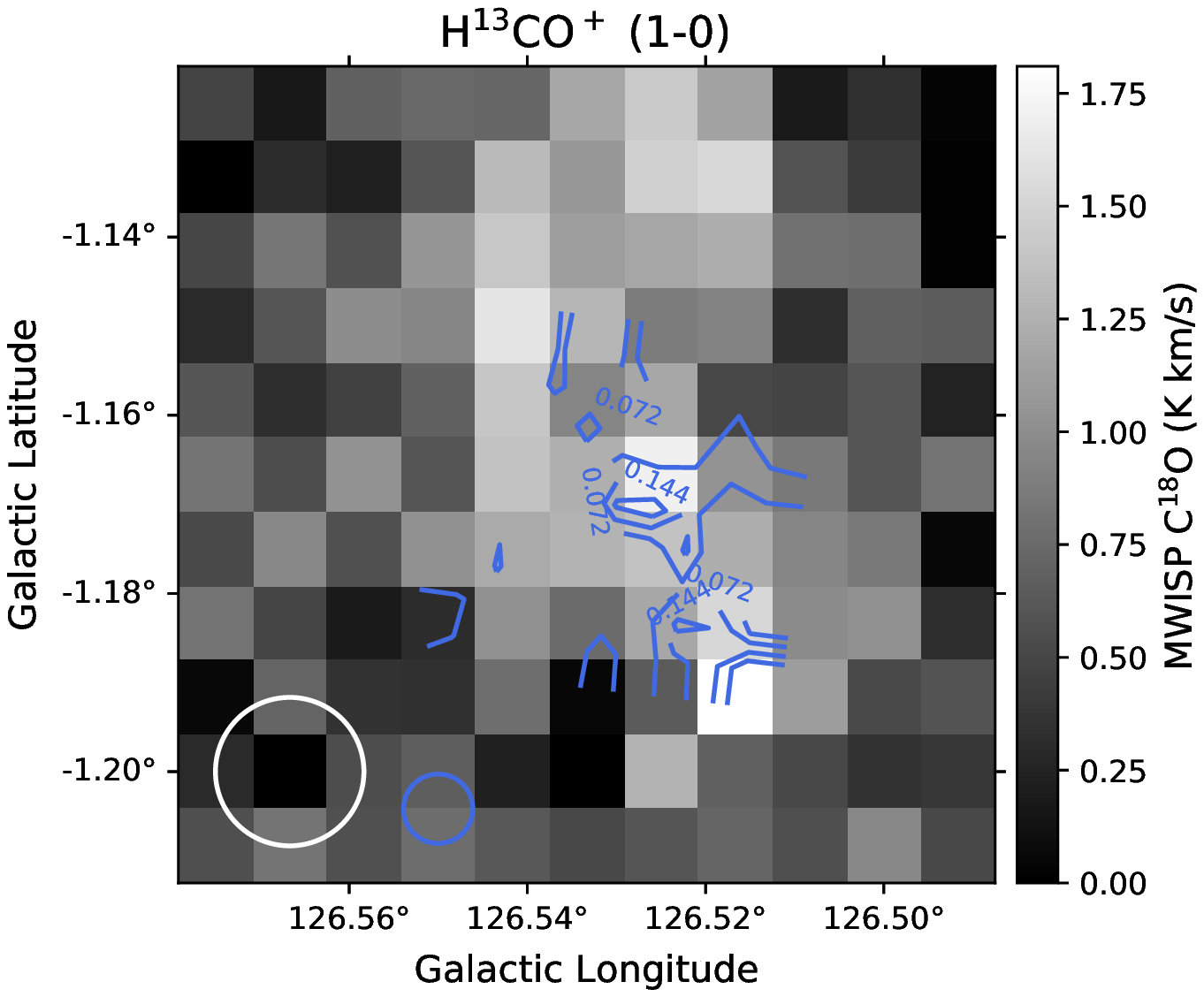}
\end{minipage}
}
\subfigure[G143.04+1.74]{
\begin{minipage}[b]{0.45\textwidth}
\includegraphics[width=1.1\textwidth]{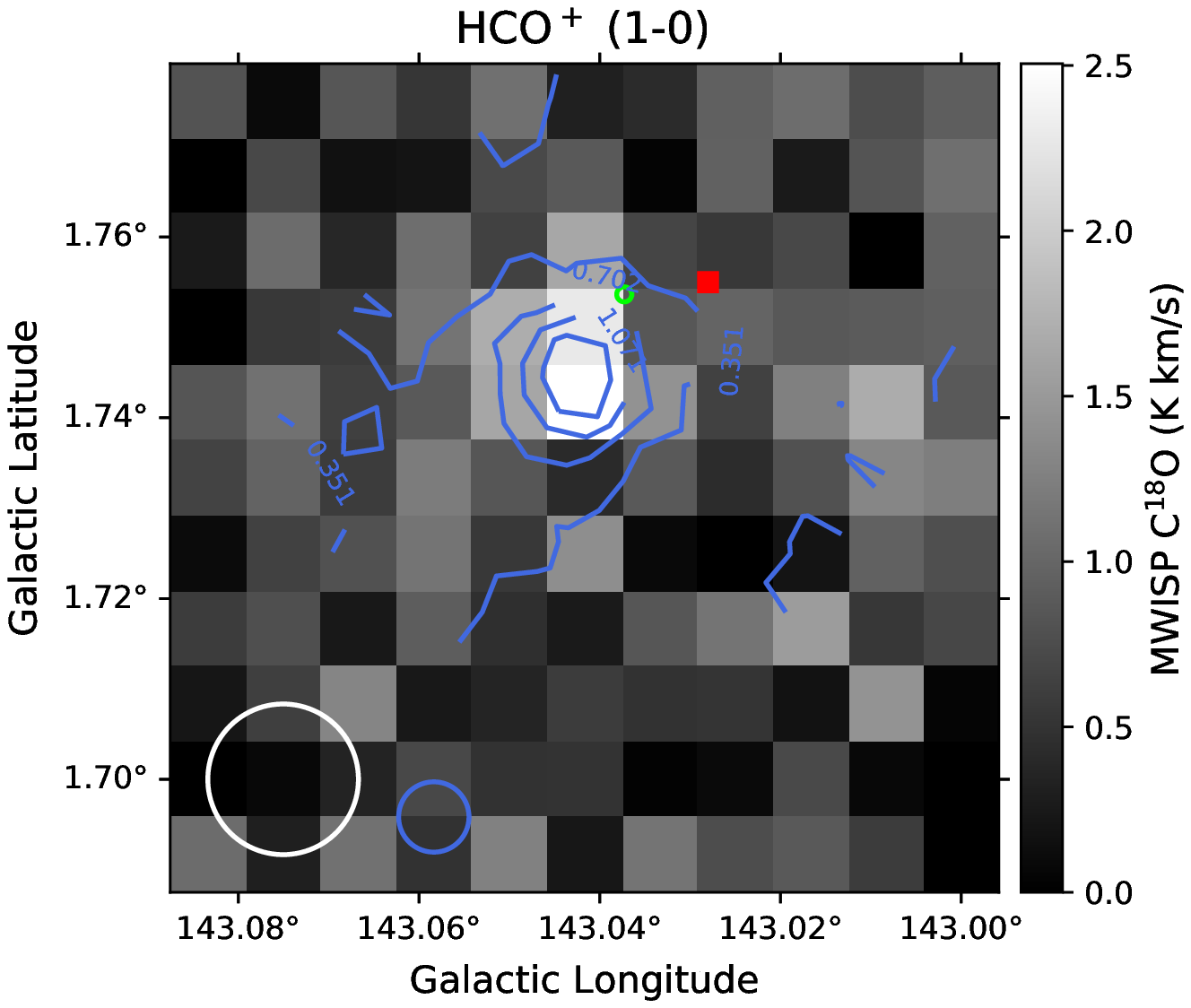} 
\end{minipage}
\begin{minipage}[b]{0.45\textwidth}
\includegraphics[width=1.1\textwidth]{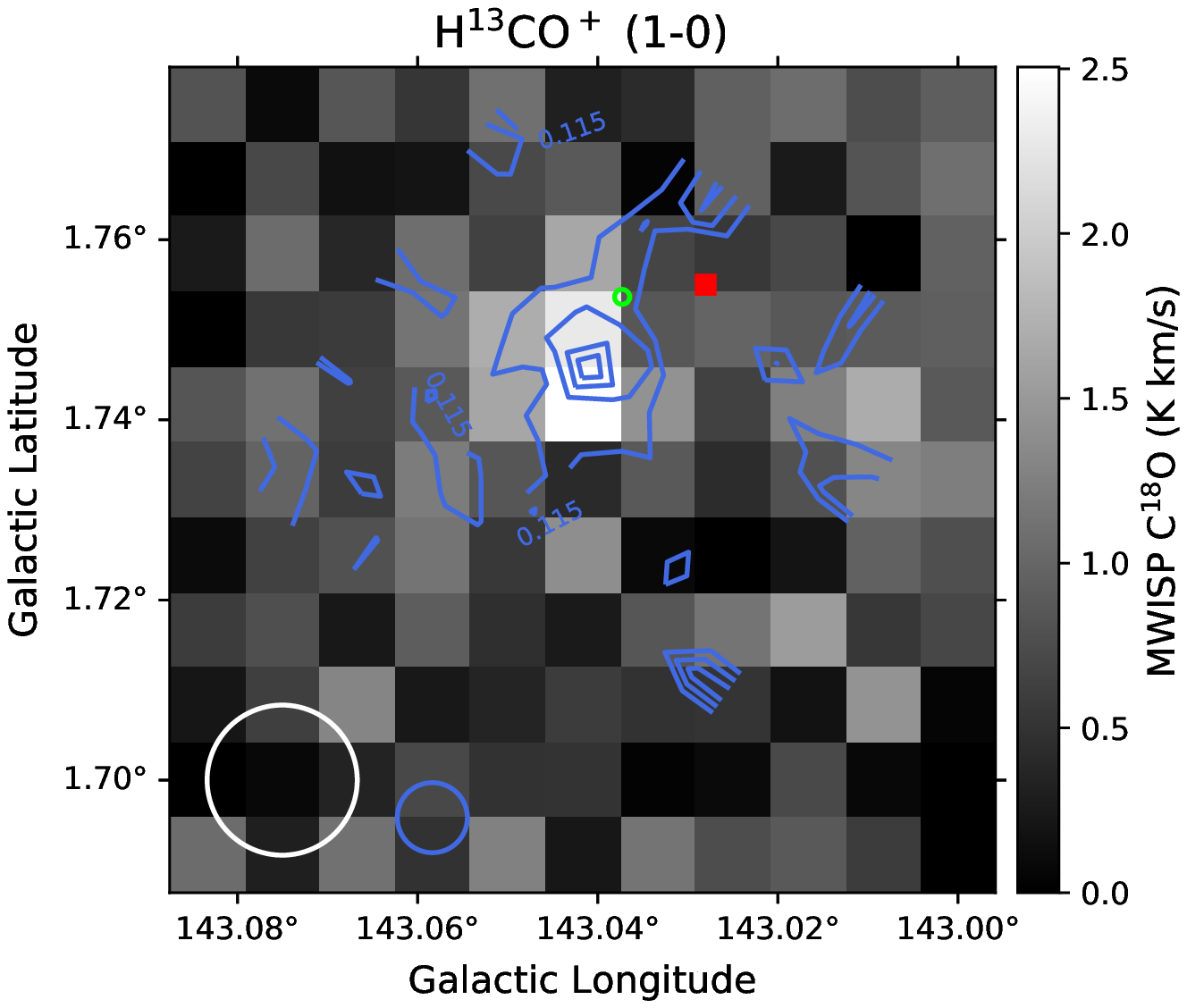}
\end{minipage}
}
\caption{Continued.}
\label{fig:map}
\end{figure}

\begin{figure}
\centering
\addtocounter{figure}{-1} 
\renewcommand{\thesubfigure} \makeatletter
\subfigure[G193.01+0.14]{
\begin{minipage}[b]{0.45\textwidth}
\includegraphics[width=1.1\textwidth]{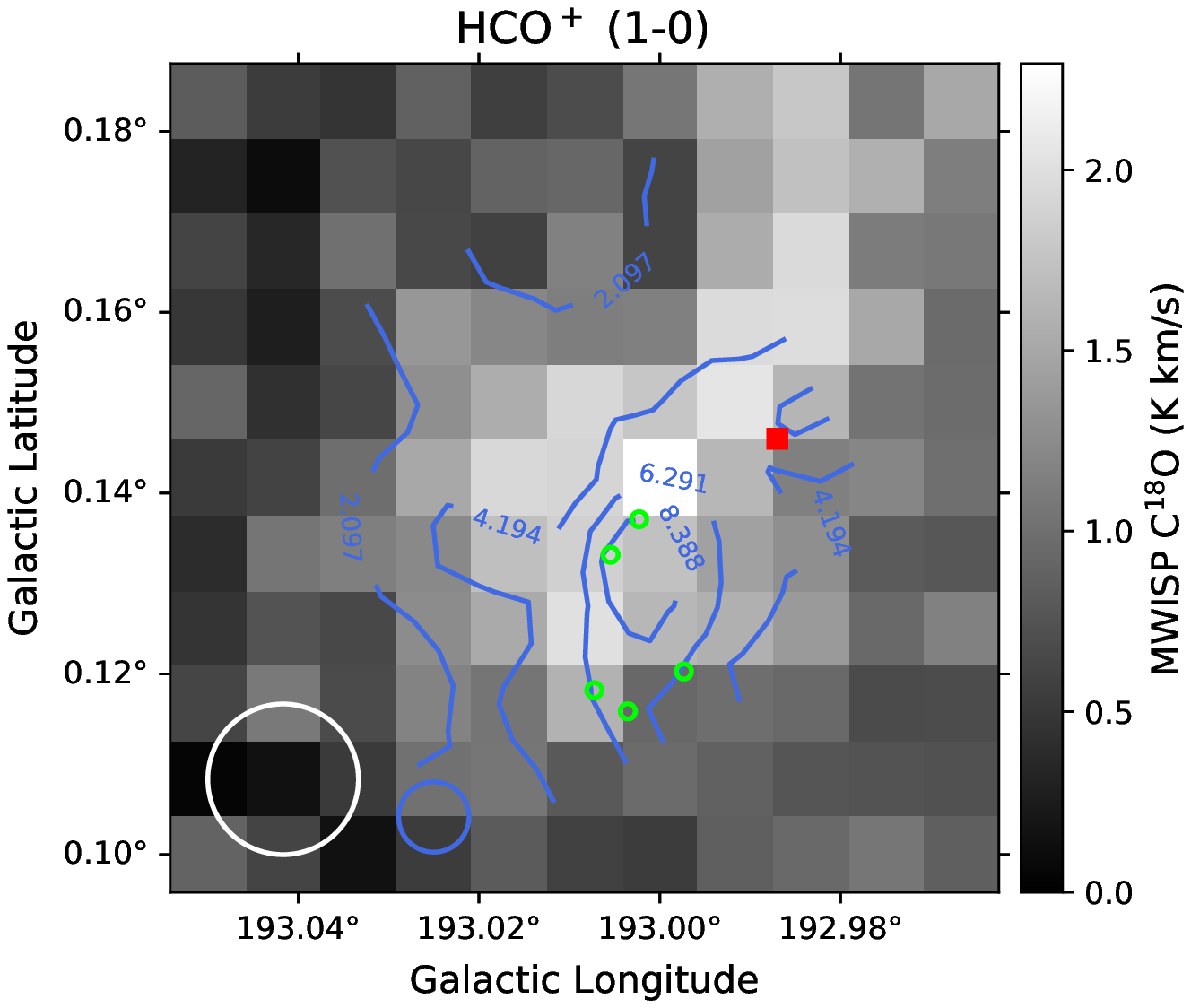}
\end{minipage}
\begin{minipage}[b]{0.45\textwidth}
\includegraphics[width=1.1\textwidth]{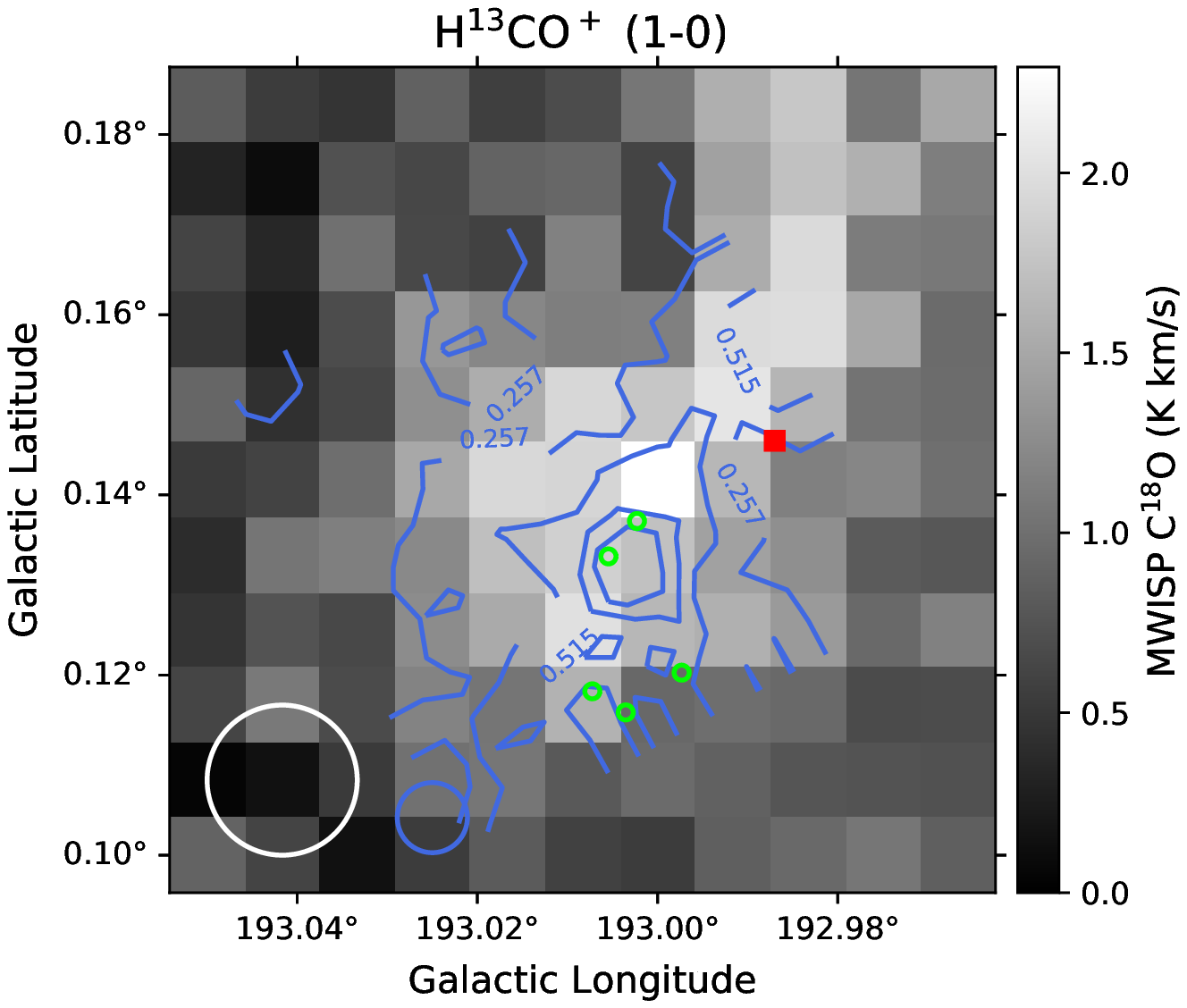}
\end{minipage}
}
\subfigure[G217.30-0.05]{
\begin{minipage}[b]{0.45\textwidth}
\includegraphics[width=1.1\textwidth]{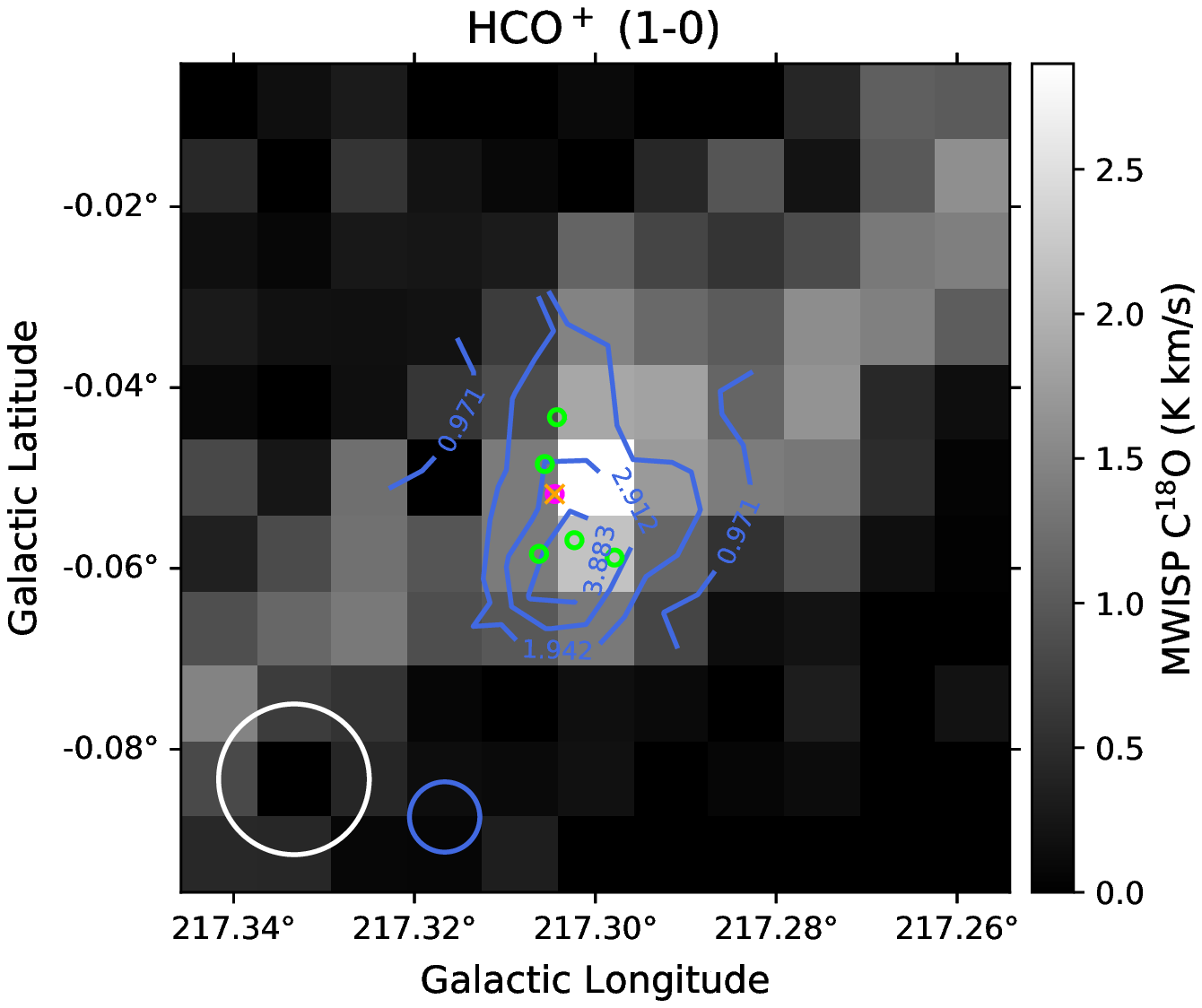} 
\end{minipage}
\begin{minipage}[b]{0.45\textwidth}
\includegraphics[width=1.1\textwidth]{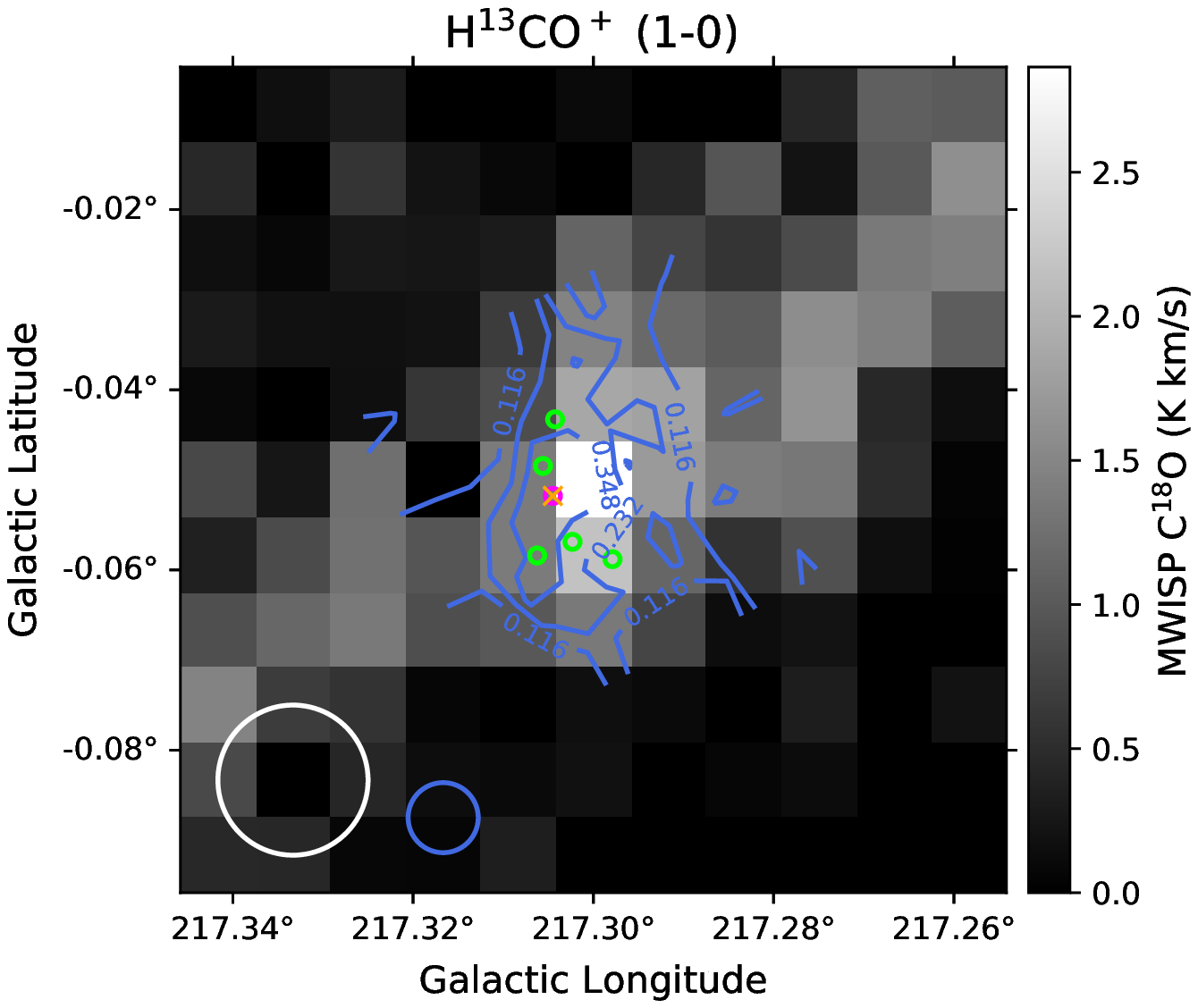}
\end{minipage}
}
\caption{Continued.}
\label{fig:map}
\end{figure}

\section{Map Grid of 24 Sources} \label{sec:Appendix2}

\begin{figure*}[h]
\gridline{\fig{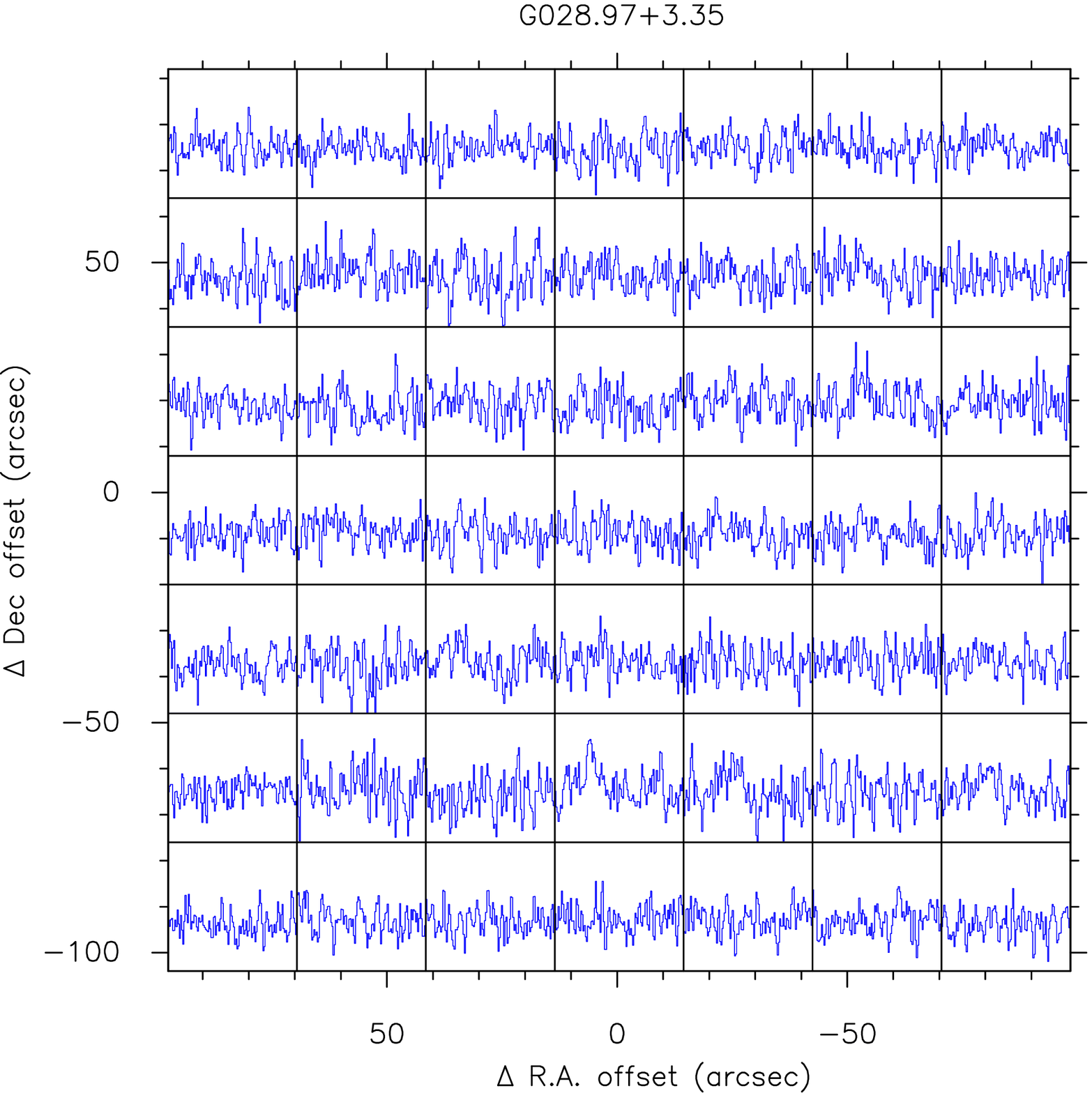}{0.4\textwidth}{}
          \fig{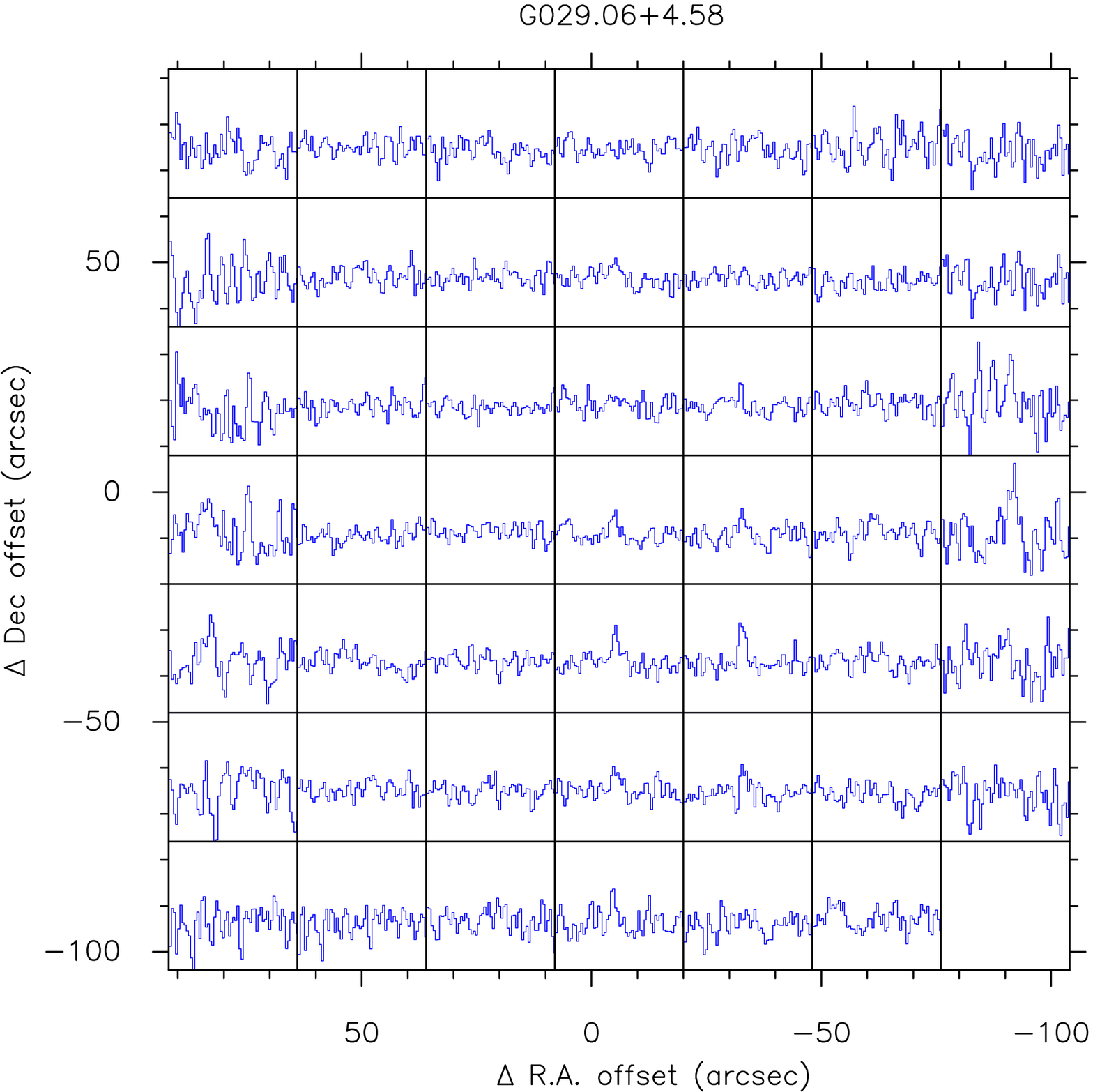}{0.4\textwidth}{}
          }
\gridline{\fig{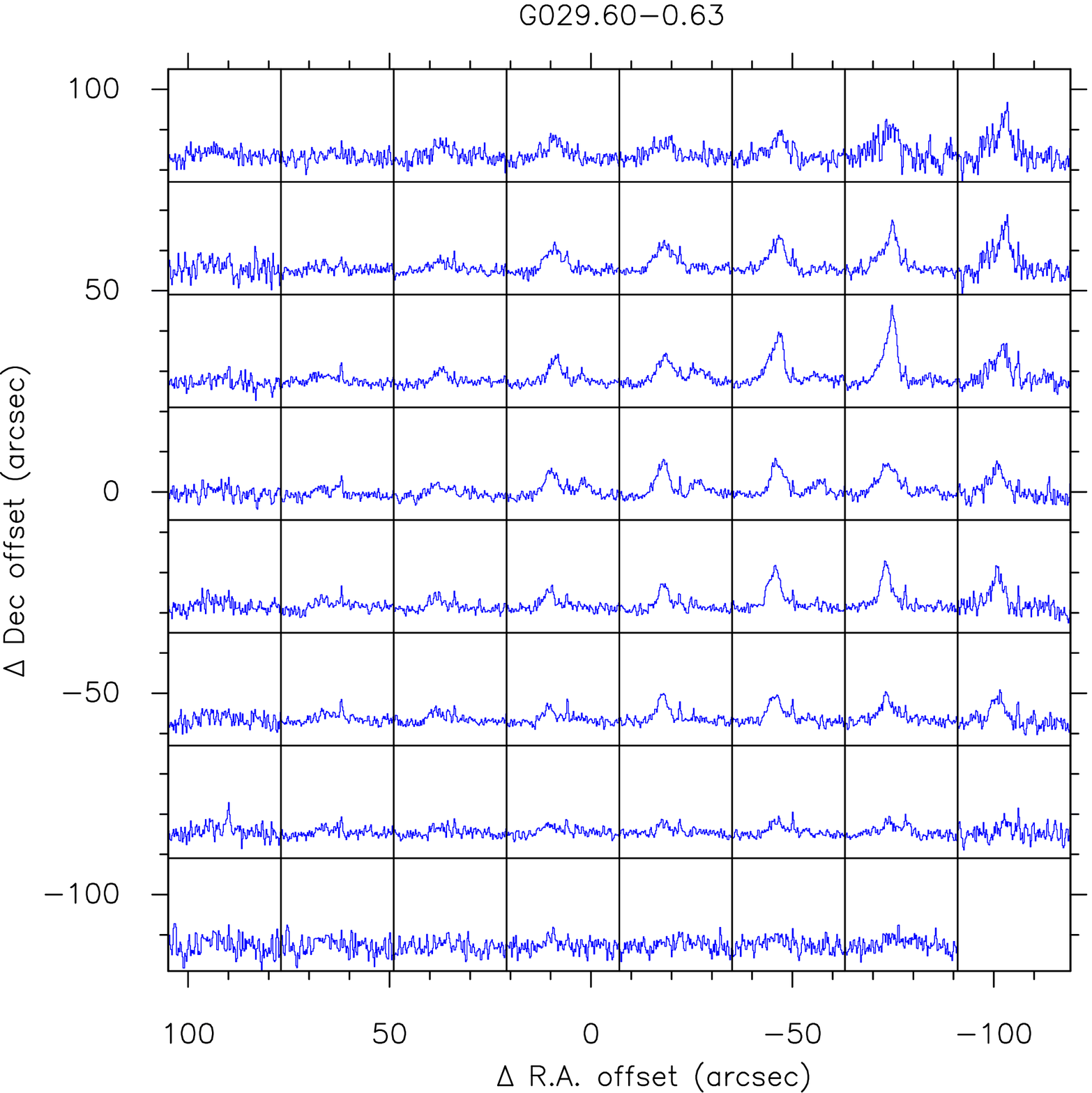}{0.4\textwidth}{}
          \fig{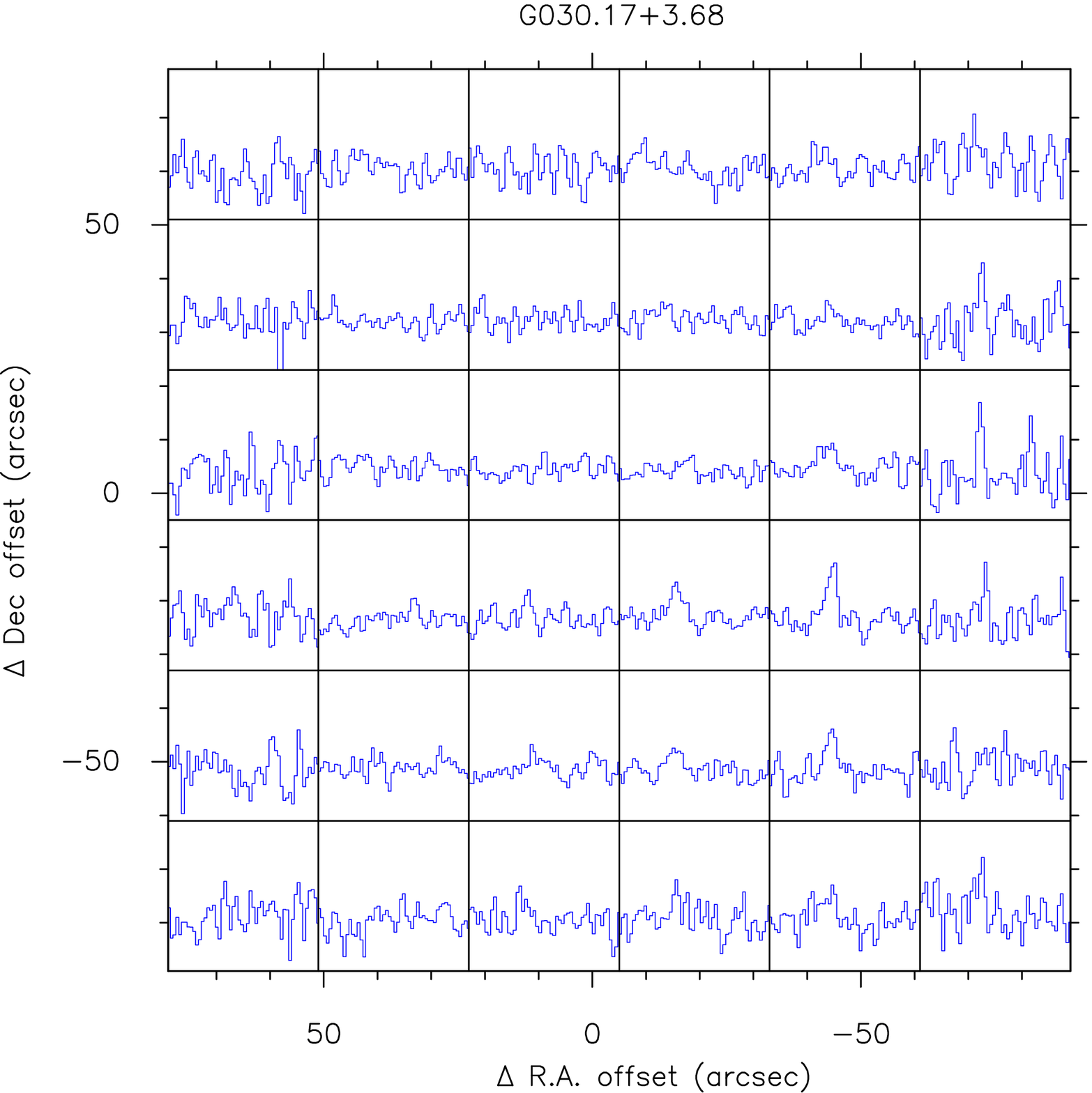}{0.4\textwidth}{}
          }
\gridline{\fig{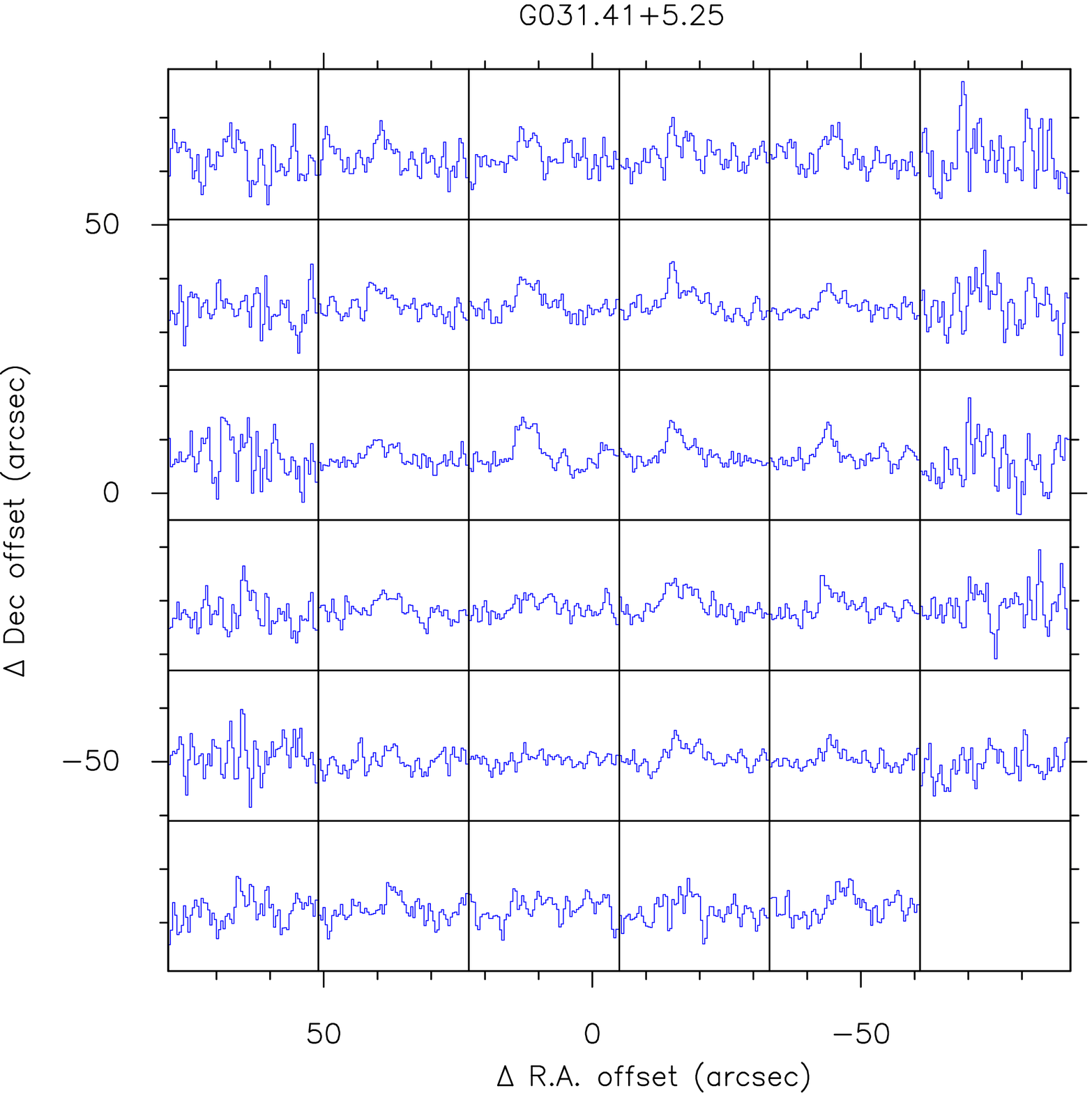}{0.4\textwidth}{}
          \fig{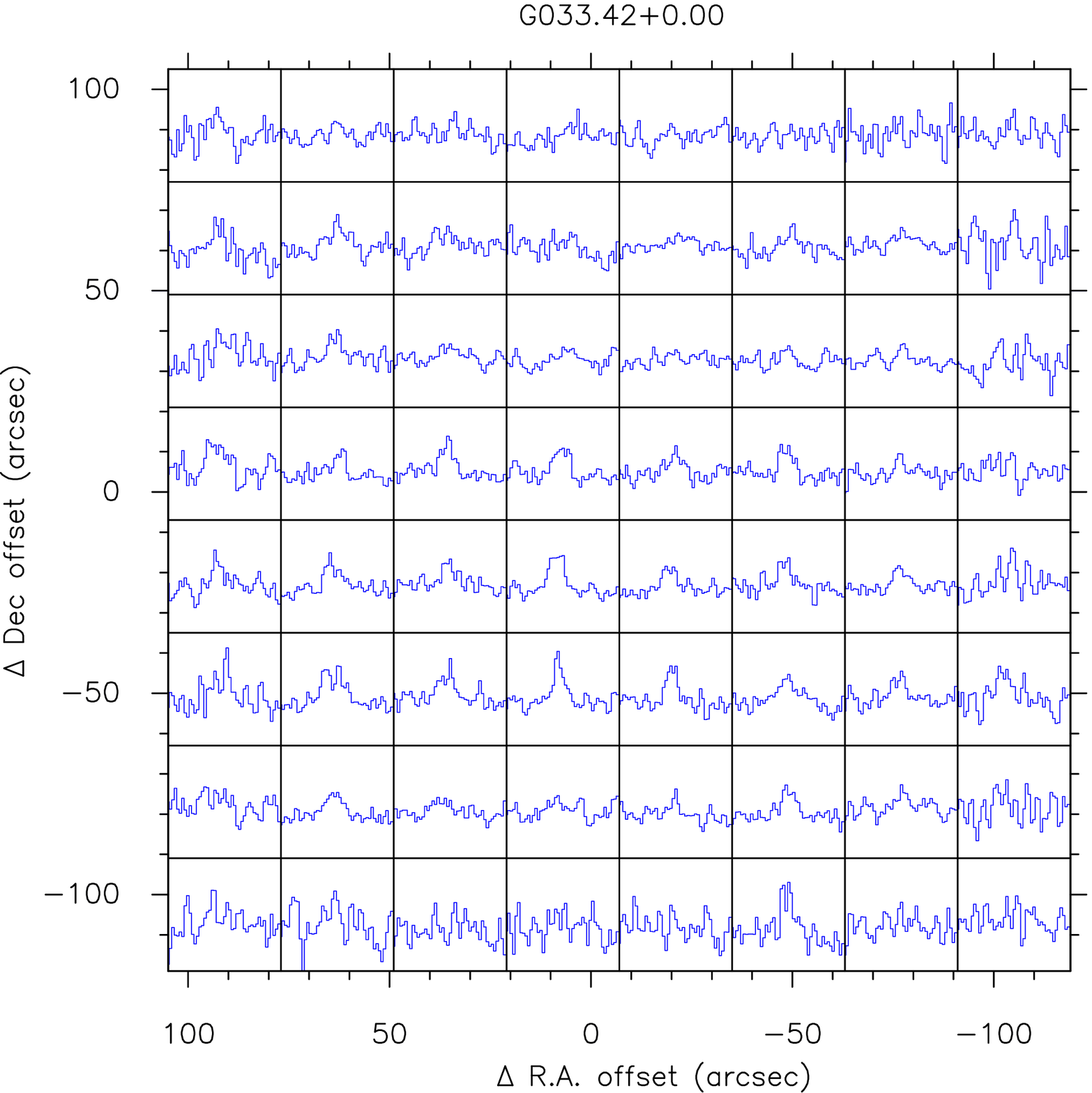}{0.4\textwidth}{}
          }
\caption{HCO$^+$ (1-0) map grid (gridded to one beam size). The axes plot the offsets $\Delta$ R.A. and $\Delta$ Dec relative to the coordinates from Table \ref{Tab:src-catalog}.
\label{fig:figA25}}
\end{figure*}

\begin{figure*}[h]
\addtocounter{figure}{-1} 
\gridline{\fig{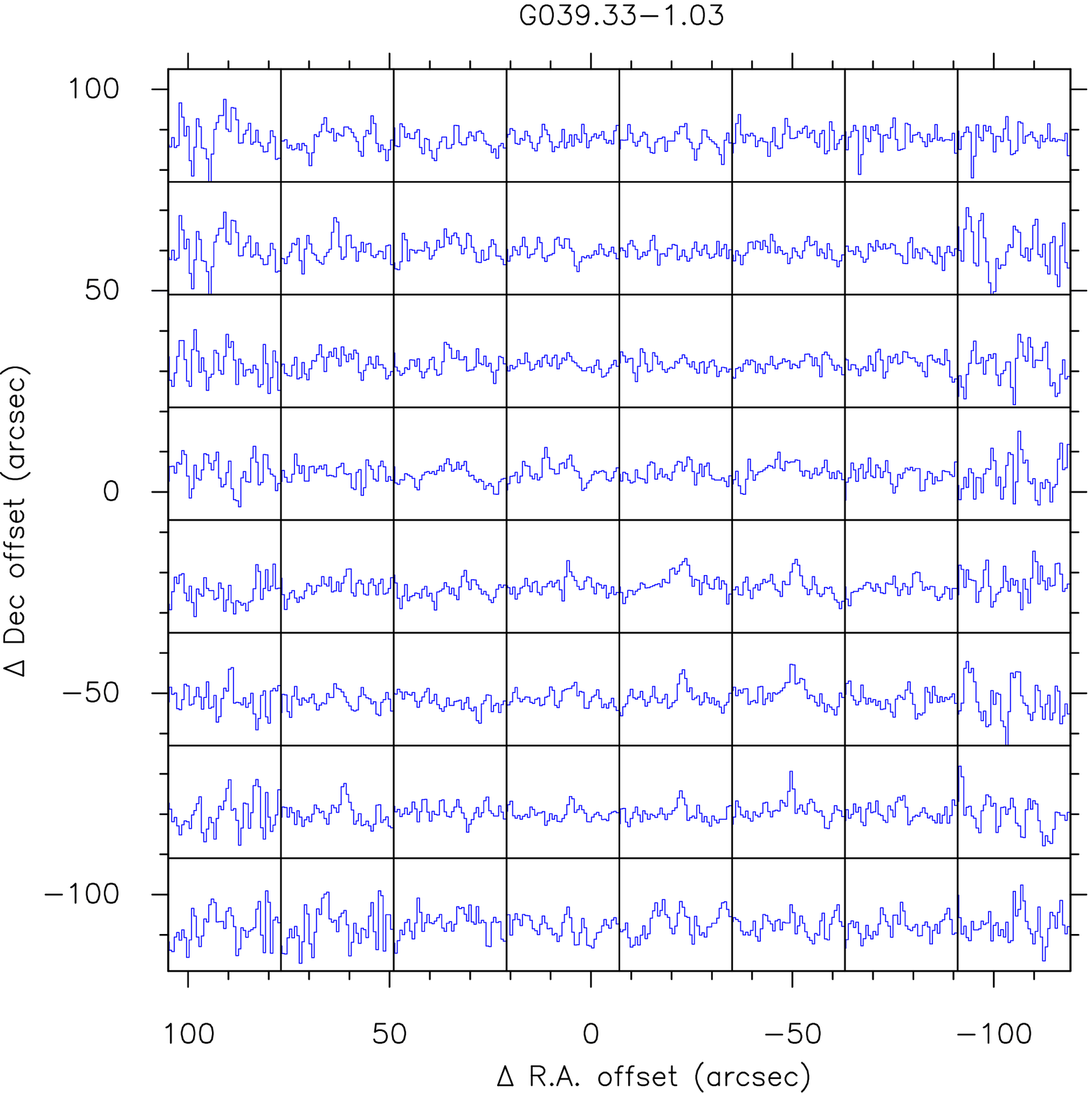}{0.4\textwidth}{}
          \fig{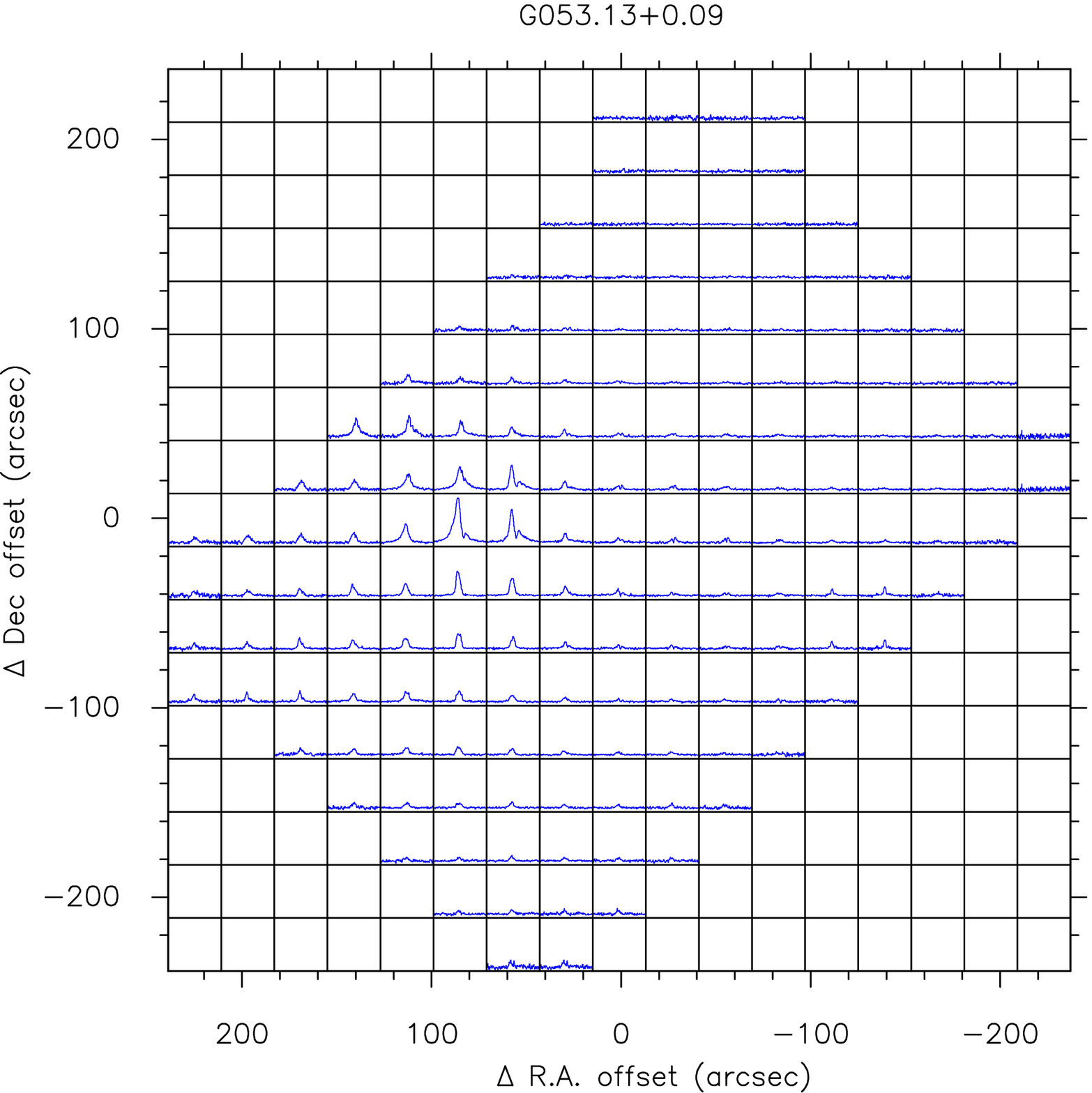}{0.4\textwidth}{}
          }
\gridline{\fig{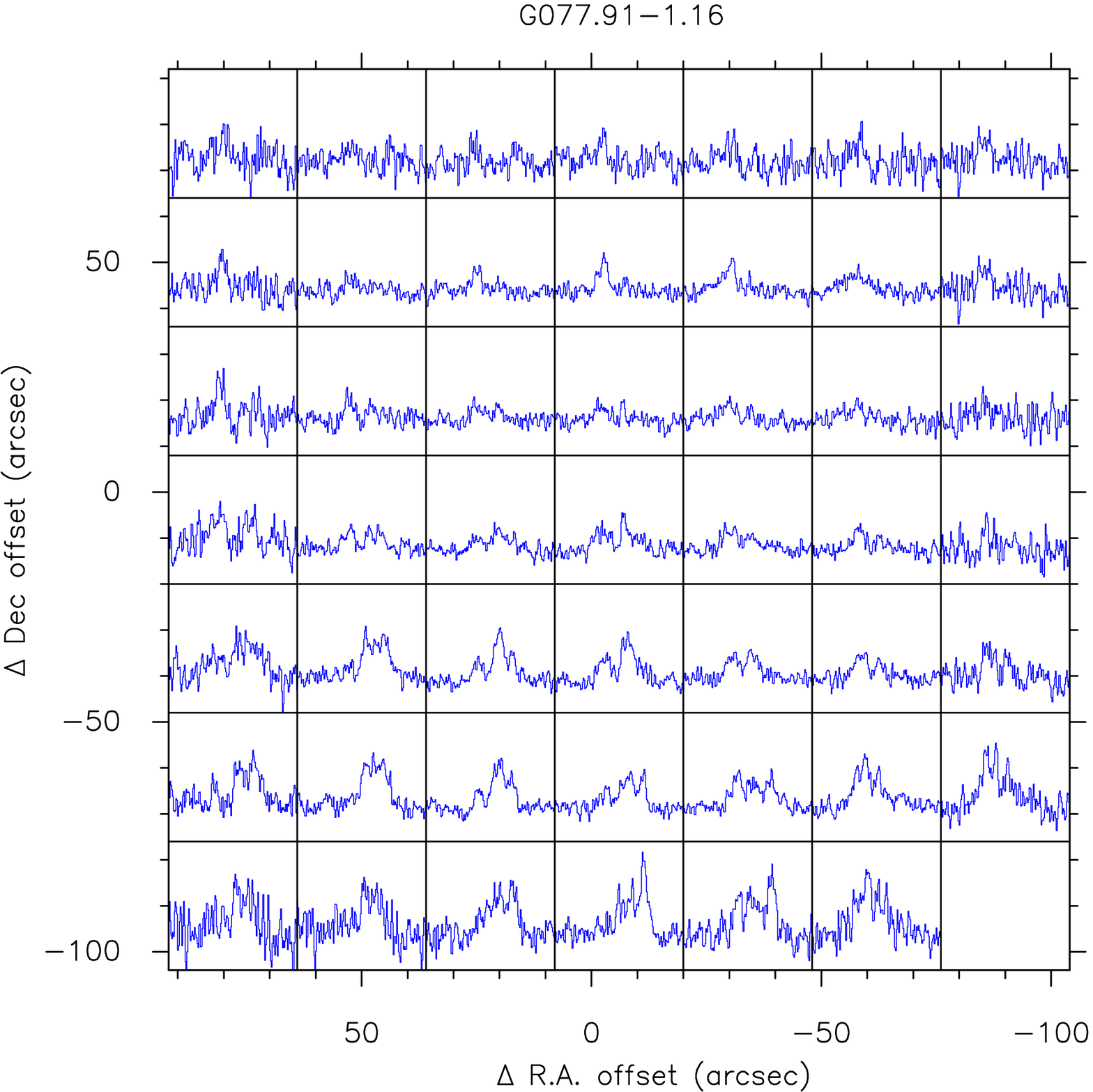}{0.4\textwidth}{}
          \fig{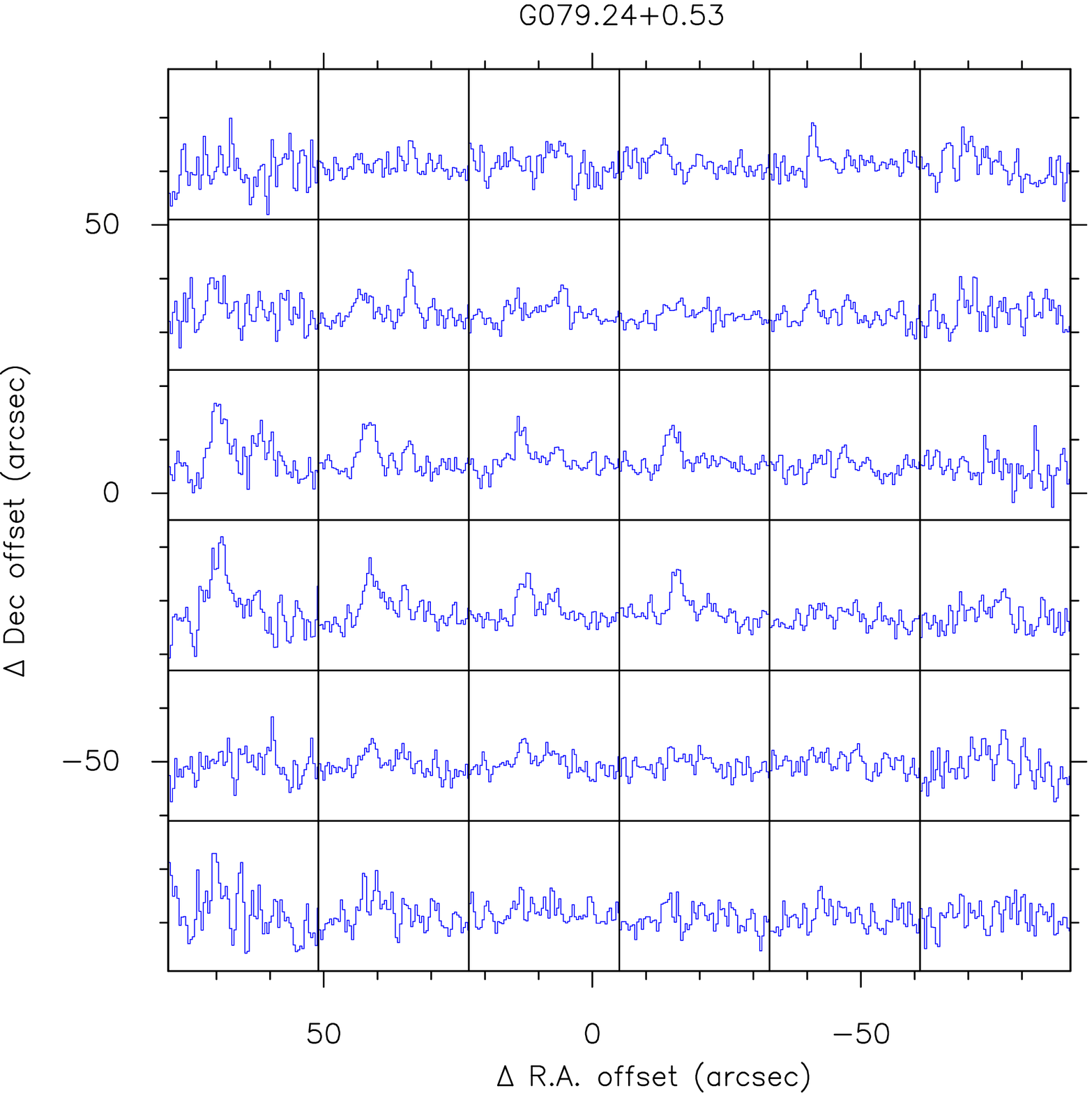}{0.4\textwidth}{}
          }
\gridline{\fig{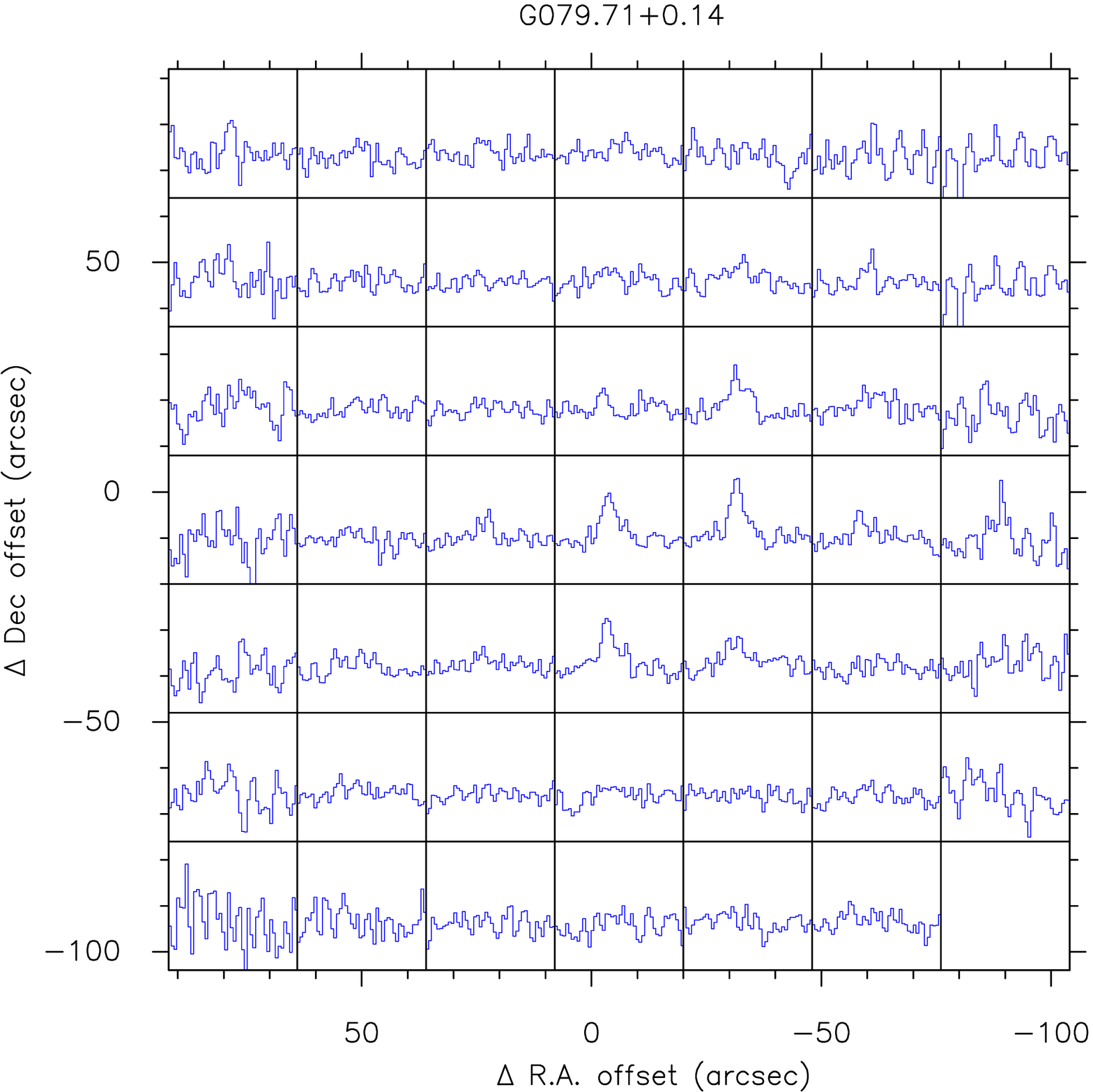}{0.4\textwidth}{}
          \fig{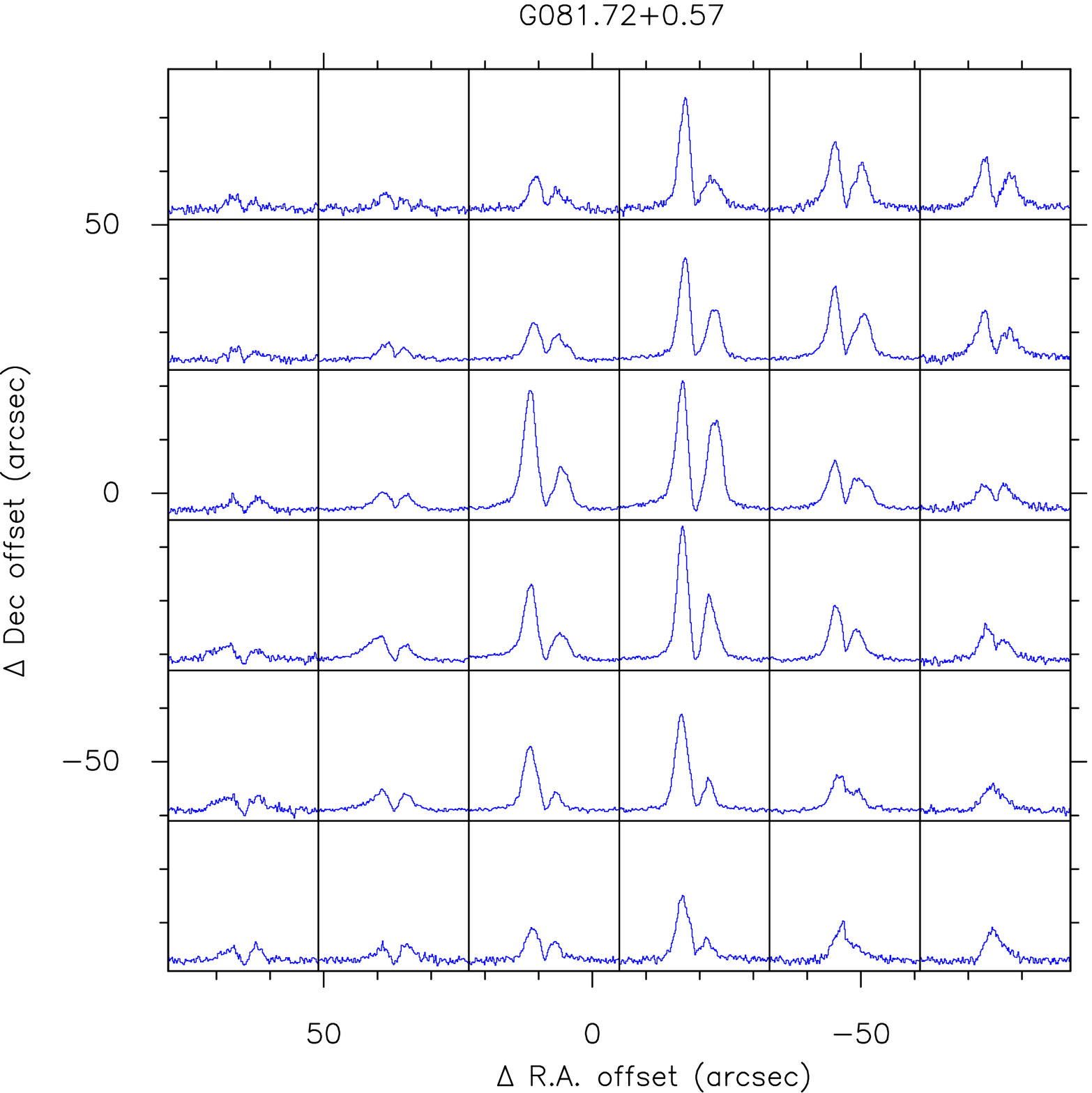}{0.4\textwidth}{}
          }
\caption{Continued.
\label{fig:figA25}}
\end{figure*}

\begin{figure*}[h]
\addtocounter{figure}{-1} 
\gridline{\fig{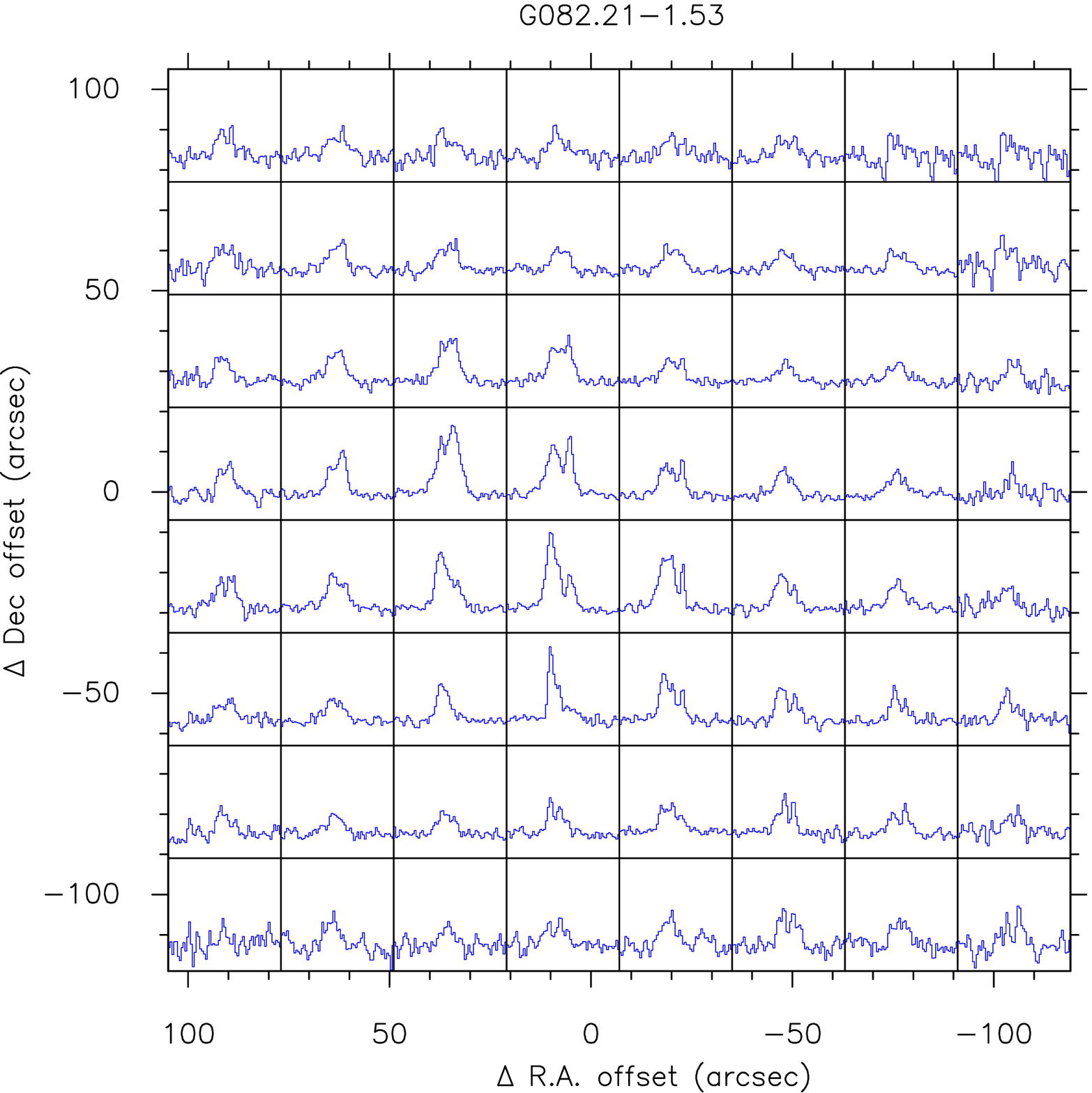}{0.4\textwidth}{}
          \fig{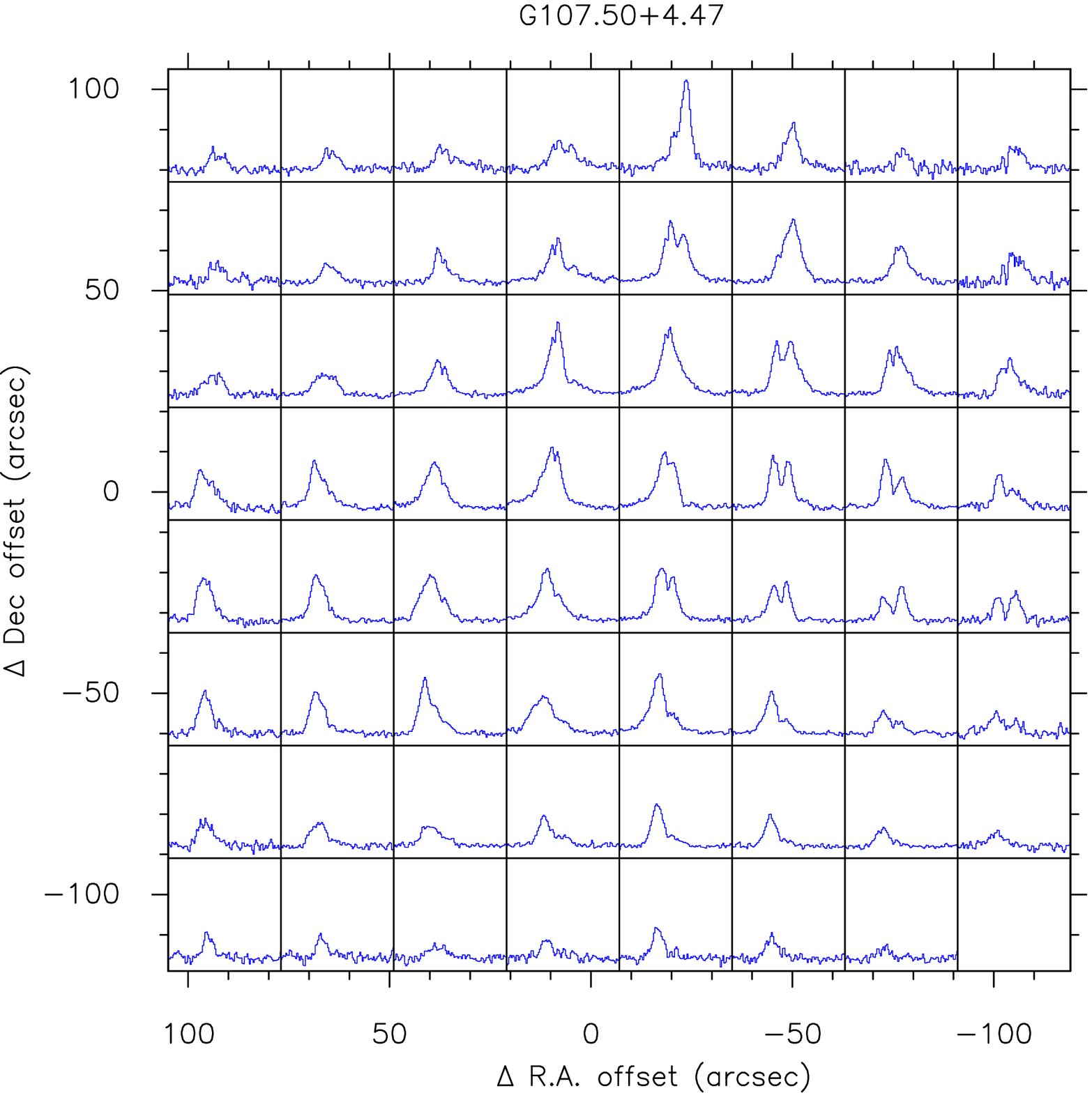}{0.4\textwidth}{}
          }
\gridline{\fig{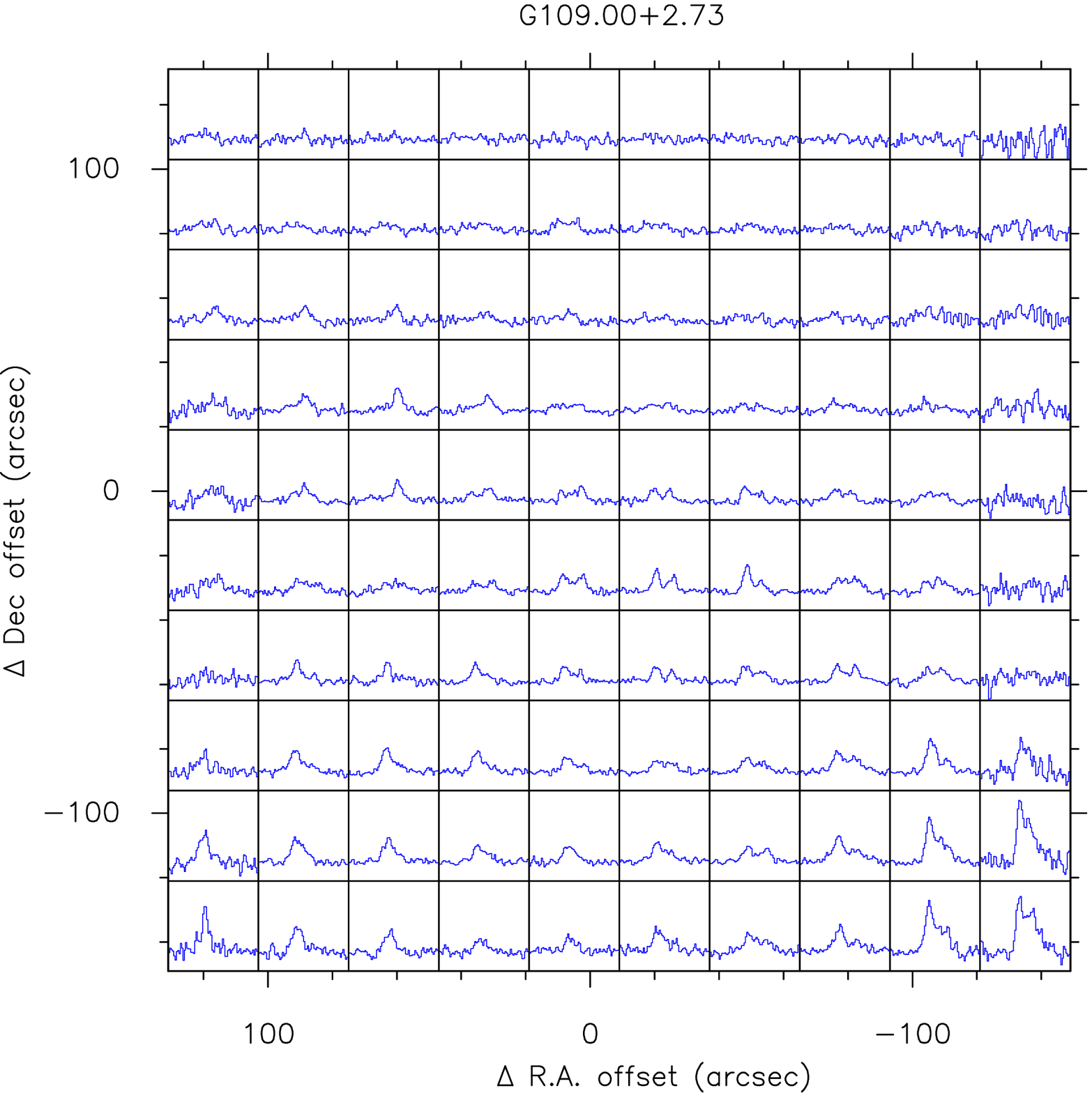}{0.4\textwidth}{}
          \fig{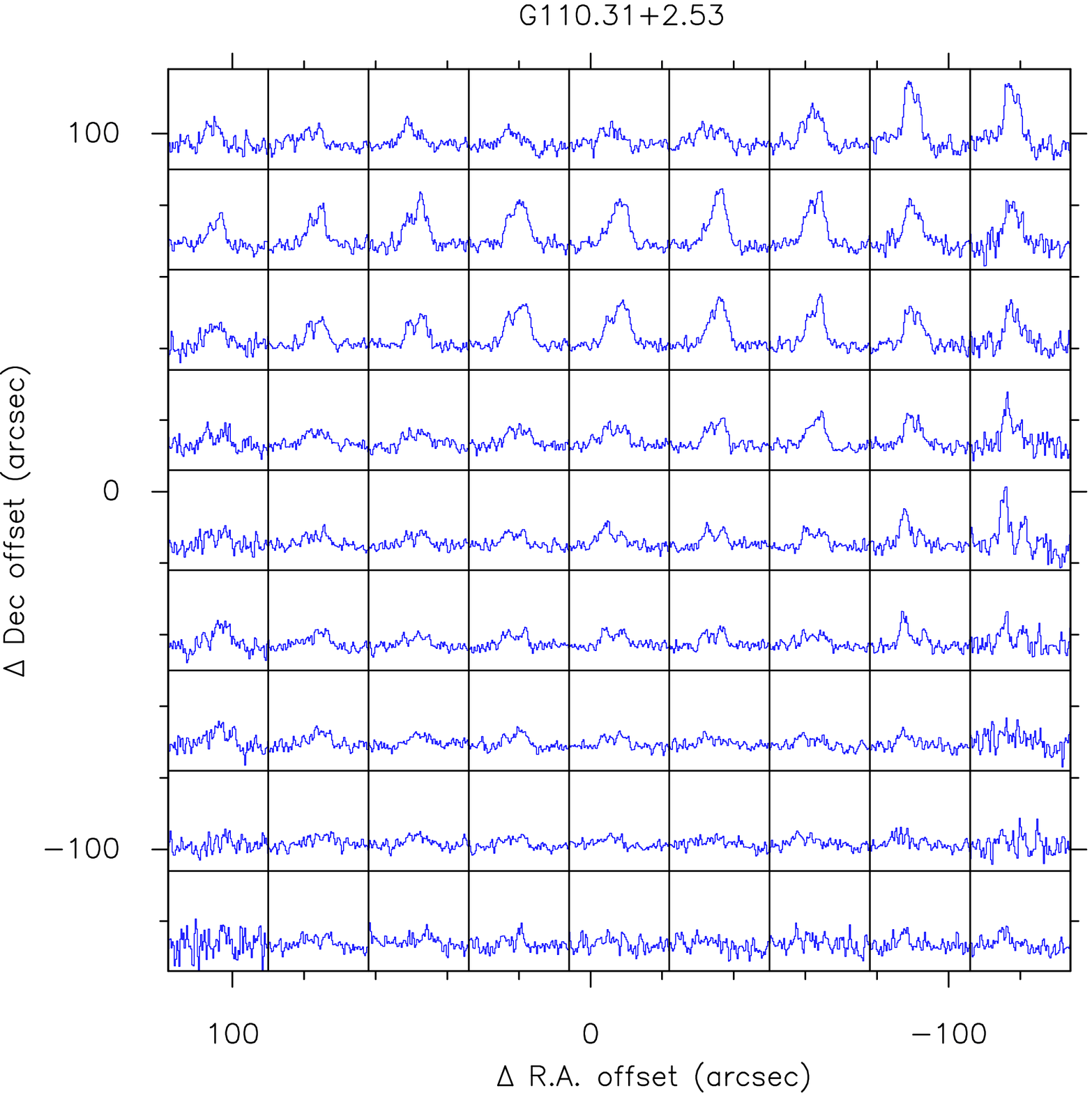}{0.4\textwidth}{}
          }
\gridline{\fig{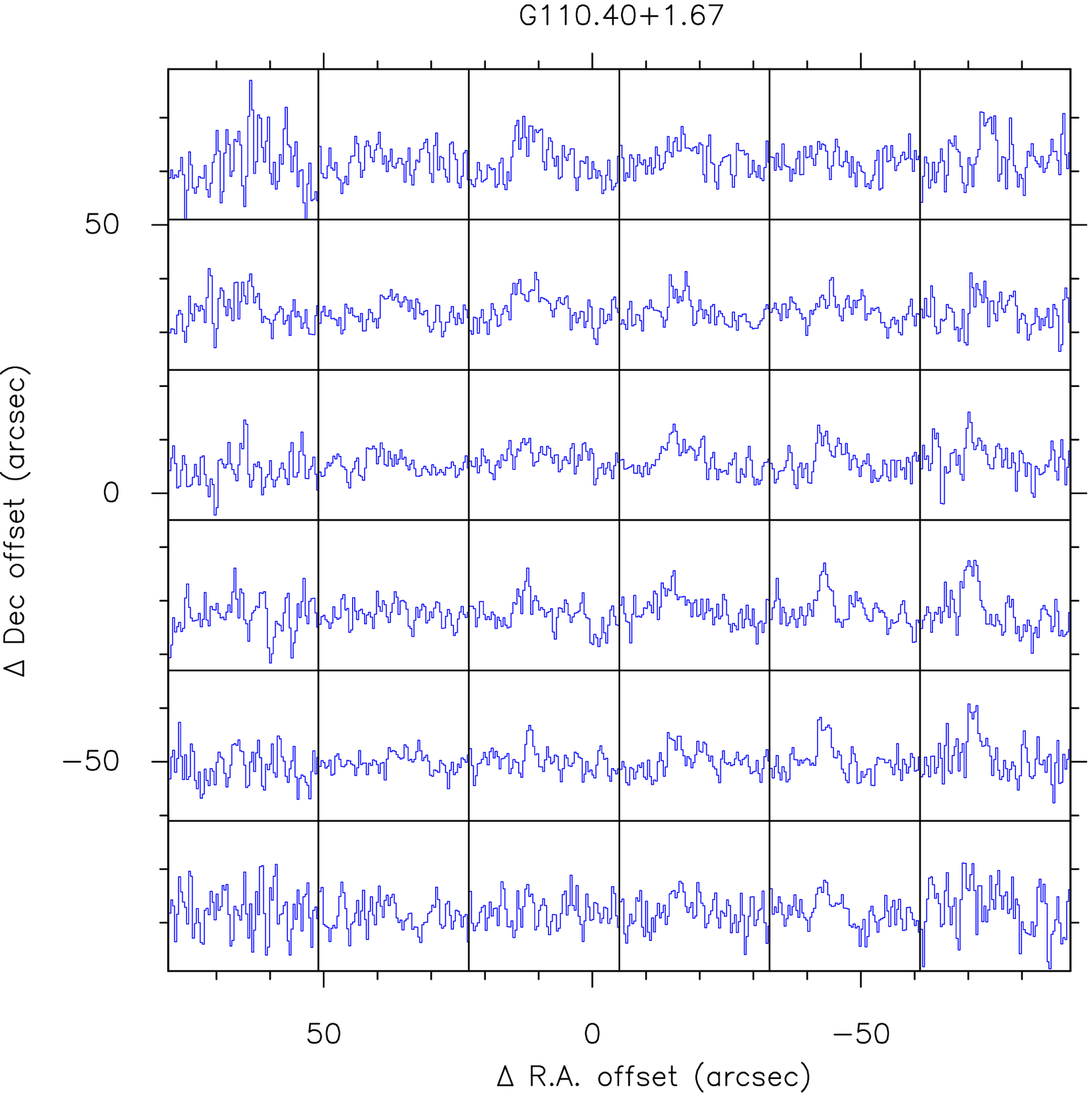}{0.4\textwidth}{}
          \fig{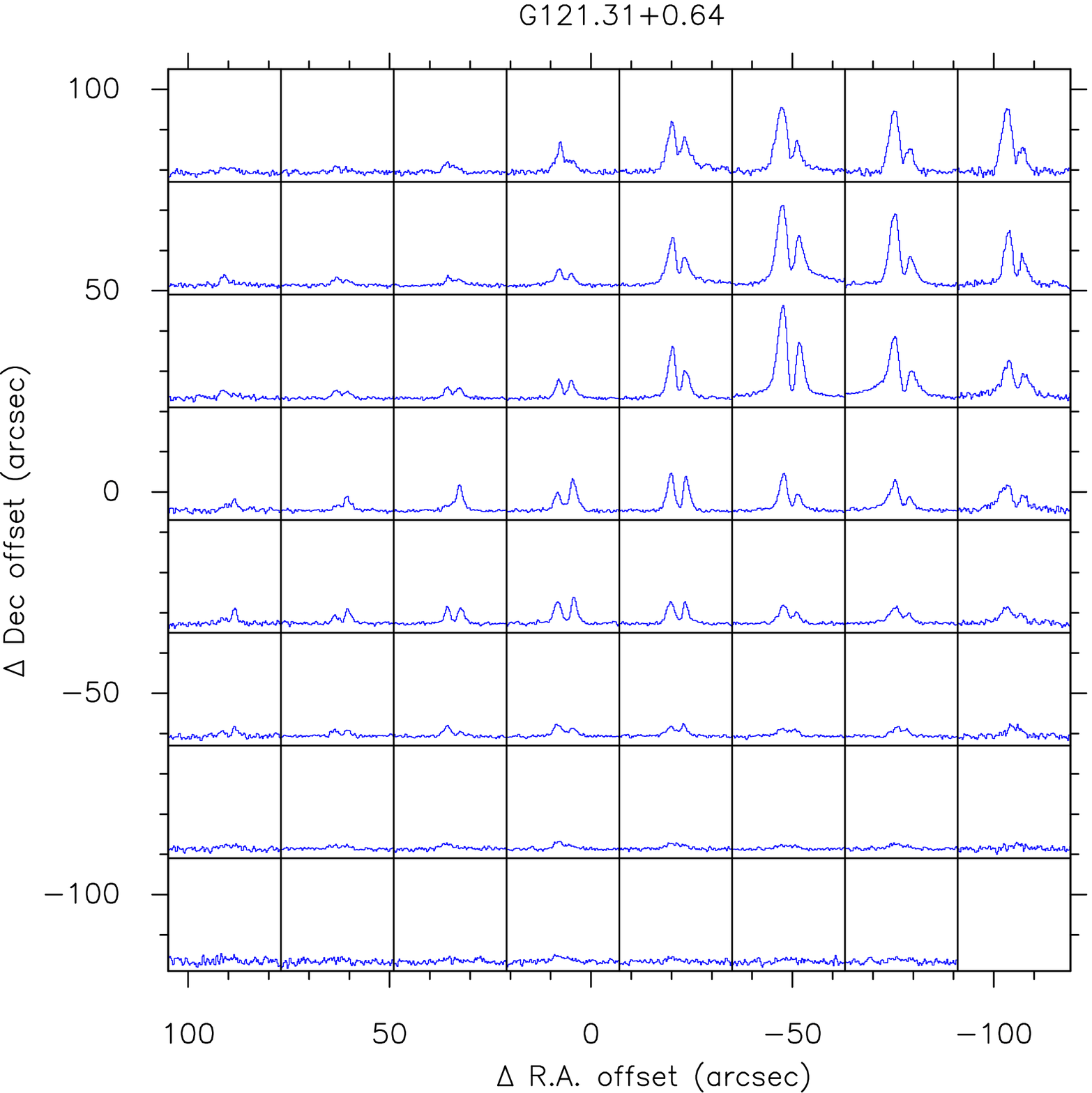}{0.4\textwidth}{}
          }
\caption{Continued.
\label{fig:figA25}}
\end{figure*}

\begin{figure*}[h]
\addtocounter{figure}{-1} 
\gridline{\fig{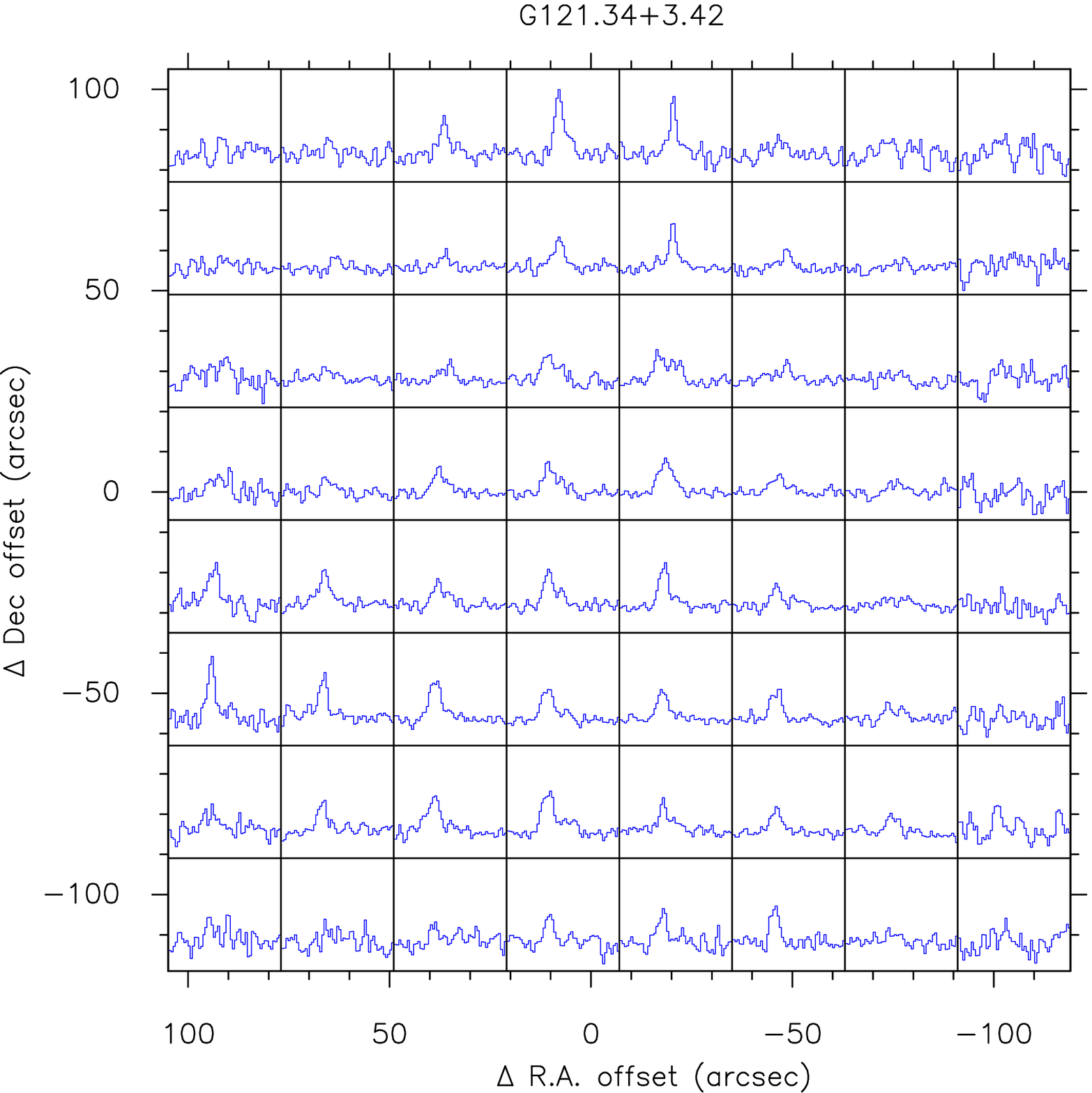}{0.4\textwidth}{}
          \fig{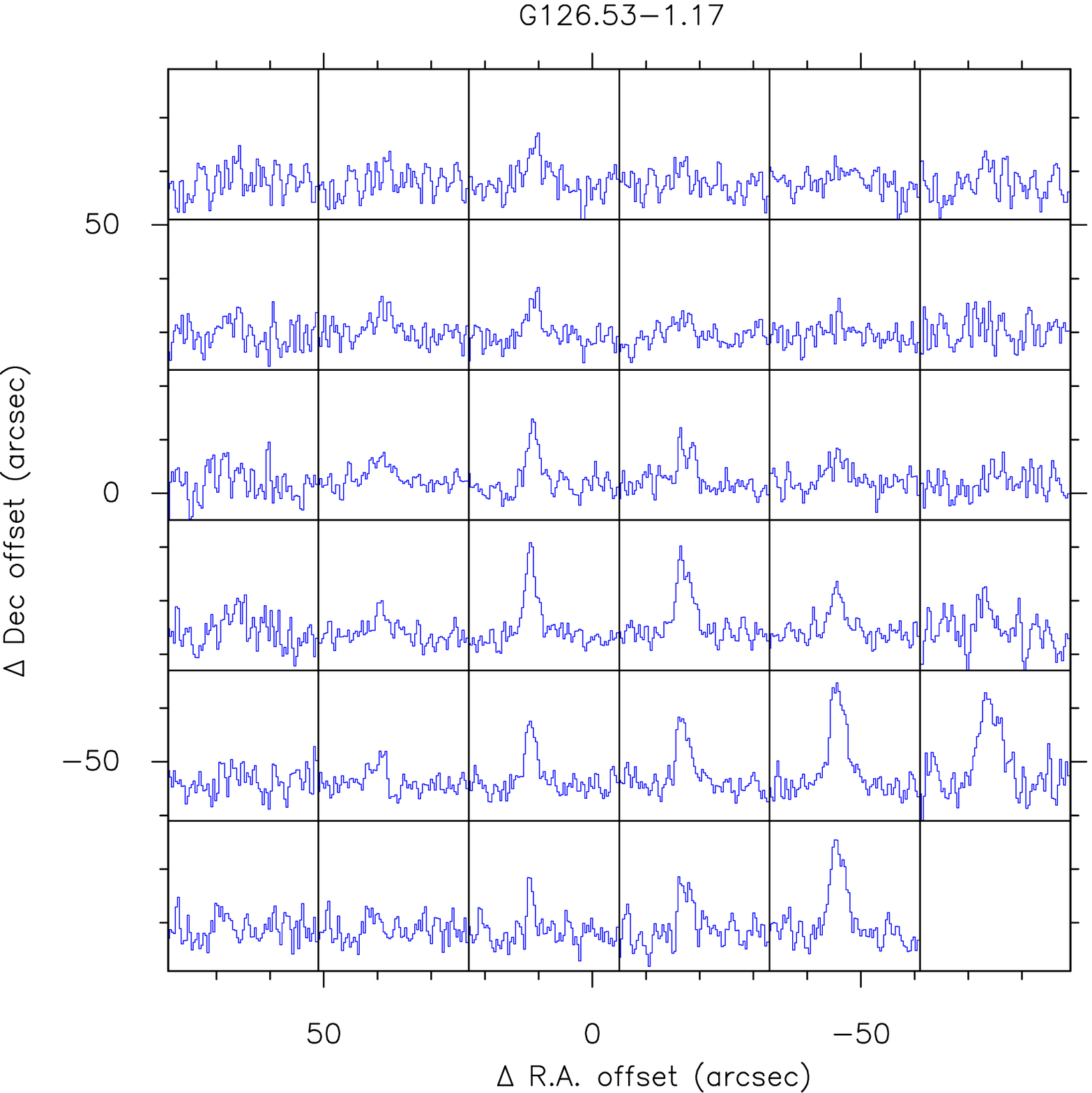}{0.4\textwidth}{}
          }
\gridline{\fig{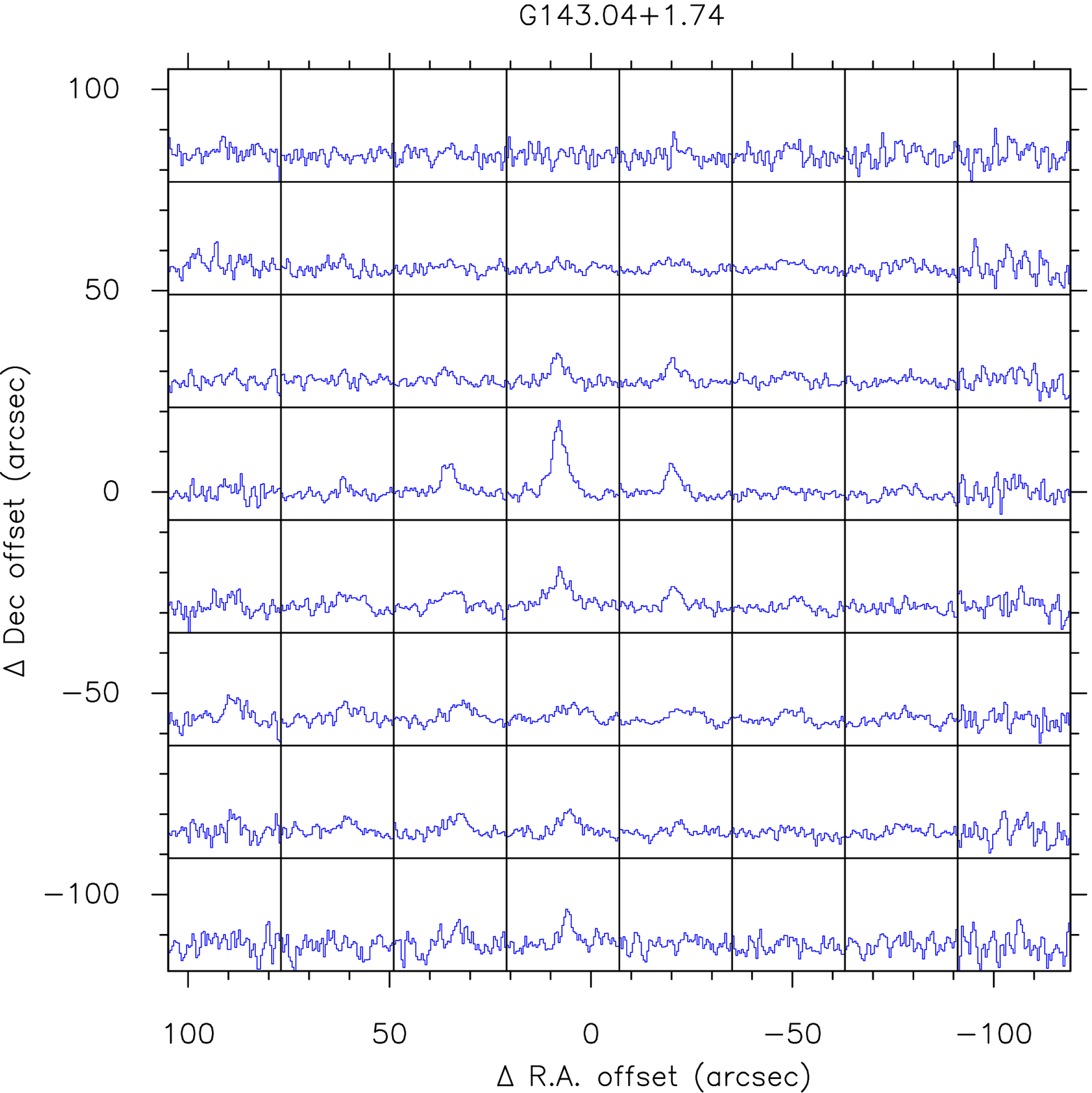}{0.4\textwidth}{}
          \fig{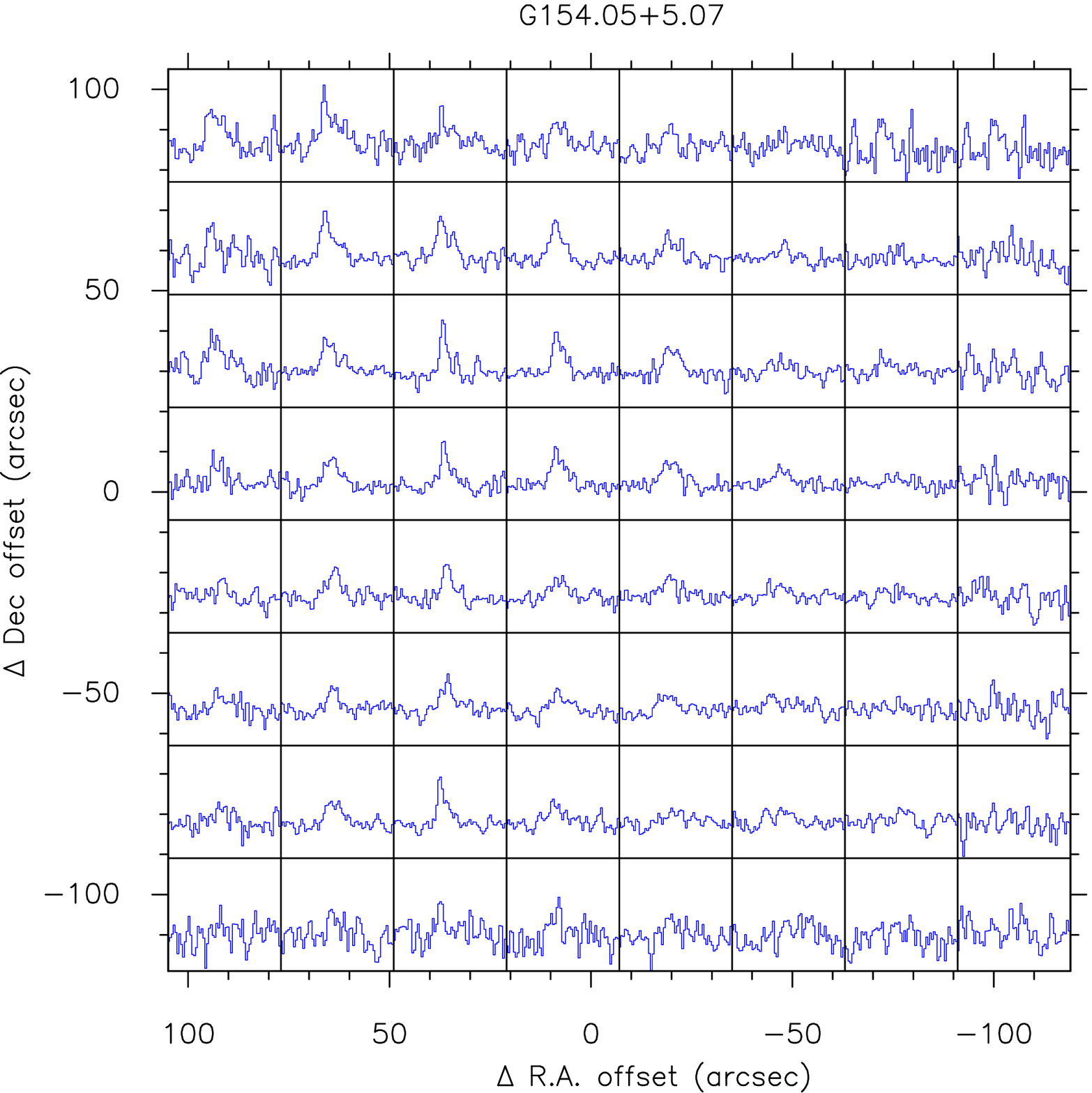}{0.4\textwidth}{}
          }
\gridline{\fig{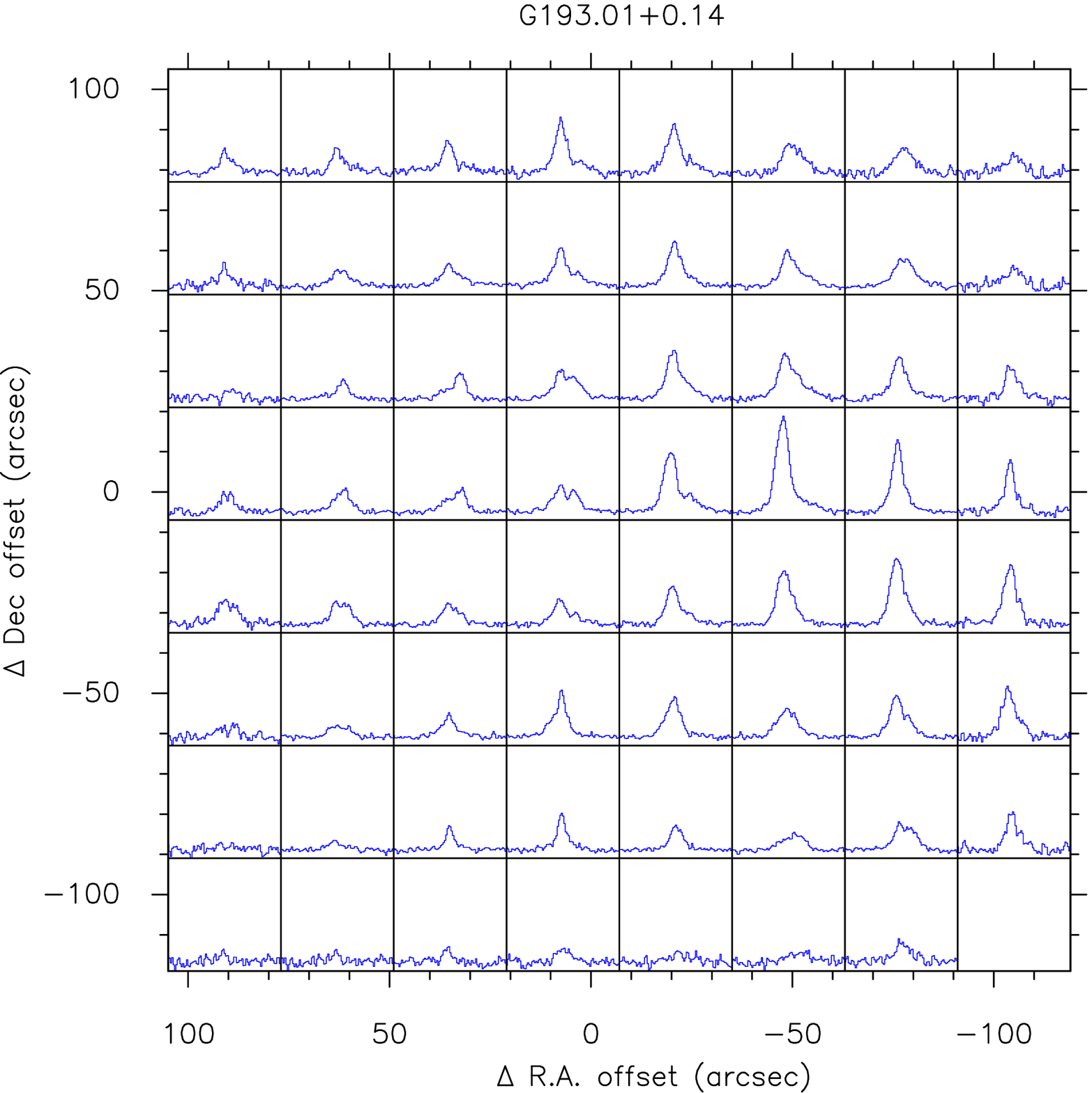}{0.4\textwidth}{}
          \fig{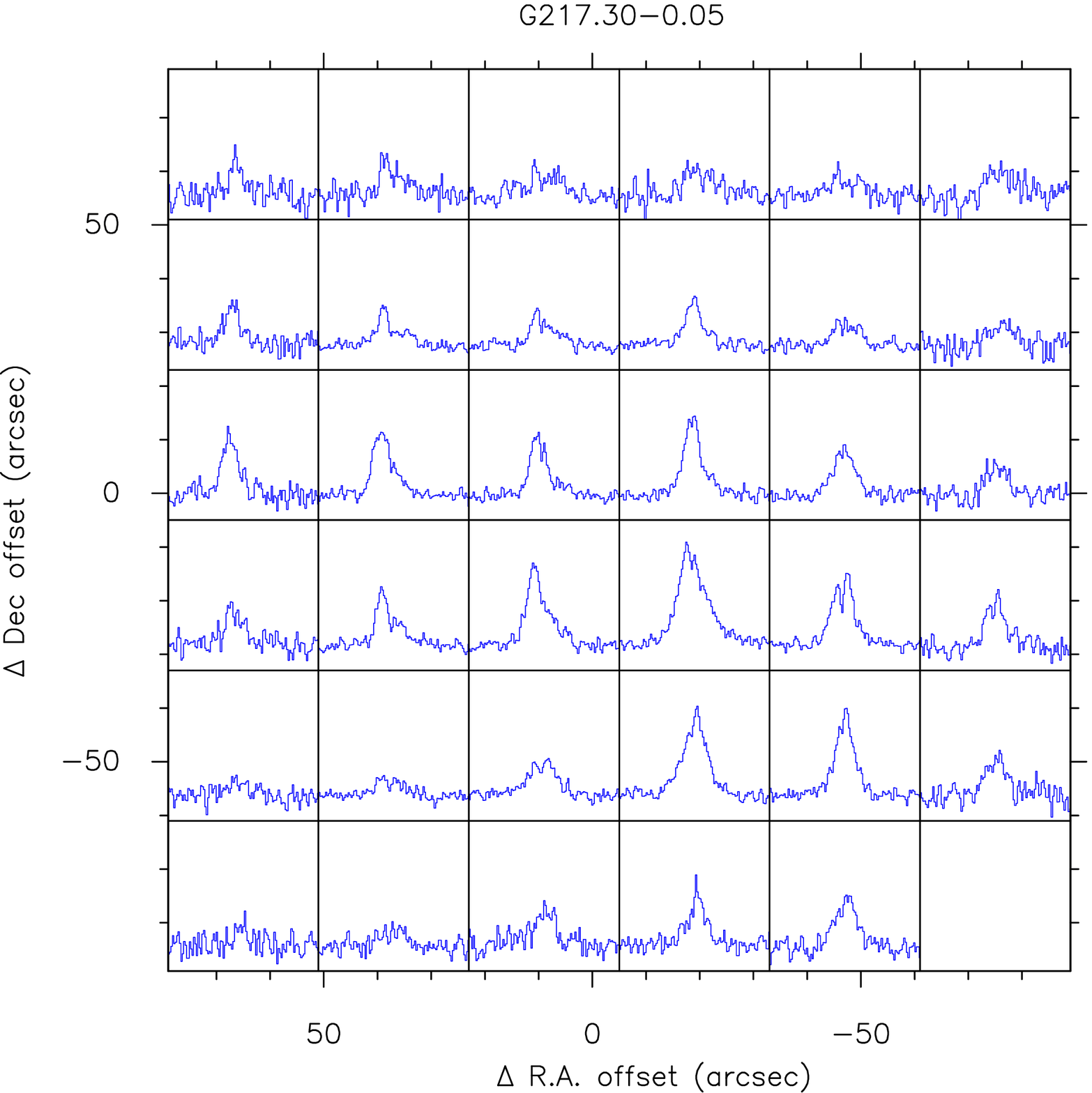}{0.4\textwidth}{}
          }
\caption{Continued.
\label{fig:figA25}}
\end{figure*}

\clearpage

\bibliography{paper_sample}{}
\bibliographystyle{aasjournal}



\end{document}